\documentclass{article}
\usepackage{epubtk}
\usepackage{amssymb}
\usepackage{graphicx}

\def\M{\mathcal{M}}
\def\R{\mathcal{R}}
\def\D{\mathcal{D}}

\begin{document}

\title{The Evolution of Compact Binary Star Systems}

\author{\epubtkAuthorData{Konstantin A.\ Postnov}
        {Sternberg Astronomical Institute \\
         13 Universitetskij Pr. \\
         119992 Moscow \\
         Russia}
        {kpostnov@gmail.com}
        {http://www.sai.msu.ru}
        \and
        \epubtkAuthorData{Lev R.\ Yungelson}
        {Institute of Astronomy of Russian Academy of Sciences \\
         48 Pyatnitskaya Str. \\
         119017 Moscow \\
         Russia}
        {lry@inasan.ru}
        {http://www.inasan.rssi.ru}}

\date{}
\maketitle


\begin{abstract}
  We review the formation and evolution of compact binary stars consisting
  of white dwarfs (WDs), neutron stars (NSs), and black holes
  (BHs). Binary NSs and BHs are thought to be the primary astrophysical
  sources of gravitational waves (GWs) within the frequency band of
  ground-based detectors, while compact binaries of WDs are important
  sources of GWs at lower frequencies to be covered by space
  interferometers (LISA). Major uncertainties in the current
  understanding of properties of NSs and BHs most relevant to the GW
  studies are discussed, including the treatment of the natal kicks
  which compact stellar remnants acquire during the core collapse of
  massive stars and the common envelope phase of binary evolution. We
  discuss the coalescence rates of binary NSs and BHs and prospects for
  their detections, the formation and evolution of binary WDs and their
  observational manifestations. Special attention is given to
  AM~CVn-stars -- compact binaries in which the Roche lobe is filled
  by another WD or a low-mass partially degenerate helium-star, as these
  stars are thought to be the best LISA verification binary GW
  sources.
\end{abstract}

\epubtkKeywords{astrophysics, binary systems, gravitational wave
  sources, supernovae, neutron stars, black holes, white dwarfs, AM
  CVn stars}

\newpage


\section{Introduction}
\label{section:introduction}

Close binary stars consisting of two compact stellar remnants (white
dwarfs (WDs), neutron stars (NSs), or black holes (BHs)) are considered
as primary targets of the forthcoming field of gravitational wave (GW) astronomy
since their orbital evolution is entirely controlled by emission of
gravitational waves and leads to ultimate coalescence (merger) of
the components. Close compact binaries can thus serve as testbeds for
theories of gravity. The double NS(BH) mergers should be the brightest
GW events in the $10\mbox{\,--\,}1000 \mathrm{\ Hz}$ frequency band of
the existing GW detectors like LIGO~\cite{Barish_Weiss99},
VIRGO~\cite{Acernese_al05}, or GEO~600~\cite{Ricci_Brillet97}. Such
mergers can be accompanied by the release of a huge amount of
electromagnetic energy in a burst and manifest themselves as short
gamma-ray bursts (GRBs). Double WDs, especially interacting binary WDs
observed as AM~CVn-stars, are potential GW sources within the
frequency band of the space GW interferometers like
LISA~\cite{fbhhs89} or future
detectors~\cite{Crowder_Cornish05}. The double WD mergers also stay
among the primary candidate mechanisms for type~Ia supernova
(SN~Ia) explosions, which are crucial in modern cosmological
studies.

Compact binaries are the end products of the evolution of binary stars, and
the main purpose of the present review is to describe the
astrophysical knowledge on their formation and evolution. We shall
discuss the present situation with the main parameters determining
their evolution and the rates of coalescence of double NSs/BHs and WDs.

About 6\% of the baryonic matter in the Universe is confined in
stars~\cite{Fukugita_Peebles04}. The typical mass of a stationary star
is close to the solar value $M_\odot \approx 2\times 10^{33} \mathrm{\
  g}$. The minimum mass of a stationary star at the main sequence (MS)
is set by the condition of stable hydrogen burning in its core
$M_\mathrm{min}\approx 0.08\,M_\odot$~\cite{Kumar63}. The maximum mass
of solar composition stars inferred observationally is close to
$150\,M_\odot$~\cite{figer_upper05}; for very low metallicity stars it
is derived by the linear analysis of pulsational stability and is
close to $300\,M_\odot$~\cite{baraffe_upper01}. Stars and stellar
systems are formed due to the development of the gravitational (Jeans)
instability in turbulized molecular clouds. The minimum protostellar
mass is dictated by the opacity conditions in the collapsing fragments
and is found to lie in the range $0.01\mbox{\,--\,}0.1\,M_\odot$ in
both analytical~\cite{Rees76} and numerical calculations
(see, e.g., \cite{Delgado-Donate_al04}). It is established from observations
that the mass distribution of main-sequence stars has a power-law
shape~\cite{salpeter55, Miller_Scalo79},
$dN/dM\sim M^{-\beta}$, with $\beta=-1.2$ for $0.08 \lesssim M/ M_\odot
\lesssim 0.5$, $\beta=-2.2$ for $0.5 \lesssim M/ M_\odot \lesssim
1.0$, and $\beta= -2.2$ to $-3.2$ for $1.0 \lesssim M/ M_\odot \lesssim
150$~\cite{kroupa_etal93, kroupa_weidner_03}.

The evolution of a single star is determined by its initial mass at
the main sequence $M_0$ and the chemical composition. If $M_0\lesssim
8\mbox{\,--\,}12\,M_\odot$, the carbon-oxygen (CO) (or oxygen-neon
(ONe) at the upper end of the range) stellar core becomes degenerate
and the evolution of the star ends up with the formation of a CO or ONe
white dwarf. The formation of a WD is accompanied by the loss of stellar
envelope by stellar wind in the red giant and asymptotic giant branch
stages of evolution and ejection of a planetary nebula. The boundary
between the masses of progenitors of WDs and NSs is not well defined and
is, probably, between $8$ and $12\,M_\odot$ (cf.~\cite{it85, iben86,
  ritossa_berro_one96, iben_ritossa_one97, berro_ritossa_one97,
  pol_98, ritossa_berro_one99, gilponsetal03, siess_agb06}).

At the upper boundary of the mass range of white dwarf progenitors,
formation of ONe WDs is possible. The masses of stars that
produce ONe WDs are still highly uncertain. However, strong
observational evidence for their existence stems from the analysis of
nova ejecta~\cite{truran_livio_one86}. This variety of WDs is
important in principle, because accretion induced collapse (AIC) of them
may result in formation of neutron stars (see~\cite{nomkondo91,
  dessart_onecoll06} and references therein), but since for the
purpose of detection of gravitational waves they are not different
from the much more numerous CO-WDs, we will, as a rule, not
consider them below as a special class.

If $M_0\gtrsim (10\mbox{\,--\,}12)\,M_\odot$, thermonuclear evolution
proceeds until iron-peak elements are produced in the core. Iron cores are
subjected to instabilities (neutronization, nuclei photodesintegration,
or pair creation for the most massive stars) that lead to
gravitational collapse. The core collapse of massive stars results in
the formation of a neutron star or, for very massive stars, a black
hole and is associated with the brightest astronomical phenomena such
as supernova explosions (of type~II, Ib, or~Ib/c, according to the
astronomical classification based on the spectra and light curves
properties). If the pre-collapsing core retains significant rotation,
powerful gamma-ray bursts lasting up to hundreds of seconds may be
produced~\cite{Woosley_Heger06}.

The boundaries between the masses of progenitors of WDs or NSs and NSs or
BHs are fairly uncertain (especially for BHs). Typically accepted masses
of stellar remnants for nonrotating solar chemical composition stars
are summarized in Table~\ref{table:ms_evolution}.

\begin{table}
  \renewcommand{\arraystretch}{1.2}
  \centering
  \begin{tabular}{ccc}
    \hline \hline
    Initial mass [$M_\odot$] &
    remnant type &
    mean remnant mass [$M_\odot$] \\
    \hline
    $ \phantom{0\mbox{\,--\,}} 0.95 < M < 8\mbox{\,--\,}10 \phantom{0} $ &
    WD &
    $ \phantom{\sim 0} 0.6 \phantom{0} $ \\
    $ \phantom{0.} 8\mbox{\,--\,}10 < M < 25\mbox{\,--\,}30 $ &
    NS &
    $ \phantom{\sim 0} 1.35 $ \\
    $ \phantom{.} 25\mbox{\,--\,}30 < M < 150 \phantom{\mbox{\,--\,}0} $ &
    BH &
    $ \sim 10 \phantom{.00} $ \\
    \hline \hline
  \end{tabular}
  \caption{\it Types of compact stellar remnants (the ranges of
    progenitor mass are shown for solar composition stars).}
  \label{table:ms_evolution}
  \renewcommand{\arraystretch}{1.0}
\end{table}

For a more detailed introduction into the physics and evolution of stars the
reader is referred to the classical
textbook~\cite{Cox_Giuli68}. Formation and physics of compact objects
is described in more detail in monographs~\cite{Shapiro_Teukolsky83,
  B-K02}. For a recent review of the evolution of massive stars and the
mechanisms of core-collapse supernovae we refer to~\cite{Woosley_al02,
  Fryer04, Kotake_al06}.

Most stars in the Galaxy are found in multiple systems, with single
stars (including our own Sun) being rather exceptions than a rule (see
for example~\cite{Duquennoy_Mayor91, Halbwachs_al03}). In the binary
stars with sufficiently large orbital separations (``wide binaries'')
the presence of the secondary component does not influence
significantly the evolution of the components. In ``close binaries'' the
evolutionary expansion of stars allows for a mass exchange between the
components. In close binaries, the initial mass of the components at
the zero-age main sequence (ZAMS) ceases to be the sole parameter
determining their evolution. Consequently, the formation of compact
remnants in binary stars differs from single stars. This is
illustrated by Figure~\ref{f:remn} which plots the type of the stellar
remnant as a function of both initial mass and the radius of a star at
the moment of the Roche-lobe overflow (RLOF). It is seen that wide
binaries evolve as single stars, while for binaries with RLOF a new
type of remnants appears -- a helium WD, whose formation from a single
star in the Hubble time is impossible\epubtkFootnote{The hydrogen burning
  time for single stars with $M_0\lesssim 0.9\,M_\odot$ exceeds the age
  of the Universe.}.

\epubtkImage{figure01.png}{%
  \begin{figure}[htbp]
    \centerline{\includegraphics[scale=0.45]{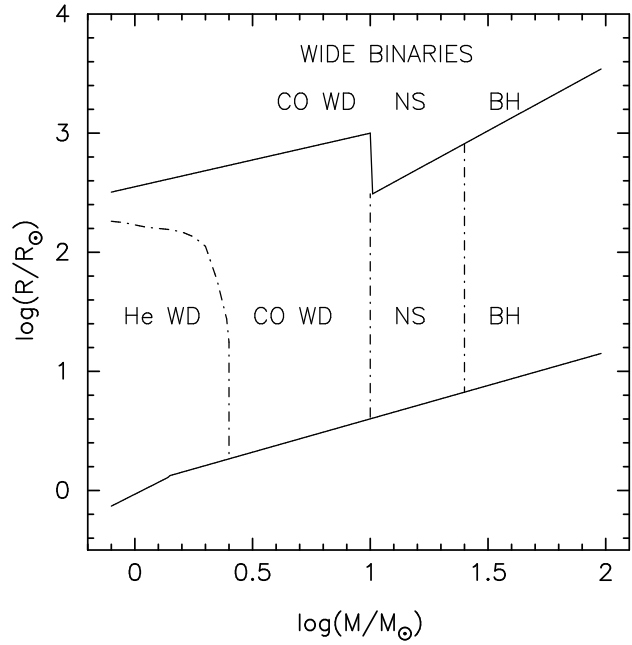}}
    \caption{\it Descendants of components of close binaries depending
      on the radius of the star at RLOF. The boundary between
      progenitors of He and CO-WDs is uncertain by several
      $0.1\,M_\odot$, the boundary between WDs and NSs by
      $\sim 1\,M_\odot$, while for the formation of BHs the lower mass
      limit may be even by $\sim 10\,M_\odot$ higher than indicated.}
    \label{f:remn}
  \end{figure}
}

Binaries with compact remnants are primary potential GW sources
(see Figure~\ref{f:GW_sources}). This figure plots the sensitivity 
of ground-based interferometer LIGO,
as well as the space laser interferometer LISA, in terms of
dimensionless GW strain $h$ measured over 1 year.
The strongest Galactic sources at all frequencies are the most compact
double NSs and BHs. Double WDs (including AM~CVn-stars) and ultra-compact
X-ray binaries (NS\,+\,WD) appear to be promising LISA sources.

\epubtkImage{figure02.png}{%
  \begin{figure}[htbp]
    \centerline{\includegraphics[scale=0.45]{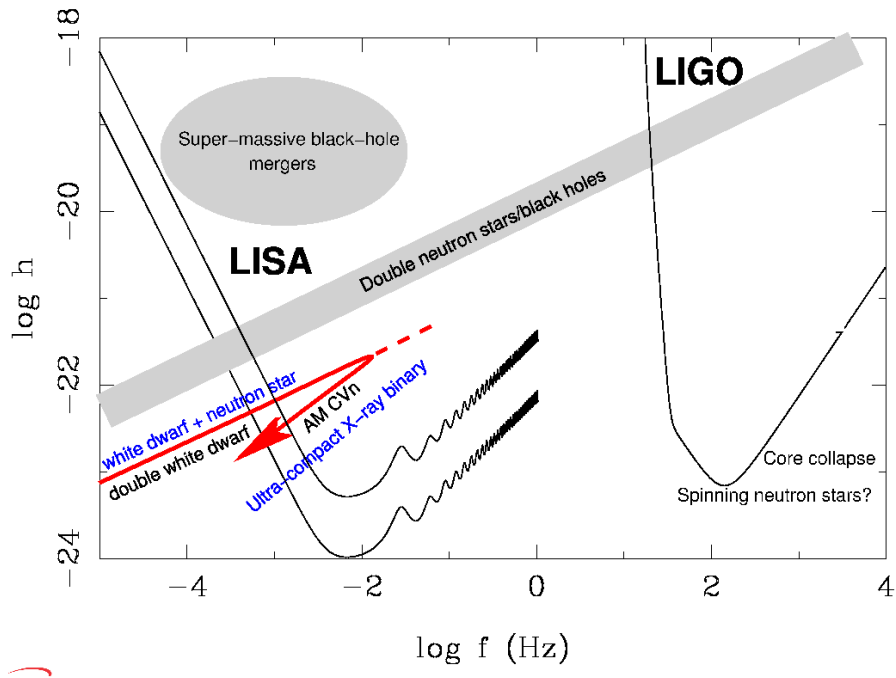}}
    \caption{\it Sensitivity limits of GW detectors and the regions of
      the $f$--$h$ diagram occupied by some of the potential GW
      sources. (Courtesy G.~Nelemans.)}
    \label{f:GW_sources}
  \end{figure}
}

Double NS/BH systems result from the evolution of initially massive
binaries, while double WDs are formed from the evolution of low-mass
binaries. We shall consider them separately.

\begin{description}
\item[Binaries with NSs and BHs] ~\\
  Binary systems with components massive enough to produce NSs or BHs at
  the end of thermonuclear evolution may remain bound after two
  supernova explosions. Then, loss of energy and momentum by GWs
  controls entirely their evolution and gradual reduction of the
  binary separation may bring the components into contact. During the
  merger process $\sim 10^{52} \mathrm{\ erg}$ are released as
  GWs~\cite{Clark_Eardley77, Clark_al79}. Such strong bursts of GWs can
  be reliably detected by the present-day ground-based GW detectors
  from distances up to several megaparsecs and are the most
  important targets for GW observatories such as LIGO, GEO, and
  VIRGO~\cite{Grishchuk_al01, Schutz03}.

  The problem is to evaluate as accurately as possible (i) the
  physical parameters of the coalescing binaries (masses of the
  components and, if possible, their spins, magnetic fields, etc.), and
  (ii) the occurrence rate of mergers in the Galaxy and in the local
  Universe. Masses of NSs in binaries are known with a rather good
  accuracy of 10\% or better from, e.g., pulsar
  studies~\cite{Thorsett_Chakrabarty99}; see
  also~\cite{Lattimer_Prakash04} for a recent update of NS mass
  measurements.

  The case is not so good with the rate of coalescence of relativistic
  binary stars. Unfortunately, there is no way to derive it from
  first principles -- neither the formation rate of the progenitor
  binaries for compact double stars nor stellar evolution are known
  well enough. However, the situation is not completely hopeless,
  especially in the case of double NS systems. Natural appearance of
  rotating NSs with magnetic fields as radio pulsars allows searching
  for binary pulsars with secondary compact companion using powerful
  methods of modern radio astronomy (for example, in dedicated pulsar
  surveys such as the Parkes multi-beam pulsar
  survey~\cite{Manchester_al01, Faulkner_al04}).

  Based on the observational statistics of the Galactic binary pulsars
  with another NS companion, one can evaluate the Galactic rate of
  binary NS formation and merging~\cite{Phinney91, nps91,
    Kim_al03}. On the other hand, a direct simulation of binary star
  evolution in the Galaxy (the \emph{population synthesis} method) can
  also predict the formation and merger rates of close compact
  binaries as a function of (numerous) parameters of binary star
  formation and evolution. It is important and encouraging that both
  estimates (observational, as inferred from recent measurements of
  binary pulsars~\cite{Burgay_al03, Kalogera_al04}, and theoretical
  from the population synthesis; see
  Section~\ref{sec:secI:DetectRate}) now give very close estimates for
  the double NS star merger rate in the Galaxy of about one event per
  10,000 years. No binary BH or NS\,+\,BH systems have been found so far,
  so merger rates of compact binaries with BHs have been evaluated as
  yet only from population synthesis studies.
\item[Binaries with WDs] ~\\
  The interest in these binaries stems from several
  circumstances. First, they are considered as testbeds for
  gravitational wave physics. Second, with them SNe~Ia are
  associated. SNe~Ia are being used as the primary standard
  candle sources for the determination of the cosmological parameters
  $\Omega$ and $\Lambda$ (see, e.g., \cite{riess_cosm98,
    perlmutter_cosm99}). A comparison of SN~Ia rates (for the
  different models of their progenitors) with observations may, in
  principle, shed light on both the star formation history and on the
  nature of the progenitors (see, e.g., \cite{yl98, madau_prog98,
    forster_delay06}). The counts of distant SNe could be used to
  constrain cosmological parameters (see, e.g.,
  \cite{ruiz_canal_cosm98}). Finally, close binaries with WDs are among
  the most promising verification binaries for LISA~\cite{stroeer06}.
\end{description}

\noindent
In this paper we shall concentrate on the formation and evolution of
binary compact stars most relevant for GW studies. The paper is
organized as follows. We start in
Section~\ref{section:observational_situation} with a review of the main
observational data on double NSs, especially measurements of masses of
NSs and BHs, which are most important for the estimate of the amplitude
of the expected GW signal. We briefly discuss the empirical methods to
determine double NS coalescence rate. The basic principles of binary
stellar evolution are discussed in
Section~\ref{section:physical_principles}. Then, in
Section~\ref{section:evolution_highmass} we describe the evolution of
massive binary stars. We then discuss the Galactic rate of formation of
binaries with NSs and BHs in
Section~\ref{section:binary_NS}. Theoretical estimates of detection
rates for mergers of binary relativistic stars are discussed in
Section~\ref{sec:secI:DetectRate}. Further we proceed to the analysis
of formation of short-period binaries with WD components in
Section~\ref{section:wd_formation}, and consider observational data on
binary white dwarfs in Section~\ref{section:observations}. A model for
the evolution of interacting double-degenerate systems is presented in
Section~\ref{section:am-evol}. In Section~\ref{section:waves} we
describe gravitational waves from compact binaries with white-dwarf
components. Sections~\ref{section:opt+x} and~\ref{section:gwr+em} are
devoted, respectively, to the model of optical and X-ray emission of
AM~CVn-stars and to their subsample potentially observed both in
electromagnetic and gravitational waves. Our conclusions follow in
Section~\ref{sec:concl}.

\newpage


\section{Observations of Double Neutron Stars}
\label{section:observational_situation}


\subsection{Compact binaries with neutron stars}

Double NSs have been discovered because one of the components of the
binary is observed as a radio pulsar. The precise pulsar timing allows
one to search for a periodic variation due to the binary motion. This
technique is reviewed in detail by Lorimer~\cite{Lorimer_LRR01};
applications of pulsar timing for general relativity tests are
reviewed by Stairs~\cite{Stairs_LRR03}.

Basically, pulsar timing provides the following Keplerian orbital
parameters of the binary system: the binary orbital period $P_\mathrm{b}$ as
measured from periodic Doppler variations of the pulsar spin, the
projected semimajor axis $x=a\sin i$ as measured from the semi-amplitude
of the pulsar radial velocity curve ($i$ is the binary inclination angle
defined such that $i=0$ for face-on systems), the orbital eccentricity
$e$ as measured from the shape of the pulsar radial velocity curve,
and the longitude of periastron $\omega$ at a particular epoch
$T_0$. The first two parameters allow one to construct the \emph{mass
  function} of the secondary companion,
\begin{equation}
  f(M_\mathrm{p},M_\mathrm{c}) =
  \frac{4\pi^2x^3}{P_\mathrm{b}^2 T_\odot} =
  \frac{(M_\mathrm{c}\sin i)^3}{(M_\mathrm{c}+M_\mathrm{p})^2}.
  \label{eq:mass_function}
\end{equation}
In this expression, $x$ is measured in light-seconds, $T_\odot\equiv
GM_\odot/c^3=4.925490947 \mathrm{\ \mu s}$, and $M_\mathrm{p}$ and
$M_\mathrm{c}$ denote masses of the
pulsar and its companion, respectively. This function gives the strict
lower limit on the mass of the unseen companion. However, assuming the
pulsar mass to have the typical value of a NS mass (for example,
confined between the lowest measured NS mass $1.25\,M_\odot$ for
PSR~J0737-303B~\cite{Lyne_al04} and the maximum measured NS mass of
$2.1\,M_\odot$ in the NS--WD binary PSR~J0751+1807~\cite{Nice_al04}), one
can estimate the mass of the secondary star even without knowing the
binary inclination angle $i$.

Long-term pulsar timing allows measurements of several relativistic 
phenomena: the advance of periastron $\dot{\omega}$, the redshift parameter
$\gamma$, the Shapiro delay within the binary system qualified through
post-Keplerian parameters $r$, $s$, and the binary orbit decay $\dot
P_\mathrm{b}$. From the post-Keplerian parameters the individual masses $M_\mathrm{p}$,
$M_\mathrm{c}$ and the binary inclination angle $i$ can be
calculated~\cite{Brumberg_al75}.

Of the post-Keplerian parameters of binary pulsars, the periastron advance rate
is usually measured most readily. Assuming it to be entirely due to general
relativity, the total mass of the system can be evaluated:
\begin{equation}
  \dot{\omega} = 3 \left( \frac{2\pi}{P_\mathrm{b}} \right)^{5/3}
  \frac{T_\odot^{2/3}(M_\mathrm{c}+M_\mathrm{p})^{2/3}}{(1-e^2)}.
  \label{eq:periastron_adv}
\end{equation}
High values of the derived total mass of the system 
($\gtrsim 2.5\,M_\odot$) suggests the presence of another NS or even
BH\epubtkFootnote{Unless the companion is directly observable and its
  mass can be estimated by other means.}.

If the individual masses, binary period, and eccentricity of a compact
binary system are known, it is easy to calculate the time it takes for
the binary companions to coalesce due to GW emission using the
quadrupole formula for GW emission~\cite{Peters64} (see
Section~\ref{A:GW_evol} for more detail):
\begin{equation}
  \tau_\mathrm{GW} \approx 4.8\times 10^{10} \mathrm{\ yr}
  \left( \frac{P_\mathrm{b}}{\mathrm{\ d}} \right)^{8/3}
  \left( \frac{\mu}{M_\odot} \right)^{-1}
  \left( \frac{M_\mathrm{c}+M_\mathrm{p}}{M_\odot} \right)^{-2/3}
  (1-e^2)^{7/2}.
  \label{eq:tau_GW}
\end{equation}
Here$\mu=M_\mathrm{p}M_\mathrm{c}/(M_\mathrm{p}+M_\mathrm{c})$ the
reduced mass of the binary. Some
observed and derived parameters of known compact binaries with NSs are
collected in Tables~\ref{table:doubleCB_o} and~\ref{table:doubleCB_d}.

\begin{table}
  \renewcommand{\arraystretch}{1.2}
  \centering
  \begin{tabular}{l|ccccccc}
    \hline \hline
    PSR & $P$ & $P_\mathrm{b}$ & $a_1 \sin i$ & $e$ &
    $\dot{\omega}$ & $\dot{P}_\mathrm{b}$ & Ref. \\
    & $[\mathrm{ms}]$ & $[\mathrm{d}]$ & $[\mbox{lt-s}]$ & &
    $[\mathrm{deg\ yr}^{-1}]$ & $[\times 10^{-12}]$ \\
    \hline
    J0737--3039A & $22.70$ & $0.102$ & $1.42$ & $0.088$ & $16.88$ & $-1.24$ &
    \cite{Burgay_al03} \\
    J0737--3039B & $2773$ & --- & --- & --- & --- & --- &
    \cite{Lyne_al04} \\
    J1518+4904 & $40.93$ & $8.634$ & $20.04$ & $0.249$ & $0.011$ & ? &
    \cite{Nice_al96} \\
    B1534+12 & $37.90$ & $0.421$ & $3.73$ & $0.274$ & $1.756$ & $-0.138$ &
    \cite{Wolszczan90, Stairs_al02} \\
    J1756--2251 & $28.46$ & $0.320$ & $2.75$ & $0.181$ & $2.585$ & ? &
    \cite{Faulkner_al05} \\
    J1811--1736 & $104.18$ & $18.779$ & $34.78$ & $0.828$ & $0.009$ & $<30$ &
    \cite{Lyne_al00} \\
    J1906+074 & $144.07$ & $0.116$ & $1.42$ & $0.085$ & $7.57$ & ? &
    \cite{Lorimer_al06} \\
    B1913+16 & $59.03$ & $0.323$ & $2.34$ & $0.617$ & $4.227$ & $-2.428$ &
    \cite{Hulse_Taylor75} \\
    B2127+11C & $30.53$ & $0.335$ & $2.52$ & $0.681$ & $4.457$ & $-3.937$ &
    \cite{Anderson_al90, Prince_al91} \\
    \hline \hline
  \end{tabular}
  \caption{\it Observed parameters of double neutron star binaries.}
  \label{table:doubleCB_o}
  \renewcommand{\arraystretch}{1.0}
\end{table}

\begin{table}
  \renewcommand{\arraystretch}{1.2}
  \centering
  \begin{tabular}{l|cccc}
    \hline \hline
    PSR & $f(m)$ & $M_\mathrm{c}+M_\mathrm{p}$ &
    $\tau_c=P/(2\dot{P})$ & $\tau_\mathrm{GW}$ \\
    & [$M_\odot$] & [$M_\odot$] & [$\mathrm{Myr}$] & [$\mathrm{Myr}$] \\
    \hline
    J0737--3039A & $0.29$ & $2.58$ & $210$ & $87$ \\
    J0737--3039B & --- & --- & $50$ & --- \\
    J1518+4904 & $0.12$ & $2.62$ & & $9.6\times 10^6$ \\
    B1534+12 & $0.31$ & $2.75$ & $248$ & $2690$ \\
    J1756--2251 & $0.22$ & $2.57$ & $444$ & $1690$ \\
    J1811--1736 & $0.13$ & $2.6$ & & $1.7\times 10^6$ \\
    J1906+074 & $0.11$ & $2.61$ & $0.112$ & $300$ \\
    B1913+16 & $0.13$ & $2.83$ & $108$ & $310$ \\
    B2127+11C & $0.15$ & $2.71$ & $969$ & $220$ \\
    \hline \hline
  \end{tabular}
  \caption{\it Derived parameters of double neutron star binaries}
  \label{table:doubleCB_d}
  \renewcommand{\arraystretch}{1.0}
\end{table}


\subsection{How frequent are double NS coalescences?}
\label{sec:ns_freq}

As it is seen from Table~\ref{table:doubleCB_d}, only six double NS
systems presently known will coalesce over a time interval shorter than
$\approx 10 \mathrm{\ Gyr}$: J0737--3039A, B1534+12, J1756--2251, J1906+074,
B1913+16, and B2127+11C. Of these six systems, one (PSR~B2127+11C) is
located in the globular cluster M15. This system may have a different
formation history, so usually it is not included in the analysis of
the coalescence rate of Galactic double compact binaries. The
formation and evolution of relativistic binaries in dense stellar
systems is reviewed elsewhere~\cite{Benacquista_LRR02}. For a recent
general review of pulsars in globular clusters see
also~\cite{Camilo_Rasio05}.

The ordinary way of estimating the double NS merger rate from binary
pulsar statistics is based on the following
extrapolation~\cite{nps91, Phinney91}. Suppose we observe $i$
classes of Galactic binary pulsars. Taking into account various
selection effects of pulsar surveys (see, e.g., \cite{Narayan87, Kim_al03}),
the Galactic number of pulsars $N_i$ in each class can be
evaluated. To compute the Galactic merger rate of double NS binaries,
we need to know the time since the birth of the NS observed as a pulsar
in the given binary system. This time is the sum of the observed
characteristic pulsar age $\tau_\mathrm{c}$ and the time required for the
binary system to merge due to GW orbit decay $\tau_\mathrm{GW}$. With the
exception of PSR~J0737--3039B and the recently discovered PSR~J1906+074,
pulsars that we observe in binary NS systems are old recycled pulsars
which were spun-up by accretion from the secondary companion to the
period of several tens of ms (see Table~\ref{table:doubleCB_o}). Thus
their characteristic ages can be estimated as the time since
termination of spin-up by accretion (for the younger pulsar
PSR~J0737--3039B this time can be also computed as the dynamical age of
the pulsar, $P/(2\dot{P})$, which gives essentially the same result).

Then the merger rate ${\cal R}_i$ can be calculated as ${\cal R}_i\sim
N_i/(\tau_\mathrm{c}+\tau_\mathrm{GW})$ (summed over all binary pulsars). The detailed
analysis~\cite{Kim_al03} indicates that the Galactic merger rate of
double NSs is mostly determined by pulsars with faint radio luminosity
and short orbital periods. Presently, it is the nearby ($600 \mathrm{\
  pc}$) double-pulsar system PSR~J0737--3039 with a short orbital
period of $2.4 \mathrm{\ hr}$~\cite{Burgay_al03} that mostly determines the
\emph{empirical} estimate of the merger rate. According to Kim
et~al.~\cite{kim_etal06}, ``the most likely values of DNS merger rate
lie in the range 3\,--\,190 per Myr depending on different pulsar
models''. The estimates by \emph{population synthesis codes} are
still plagued by uncertainties in statistics of binaries, in modeling
binary evolution and supernovae. The most optimistic ``theoretical''
predictions amount to $\simeq 300 \mathrm{\ Myr}^{-1}$~\cite{ty93b, bel_kal02}.

Recently, the bursting radio source GCRT J1745--3009 in the direction
to the Galactic centre was proposed to be a possible double NS
binary~\cite{Turolla_al05}. The source was found to emit a series of
radio bursts with high brightness temperature, of typical duration
$\sim 10 \mathrm{\ min}$, with an apparent periodic pattern of $\sim
77 \mathrm{\ min}$~\cite{Hyman_al05}. Confirmation of the binary NS
nature of transient radio sources like GCRT J1745--3009 would be
important to get a more precise estimate of the Galactic coalescence
rate of double NS.

The extrapolation beyond the Galaxy is usually done by scaling the
Galactic merger rate to the volume from which the merger events can be
detected with given GW detector's sensitivity. The scaling factor
widely used is the ratio between the B-band luminosity density in the
local Universe, correlating with the star-formation rate, and the
B-band luminosity of the Galaxy~\cite{Phinney91, Kalogera_al01}. One
can also use for this purpose the direct ratio of the Galactic star
formation rate $\mathrm{SFR}_\mathrm{G}\simeq 3\,M_\odot \mathrm{\ yr}^{-1}$~\cite{Miller_Scalo79,
  Timmes_al97} to the star formation rate on the local Universe
$\mathrm{SFR}_\mathrm{loc}\simeq 0.03\,M_\odot \mathrm{\ yr}^{-1}$~\cite{Perez-Gonzalez_al03,
  Schiminovich_al05}. These estimates yield the relation 
\begin{equation}
  {\cal R}_\mathrm{V} \approx 0.01 {\cal R}_\mathrm{G} [\mathrm{Mpc}^{-3}].
  \label{R_V}
\end{equation}
So for the Galactic merger rate ${\cal R}_\mathrm{G}\sim 10^{-4}
\mathrm{\ yr}^{-1}$ a
very optimistic detection rate for binary NSs of about once per 1\,--\,2
years of observations by the first-generation GW detectors is
predicted~\cite{Burgay_al03}. This estimate is still uncertain, mostly
due to poor knowledge of the luminosity function for faint radio
pulsars~\cite{kim_etal06}.

Recently, the first results of the search for GWs from coalescing
binary systems in the Milky Way and the Magellanic Clouds using data
taken by two of the three LIGO interferometers~\cite{Abbott_al04}
established an observational upper limit to the Galactic binary NS
coalescence rate of $\R_\mathrm{G} < 1.7 \times 10^2 \mathrm{\ yr}^{-1}$. With increasing
sensitivity of GW detectors, this limit will be much improved in the
nearest future.


\subsection{On the connection between GRBs and compact binary mergers}

Here we wish to mention the possibility of observational
manifestations of NS/BH binary mergers other than the violent GW
emission. First of all, it is the connection of relativistic binary
mergers to some subclasses of cosmic GRBs. That
catastrophic events like the coalescence of binary NSs or BHs can be
related to GRBs has been suggested for quite a long time. First ideas can
be found in the papers of Sergei Blinnikov and his
coauthors~\cite{Blinnikov_al84} and Bohdan Paczy\'{n}ski
\cite{Paczynski86} with subsequent studies in~\cite{Eichler_al89,
  Paczynski91, Mochkovitch_al93, Katz_Canel96}, etc.

Now these ideas gained strong observational support from the accurate
localization of short GRBs with hard spectrum by the Swift and
HETE-II space missions~\cite{Gehrels_al05, Bloom_al05,
  Berger_al05, Fox_al05}. The short-hard subclass of GRBs includes up
to 30\% of all GRBs~\cite{Paciesas_al99}. The most important recent
discovery is that these GRBs occur both in late-type~\cite{Covino_al05,
  Hjorth_al05} and early-type galaxies~\cite{Berger_al05,
  Barthelmy_al05}, suggesting old stellar population progenitors. This
is in sharp contrast to long GRBs, some of which are definitely
associated with peculiar type~Ib/c supernovae produced by the core
collapse of massive stars~\cite{Hjorth_al03, Stanek_al03}.

The principal observational facts about several well-localized 
short GRBs (see~\cite{Berger06} for more discussion) are:

\begin{enumerate}
\item They occur at cosmological redshifts from 0.160 to 1.8 and may
  constitute up to 20\% of the local short GRB population detected by
  the BATSE\epubtkFootnote{Burst And Transient Source
    Experiment~\cite{url01}.} experiment on board of the Compton
  Gamma-ray Observatory.
\item The isotropic energy release is typically lower than in long
  GRBs (from $\sim 10^{48} \mathrm{\ erg}$ to $\sim 4 \times 10^{51}
  \mathrm{\ erg}$).
\item The opening angle of the ejecta in these GRBs is on average
  larger than in long GRBs~\cite{Burrows_al06}.
\item Short GRBs are found both in elliptical and star forming
  galaxies. Statistical analysis suggests that the occurrence rate of
  short bursts is roughly equal in early-type and late-type
  galaxies~\cite{Berger06}.
\end{enumerate}

The short GRB rate inferred from these
observations~\cite{Piran_Guetta06}, $\R_\mathrm{sGRB}\sim (10\mbox{\,--\,}30)
\, _{70}^3 \mathrm{\ Gpc}^{-3} \mathrm{\ yr}^{-1}$ agrees with the double NS
merger rate derived from binary pulsar statistics. Depending on the unknown
beaming factor a possible upper limit of about $10^5 \mathrm{\ events\ yr}^{-1}
\mathrm{\ Gpc}^{-3}$ was obtained in~\cite{Nakar_al06}. That paper also
extensively discusses the application of the rate of short GRBs to
LIGO/VIRGO detections of double NS binary mergings if they are
associated with short GRBs, and gives very good prospects for the
Advanced LIGO sensitivity (up to hundreds detections per
year). However, recent deep optical observations of several short GRBs
provide evidence for their association with very faint
galaxies~\cite{Berger_al06}, suggesting the intrinsic luminosity of a
significant part of short GRBs to be much higher than
$10^{48}\mbox{\,--\,}10^{49} \mathrm{\ erg}$, as inferred from observations of
close short GRBs by~\cite{Nakar_al06}, close to that of classical long
GRBs $\sim 10^{51}-10^{52} \mathrm{\ erg}$. Taking this finding into account
decreases the expected detection rate of NS mergers (if they are
associated with short GRBs) down to several events per year by the
Advanced LIGO detector~\cite{Berger_al06}.

We also emphasize the agreement of the observational estimates with
population synthesis calculations of binary mergers in galaxies
of different types~\cite{bag_spz_lry_grb98, Panchenko_al99,
  Belczynski_al02, Bulik_Belczynski04, belcz_short_grb06}. The
analysis of luminosity function and statistics of short GRBs from the
BATSE catalog~\cite{Piran_Guetta06} implies a delay relative to the
star formation history, which can favour double NS systems
dynamically formed in stellar clusters as
progenitors~\cite{Hopman_al06, Grindlay_al06}. Theoretical issues
related to the generation of short hard GRBs from binary NS and NS--BH
mergers are discussed in~\cite{Lee_al05a, Lee_al05b,
  Oechslin_Janka06}.

Now let us see what theory says about the formation, evolution, and
detection rates of close compact binaries and their properties.

\newpage


\section{Basic Principles of Binary Star Evolution}
\label{section:physical_principles}

Beautiful general reviews of the topic can be found
in~\cite{Bhattacharya_vandenHeuvel91, vandenHeuvel83}. Here we
restrict ourselves to recalling several facts about binary evolution
which are most relevant to the formation and evolution of compact
binaries. The readers who have experience in the field can skip this
section.


\subsection{Keplerian binary system and radiation back reaction}
\label{sec:appA}

We start with some basic facts about Keplerian motion in a binary
system and the simplest case of evolution of two point masses due to
gravitational radiation losses. The stars are highly condensed
objects, so their treatment as point masses is usually adequate for
the description of their interaction in the binary. Furthermore,
Newtonian gravitation theory is sufficient for this purpose as long as
the orbital velocities are small in comparison with the speed of light
$c$. The systematic change of the orbit caused by the emission of
gravitational waves will be considered in a separate paragraph below.


\subsubsection{Keplerian motion}

Let us consider two point masses $M_1$ and $M_2$ orbiting each other 
under the force of gravity.
It is well known (see~\cite{L_L_v1}) that this problem
is equivalent to the problem of a single body with mass $\mu$ moving 
in an external gravitational potential. The value of the external
potential is determined by the total mass of the system 
\begin{equation}
  M = M_1 + M_2.
  \label{B:M}
\end{equation}
The reduced mass $\mu$ is
\begin{equation}
  \mu = \frac{M_1M_2}{M}.
  \label{B:mu}
\end{equation}
The body $\mu$ moves in an elliptic orbit with eccentricity $e$
and major semi-axis $a$. 
The orbital period $P$ and orbital frequency $\Omega=2\pi/P$
are related to $M$ and $a$ by Kepler's third law
\begin{equation}
  \Omega^2 = \left( \frac{2\pi}{P} \right)^2 = \frac{GM}{a^3}.
  \label{B:3Kepl}
\end{equation}
This relationship is valid for any eccentricity $e$.

Individual bodies $M_1$ and $M_2$ move around the 
barycentre of the system in elliptic orbits with the same
eccentricity $e$. The major semi-axes $a_i$ of the two ellipses
are inversely proportional to the masses
\begin{equation}
  \frac{a_1}{a_2} = \frac{M_2}{M_1},
  \label{B:a1}
\end{equation}
and satisfy the relationship $a=a_1+a_2$. 
The position vectors of the bodies from the system's barycentre are 
$\vec{r}_1 = M_2 \vec{r} /(M_1+M_2)$ and 
$\vec{r}_2 = - M_1 \vec{r} /(M_1+M_2)$, where $\vec{r} = \vec{r}_1 - \vec{r}_2$
is the relative position vector. Therefore, 
the velocities of the bodies with respect to the system's
barycentre are related by 
\begin{equation}
  -\frac{\vec{V}_1}{\vec{V}_2} = \frac{M_2}{M_1},
  \label{B:v1}
\end{equation}
and the relative velocity is $\vec{V}= \vec{V}_1-\vec{V}_2$.

The total conserved energy of the binary system is 
\begin{equation}
  E = \frac{M_1 \vec{V}_1^2}{2} + \frac{M_2 \vec{V}_2^2}{2} - 
  \frac{GM_1M_2}{r} =
  \frac{\mu \vec{V}^2}{2} - \frac{GM_1M_2}{r} =
  -\frac{GM_1M_2}{2a},
  \label{B:E}
\end{equation}
where $r$ is the distance between the bodies.
The orbital angular momentum vector is perpendicular to the orbital plane
and can be written as
\begin{equation}
  \vec{J}_\mathrm{orb} =
  M_1\vec{V}_1\times\vec{r}_1 + M_2\vec{V}_2\times\vec{r}_2 =
  \mu\vec{V}\times\vec{r}.
  \label{B:vecJ}
\end{equation}
The absolute value of the orbital angular momentum is
\begin{equation}
  |\vec{J}_\mathrm{orb}| = \mu\sqrt{GMa(1-e^2)}.
  \label{B:Je}
\end{equation}

For circular binaries with $e=0$
the distance between orbiting bodies does not depend on time,
\begin{displaymath}
  r(t,e=0) = a,
\end{displaymath}
and is usually referred to as orbital separation.
In this case, the velocities of the bodies, as well as their 
relative velocity, are also time-independent,
\begin{equation}
  V \equiv |\vec{V}| = \Omega a = \sqrt{GM/a},
  \label{B:Vorb}
\end{equation}
and the orbital angular momentum becomes
\begin{equation}
  |\vec{J}_\mathrm{orb}| = \mu Va = \mu \Omega a^2.
  \label{B:J}
\end{equation}


\subsubsection{Gravitational radiation from a binary}

The plane of the orbit is determined by the orbital
angular momentum vector $\vec{J}_\mathrm{orb}$. The
line of sight is defined by a unit vector $\vec{n}$. The
binary inclination angle $i$ is defined by the relation
$\cos i = (\vec{n},\vec{J}_\mathrm{orb}/J_\mathrm{orb})$
such that $i=90^\circ$ corresponds to a system visible edge-on.

Let us start from two point masses $M_1$ and $M_2$ in a circular
orbit. In the quadrupole approximation~\cite{L_L_v2}, the two polarization
amplitudes of GWs at a distance $r$ from the source are given by
\begin{eqnarray}
  h_+ &=& \frac{G^{5/3}}{c^4}
  \frac{1}{r}\,2(1+\cos^2i)(\pi f M )^{2/3}\mu\cos(2\pi f t),
  \label{A:h_+}
  \\
  h_\times &=& \pm \frac{G^{5/3}}{c^4}
  \frac{1}{r}\,4 \cos i (\pi f M )^{2/3}\mu\sin(2\pi f t).
  \label{A:h_x}
\end{eqnarray}%
Here $f=\Omega/\pi$ is the frequency of the emitted GWs (twice the
orbital frequency). Note that for a fixed distance $r$ and a given
frequency $f$, the GW amplitudes are fully determined by $\mu M^{2/3}
= {\M}^{5/3}$, where the combination
\begin{displaymath}
  \M\equiv\mu^{3/5}M^{2/5}
\end{displaymath}
is called the ``chirp mass'' of the binary. 
After averaging over the orbital period (so that
the squares of periodic functions are replaced by 1/2)
and the orientations of the binary orbital plane, 
one arrives at the averaged (characteristic) GW amplitude
\begin{equation}
  h(f,\M,r) =
  \left( \langle h_+^2 \rangle + \langle h_\times^2 \rangle \right)^{1/2} =
  \left( \frac{32}{5} \right)^{1/2} \frac{G^{5/3}}{c^4}
  \frac{\M^{5/3}}{r} (\pi f)^{2/3}.
  \label{A:meanh}
\end{equation}


\subsubsection{Energy and angular momentum loss}
\label{section:AML}

In the approximation and under the choice of coordinates that we are
working with, it is sufficient to use the Landau--Lifshitz
gravitational pseudo-tensor~\cite{L_L_v2} when calculating the
gravitational waves energy and flux. (This calculation can be
justified with the help of a fully satisfactory gravitational
energy-momentum tensor that can be derived in the field theory
formulation of general relativity~\cite{Babak_Grishchuk00}). The energy
$dE$ carried by a gravitational wave along its direction of
propagation per area $dA$ per time $dt$ is given by 
\begin{equation}
  \frac{dE}{dA\,dt} \equiv F =
  \frac{c^3}{16\pi G}
  \left[ \left( \frac{\partial h_+}{\partial t} \right)^2 +
  \left( \frac{\partial h_\times}{\partial t} \right)^2 \right].
  \label{A:flux} 
\end{equation}
The energy output $dE/dt$ from a localized source in all directions is
given by the integral
\begin{equation}
  \frac{dE}{dt} = \int F(\theta,\phi) r^2 \, d\Omega.
  \label{A:loss}
\end{equation}
Replacing
\begin{displaymath}
  \left( \frac{\partial h_+}{\partial t} \right)^2 +
  \left( \frac{\partial h_\times}{\partial t} \right)^2 = 
  4\pi^2 f^2 h^2(\theta, \phi)
\end{displaymath}
and introducing
\begin{displaymath}
  h^2 = \frac{1}{4 \pi} \int h^2(\theta, \phi) \, d\Omega,
\end{displaymath}
we write Equation~(\ref{A:loss}) in the form
\begin{equation}
  \frac{dE}{dt} = \frac{c^3}{G} (\pi f)^2 h^2 r^2.
  \label{A:loss2}
\end{equation}

Specifically for a binary system in a circular orbit, one finds
the energy loss from the system (sign minus) with the help of 
Equations~(\ref{A:loss2}) and~(\ref{A:meanh}):
\begin{equation}
  \frac{dE}{dt} = 
  - \left( \frac{32}{5}\right)\frac{G^{7/3}}{c^5} (\M\pi f)^{10/3}.
  \label{A:dEdt}
\end{equation}
This expression is exactly the same one that can be obtained directly from
the quadrupole formula~\cite{L_L_v2},
\begin{equation}
  \frac{dE}{dt} =
  -\frac{32}{5} \frac{G^4}{c^5} \frac{M_1^2M_2^2M}{a^5},
  \label{A:GW:dEdt}
\end{equation}
rewritten using the definition of the chirp mass and
Kepler's law. Since energy and angular momentum are continuously
carried away by gravitational radiation, the two masses in orbit spiral
towards each other, thus increasing their orbital frequency $\Omega$. 
The GW frequency $f=\Omega/\pi$ and the GW amplitude $h$ 
are also increasing functions of time. The rate of the frequency 
change is\epubtkFootnote{A signal with such an increasing frequency is 
reminiscent of the chirp of a bird. This explains the
origin of the term ``chirp mass'' for the 
parameter $\M$ which fully determines the GW frequency and 
amplitude behaviour.}
\begin{equation}
  \dot{f} = \left( \frac{96}{5} \right)
  \frac{G^{5/3}}{c^5}\pi^{8/3}\M^{5/3}f^{11/3}.
  \label{A:dotf}
\end{equation}

In spectral representation, the flux of energy per unit area
per unit frequency interval is given by the right-hand-side of the
expression 
\begin{equation}
  \frac{dE}{dA\,df} = \frac{c^3}{G} \frac{\pi f^2}{2}
  \left( \left| \tilde h(f)_+ \right|^2 +
  \left| \tilde h(f)_\times \right|^2 \right)
  \equiv \frac{c^3}{G} \frac{\pi f^2}{2} S_h^2(f),
  \label{A:S_h}
\end{equation}
where we have introduced the spectral density $S_h^2(f)$ of the
gravitational wave field $h$. In the case of a binary system, the
quantity $S_h$ is calculable from Equations~(\ref{A:h_+}, \ref{A:h_x}):
\begin{equation}
  S_h = \frac{G^{5/3}}{c^3} \frac{\pi}{12} \frac{\M^{5/3}}{r^2}
  \frac{1}{(\pi f)^{7/3}}.
  \label{A:S_h:2}
\end{equation}


\subsubsection{Binary coalescence time}
\label{A:GW_evol}

A binary system in a circular orbit loses energy according to
Equation~(\ref{A:dEdt}). For orbits with non-zero eccentricity $e$, the
right-hand-side of this formula should be multiplied by the factor 
\begin{displaymath}
  f(e) = \left( 1+\frac{73}{24}e^2+\frac{37}{96}e^4 \right)(1-e^2)^{-7/2}
\end{displaymath}
(see~\cite{Peters64}). The initial binary separation $a_0$ decreases
and, assuming Equation~(\ref{A:GW:dEdt}) is always valid, it should vanish
in a time
\begin{equation}
  t_0 = \frac{c^5}{G^3} \frac{5 a_0^4}{256 M^2\mu} =
  \frac{5c^5}{256} \frac{(P_0/2\pi)^{8/3}}{(G\M)^{5/3}} \approx
  (9.8 \times 10^6 \mathrm{\ yr})
  \left( \frac{P_0}{1 \mathrm{\ h}} \right)^{8/3}
  \left( \frac{\M}{M_\odot} \right)^{-5/3}\!\!\!\!\!\!\!\!\!\!\!.
  \label{A:GW:t_0}
\end{equation}
As we noted above, gravitational radiation from the binary depends on
the chirp mass $\M$, which can also be written as $\M\equiv
M\eta^{3/5}$, where $\eta$ is the dimensionless ratio
$\eta=\mu/M$. Since $\eta\le 1/4$, one has $\M\lesssim 0.435 M$. For
example, for two NS with equal masses $M_1=M_2=1.4\,M_\odot$, the chirp
mass is $\M\approx 1.22\,M_\odot$. This explains the choice of
normalization in Equation~(\ref{A:GW:t_0}).

\epubtkImage{figure03.png}{
  \begin{figure}[htbp]
    \centerline{\includegraphics[scale=0.5]{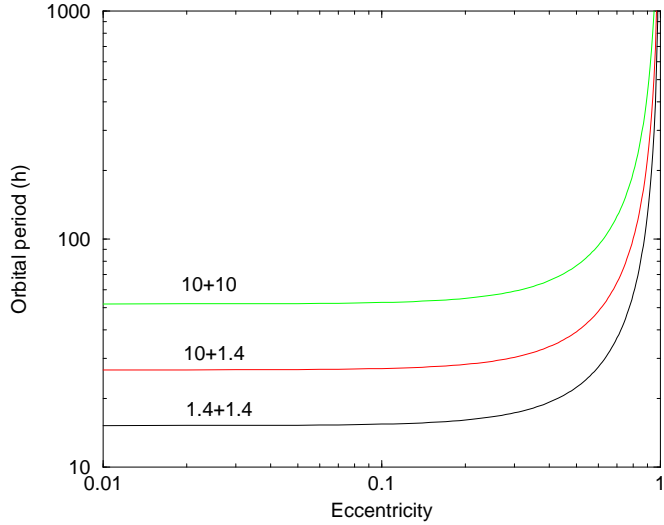}}
    \caption{\it The maximum initial orbital period (in hours) of two
      point masses which will coalesce due to gravitational wave
      emission in a time interval shorter than $10^{10} \mathrm{\  yr}$, as a
      function of the initial eccentricity $e_0$. The lines are
      calculated for $10\,M_\odot + 10\,M_\odot$ (BH\,+\,BH),
      $10\,M_\odot + 1.4\,M_\odot$ (BH\,+\,NS), and
      $1.4\,M_\odot + 1.4\,M_\odot$ (NS\,+\,NS).}
    \label{A:GW:p-e}
  \end{figure}
}

The coalescence time for an initially eccentric orbit with $e_0 \ne 0$
and separation $a_0$ is shorter than the coalescence time for a
circular orbit with the same initial separation $a_0$~\cite{Peters64}:
\begin{equation}
  t_\mathrm{c}(e_0) = t_0 \, f(e_0),
  \label{A:GW:t_c}
\end{equation}
where the correction factor $f(e_0)$ is 
\begin{equation}
  f(e_0) = \frac{48}{19} \frac{(1-e_0^2)^4}{e_0^{48/19}
  \left( 1+\frac{121}{304}e_0^2 \right)^{3480/2299}}
  \int_0^{e_0} \frac{\left( 1+\frac{121}{304}e^2 \right)^{1181/2299}}
  {(1-e^2)^{3/2}} \, e^{29/19} \, de.
  \label{A:GW:f(e)}
\end{equation}
To merge in a time interval shorter than $10 \mathrm{\ Gyr}$ the
binary should have a small enough initial orbital period $P_0\le
P_{cr}(e_0,\M)$ and, accordingly, a small enough initial semimajor
axis $a_0\le a_{cr}(e_0,\M)$. The critical orbital period is plotted
as a function of the initial eccentricity $e_0$ in
Figure~\ref{A:GW:p-e}. The lines are plotted for three typical sets of
masses: two neutron stars with equal masses ($1.4\,M_\odot +
1.4\,M_\odot$), a black hole and a neutron star ($10\,M_\odot +
1.4\,M_\odot$), and two black holes with equal masses ($10\,M_\odot +
10\,M_\odot$). Note that in order to get a significantly shorter
coalescence time, the initial binary eccentricity should be $e_0\geq
0.6$.


\subsubsection{Magnetic stellar wind}
\label{sec:msw}

In the case of low-mass binary evolution, there is another important
physical mechanism responsible for the removal of orbital angular
momentum, in addition to GW emission discussed above. This is the
magnetic stellar wind (MSW), or magnetic braking, which is thought to
be effective for main-sequence G-M dwarfs with convective envelopes,
i.e.\ in the mass interval $0.3\mbox{\,--\,}1.5\,M_\odot$. The upper
mass limit corresponds to the disappearance of a deep convective zone,
while the lower mass limit stands for fully convective stars. In both
cases a dynamo mechanism, responsible for enhanced magnetic activity, is
thought to be ineffective. The idea behind angular momentum loss (AML)
by magnetically coupled stellar wind is that the stellar wind is compelled by
magnetic field to corotate with the star to rather large distances where
it carries away large specific angular
momentum~\cite{schatz_msw62}. Thus, it appears possible to take away
substantial angular momentum without evolutionary significant
mass-loss in the wind. The concept of an MSW was introduced into analyses of
the evolution of compact binaries by Verbunt and Zwaan~\cite{vz81} when
it became evident that momentum loss by GWs alone is unable to explain
the observed mass-transfer rates in cataclysmic variables.
The latter authors based their reasoning on observations of the spin-down
of single G-dwarfs in stellar clusters with age~\cite{Skumanich72}
$V\propto t^{-1/2}$ (the Skumanich law). Applying this to a binary
component and assuming tidal locking between the star's axial rotation and
orbital revolution, one arrives at the rate of angular momentum loss
via an MSW,
\begin{equation}
  \frac{\dot{J}_\mathrm{MSW}}{J_\mathrm{orb}} \sim
  -\frac{R_\mathrm{o}^4}{M_\mathrm{c}}\frac{GM^2}{a^{5}},
  \label{MSW}
\end{equation}
where $R_\mathrm{o}$ is the radius of the optical companion and
$M_\mathrm{c}$ is the mass of the compact star.

Radii of stars filling their Roche lobes should be proportional to
binary separations,  $R_\mathrm{o}\propto a$, which means that the
characteristic time of orbital angular momentum removal by an MSW is
$\tau_\mathrm{MSW}\equiv (\dot{J}_\mathrm{MSW}/J_\mathrm{orb})^{-1}\propto a$. This should
be compared with AML by GWs with $\tau_\mathrm{GW} \propto
a^4$. Clearly, the MSW
(if it operates) is more effective in removing angular momentum from
binary system at larger separations (orbital periods), and at small
orbital periods GWs always dominate. Magnetic braking is especially
important in CVs and in LMXBs with orbital periods exceeding several
hours and is the driving mechanis for mass accretion onto the compact
component. 

We should note that the above prescription for an MSW is still debatable, 
since it is based on extrapolation of stellar rotation rates 
over several orders of magnitude -- from slowly rotating field 
stars to rapidly spinning components of close binaries. 
There are strong indications that actual magnetic braking 
for compact binaries may be much weaker than predictions 
based on the Skumanich law
(see, e.g., \cite{Politano_Weiler06} for recent discussion and references).


\subsection{Mass transfer modes and mass loss in binary systems}
\label{A:mass_transf}

GW emission is the sole factor responsible for 
the change of orbital parameters of a pair of compact (degenerate) stars. 
However, at the early stages of binary evolution, it is the 
mass transfer between the components and the loss of matter and its 
orbital momentum that play a dominant dynamical role. 
Strictly speaking, these processes should be treated hydro-dynamically
and they require complicated numerical calculations. However,
binary evolution can also be described semiqualitatively,
using a simplified description in terms of point-like bodies. 
The change of their integrated physical quantities, such as masses,
orbital angular momentum, etc.\ governs the evolution of the orbit.
This description turns out to be successful in reproducing the
results of more rigorous numerical calculations
(see, e.g., \cite{Shore_al94} for more detail and references). In this approach, 
the key role is allocated to the total orbital angular momentum $J_\mathrm{orb}$
of the binary. 

Let star 2 lose matter at a rate $\dot{M}_2<0$ and let $\beta$
$(0\le\beta\le1)$ be the fraction of the ejected matter which leaves
the system (the rest falls on the first star),
i.e.\ $\dot{M}_1=-(1-\beta)\dot{M}_2\ge 0$.
Consider circular orbits with orbital angular momentum given by
Equation~(\ref{B:J}). Differentiate both parts of Equation~(\ref{B:J}) by time $t$ 
and exclude $d\Omega/dt$ with the help of Kepler's third
law (\ref{B:3Kepl}). This gives us the rate of change of the
orbital separation:
\begin{equation}
  \frac{\dot{a}}{a} = -2 \left( 1+(\beta-1)\frac{M_2}{M_1} -
  \frac{\beta}{2} \frac{M_2}{M} \right)
  \frac{\dot{M}_2}{M_2} + 2\frac{\dot{J}_\mathrm{orb}}{J_\mathrm{orb}}.
  \label{A:dotaa}
\end{equation}
In the previous equation $\dot{a}$ and $\dot{M}$ are not independent
variables if the donor fills its Roche lobe. One defines the mass transfer
as conservative if both $\beta=0$ and $\dot{J}_\mathrm{orb}=0$. The mass
transfer is called non-conservative if at least one of these
conditions is violated.

It is important to distinguish some specific cases (modes) of mass
transfer:
\begin{enumerate}
\item conservative mass transfer,
\item non-conservative Jeans mode of mass loss (or fast wind mode),
\item non-conservative isotropic re-emission,
\item sudden mass loss from one of the component during supernova
  explosion, and
\item common-envelope stage.
\end{enumerate}

As specific cases of angular momentum loss we consider GW emission
(see Section~\ref{section:AML} and~\ref{A:GW_evol}) and a magnetically
coupled stellar wind (see Section~\ref{sec:msw})
which drive the orbital evolution for short-period binaries. For
non-conservative modes, one can also introduce some sub-cases, such as,
for example, a ring-like mode in which a circumbinary ring of
expelled matter is being formed (see, e.g., \cite{Soberman_al97}). Here, we
will not go into the details of such sub-cases.


\subsubsection{Conservative accretion}

In the case of conservative accretion, matter from $M_2$ is fully
deposited onto $M_1$. The transfer process preserves the total mass
($\beta=0$) and the orbital angular momentum of the system. It follows
from Equation~(\ref{A:dotaa}) that 
\begin{displaymath}
  M_1 M_2 \sqrt{a} = \mathrm{const},
\end{displaymath}
so that the initial and final binary separations are related as
\begin{equation}
  \frac{a_\mathrm{f}}{a_\mathrm{i}} =
  \left( \frac{M_{1\mathrm{i}}\,M_{2\mathrm{i}}}
  {M_{1\mathrm{f}}\,M_{2\mathrm{f}}} \right)^2\!\!\!.
  \label{A:conserv}
\end{equation}
The well-known ``rule of thumb'' for this case says that the orbit
shrinks when the more massive component loses matter, and the orbit
widens in the opposite situation. During such a mass exchange, the
orbital separation passes through a minimum, if the masses become
equal in course of mass transfer.


\subsubsection{The Jeans (fast wind) mode}

In this mode the ejected matter completely escapes from the system,
that is, $\beta =1$. The escape of matter can take place either in a
spherically symmetric way or in the form of bipolar jets moving from
the system at high velocity. In both cases, matter carries away some
amount of the total orbital momentum proportional to the orbital
angular momentum of the mass losing star $J_2 = (M_1/M) J_\mathrm{orb}$ (we
neglect a possible proper rotation of the star,
see~\cite{vandenHeuvel83}). For the loss of orbital momentum
$\dot{J}_\mathrm{orb}$ it is reasonable to take
\begin{equation}
  \dot{J}_\mathrm{orb} = \frac{\dot{M}_2}{M_2} J_2.
  \label{A:jspecif}
\end{equation}
In the case $\beta =1$, Equation~(\ref{A:dotaa}) can be written as
\begin{equation}
  \frac{\dot{(\Omega a^2)}}{\Omega a^2} =
  \frac{\dot{J}_\mathrm{orb}}{J_\mathrm{orb}} -
  \frac{M_1\dot{M}_2}{M M_2}.
  \label{A:dotaa2} 
\end{equation}
Then Equation~(\ref{A:dotaa2}) in conjunction with Equation~(\ref{A:jspecif}) 
gives $\Omega a^2 = \mathrm{const}$, that is, 
$\sqrt{GaM}=\mathrm{const}$. Thus, as a result of such a mass loss,
the change in orbital separation is 
\begin{equation}
  \frac{a_\mathrm{f}}{a_\mathrm{i}} =
  \frac{M_\mathrm{i}}{M_\mathrm{f}}.
  \label{A:Jeans}
\end{equation}
Since the total mass decreases, the orbit always widens.


\subsubsection{Isotropic re-emission}
\label{sec:reemission}

The matter lost by star 2 can first accrete to star 1, and then, a
fraction $\beta$ of the accreted matter, can be expelled from the
system. This happens, for instance, when a massive star transfers
matter to a compact star on the thermal timescale ($<10^6$ years). The
accretion luminosity may exceed the Eddington luminosity limit, and
the radiation pressure pushes the infalling matter away from the
system, in a manner similar to the spectacular example of the SS~433
binary system. Another examples may be systems with helium stars
transferring mass onto relativistic objects~\cite{lyhnp05,
  ghosh_cyg3analog06} or hot white dwarf components of cataclysmic
variables losing mass by optically thick winds~\cite{kathac94}. The
same algorithm may be applied to the time-averaged mass loss from
novae~\cite{y73a}. In this mode of mass-transfer, the binary orbital
momentum carried away by the expelled matter is determined by the
orbital momentum of the accreting star $M_1$, rather than by the
orbital momentum of the mass-losing star $M_2$. The orbital momentum
loss can be written as
\begin{equation}
  \dot{J}_\mathrm{orb} = \beta \frac{\dot{M}_2}{M_1} J_1,
  \label{A:jspecif_1}
\end{equation}
where $J_1=(M_2/M)J_\mathrm{orb}$ is the orbital momentum of the star $M_1$.
In the limiting case when all the mass attracted
by $M_1$ is expelled from the system, $\beta=1$, Equation~(\ref{A:jspecif_1})
simplifies to 
\begin{equation}
  \frac{\dot{J}_\mathrm{orb}}{J_\mathrm{orb}} =
  \frac{\dot{{}M_2} M_2}{M_1 M}.
  \label{A:jreemiss}
\end{equation}
After substitution of this formula into Equation~(\ref{A:dotaa})
and integration over time, one arrives at
\begin{equation}
  \frac{a_\mathrm{f}}{a_\mathrm{i}} =
  \frac{M_\mathrm{i}}{M_\mathrm{f}}
  \left( \frac{M_{2\mathrm{i}}}{M_{2\mathrm{f}}} \right)^2
  \exp{\left( -2 \, \frac{M_{2\mathrm{i}}-M_{2\mathrm{f}}}{M_1} \right)}.
  \label{A:re-em}
\end{equation}
The exponential term makes this mode of the mass transfer
very sensitive to the components mass ratio.
If $M_1/M_2\ll 1$, the separation $a$ between the stars
may decrease so much that the approximation of point masses becomes
invalid. The tidal orbital instability
(Darwin instability) may set in, and the compact star
may start spiraling toward the companion star center
(the common envelope stage; see Section~\ref{A:CE} below).


\subsection{Supernova explosion}
\label{A:SN}

A supernova explosion in a massive binary system occurs on a timescale
much shorter than the orbital period, so the loss of mass is
practically instantaneous. This case can be treated analytically
(see, e.g., \cite{Blaauw61, Flannery_vdHeuvel75, mitalas76, Sutantyo78,
  hills83, Yamaoka_al93, bp95, kalogera96, tautak98}).

In general, the loss of matter and radiation is non-spherical, so that
the remnant of the supernova explosion (neutron star or black hole)
acquires some recoil velocity called kick velocity $\vec{w}$. In a
binary, the kick velocity should be added to the orbital velocity of the
pre-supernova star.

The usual treatment proceeds as follows. Let us consider a pre-SN
binary with initial masses $M_1$ and $M_2$. The stars move in a
circular orbit with orbital separation $a_\mathrm{i}$ and relative velocity
$\vec{V}_\mathrm{i}$. The star $M_1$ explodes leaving a compact remnant of mass
$M_\mathrm{c}$. The total mass of the binary decreases by the amount $\Delta M
= M_1 - M_\mathrm{c}$. It is usually assumed that the compact star acquires
some additional velocity (kick velocity) $\vec{w}$ (see detailed
discussion in Section~\ref{kicks}). Unless the binary is disrupted, it will end up in
a new orbit with eccentricity $e$, major semi axis $a_\mathrm{f}$, and
angle $\theta$ between the orbital planes before and after the
explosion. In general, the new barycentre will also receive some
velocity, but we neglect this motion. The goal is to evaluate the
parameters $a_\mathrm{f}$, $e$, and $\theta$.
 
It is convenient to work in an instantaneous 
reference frame centered on $M_2$ right at the time of explosion. 
The $x$-axis is the line from $M_2$ to $M_1$, the $y$-axis points in
the direction of $\vec{V}_\mathrm{i}$, and the $z$-axis is perpendicular to the 
orbital plane. In this frame, the pre-SN relative velocity is
$\vec{V}_\mathrm{i} = (0, V_\mathrm{i}, 0)$, where 
$V_\mathrm{i}=\sqrt{G(M_1+M_2)/a_\mathrm{i}}$ (see Equation~(\ref{B:Vorb})).
The initial total orbital momentum is $\vec{J}_\mathrm{i} = \mu_\mathrm{i} a_\mathrm{i} (0, 0, -V_\mathrm{i})$.
The explosion is considered to be instantaneous. 
Right after the explosion, the position vector of the exploded 
star $M_1$ has not changed: $\vec{r}= (a_\mathrm{i}, 0, 0)$. However,
other quantities have changed: $\vec{V}_\mathrm{f}= (w_x, V_\mathrm{i}+w_y, w_z)$
and $\vec{J}_\mathrm{f} = \mu_\mathrm{f} a_\mathrm{i} (0, w_z, -(V_\mathrm{i}+w_y))$, where 
$\vec{w} = (w_x, w_y, w_z)$ is the kick velocity and
$\mu_\mathrm{f}= M_\mathrm{c}M_2/(M_\mathrm{c}+M_2)$ is the reduced mass of the system after explosion.
The parameters $a_\mathrm{f}$ and $e$ are being found from equating the total
energy and the absolute value of orbital momentum of the initial
circular orbit to those of the resulting elliptical orbit
(see Equations~(\ref{B:E}, \ref{B:J}, \ref{B:Je})): 
\begin{eqnarray}
  \mu_\mathrm{f} \frac{V_\mathrm{f}^2}{2} -
  \frac{GM_\mathrm{c}M_2}{a_\mathrm{i}} &=&
  -\frac{GM_\mathrm{c}M_2}{2a_\mathrm{f}},
  \label{A:SN:E}
  \\
  \mu_\mathrm{f} a_\mathrm{i}
  \sqrt{w_z^2+(V_\mathrm{i}+w_y)^2} &=&
  \mu_\mathrm{f} \sqrt{G(M_\mathrm{c}+M_2)a_\mathrm{f}(1-e^2)}.
  \label{A:SN:J}
\end{eqnarray}%
For the resulting $a_\mathrm{f}$ and $e$ one finds 
\begin{equation}
  \frac{a_\mathrm{f}}{a_\mathrm{i}} =
  \left[ 2-\chi \left( \frac{w_x^2+w_z^2+(V_\mathrm{i}+w_y)^2}
  {V_\mathrm{i}^2} \right) \right]^{-1}
  \label{A:SN:afai}
\end{equation}
and
\begin{equation}
  1-e^2 = \chi \frac{a_\mathrm{i}}{a_\mathrm{f}}
  \left( \frac{w_z^2+(V_\mathrm{i}+w_y)^2}{V_\mathrm{i}^2} \right),
  \label{A:SN:ecc}
\end{equation}
where $\chi \equiv (M_1+M_2)/(M_\mathrm{c}+M_2) \ge 1$. The angle $\theta$ is 
defined by
\begin{displaymath}
  \cos \theta =
  \frac{\vec{J}_\mathrm{f} \cdot \vec{J}_\mathrm{i}}
  {|\vec{J}_\mathrm{f}|~|\vec{J}_\mathrm{i}|},
\end{displaymath}
which results in
\begin{equation}
  \cos\theta = \frac{V_\mathrm{i}+w_y}{\sqrt{w_z^2+(V_\mathrm{i}+w_y)^2}}.
  \label{A:SN:theta} 
\end{equation}

The condition for disruption of the binary system depends on the absolute
value $V_\mathrm{f}$ of the final velocity, and on the parameter $\chi$.
The binary disrupts if its total energy defined by the left-hand-side
of Equation~(\ref{A:SN:E}) becomes non-negative or, equivalently, if its eccentricity
defined by Equation~(\ref{A:SN:ecc}) becomes $e \ge 1$. From either of these
requirements one derives the condition for disruption:
\begin{equation}
  \frac{V_\mathrm{f}}{V_\mathrm{i}} \ge \sqrt{\frac{2}{\chi}}.
  \label{A:disrupt}
\end{equation}
The system remains bound if the opposite inequality is
satisfied. Equation~(\ref{A:disrupt}) can also be written in terms
of the escape (parabolic) velocity $V_\mathrm{e}$ defined by the
requirement
\begin{displaymath}
  \mu_\mathrm{f} \frac{V_\mathrm{e}^2}{2} -
  \frac{GM_\mathrm{c}M_2}{a_\mathrm{i}} = 0. 
\end{displaymath}
Since $\chi = M/(M - \Delta M)$ and $V_\mathrm{e}^2 = 2 G(M - \Delta M)/ a_\mathrm{i} =
2V_\mathrm{i}^2 / \chi$, one
can write Equation~(\ref{A:disrupt}) in the form
\begin{equation}
  V_\mathrm{f} \ge V_\mathrm{e}.
  \label{A:disrupt2}
\end{equation}
The condition of disruption simplifies 
in the case of a spherically symmetric SN explosion, that is, when there is
no kick velocity, $\vec{w}= 0$, and, therefore, $V_\mathrm{f} = V_\mathrm{i}$. In this case,
Equation~(\ref{A:disrupt}) reads $\chi \ge 2$, which is equivalent to
$\Delta M \ge M/2$. Thus, the system unbinds if more
than a half of the mass of the binary is lost. In other words, 
the resulting eccentricity 
\begin{equation}
  e = \frac{M_1-M_\mathrm{c}}{M_\mathrm{c}+M_2}
  \label{A:SN:symm-ecc}
\end{equation}
following from Equations~(\ref{A:SN:afai}, \ref{A:SN:ecc}), and $\vec{w} = 0$
becomes larger than 1, if $\Delta M > M/2$. 

So far, we have considered
an originally circular orbit. If the pre-SN star moves in an 
originally eccentric orbit, the condition of disruption of the system 
under symmetric explosion reads
\begin{displaymath}
  \Delta M = M_1-M_\mathrm{c} > \frac{1}{2} \frac{r}{a_\mathrm{i}},
\end{displaymath}
where $r$ is the distance between the components at the moment of explosion.


\subsection{Kick velocity of neutron stars}
\label{kicks}

The kick velocity imparted to a NS at birth is one of the 
major problems in the theory of stellar evolution. By itself it is an
additional parameter, the introduction of which has been motivated first of
all by high space velocities of radio pulsars inferred from the
measurements of their proper motions and distances. Pulsars were
recognized as a high-velocity Galactic population soon after their
discovery in
1968~\cite{Gunn_Ostriker70}. Shklovskii~\cite{Shklovskii70} put
forward the idea that high pulsar velocities may result from
asymmetric supernova explosions. Since then this hypothesis has been
tested by pulsar observations, but no definite conclusions on its
magnitude and direction have been obtained as yet.

Indeed, the distance to a pulsar is usually derived from the dispersion
measure evaluation and crucially depends on the assumed model of the electron
density distribution in the Galaxy. In the middle of the 1990s, Lyne and
Lorimer~\cite{Lyne_Lorimer94} derived a very high mean space velocity of
pulsars with known proper motion of about $450 \mathrm{\ km\ s}^{-1}$. This value was
difficult to adopt without invoking an additional natal kick velocity of
NSs. 
It was suggested~\cite{LPP96} that the observed 2D pulsar velocity distribution
found by Lyne and Lorimer~\cite{Lyne_Lorimer94} is reproduced if the
absolute value of the (assumed to be isotropic) NS kick has a power-law shape,
\begin{equation}
  f_\mathrm{LL}(|\vec{w}|) \propto
  \frac{(|\vec{w}|/w_0)^{0.19}} {(1+(|\vec{w}|/w_0)^{6.72})^{0.5}},
  \label{bin:LL}
\end{equation}
with $w_0 \approx 400 \mathrm{\ km\ s}^{-1}$. However, using this
formula or a Maxwellian
distribution for NS kicks in population synthesis
calculations~\cite{LPP96} give similar results, and later we shall not
distinguish these kicks.
The high mean space velocity of pulsars, consistent with earlier
results by Lyne and Lorimer, was confirmed by the analysis of a
larger sample of pulsars ~\cite{Hobbs_al05}. The recovered
distribution of 3D velocities is well fit by a Maxwellian distribution
with the mean value $w_0=400\pm 40 \mathrm{\ km\ s}^{-1}$ and a 1D rms $\sigma=265 \mathrm{\ km\ s}^{-1}$.

Possible physical reasons for natal NS kicks due to hydrodynamic
effects in core-collapse supernovae are summarized in~\cite{Lai_al01,
  Lai01}. Neutrino effects in the strong magnetic field of a young NS
may be also essential in explaining kicks up to $\sim 100 \mathrm{\ km\ s}^{-1}$~\cite{Chugai84, Dorofeev_al85, Kusenko04}.
Astrophysical arguments favouring a kick velocity are also summarized
in~\cite{Tauris_vdHeuvel00}. To get around the theoretical difficulty
of insufficient rotation of pre-supernova cores in single stars to
produce rapidly spinning young pulsars, Spruit and
Phinney~\cite{Spruit_Phinney98} proposed that random off-center kicks
can lead to a net spin-up of proto-NSs. In this model, correlations
between pulsar space velocity and rotation are possible and can be
tested in further observations.

Here we should note that the existence of some kick follows not only
from the measurements of radio pulsar space velocities, but also from
the analysis of binary systems with NSs. The impact of a kick
velocity $\sim 100 \mathrm{\ km\ s}^{-1}$ explains the precessing binary pulsar orbit
in PSR~J0045--7319~\cite{Kaspi_al96}. The evidence of the kick velocity
is seen in the inclined, with respect to the orbital plane,
circumstellar disk around the Be star SS~2883 -- an optical component
of binary PSR~B1259--63~\cite{Prokhorov_Postnov97}.

Long-term pulse profile changes interpreted as geodetic precession are
observed in the relativistic binary pulsars
PSR~1913+16~\cite{Weisberg_Taylor02}, PSR~B1534+12~\cite{Stairs_al04},
PSR~J1141--6545~\cite{Hotan_al05}, and
PSR~J0737--3039B~\cite{Burgay_al05}. These observations indicate that
in order to produce the misalignment between the orbital angular momentum
and the neutron star spin, a component of the kick velocity
perpendicular to the orbital plane is required~\cite{Wex_al00,
  Willems_Kalogera04, Willems_al04}. This idea seems to gain
observational support from recent thorough polarization
measurements~\cite{Johnston_al05} suggesting alignment of the
rotational axes with pulsar's space velocity. Such an alignment
acquired at birth may indicate the kick velocity directed preferably
along the rotation of the proto-NS. For the first SN explosion in a close
binary system this would imply the kick to be perpendicular to the orbital
plane.

It is worth noticing that the analysis of the formation of the double
relativistic pulsar PSR~J0737--3039~\cite{Podsiadlowski_al05} may
suggest, from the observed low eccentricity of the system $e\simeq
0.09$, that a small (if any) kick velocity may be acquired if the
formation of the second NS in the system is associated with the collapse
of an ONeMg WD due to electron-captures. The symmetric nature of
electron-capture supernovae was discussed in~\cite{Podsiadlowski_al04}
and seems to be an interesting issue requiring further studies (see,
e.g., \cite{Pfahl_al02a, Kuranov_Postnov06} for the analysis of the
formation of NSs in globular clusters in the frame of this
hypothesis). Note that electron-capture SNe are expected to be weak
events, irrespective of whether they are associated with the core-collapse
of a star which retained some original envelope or with the AIC of a
WD~\cite{ritossa_berro_one99, kitaura_one06, dessart_onecoll06}.

We also note the hypothesis of Pfahl et al.~\cite{prps02}, based
on observations of high-mass X-ray binaries with long orbital
periods ($\gtrsim 30 \mathrm{\ d}$) and low eccentricities ($e<0.2$), that
rapidly rotating precollapse cores may produce neutron stars with
relatively small kicks, and vice versa for slowly rotating
cores. Then, large kicks would be a feature of stars that retained
deep convective envelopes long enough to allow a strong magnetic
torque, generated by differential rotation between the core and the
envelope, to spin the core down to the very slow rotation rate of
the envelope. A low kick velocity imparted to the second (younger)
neutron star ($<50 \mathrm{\ km\ s}^{-1}$) was inferred from the analysis of
large-eccentricity binary pulsar
PSR~J1811--1736~\cite{Corongiu_al06}. The large orbital period of
this binary pulsar ($18.8 \mathrm{\ d}$) then may suggest an evolutionary
scenario with inefficient (if any) common envelope
stage~\cite{Dewi_Pols03}, i.e.\ the absence of deep convective shell
in the supernova progenitor (a He-star). This conclusion can be
regarded as supportive to ideas put forward in~\cite{prps02}.

In principle, it is possible to assume some kick velocity during BH
formation as well~\cite{Lipunov_al97, fbb98, py98,
  Postnov_Prokhorov00, nyp01, bel_kal02, yungelson_bh06}. The similarity
of NS and BH distribution in the Galaxy suggesting BH kicks was noted
in~\cite{jonker_nelemans04}. A recent analysis of the space velocity of some
BH binary systems~\cite{Willems_al05} put an upper limit on the BH
kick velocity of less than $\sim 200 \mathrm{\ km\ s}^{-1}$. However, no kick seems to
be required to explain the formation of other BH candidates (Cyg X-1,
X-Nova Sco, etc.)~\cite{Nelemans_al99}.

To summarize, the kick velocity remains one of the 
important unknown parameters of binary evolution with NSs and BHs.
Further constraining this parameter from various observations and
theoretical understanding of possible asymmetry of 
core-collapse supernovae seem to be of paramount importance for 
the formation and evolution of close compact binaries.


\subsubsection{Effect of the kick velocity on the evolution of a binary system}

The collapse of a star to a BH, or its explosion leading to the
formation of a NS, are normally considered as instantaneous. 
This assumption is well justified in binary systems, 
since typical orbital velocities before the explosion 
do not exceed a few hundred km/s, while most of the mass 
is expelled with velocities about several thousand km/s. 
The exploding star $M_1$ leaves
the remnant $M_\mathrm{c}$, and the binary loses a portion of its mass: 
$\Delta M = M_1 - M_\mathrm{c}$. The
relative velocity of stars before the event is 
\begin{equation}
  V_\mathrm{i} = \sqrt{G(M_1+M_2)/a_\mathrm{i}}.
\end{equation}
Right after the event, the relative velocity is
\begin{equation}
  \vec{V}_\mathrm{f} = \vec{V}_\mathrm{i}+\vec{w}.
  \label{V+W}
\end{equation}
Depending on the direction of the kick velocity vector $\vec{w}$, the
absolute value of $\vec{V}_\mathrm{f}$ varies in the interval from
the smallest $V_\mathrm{f} = |V_\mathrm{i} - w|$ to the largest
$V_\mathrm{f} = V_\mathrm{i} + w$. 
The system gets disrupted
if $V_\mathrm{f}$ satisfies the condition (see Section~\ref{A:SN})
\begin{equation}
  V_\mathrm{f} \ge V_\mathrm{i} \sqrt{\frac{2}{\chi}},
  \label{I:disrupt}
\end{equation}
where $\chi \equiv (M_1 + M_2)/(M_\mathrm{c}+M_2)$.

Let us start from the limiting case when the mass loss is 
practically zero ($\Delta M = 0$, $\chi =1$), while a non-zero kick 
velocity can still be present. This situation can be relevant to 
BH formation.
It follows from Equation~(\ref{I:disrupt}) that, for relatively small
kicks, $w< (\sqrt{2}-1)V_\mathrm{i}$, the system always (independently of the
direction of $\vec{w}$) remains bound, while for $w> (\sqrt{2}+1)V_\mathrm{i}$
the system always unbinds. By averaging over equally probable
orientations of $\vec{w}$ with a fixed amplitude $w$, one can show that
in the particular case $w= V_\mathrm{i}$ the system disrupts or survives with
equal probabilities. If $V_\mathrm{f} < V_\mathrm{i}$, the semimajor axis of the system
becomes smaller than the original binary separation, $a_\mathrm{f} < a_\mathrm{i}$ (see
Equation~(\ref{A:SN:afai})). This means that the system becomes more hard
than before, i.e.\ it has a greater negative total energy than the
original binary. If $V_\mathrm{i} <V_\mathrm{f}
<\sqrt{2}V_\mathrm{i}$, the system remains bound,
but $a_\mathrm{f} > a_\mathrm{i}$. For small and moderate kicks $w
\gtrsim V_\mathrm{i}$, the
probabilities for the system to become more or less bound are
approximately equal.

In general, the binary system loses some fraction of its mass
$\Delta M$. 
In the absence of the kick, the system remains bound if $\Delta M <
M/2$ and gets disrupted if $\Delta M \ge M/2$ (see
Section~\ref{A:SN}). Clearly, a properly oriented kick velocity
(directed against the vector $\vec{V}_\mathrm{i}$) can keep the system bound,
even if it would have been disrupted without the kick. And, on the
other hand, an unfortunate direction of $\vec{w}$ can disrupt the
system, which otherwise would stay bound.

Consider, first, the case $\Delta M < M/2$. The parameter $\chi$ varies
in the interval from 1 to 2, and the escape velocity $V_\mathrm{e}$ varies in
the interval from $\sqrt{2} V_\mathrm{i}$ to $V_\mathrm{i}$. 
It follows from Equation~(\ref{A:disrupt2}) that the binary always remains
bound if $w< V_\mathrm{e} - V_\mathrm{i}$, and always unbinds if $w>
V_\mathrm{e} + V_\mathrm{i}$.
This is a generalization of the formulae 
derived above for the limiting case $\Delta M = 0$. Obviously, 
for a given $w$, the probability for the system to disrupt or become 
softer increases when $\Delta M$ becomes larger. Now turn 
to the case $\Delta M > M/2$. The escape velocity of the compact 
star becomes $V_\mathrm{e}<V_\mathrm{i}$. The binary is always disrupted if the
kick velocity is too large or too small: $w > V_\mathrm{i} + V_\mathrm{e}$ or
$w < V_\mathrm{i} - V_\mathrm{e}$. However, for all intermediate values of $w$,
the system can remain bound, and sometimes even more bound than before,
if the direction of $\vec{w}$ happened to be approximately
opposite to $\vec{V}_\mathrm{i}$. A detailed calculation of the probabilities
for the binary survival or disruption requires integration
over the kick velocity distribution function $f (\vec{w})$ 
(see, e.g., \cite{Brandt_Podsiadlowski95}).


\subsection{Common envelope stage}
\label{A:CE}

This is a very important stage in the evolution of binaries of all
masses. In different contexts, the evolution of binary components in an
envelope engulfing both of them was considered, for instance,
in~\cite{rrw74, sparks-74, alexander-76}, but the importance of common
envelopes was really recognized after Paczy\'{n}ski~\cite{pac76} and
Ostriker~\cite{Ostriker75} applied them to explain the
formation of cataclysmic variables and massive X-ray binaries,
respectively. A detailed review of problems related to common
envelopes may be found, e.g., in~\cite{il93, taam_ce00}.

Generally, common envelopes form in binary systems where the mass
transfer from the mass-losing star is high, and the companion cannot
accommodate all the accreting matter. The common envelope stage
appears unavoidable on observational grounds. The evidence for a
dramatic orbital angular momentum decrease in some preceding
evolutionary stage follows from observations of certain types of
close binary stars. They include cataclysmic variables, in which a
white dwarf accretes matter from a small red dwarf main-sequence
companion, planetary nebulae with double cores, low-mass X-ray
binaries and X-ray transients (neutron stars and black holes accreting
matter from low-mass main-sequence dwarfs). The radii of progenitors
of compact stars in these binaries typically should have been
$100\mbox{\,--\,}1000$ solar radii, that is, much larger than the currently
observed binary separations. This testifies of some dramatic reduction
of the orbital momentum in the earlier stages of evolution and
eventual removal of the common envelope. Additional indirect evidence
for reality of the common envelope stage in the typical
pre-cataclysmic binary V471 Tau has recently been obtained from X-ray
Chandra observations~\cite{Drake_Sarna03} showing anomalous C/N
contamination of the K-dwarf companion. Recent studies also indicate
that many planetary nebulae are actually binaries, which may suggest
that most of them result from common envelope interaction~\cite{ytl93,
  deMarco_Moe05}.

Exact criteria for the formation of a common envelope are
absent. However, a high rate of mass overflow onto a compact star from
a normal star is always expected when the normal star goes off the
main sequence and develops a deep convective envelope. The physical
reason for this is that convection tends to make entropy constant
along the radius, so the radial structure of convective stellar
envelopes is well described by a polytrope (i.e.\ the equation of
state can be written as $P=K\rho^{1+1/n}$) with an index $n=3/2$. The
polytropic approximation with $n=3/2$ is also valid for degenerate
white dwarfs with masses not too close to the Chandrasekhar limit. For
a star in hydrostatic equilibrium, this results in the well known
inverse mass-radius relation, $R\propto M^{-1/3}$, first measured for
white dwarfs. Removing mass from a star with a negative power of the
mass-radius relation increases its radius. On the other hand, the
Roche lobe of the more massive star should shrink in response to the
conservative mass exchange between the components. This further
increases the mass loss rate from the Roche-lobe filling star leading to
a continuation of an unstable mass loss and eventual formation of a
common envelope.
The critical mass ratio for the unstable Roche lobe overflow depends
on specifics of the stellar structure and mass ratio of components;
typically, mass loss is unstable for stars with convective envelopes,
stars with radiative envelopes if $q \gtrsim2$, and white dwarfs if $q
\gtrsim 2/3$.

As other examples for the formation of a common envelope one may
consider, for instance, direct penetration of a compact star into the
dense outer layers of the companion, which can happen as a result of
the Darwin tidal orbital instability in binaries~\cite{Counselman73,
  Bagot96}; it is possible that a compact remnant of a supernova
explosion with appropriately directed kick velocity finds itself in an
elliptic orbit whose minimum periastron distance $a_\mathrm{f}(1-e)$ is smaller
than the stellar radius of the companion; a common envelope
enshrouding both components of a binary may form due to unstable
thermonuclear burning in the surface layers of an accreting WD.

The common envelope stage is, usually, treated in the following simplified
way~\cite{Webbink84, dek90}. The orbital evolution of the compact star
$m$ inside the envelope of the normal star $M_1$ is driven by the
dynamical friction drag. This leads to a gradual spiral-in process of
the compact star. The released orbital energy $\Delta E_\mathrm{orb}$, or a
fraction of it, can become numerically equal to the binding energy
$E_\mathrm{bind}$ of the envelope with the rest of the binary system. It is
generally assumed that the orbital energy of the binary is used to
expel the envelope of the donor with an efficiency
$\alpha_\mathrm{ce}$: 
\begin{displaymath}
  E_\mathrm{bind} = \alpha_\mathrm{ce} \Delta E_\mathrm{orb},
\end{displaymath}
where $E_\mathrm{bind}$ is the total binding energy of the envelope
and $\Delta E_\mathrm{orb}$ is the orbital energy released in the
spiral-in. What remains of the normal star $M_1$ is its stellar core $M_\mathrm{c}$. 
The above energy condition reads
\begin{equation}
  \frac{GM_1(M_1-M_{c})}{\lambda R_\mathrm{L}} =
  \alpha_\mathrm{ce}
  \left( \frac{GmM_{c}}{2a_\mathrm{f}}-\frac{GM_1m}{2a_\mathrm{i}} \right),
  \label{A:CE:eq}
\end{equation}
where $a_\mathrm{i}$ and $a_\mathrm{f}$ are the initial and the final orbital
separations, and $\lambda$ is a numerical coefficient that depends on the
structure of the donor's envelope. $R_\mathrm{L}$ is the Roche lobe radius of
the normal star that can be approximated as~\cite{Eggleton83} 
\begin{equation}
  \frac{R_\mathrm{L}}{a_\mathrm{i}} =
  \frac{0.49}{0.6+q^{2/3} \ln (1+q^{1/3})}
  \label{A:Roche}
\end{equation}
and $q\equiv M_1/m$. From Equation~(\ref{A:CE:eq}) one derives 
\begin{equation}
  \frac{a_\mathrm{f}}{a_\mathrm{i}} =
  \frac{M_{c}}{M_1} \left( 1+\frac{2a_\mathrm{i}}
  {\lambda\alpha_\mathrm{ce}R_\mathrm{L}}
  \frac{M_1-M_{c}}{m} \right)^{-1} \lesssim
  \frac{M_{c}}{M_1} \frac{M_\mathrm{c}}{\Delta M},
  \label{A:CE:afai}
\end{equation}
where $\Delta M=M_1-M_{c}$ is the mass of the ejected envelope.
The mass $M_{c}$ of a helium core of a massive star may be
approximated as~\cite{ty73}
\begin{equation}
  M_\mathrm{He} \approx 0.1 (M_1/M_\odot)^{1.4},
  \label{A:MHe}
\end{equation}
so the orbital separation during the common envelope stage may 
decrease as much as by factor 100 and more. 

The above formalism for the common envelope stage depends in fact on the
product of two parameters: $\lambda$, which is the measure of the binding
energy of the envelope to the core prior to mass transfer in a binary
system, and $\alpha_\mathrm{ce}$, which is the common envelope efficiency itself.
Numerical calculations of evolved giant stars with masses
$3\mbox{\,--\,}10\,M_\odot$~\cite{Dewi_Tauris00} showed that the value
of the $\lambda$-parameter is typically between 0.2 and 0.8; however,
it can be as high as 5 on the asymptotic giant branch. For more
massive primaries ($>20\,M_\odot$), which are appropriate for the
formation of BH binaries, the $\lambda$-parameter was found
to depend on the mass of the star and vary within a wide range
0.01\,--\,0.5~\cite{Podsiadlowski_al03}. Some hydrodynamical
simulations~\cite{Rasio_Livio96} indicated that $\alpha_\mathrm{ce}\simeq 1$,
while in others~\cite{sandq_binhedw00} a wider range for values of
$\alpha_\mathrm{ce}$ was obtained. There are debates in the literature as to
should additional sources of energy (e.g., ionization energy in the
envelope~\cite{Han_al02}) should be included in the ejection criterion of
common envelopes~\cite{Soker_Harpaz03}.

There is another approach, different from the standard Webbink
formalism, which is used to estimate the common envelope efficiency
$\alpha_\mathrm{ce}$. In the case of systems with at least one
white-dwarf component one can try to reconstruct the evolution of
double compact binaries with known masses of both components, since
there is a unique relation between the mass of a white dwarf and the
radius of its red giant progenitor. Close binary white dwarfs should
definitely result from the spiral-in phase in the common envelope that
appears inevitable during the second mass transfer (i.e.\ from the red
giant to the white dwarf remnant of the original primary in the
system). Such an analysis~\cite{Nelemans_al00}, extended
in~\cite{nelemans_tout05, sluys_wd06}, suggests that the standard
energy prescription for the treatment of the common envelope stage
cannot be applied to the first mass transfer episode. Instead, the
authors proposed to apply the so-called $\gamma$-formalism for the
common envelope, in which not the energy but the angular momentum $J$
is balanced and conservation of energy is implicitly implied:
\begin{equation}
  \frac{\delta J}{J} = \gamma \frac{\Delta M}{M_\mathrm{tot}},
  \label{ce_gamma}
\end{equation}
where $\Delta M$ is the mass of the common envelope, $M_\mathrm{tot}$ is the total
mass of the binary system before the common envelope, and $\gamma$ is a
numerical coefficient. For all binary systems considered, the parameter 
$\gamma$ was found to be within the range 1.1\,--\,4, with the mean value around
1.5. (Notice that $\gamma=1$ corresponds to loss of the angular momentum
through stellar wind, as considered above, which always increases the 
orbital separation of the binary.) The applicability of this algorithm should 
be investigated further. 

Note also, that formulations of the common envelope equation different
from Equation~(\ref{A:CE:eq}) are met in the literature (see, e.g., \cite{ty79,
  il93}); $a_\mathrm{f}/a_\mathrm{i}$ similar to the values produced by
Equation~(\ref{A:CE:eq}) are then obtained for different $\alpha_{\rm
  ce}\lambda$ values.

\newpage


\section{Evolutionary Scenario for Compact Binaries with Neutron Star
  or Black Hole Components}
\label{section:evolution_highmass}


\subsection{Compact binaries with NSs}

Compact binaries with NS and BH components are descendants of
initially massive binaries with $M_1 \gtrsim (8\mbox{\,--\,}10)
M_\odot$. The evolutionary scenario of massive binaries was elaborated
shortly after the discovery of binary X-ray sources~\cite{ty73,
  Tutukov_Yungelson73b, vdHeuvel_deLoore73} and is depicted in
Figure~\ref{figure:massive_flow}.

\epubtkImage{figure04.png}{
  \begin{figure}[htbp]
    \centerline{\includegraphics[scale=0.7]{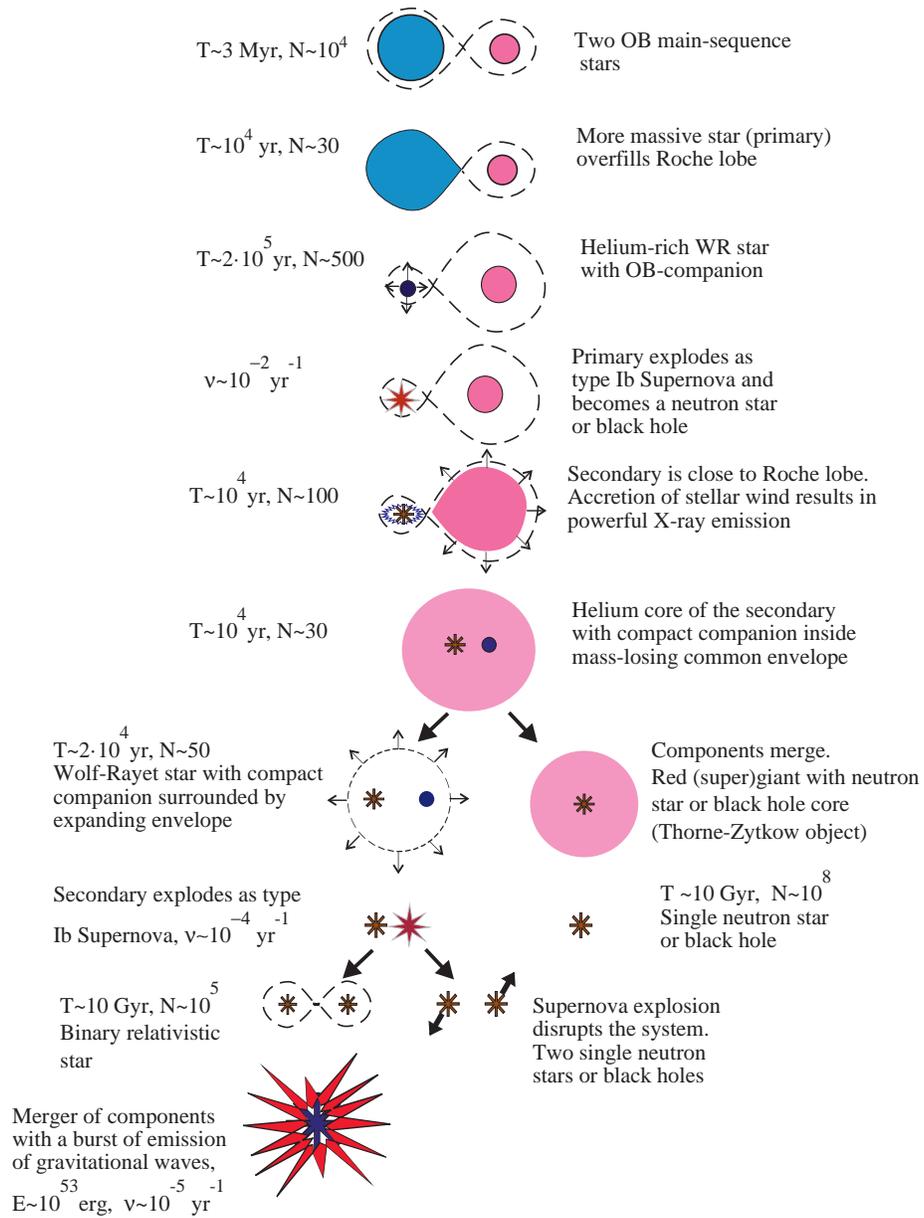}}
    \caption{\it Evolutionary scenario for the formation of neutron
      stars or black holes in close binaries.}
    \label{figure:massive_flow}
  \end{figure}
}

This scenario is fully confirmed by more than 30-years' history of
astronomical observations and is now considered as standard. A massive
X-ray binary is an inevitable stage preceding the formation of a
double compact system after the second supernova explosion of the
helium-rich companion in such a stellar system. The formation scenario
for binary pulsars proposed immediately after the discovery of
PSR~1913+16~\cite{Flannery_vdHeuvel75, mty76} has also been tested by
subsequent observations of binary pulsars. In fact, the scenario for
binary pulsars was proposed even earlier in~\cite{ty73}, but because
no binary pulsars were known at that time, it was suggested that all
pairs of NS are disrupted at the second NS formation.

It is convenient to separate the evolution of a massive binary into
several stages according to the physical state of the binary
components, including phases of mass exchange between them. The
simplest evolutionary scenario can be schematically described as
follows (see Figure~\ref{figure:massive_flow}).
\begin{enumerate}
\item Initially, two high-mass OB main-sequence stars are separated
  and are inside their Roche lobes. Tidal interaction is very effective
  so that a possible initial eccentricity vanishes before the primary
  star $M_1$ fills its Roche lobe. The duration of this stage is
  determined by the hydrogen burning time of the primary and typically
  is several million years (for massive main-sequence stars, the time
  of core hydrogen burning is $t_\mathrm{nucl}\propto M^{-2}$). The star burns
  out hydrogen in its central parts, so that a dense central helium
  core with a mass $M_\mathrm{He}\simeq 0.1 (M/M_\odot)^{1.4}$ forms by the
  time when the star leaves the main sequence. The expected number of
  such binaries in the Galaxy is around $10^4$.
\item After core hydrogen exhaustion, the primary leaves the main
  sequence and starts to expand rapidly. When its radius approaches
  the Roche lobe (see Equation~(\ref{A:Roche})), mass transfer onto the
  secondary, less massive star which still resides on the main
  sequence begins. The mass-transfer rate can be crudely estimated as
  $\dot{M}\sim M_1/\tau_\mathrm{KH}$, where $\tau_\mathrm{KH}=GM_1^2/R_1L_1$ is the
  primary's thermal time scale.
  
  The mass transfer ends when most of the primary's hydrogen envelope is
  transferred onto the secondary, so a naked helium core is left. This
  core can be observed as a Wolf--Rayet (WR) star with intense stellar
  wind if its mass exceeds $(7\mbox{\,--\,}8)\,M_\odot$~\cite{nl00,
    fadeyev03a, fadeyev04}.
  
  While the mass of the primary star reduces, the mass of the secondary
  star increases, since the mass transfer at this stage is thought to be
  quasi-conservative. For not too massive main-sequence stars, $M \lesssim
  40\,M_\odot$, no significant stellar wind mass loss occurs which could,
  otherwise, remove too much matter from the binary, thereby increasing
  binary separation. The secondary star acquires large angular
  momentum due to the infalling material, so that its outer envelope can
  be spun up to an angular velocity close to the limiting (Kepler orbit)
  value. Such massive rapidly rotating stars are observed as
  Be-stars. During the conservative stage of mass transfer, the
  semimajor axis of the orbit first decreases, reaches a minimum when
  the masses of the binary components become equal to each other, and
  then increases. This behavior is dictated by the angular momentum
  conservation law~(\ref{A:conserv}). After the completion of the
  conservative mass transfer, the initially more massive star becomes
  less massive than its initially lighter companion.
  
  For the typical parameters the duration of the first RLOF is rather
  short, of the order of $10^4 \mathrm{\ yr}$, so only several dozens of
  such binaries are expected to be in the Galaxy.
  
\item The duration of the WR stage is about several $10^5 \mathrm{\ yr}$, so
  the Galactic number of such binaries should be several hundreds.
\item At the end of the thermonuclear evolution, the WR star explodes as
  Ib (or Ic) supernova to become a NS or BH. The inferred Galactic
  type~Ib SN rate is around $10^{-2}$ per year, at least half of them
  should be in binaries. At this stage the disruption of the binary is
  possible (e.g., if the mass lost during the symmetric SN explosion
  exceeds 50\% of the total mass of the pre-SN binary, or even smaller
  in the presence of the kick velocity; see Section~\ref{A:SN}
  above). Some runaway Galactic OB-stars must have been formed in this
  way.
\item If the system survives the first SN explosion, a rapidly
  rotating Be star in pair with a young NS appears. Orbital evolution
  following the SN explosion is described above by
  Equations~(\ref{A:SN:afai}--\ref{A:SN:symm-ecc}). The orbital
  eccentricity after the SN explosion is high, so enhanced accretion onto
  the NS occurs at the periastron passages. Most of about 100 Galactic
  Be/X-ray binaries~\cite{Raguzova_Popov05} are formed in this way.
  The duration of this stage depends on the binary parameters, but in
  all cases it is limited by the time left for the (now more massive)
  secondary to burn hydrogen in its core.
  
  An important parameter of NS evolution is the surface magnetic field
  strength. In binary systems, magnetic field, in combination with NS
  spin period and accretion rate onto the NS surface, determines the
  observational manifestation of the neutron star (see~\cite{Lipunov92}
  for more detail). Accretion of matter onto the NS can reduce the surface
  magnetic field and spin-up the NS rotation (pulsar
  recycling)~\cite{BK_Komberg74, Romani90, Romani95, B-K06}.
\item The secondary expands to engulf the NS. The common envelope
  stage begins and after $\sim 10^3 \mathrm{\ yr}$ ends up with the
  formation of a WR star with a compact companion surrounded by an
  expanding envelope (Cyg X-3 may serve as an example), or the NS
  merges with the helium core during the common envelope to form a
  still hypothetic Thorne--\.{Z}ytkow (TZ) object. The fate of TZ stars
  remains unclear (see~\cite{Barkov_al01} for the recent
  study). Single (possibly, massive) NS or BH should descend from
  them.

  A note should be made concerning the phase when a common envelope
  engulfs the first-formed NS and the core of the
  secondary. Colgate~\cite{colgate71} and Zel'dovich
  et~al.~\cite{zeld72} have shown that hyper-Eddington accretion onto
  a neutron star is possible if the gravitational energy released in
  accretion is lost by neutrinos. Chevalier~\cite{chevalier93}
  suggested that this may be the case for the accretion in common
  envelopes. Since the accretion rates in this case may be as high as
  $\sim 0.1 \, M_\odot \mathrm{\ yr}^{-1}$, the NS may collapse
  into a BH inside
  the common envelope. An essential caveat is that the accretion in
  the hyper-Eddington regime may be prevented by the angular momentum of
  the captured matter. The magnetic field of the NS may also be
  a complication. The possibility of hyper-critical accretion still
  has to be studied. Nevertheless, implications of this hypothesis for
  different types of relativistic binaries were explored in great
  detail by H.~Bethe and G.~Brown and their coauthors (see,
  e.g., \cite{brown_bh00} and references therein). Also, the possibility
  of hyper-Eddington accretion was included in several population
  synthesis studies with evident result of diminishing the population
  of NS\,+\,NS binaries in favour of neutron stars in pairs with low-mass
  black holes (see, e.g., \cite{py98, bel_kal02}).
\item The secondary WR ultimately explodes as a type~Ib supernova
  leaving behind a double NS binary, or the system is disrupted to
  form two single high-velocity NSs or BHs. Even for a symmetric SN
  explosion the disruption of binaries after the second SN explosion could
  result in the observed high average velocities of radiopulsars (see
  Section~\ref{kicks} above). In the surviving close binary NS system,
  the older NS is expected to have faster rotation velocity (and
  possibly higher mass) than the younger one because of the recycling
  at the preceding accretion stage. The subsequent orbital evolution
  of such double NS systems is entirely due to GW emission (see
  Section~\ref{A:GW_evol}) and ultimately leads to the coalescence of
  the components.
\end{enumerate}
Detailed studies of possible evolutionary channels which produce 
merging binary NS can be found in the literature
(see, e.g., \cite{Tutukov_Yungelson93a, ty93b, 
Lipunov_al97,
py98, Bagot97, Wettig_Brown96,
bel_kal02,
Ivanova_al03, Dewi_vdHeuvel04, Willems_Kalogera04}).

We emphasize that this scenario applies only to initially massive
binaries. There exists also a population of NSs accompanied by low-mass
$[\sim (1\mbox{\,--\,}2)\,M_\odot]$ companions. A scenario similar to the one
presented in Figure~\ref{figure:massive_flow} may be sketched for them
too, with the difference that the secondary component stably transfers
mass onto the companion (see, e.g., \cite{ity95b, kw96, kw98,
  tf_bh02}). This scenario is similar to the one for low- and
intermediate-mass binaries considered in
Section~\ref{section:wd_formation}, with the WD replaced by a NS or a
BH. Compact low-mass binaries with NSs may be dynamically formed in
dense stellar environments, for example in globular clusters. The
dynamical evolution of binaries in globular clusters is beyond the
scope of this review; see~\cite{Benacquista_LRR02} and~\cite{B-K06}
for more detail and further references.


\subsection{Black hole formation parameters}

So far, we have considered the formation of NSs and binaries with
NSs. It is believed that very massive stars end up their evolution with
the formation of stellar mass black holes. We will discuss now their
formation.

In the analysis of BH formation, new important parameters appear. The first
one is the threshold mass $M_\mathrm{cr}$ beginning from which a main-sequence star,
after the completion of its nuclear evolution, can collapse into a BH. This
mass is not well known; different authors suggest different values: van den
Heuvel and Habets~\cite{Heuvel_Habets84} -- $40\,M_\odot$; Woosley
et~al.~\cite{Woosley_al95} -- $60\,M_\odot$; Portegies Zwart, Verbunt,
and Ergma~\cite{PortegiesZwart_al97} -- more than $20\,M_\odot$. A simple physical
argument usually put forward in the literature is that the mantle of the
main-sequence star with $M>M_\mathrm{cr}\approx 30\,M_\odot$ before the collapse has
a binding energy well above $10^{51} \mathrm{\ erg}$ (the typical supernova
energy observed), so that the supernova shock is not strong enough to
expel the mantle~\cite{Fryer99, Fryer04}.

The upper mass limit for BH formation (with the caveat that the
role of magnetic-field effects is not considered) is, predominantly,
a function of stellar-wind mass loss in the core-hydrogen,
hydrogen-shell, and core-helium burning stages. For a specific
combination of winds in different evolutionary stages and
assumptions on metallicity it is possible to find the types of
stellar remnants as a function of initial mass (see, for
instance~\cite{heger_death03}). Since stellar winds are mass (or
luminosity) and metallicity-dependent, a peculiar consequence of
mass-loss implementation in the latter study is that for $Z \simeq
Z_\odot$ the mass-range of precursors of black holes is constrained
to $M\approx (25\mbox{\,--\,}60)\,M_\odot$, while more massive stars
form NSs because of heavy mass loss. The recent discovery of the possible
magnetar in the young stellar cluster
Westerlund~1~\cite{muno_wester_magn06} hints to the reality of such
a scenario. Note, however, that the estimates of $\dot{M}$ are rather
uncertain, especially for the most massive stars, mainly because of
clumping in the winds (see, e.g., \cite{kudr_urban_winds06,
  crowther_araa06, hamann_wn06}). Current reassessment of the role of
clumping generally results in the reduction of previous mass-loss
estimates. Other factors that have to be taken into account in the
estimates of the masses of progenitors of BHs are rotation and
magnetic fields.

The second parameter is the mass $M_\mathrm{BH}$ of the nascent BH. There are
various studies as for what the mass of the BH should be
(see, e.g., \cite{Timmes_al96, Bethe_Brown98, Fryer99,
  Fryer_Kalogera01}). In some papers a typical BH mass was found to be
not much higher than the upper limit for the NS mass
(Oppenheimer--Volkoff limit $\sim(1.6\mbox{\,--\,}2.5)\,M_\odot$, depending on
the unknown equation of state for NS matter) even if the
fall-back accretion onto the supernova remnant is
allowed~\cite{Timmes_al96}. Modern measurements of black hole
masses in binaries suggest a broad range of BH masses of the order
of $4\mbox{\,--\,}17\,M_\odot$~\cite{orosz_bh03, McClintock_Remillard03,
  Remillard_McClintock06}. 
A continuous range of BH masses up to $10\mbox{\,--\,}15\,M_\odot$ was derived in
calculations~\cite{Fryer_Kalogera01}. Since present day
calculations are still unable to reproduce self-consistently even the
supernova explosion, in the further discussion we have parameterized
the BH mass $M_\mathrm{BH}$ by the fraction of the pre-supernova mass $M_*$
that collapses into the BH: $k_\mathrm{BH}=M_\mathrm{BH}/M_*$. In fact, the
pre-supernova mass $M_*$ is directly related to $M_\mathrm{cr}$, but the form
of this relationship is somewhat different in different scenarios for
massive star evolution, mainly because of different mass-loss
prescriptions. According to our parameterization, the minimal BH mass
can be $M_\mathrm{BH}^\mathrm{min}=k_\mathrm{BH}M_*$, where $M_*$ itself depends on
$M_\mathrm{cr}$. The parameter $k_\mathrm{BH}$ can vary in a wide range.

The third parameter, similar to the case of NS formation, is the
possible kick velocity ${\bf w}_\mathrm{BH}$ imparted to the newly formed BH
(see the end of Section~\ref{kicks}). In general, one expects
that the BH should acquire a smaller kick velocity than a NS, as black
holes are more massive than neutron stars. A possible relation (as
adopted, e.g., in calculations~\cite{Lipunov_al97}) reads
\begin{equation}
  \frac{w_\mathrm{BH}}{w_\mathrm{NS}} =
  \frac{M_*-M_\mathrm{BH}}{M_*-M_\mathrm{OV}} =
  \frac{1-k_\mathrm{BH}}{1-M_\mathrm{OV}/M_*},
  \label{BH_kick}
\end{equation}
where $M_\mathrm{OV}=2.5 \,M_\odot$ is the maximum NS mass.
When $M_\mathrm{BH}$ is close to $M_\mathrm{OV}$, the ratio $w_\mathrm{BH}/w_\mathrm{NS}$ 
approaches 1, and
the low-mass black holes acquire kick velocities similar to those of 
neutron stars. When $M_\mathrm{BH}$ is significantly larger than $M_\mathrm{OV}$,
the parameter $k_\mathrm{BH}=1$, and the BH kick velocity becomes vanishingly
small.
The allowance for a quite moderate $w_\mathrm{BH}$ can 
increases the coalescence rate of binary BH~\cite{Lipunov_al97}.

The possible kick velocity imparted to newly born black holes
makes the orbits of survived systems highly eccentric. It is
important to stress that some fraction of such binary BH can retain
their large eccentricities up to the late stages of their
coalescence. This signature should be reflected in their emitted
waveforms and should be modeled in templates.

Asymmetric explosions accompanied by a kick change the space
orientation of the orbital angular momentum. On the other hand, the
star's spin axis remains fixed (unless the kick was off-center). As
a result, some distribution of the angles between the BH spins and
the orbital angular momentum (denoted by $J$) will be
established~\cite{Postnov_Prokhorov00}. It is interesting that even
for small
kicks of a few tens of km/s an appreciable fraction (30\,--\,50\%) of
the merging binary BH can have $\cos J<0$. This means that in these
binaries the orbital angular momentum vector is oriented almost
oppositely to the black hole spins. This is one more signature of
imparted kicks that can be tested observationally. These effects are
also discussed in~\cite{Kalogera00}.

\newpage


\section{Formation of Double Compact Binaries}
\label{section:binary_NS}


\subsection{Analytical estimates}

A rough estimate of the formation rate of double compact binaries can
be obtained ignoring many details of binary evolution. To do this, we
shall use the observed initial distribution of binary orbital
parameters and assume the simplest conservative mass transfer
($M_1 + M_2= \mathrm{const}$) without kick velocity imparted to the
nascent compact stellar remnants during SN explosions.

\noindent
{\bf Initial binary distributions.~~}
From observations of spectroscopic binaries it is possible to derive
the formation rate of binary stars with initial masses $M_1$, $M_2$
(with mass ratio $q=M_2/M_1\le 1$), orbital semimajor axis $A$, and
eccentricity $e$. According to~\cite{Popova_al82}, the present
birth rate of binaries in our Galaxy can be written in factorized
form as 
\begin{equation}
  \frac{dN}{dA\,dM_1\,dq\,dt} \approx
  0.087 \left( \frac{A}{R_\odot} \right)^{-1}
  \left( \frac{M_1}{M_\odot} \right)^{-2.5} f(q),
  \label{E:bin_brate}
\end{equation}
where $f(q)$ is a poorly constrained distribution over the initial mass
ratio of binary components.

One usually assumes a mass ratio distribution law in the form 
$f(q)\sim q^{-\alpha_q}$ where $\alpha_q$ is a
free parameter; another often used form of the $q$-distribution was
suggested by Kuiper~\cite{Kuiper35}:
\begin{displaymath}
  f(q) = 2/(1+q)^2.
\end{displaymath} 
The range of $A$ is $10 \leq A/R_\odot \leq 10^6$. In deriving the above
Equation~(\ref{E:bin_brate}), the authors of~\cite{Popova_al82} took
into account selection effects to convert the ``observed'' distribution of
stars into the true one. An almost flat logarithmic distribution of
semimajor axes was also found in~\cite{Abt83}. Integration of
Equation~(\ref{E:bin_brate}) yields one binary system with $M_1\gtrsim
0.8M_\odot$ and $10\,R_\odot < A < 10^6\,R_\odot$ per year in the
Galaxy, which is in reasonable agreement with the Galactic star
formation rate estimated by various methods; the present-day star
formation rate is about several $M_\odot$ per year (see, for
example,~\cite{Miller_Scalo79, Timmes_al97}).

\noindent
{\bf Constraints from conservative evolution.~~}
To form a NS at the end of thermonuclear evolution, the primary mass
should be at least $10\,M_\odot$. Equation~(\ref{E:bin_brate}) says
that the formation rate of such binaries is about 1 per 50 years. We
shall restrict ourselves by considering only close binaries, in which
mass transfer onto the secondary is possible. This narrows the binary
separation interval to $10\mbox{\,--\,}1000\,R_\odot$ (see
Figure~\ref{f:remn}); the birth rate of close massive
($M_1>10\,M_\odot$) binaries is thus $1/50 \times 2/5 \mathrm{\ yr}^{-1}= 1/125 \mathrm{\
  yr}^{-1}$. The mass ratio $q$ should not be very small to make the
formation of the second NS possible. The lower limit for $q$ is
derived from the condition that after the first mass transfer stage,
during which the mass of the secondary increases, $M_2+\Delta M\ge
10\,M_\odot$. Here $\Delta M=M_1-M_\mathrm{He}$ and the mass of the helium
core left after the first mass transfer is $M_\mathrm{He}\approx 0.1
(M_1/M_\odot)^{1.4}$. This yields
\begin{displaymath}
  m_2+(m_1-0.1m_1^{1.4}) > 10,
\end{displaymath}
where we used the notation $m=M/M_\odot$, or in terms of $q$:
\begin{equation}
  q \ge 10/m_1+0.1 m_1^{0.4}-1.
  \label{E:ll_q}
\end{equation}

An upper limit for the mass ratio is obtained from the requirement
that the binary system remains bound after the sudden mass loss in the
second supernova explosion\epubtkFootnote{For the first supernova
  explosion without kick this is always satisfied.}. From
Equation~(\ref{A:SN:symm-ecc}) we obtain
\begin{displaymath}
  \frac{0.1[m_2+(m_1-0.1m_1^{1.4})]^{1.4}-1.4}{2.8} < 1,
\end{displaymath}
or in terms of $q$:
\begin{equation}
  q \le 14.4/m_1+0.1m_1^{0.4}-1.
  \label{E:ul_q}
\end{equation}

Inserting $m_1=10$ in the above two equations yields the appropriate
mass ratio range $0.25< q< 0.69$, i.e.\ 20\% of the binaries for Kuiper's
mass ratio distribution. So we conclude that the birth rate of binaries
which potentially can produce double NS system is $\lesssim 0.2 \times
1/125 \mathrm{\ yr}^{-1} \simeq 1/600 \mathrm{\ yr}^{-1}$.

Of course, this is a very crude upper limit -- we have not taken into
account the evolution of the binary separation, ignored initial binary
eccentricities, non-conservative mass loss, etc. However, it is not
easy to treat all these factors without additional knowledge of
numerous details and parameters of binary evolution (such as the
physical state of the star at the moment of the Roche lobe overflow,
the common envelope efficiency, etc.). All these factors should
decrease the formation rate of double NS. The coalescence rate of
compact binaries (which is ultimately of interest for us) will be even
smaller -- for the compact binary to merge within the Hubble time, the
binary separation after the second supernova explosion should be less
than $\sim 100\,R_\odot$ (orbital periods shorter than $\sim 40
\mathrm{\ d}$) 
for arbitrary high orbital eccentricity $e$ (see Figure~\ref{A:GW:p-e}).
The model-dependent distribution of NS kick velocities provides another
strong complication. We also stress that this upper limit was obtained
assuming a constant Galactic star-formation rate and normalization 
of the binary formation by Equation~(\ref{E:bin_brate}).

Further (semi-)analytical investigations of the parameter space of binaries
leading to the formation of coalescing binary NSs are still possible
but technically very difficult, and we shall not reproduce them here.
The detailed semi-analytical approach to the problem of formation of NSs in
binaries and evolution of compact binaries has been developed by Tutukov and
Yungelson~\cite{Tutukov_Yungelson93a, ty93b}.


\subsection{Population synthesis results}

A distinct approach to the analysis of binary star evolution is based
on the population synthesis method -- a Monte-Carlo simulation of the
evolution of a sample of binaries with different initial
parameters. This approach was first applied to model various
observational manifestations of magnetized NSs in massive binary
systems~\cite{Kornilov_Lipunov83a, Kornilov_Lipunov83b,
  Dewey_Cordes87} and generalized to binary systems of arbitrary mass
in~\cite{Lipunov_al96} (The Scenario Machine code). To achieve a
sufficient statistical significance, such simulations usually involve
a large number of binaries, typically of the order of a million. The
total number of stars in the Galaxy is still four orders of magnitude
larger, so this approach cannot guarantee that rare stages of the
binary evolution will be adequately reproduced\epubtkFootnote{Instead
  of Monte-Carlo simulations one may use a sufficiently dense grid in
  the 3D space of binary parameters and integrate over this grid (see,
  e.g., \cite{ty02} and references therein).}.

Presently, there are several population synthesis codes used for
massive binary system studies, which take into account with different
degree of completeness various aspects of binary stellar evolution
(e.g., the codes by Portegies Zwart et
al.~\cite{py98, yungelson_bh06}, Bethe and
Brown~\cite{Bethe_Brown98}, Hurley, Tout, and Pols~\cite{Hurley_al02},
Belczynski et al.~\cite{Belczhynski_al_startreck}, Yungelson and
Tutukov~\cite{ty02}). A review of applications of the population
synthesis method to various types of astrophysical sources and further
references can be found in~\cite{Popov_Prokhorov04,
  yungelson05b}. Some results of population synthesis
calculations of compact binary mergers carried out by different groups
are presented in Table~\ref{table:mergings}.

\begin{table}
  \renewcommand{\arraystretch}{1.2}
  \centering
  \begin{tabular}{l|rrrr}
    \hline \hline
    Authors & Ref. & NS\,+\,NS & NS\,+\,BH & BH\,+\,BH \\
    & & \multicolumn{1}{c}{[$\mathrm{yr}^{-1}$]} &
    \multicolumn{1}{c}{[$\mathrm{yr}^{-1}$]} &
    \multicolumn{1}{c}{[$\mathrm{yr}^{-1}$]} \\
    \hline
    Tutukov and Yungelson (1993) & \cite{ty93b} &
    $3\times 10^{-4}$ & $2\times 10^{-5}$ & $1\times 10^{-6}$ \\
    Lipunov et~al.\ (1997) & \cite{LPP97} &
    $3\times 10^{-5}$ & $2\times 10^{-6}$ & $3\times 10^{-7}$ \\
    Portegies Zwart and Yungelson (1998) & \cite{py98} &
    $2\times 10^{-5}$ & $10^{-6}$ \\
    Nelemans et~al.\ (2001) & \cite{nyp01} &
    $2\times 10^{-5}$ & $4\times 10^{-6}$ & \\
    Voss and Tauris (2003) & \cite{voss_taurisns03} &
    $2\times 10^{-6}$ & $6\times 10^{-7}$ & $10^{-5}$ \\
    O'Shaughnessy et al.\ (2005) & \cite{Oshaughnessy_al05} &
    $7\times 10^{-6}$ & $1\times 10^{-6}$ & $1\times 10^{-6}$ \\
    de Freitas Pacheco et al.\ (2006) & \cite{freitas_ns06} &
    $2\times 10^{-5}$ \\
    \hline \hline
  \end{tabular}
  \caption{\it Examples of the estimates for Galactic merger rates of
    relativistic binaries calculated under different assumptions on
    the parameters entering population synthesis.}
  \label{table:mergings}
  \renewcommand{\arraystretch}{1.0}
\end{table}

Actually, the authors of the studies mentioned in
Table~\ref{table:mergings} make their simulations for a range of
parameters. We list in the table the rates for the models which the
authors themselves consider as ``standard'' or ``preferred'' or
``most probable''. Generally, for the NS\,+\,NS merger rate
Table~\ref{table:mergings} shows the scatter within a factor $\sim 4$,
which may be considered quite reasonable, having in mind the
uncertainties in input parameters. There are two clear outliers,
\cite{ty93b} and~\cite{voss_taurisns03}. The high rate
in~\cite{ty93b} is due to the assumption that kicks to
nascent neutron stars are absent. The low rate in~\cite{voss_taurisns03}
is due to the fact that these authors apply in the common envelope
equation an evolutionary-stage-dependent structural constant
$\lambda$. Their range for $\lambda$ is $0.006\mbox{\,--\,}0.4$, to
be compared with the ``standard'' $\lambda=0.5$ applied in most of the
other studies. A low $\lambda$ favours mergers in the first critical
lobe overflow episode and later mergers of the first-born neutron
stars with their non-relativistic companions\epubtkFootnote{Note that
  similar low values of $\lambda$ for $20$ to $50\,M_\odot$ stars were
  obtained also by~\cite{Podsiadlowski_al03}. If confirmed, these
  results may have major impact on the estimates of merger rates for
  relativistic binaries.}. A considerable scatter in the rates of
mergers of systems with BH companions is due, mainly, to
uncertainties in stellar wind mass loss for the most massive
stars. For instance, the implementation of winds in the code used
in~\cite{py98, nyp01} resulted in the absence
of merging BH\,+\,BH systems, while a rather low $\dot{M}$ assumed
in~\cite{voss_taurisns03} produced a high merger rate of BH\,+\,BH
systems.

A word of caution should be said here. It is hardly possible to trace
a detailed evolution of each binary, so one usually invokes the
approximate approach to describe the change of evolutionary stages of
the binary components (the so-called evolutionary track), their
interaction, effects of supernovae, etc. Thus, fundamental uncertainties
of stellar evolution mentioned above are complemented with (i)
uncertainties of the scenario and (ii) uncertainties in the
normalization of the calculations to the real galaxy (such as the
fraction of binaries among allstars, the star formation history,
etc.). The intrinsic uncertainties in the population synthesis results
(for example, in the computed event rates of binary mergers etc.) are
in the best case not less than of the order of factor two or
three. This should always be born in mind when using the population
synthesis calculations. However, we emphasize again the fact that the
double NS merger rate, as inferred from binary pulsar statistics with
account for the double pulsar observations~\cite{Burgay_al03,
  Kalogera_al04}, is very close to the population syntheses estimates
with a kick of about $(250\mbox{\,--\,}300) \mathrm{\ km\ s}^{-1}$.

\newpage


\section{Detection Rates} 
\label{sec:secI:DetectRate}

The detection of a gravitational wave signal from merging close
binaries is characterized by the signal-to-noise ratio $S/N$, which
depends on the binary masses, the distance to the binary, the
frequency, and the noise characteristics. A pedagogical derivation of
the signal-to-noise ratio and its discussion for different detectors
is given, for example, in Section~8 of the review~\cite{Grishchuk_al01}.

In this section we focus of two particular points: the plausible enhancement 
of the detection of merging binary black holes with respect to 
binary neutron stars and the way how absolute detection rates
of binary mergings can be calculated.


\subsection{Enhancement of the detection rate for binary BH mergers}

Coalescing binaries emit gravitational wave signals with a well known
time-dependence (waveform) (see Section~\ref{sec:appA} above). This
allows one to use the technique of matched
filtering~\cite{Thorne87}. The signal-to-noise ratio $S/N$ for a
particular detector, which is characterized by the dimensionless noise
rms amplitude $h_\mathrm{rms}$ at a given frequency $f$, depends
mostly on the ``chirp'' mass of the binary system
$\M=(M_1+M_2)^{-1/5}(M_1M_2)^{3/5}$ and its distance $r$. Here, we
will use the simplified version for $S/N$ (\cite{Thorne87}; see
also~\cite{Flanagan_Hughes98}):
\begin{equation}
  \frac{S}{N} = 3^{-1/2}\pi^{-2/3} \frac{G^{5/6}}{c^{3/2}}
  \frac{\M^{5/6}}{r} \frac{f^{-1/6}}{h_\mathrm{rms}(f)}.
  \label{D:S/N}
\end{equation}
At a fixed level of $S/N$, the detection volume is proportional to
$r^3$ and therefore it is proportional to $\M^{5/2}$. The detection
rate $\D$ for binaries of a given class (NS\,+\,NS, NS\,+\,BH or BH\,+\,BH) is the
product of their coalescence rate $\R_\mathrm{V}$ with the detector's
registration volume $\propto \M^{5/2}$ for these binaries.

It is seen from Table~\ref{table:mergings} that the model Galactic
rate $\R_\mathrm{G}$ of NS\,+\,NS coalescences is typically higher than the rate of
NS\,+\,BH and BH\,+\,BH coalescences. However, the BH mass can be
significantly larger than the NS mass. So a binary involving one or
two black holes, placed at the same distance as a NS\,+\,NS binary,
produces a significantly larger amplitude of gravitational waves. With
the given sensitivity of the detector (fixed $S/N$ ratio), a BH\,+\,BH
binary can be seen at a greater distance than a NS\,+\,NS binary. Hence,
the registration volume for such bright binaries is significantly
larger than the registration volume for relatively weak binaries. The
detection rate of a given detector depends on the interplay between
the coalescence rate and the detector's response to the sources of one
or another kind.

If we assign some characteristic (mean) chirp mass to different types
of double NS and BH systems, the expected ratio of their detection
rates by a given detector is
\begin{equation}
  \frac{\D_\mathrm{BH}}{\D_\mathrm{NS}} =
  \frac{\R_\mathrm{BH}}{\R_\mathrm{NS}}
  \left( \frac{\M_\mathrm{BH}}{\M_\mathrm{NS}} \right)^{5/2}\!\!\!\!\!\!\!\!,
\label{DBH/DNS}
\end{equation}
where $\D_\mathrm{BH}$ and $\D_\mathrm{NS}$ refer to BH\,+\,BH and NS\,+\,NS pairs,
respectively. Here, we discuss the ratio of the detection rates,
rather than their absolute values. The derivation of absolute 
values requires detailed evolutionary calculations, as we discussed
above.
Taking $\M_\mathrm{BH}=8.7\,M_\odot$ (for $10M_\odot + 10\,M_\odot$) 
and $\M_\mathrm{NS}=1.22\,M_\odot$ (for $1.4\,M_\odot + 1.4\,M_\odot$), 
Equation~(\ref{DBH/DNS}) yields
\begin{equation}
   \frac{\D_\mathrm{BH}}{\D_\mathrm{NS}} \approx
   140 \frac{\R_\mathrm{BH}}{\R_\mathrm{NS}}.
   \label{DBH/DNS.2}
\end{equation}
As $\frac{\R_\mathrm{BH}}{\R_\mathrm{NS}}$ is typically 0.1\,--\,0.01
(see Table~\ref{table:mergings}), this relation suggests that the
registration rate of BH mergers can be {\it higher} than that of NS
mergers. This estimate is, of course, very rough, but it can serve as
an indication of what one can expect from detailed calculations. We
stress that the effect of an enhanced detection rate of BH binaries is
independent of the desired $S/N$ and other characteristics of the
detector; it was discussed, for example,
in~\cite{ty93b, LPP97, Grishchuk_al01}.


\subsection{Note on realistic calculation of the detection rates of
  binary mergings}

Now we shall briefly discuss how the detection rates of binary mergings
can be calculated for a given gravitational wave detector. For a secure
detection, the $S/N$ ratio is usually raised up to 7\,--\,8 to avoid false
alarms over a period of a year (assuming Gaussian
noise)\epubtkFootnote{If a network of three detectors, such as two
  LIGOs and VIRGO, runs simultaneously, the $S/N$ ratio in an
  individual detector should be $>7/\sqrt{3}\approx 4$)}. This
requirement determines the maximum distance from which an event can be
detected by a given interferometer. The distance ranges of
LIGO~I/VIRGO (LIGO~II) interferometers for relativistic binary
inspirals are given in~\cite{Cutler_Thorne02}: $20 (300) \mathrm{\
  Mpc}$ for NS\,+\,NS ($1.4\,M_\odot + 1.4\,M_\odot$), $43 (650) \mathrm{\
  Mpc}$ for NS\,+\,BH ($1.4\,M_\odot + 10\,M_\odot$) Note that the
distances increase for a network of detectors.

To calculate a realistic detection rate of binary mergers the
distribution of galaxies should be taken into account within the
volume bounded by the distance range of the detector (see, for
example, the earlier attempt to take into account only bright galaxies
from Tully's catalog of nearby galaxies in~\cite{Lipunov_al95_gwsky},
and the use of LEDA database of galaxies to estimate the detection
rate of supernova explosions~\cite{Baryshev_Paturel01}). However, not
only the mass and type of a given galaxy, but also the star formation
rate and, better, the history of the star formation rate in that
galaxy (since the binary merger rate in galaxies strongly evolves
with time~\cite{Lipunov_al95_evol}) are needed to estimate the
expected detection rate $\cal D$. But this is a tremendous problem --
even the sample of galaxies (mostly, dwarfs) within the Local Volume
($<8 \mathrm{\ Mpc}$) is only 70\%--80\%
complete~\cite{Karachentsev_al04}, and the number of new nearby
galaxies continues to increase. So to assess the merger rate from a
large volume based on the Galactic values, the best one can do at
present appears to be using formulas like Equation~(\ref{R_V}) given
earlier in Section~\ref{sec:ns_freq}. This, however, adds another
factor two of uncertainty in the estimates. Clearly, a more accurate
treatment of the transition from Galactic rates to larger volumes with
an account of the galaxy distribution is required.


\subsection{Going further}
\label{gf:observations}

The most important future observations include: 
\begin{enumerate}
\item Routine increase of the statistics of binary pulsars, especially
  with low radio luminosity. This will allow one to put stronger
  constraints on the NS\,+\,NS merger rate as directly inferred from the
  binary pulsars statistics. More indirectly, a larger sample of NS
  parameters in binary pulsars would be useful for constraining the
  range of parameters of scenarios of formation for double NSs and, hence,
  a better understanding their origin (see, for example, a recent
  attempt of such an analysis in~\cite{Oshaughnessy_al05}).
\item Discovery of a possible NS\,+\,BH binary. Measurements of its
  parameters  would be crucial for models of formation and evolution
  of BHs in binary systems in general. Current estimates of the number
  of such binaries in the Galaxy, obtained by the population synthesis
  method, range from one per several thousand ordinary
  pulsars~\cite{Lipunov_al94, Lipunov_al05} to much smaller values of
  about 0.1\,--\,1\% of the number of double NSs in the Galactic
  disk~\cite{Pfahl_al05}.
\item Search for unusual observational manifestations of relativistic
  binaries (e.g., among some new radio transient sources like GCRT
  J1745-3009, firm identifications with some GRBs, etc.)
\item Improving the estimates of binary merger rate limits from data
  taken by GW detectors.
\end{enumerate}
The most important theoretical issues include:
\begin{enumerate}
\item Stellar physics: post-helium burning evolution of massive stars,
  supernova explosion mechanism, masses of compact stars formed in the
  collapse, mechanism(s) of kick velocity imparted to nascent compact
  remnants (neutron stars and black holes), stellar winds from
  hydrogen- and helium-rich stars.
\item Binary evolution: treatment of the common envelope stage,
  magnetic braking for low-mass binaries, observational constraints on
  the initial distributions of orbital parameters of binary stars
  (masses, semimajor axes, eccentricities).
\item Last but not least, it is very important to improve our
  knowledge of such ``traditional'' topics of stellar astronomy as the
  fraction of binary stars among the total population, distributions
  of binary stars over separations of components, and their mass
  ratios.
\end{enumerate}

\newpage


\section{Formation of Short-Period Binaries with a White-Dwarf Components}
\label{section:wd_formation}

Binary systems with white dwarf components that are interesting for general 
relativity and cosmology come in several flavours:
\begin{itemize}
\item Detached binary white dwarfs or ``double
  degenerates'' (DDs, we shall use both terms as synonyms below).
\item Cataclysmic variables (CVs) -- a class of semidetached
  binary stars containing a white dwarf and a companion star that is
  usually a red dwarf or a slightly evolved star, a subgiant.
\item A subclass of the former systems in which the Roche lobe
  is filled by another white dwarf or low-mass partially degenerate
  helium star (AM~CVn-type stars or ``interacting
  double-degenerates'', IDDs). They appear to be important LISA
  verification sources.
\item Detached systems with a white dwarf accompanied by a
  low-mass nondegenerate helium star (SD\,+\,WD systems).
\item Ultracompact X-ray binaries (UCXBs) where one of the
  components is a NS, while the Roche lobe overflowing component is a
  WD.
\end{itemize} 
As Figure~\ref{f:GW_sources} shows, compact binary stars emit
gravitational waves within the sensitivity limits for space-based
detectors if their orbital periods are from $\sim 20 \mathrm{\ s}$ to
$20\!,000 \mathrm{\ s}$. This means that in principle all AM~CVn-stars,
UCXBs, a considerable fraction of all CVs with measured orbital periods,
and some DD and SD\,+\,WD systems would be observable in GWs in the absence
of confusion noise and sufficient sensitivity of detectors.

Though general relativity predicted that binary stars have to be the
sources of gravitational waves in the 1920s, this prediction became a
matter of actual interest only with the discovery of the $P_\mathrm{orb} \approx
81.5 \mathrm{\ min}$ cataclysmic variable WZ~Sge by Kraft, Mathews, and
Greenstein in 1962~\cite{kmg62}, who immediately recognized the
significance of short-period binary stars as testbeds for
gravitational waves physics. Another impetus to the study of binaries
as sources of gravitational wave radiation (GWR) was imparted by the
discovery of the ultra-short period
binary HZ~29\,=\,AM~CVn ($P_\mathrm{orb} \approx 18 \mathrm{\ min}$) by Smak in
1967~\cite{sma67}. Smak~\cite{sma67} and Paczy\'{n}ski~\cite{pac67a}
speculated that the latter system is a close pair of white dwarfs,
without specifying whether it is detached or semidetached. Faulkner et
al.~\cite{ffw72} inferred the status of AM~CVn as a
``double-white-dwarf semidetached'' nova. AM~CVn was later classified
as a cataclysmic variable after flickering typical for CVs was found for
AM~CVn by Warner and Robinson~\cite{wr72}\epubtkFootnote{Flickering is
  a fast intrinsic brightness scintillation occurring on time scales
  from seconds to minutes with amplitudes of $0.01\mbox{\,--\,} 1
  \mathrm{\ mag}$ and
  suggesting an ongoing mass transfer.} and became a prototype of a
subclass of binaries\epubtkFootnote{Remarkably, a clear-cut confirmation
  of the binary nature of AM~CVn and the determination of its true period
  awaited for almost 25 years~\cite{nsg_am00}.}.

The origin of all above mentioned classes of short-period binaries was
understood after the notion of common envelopes and the formalism for
their treatment were suggested in 1970s (see Section~\ref{A:CE}). A
spiral-in of components in common envelopes allowed to explain how
white dwarfs -- former cores of highly evolved stars with radii of
$\sim 100\,R_\odot$ -- may acquire companions separated by $\sim\,R_\odot$
only (for pioneering work see~\cite{pac76, webbink-79, ty79a, ty81,
  it84a, Webbink84}). We recall, however, that most studies of the formation
of compact objects through common envelopes are based on a simple
formalism of comparison of binding energy of the envelope with the orbital
energy of the binary, thought to be the sole source of energy for the loss
of the envelope as described in Section~\ref{A:CE}. Though full-scale
hydrodynamic calculations of a common-envelope evolution exist, for
instance a series of papers by Taam and coauthors published over more
than two decades (see~\cite{sandq_ce98, sandq_binhedw00} and
references therein), the process is still very far from comprehension.

We recall also that the stability and timescale of mass-exchange in a
binary depends on the mass ratio of components $q$, the structure of
the envelope of Roche-lobe filling star, and possible stabilizing
effects of mass and momentum loss from the system~\cite{tfy82, hw87,
  yl98, hknu99, hpmmi02, ivanova_taam03, fty04,
  bunning_ritter_me06}. For stars with radiative envelopes, to the
first approximation, mass exchange is stable if $q \lesssim 1.2$; for
$1.2 \lesssim q \lesssim 2$ it proceeds in the thermal time scale of the
donor; for $q \gtrsim 2$ it proceeds in the dynamical time scale. Mass
loss occurs on a dynamical time scale if the donor has a deep convective
envelope or if it is degenerate and conditions for stable mass
exchange are not satisfied. It is currently commonly accepted, despite
a firm observational proof is lacking, that the distribution of
binaries over $q$ is even or rises to small $q$ (see
Section~\ref{section:binary_NS}). Since typically the accretion rate is
limited either by the rate that corresponds to the thermal time scale
of the accretor or its Eddington accretion rate, both of which are lower
than the mass-loss rate by the donor, the overwhelming majority ($\sim
90\%$) of close binaries pass in their evolution through one to four
stages of a common envelope.

An ``initial donor mass -- donor radius at RLOF'' diagram showing
descendants of stars after mass-loss in close binaries is presented in
Figure~\ref{f:remn}. We should remember here that solar metallicity
stars with $M\lesssim 0.95\,M_\odot$ do not evolve past the core-hydrogen
burning stage in Hubble time.

\epubtkImage{figure05.png}{
  \begin{figure}[htbp]
    \centerline{\includegraphics[scale=0.55]{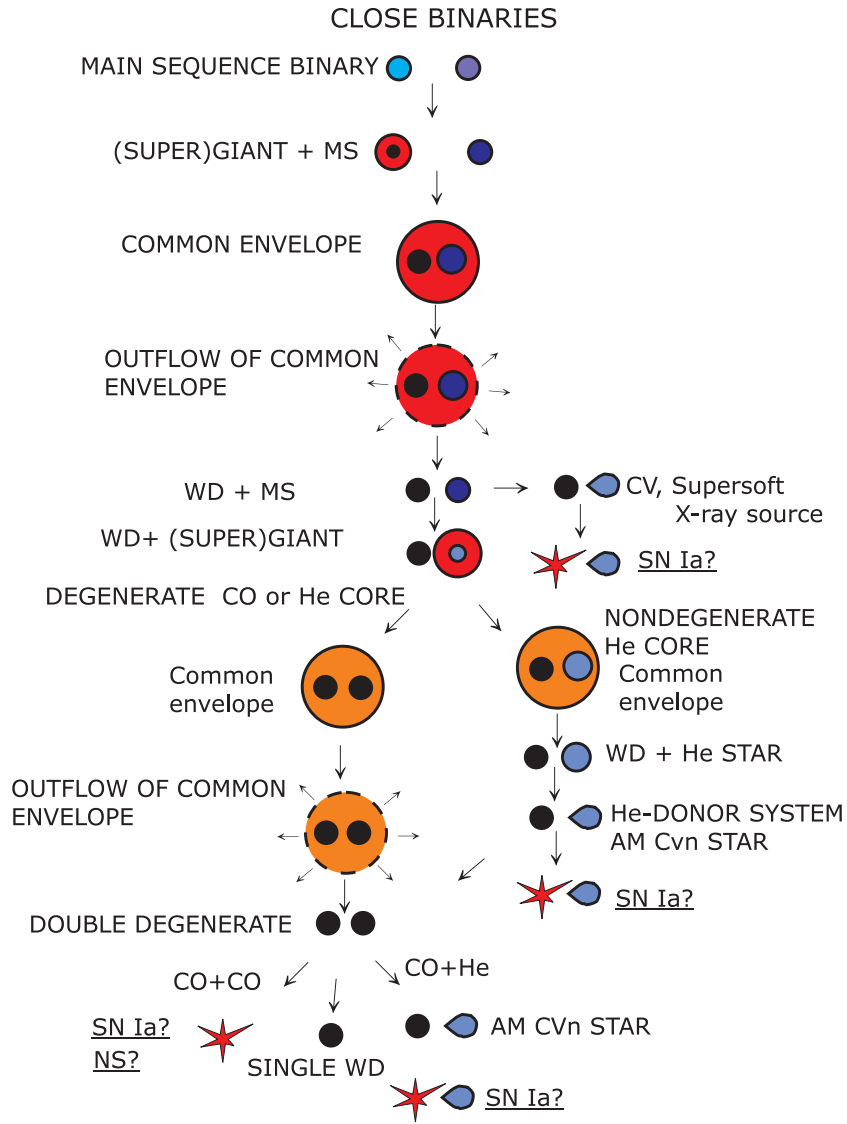}}
    \caption{\it Formation of close binary dwarfs and their
      descendants (scale and colour-coding are arbitrary).}
    \label{figure:wd_flow}
  \end{figure}
}

\noindent
{\bf Formation of compact binaries with WDs.~~}
A flowchart schematically presenting the typical scenario for formation of
low-mass compact binaries with white-dwarf components and some
endpoints of evolution is shown in Figure~\ref{figure:wd_flow}. Of
course, not all possible scenarios are plotted, but only the most probable
routes to SNe~Ia and systems that may emit potentially detectable
gravitational waves. For simplicity, we consider only the most general
case when the first RLOF results in the formation of a common envelope.

The overwhelming majority of stars overflow their Roche lobes when they
have He- or CO-cores. In stars with a mass below $(2.0\mbox{\,--\,}2.5)
M_\odot$, helium cores are degenerate and if these stars overflow the Roche
lobe prior to He-ignition, they produce helium white dwarfs. Binaries
with non-degenerate He-core donors ($M\gtrsim (2.0\mbox{\,--\,}2.5)
 \,M_\odot$) first form a He-star\,+\,MS-star pair that may be observed
as a subdwarf (sdB or sdO) star with MS companion.  When the He-star
completes its evolution, a pair harbouring a CO white dwarf and a
MS-star appears.

If after the first common-envelope stage the orbital separation of the
binary $a\simeq \mathrm{several\ } R_\odot$ and the WD has a low-mass ($\lesssim
1.5\,M_\odot$) MS companion the pair may evolve into contact during
the hydrogen-burning stage or shortly after because of loss of angular
momentum by a magnetically coupled stellar wind and/or GW
radiation. If, additionally, the mass-ratio of components is
favourable for stable mass transfer a cataclysmic variable may
form. If the WD belongs to the CO-variety and accreted hydrogen
burns at the surface of the WD stably, the WD may accumulate enough mass to
explode as a type~Ia supernova; the same may happen if in the recurrent
outbursts less mass is ejected than accreted (the so-called
SD scenario for SNe~Ia originally suggested by
Whelan and Iben~\cite{wi73}; see, e.g., \cite{lh97, yl98, hachisu_rec01,
  hp04, fty04, hp06, lesaffre_sn06} and references therein for later
studies).

Some CV systems that burn hydrogen stably or are in the stage of
residual hydrogen burning after an outburst may be also observed as
supersoft X-ray sources (see, e.g., \cite{hbnr92, rdss94, kahabka_sss95,
  ylttf96, kh97, kahabka02}).

If the WD belongs to the ONe-variety, it may experience an AIC into a
neutron star due to electron captures on
Ne and Mg, and a low-mass X-ray binary may be formed.

The outcome of the evolution of a CV is not completely clear. It was
hypothesized that the donor may be disrupted when its mass decreases
below several hundredth of $M_\odot$~\cite{rs83}. Note, that for
$q\lesssim0.02$, that may be attained in Hubble time, and the conventional
picture of mass exchange may become non-valid, since the
circularization radius becomes greater than the outer radius of the
disk. Matter flowing in from the companion circularises onto unstable
orbits. At $q\approx 0.02$, matter is added at $R_\mathrm{circ}$ onto
orbits that can become eccentric due to the 3:1 resonance. At
$q\approx 0.005$ the circularization radius approaches the 2:1
Lindblad resonance. This can efficiently prevent mass being
transferred onto the compact object. These endpoints of the evolution
of binaries with low-mass donors, were, in fact, never studied.

The second common envelope may form when the companion to the WD overfills
its Roche lobe. If the system avoids merger and the donor had a
degenerate core, a close binary WD (or double-degenerate, DD) is
formed. The fate of the DD is solely defined by GWR. The closest of them
may be brought into contact by AML via GWR. The outcome of the contact
depends on the chemical composition of the stars and their masses. The
lighter of the two stars fills the Roche lobe first (by virtue of the mass--radius
relation $R \propto M^{-1/3}$). For a zero-temperature WD the condition of
stable mass transfer is $q<2/3$ (but see the more detailed discussion in
Section~\ref{section:am-evol}). The merger of the CO-WD pair with a total
mass exceeding $M_\mathrm{Ch}$ may result in a SN~Ia leaving no remnant
(``double-degenerate SN~Ia scenario'' first suggested by
Webbink~\cite{webbink-79} and independently by Tutukov and
Yungelson~\cite{ty81}) or in an AIC with formation of a single neutron
star~\cite{mochko90}. The issue of the merger outcome for $M_\mathrm{tot} >
M_\mathrm{Ch}$ still remains an unsolved issue and a topic of fierce
discussion, see below. For total masses lower than $M_\mathrm{Ch}$ the
formation of a single WD is expected.

If in a CO\,+\,He WD dwarfs pair the conditions for stable mass exchange are
fulfilled, an AM~CVn system forms (see for details~\cite{ty79a, ty96,
  nyp01a, mns04, gokhale_stab_am06}).

\epubtkImage{figure06.png}{
  \begin{figure}[htbp]
    \centerline{\includegraphics[scale=0.4]{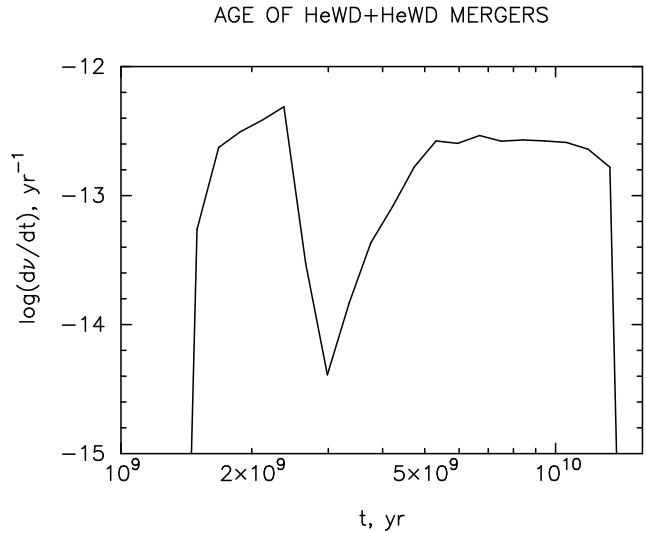}}
    \caption{\it The age of merging pairs of helium WDs. Two
      components of the distribution correspond to the systems that
      experienced in the course of formation two or one common
      envelope episodes, respectively.}
    \label{figure:hewdmerg}
  \end{figure}
}

The current Galactic merger rate of close binary WDs is about 50\%
of their current birth rate~\cite{nyp01, ty02}. It is not yet clear
how the merger proceeds; it is possible that for He\,+\,He or CO\,+\,He pairs
a helium star is an intermediate stage (see, e.g., \cite{guerrero04}). It
is important in this respect that binary white dwarfs at birth have
a wide range of separations and merger of them may occur gigayears
after formation. Formation of helium stars via merger may be at
least partially responsible for the ultraviolet flux from the giant
elliptical galaxies, where all hot stars finished their evolution
long ago. This is illustrated by Figure~\ref{figure:hewdmerg} which
shows the occurrence rate of mergers of pairs of He-WDs vs.\ age.

If the donor has a nondegenerate He-core ($M \gtrsim
  (2.0\mbox{\,--\,}2.5)\,M_\odot$) and the system does not merge, after
the second CE-stage a helium subdwarf\,+\,WD system may emerge. If the
separation of components is sufficiently small, AML via GWR may bring
the He-star into contact while He is still burning in its core. If $M_{\rm
  He}/M_\mathrm{wd} \lesssim 1.2$, stable mass exchange is possible
with a typical $\dot{M} \sim 10^{-8} \,M_\odot \mathrm{\
  yr}^{-1}$~\cite{skh86}. Mass loss quenches
nuclear burning and the helium star becomes ``semidegenerate''. An
AM~CVn-type system may be formed in this way (the ``nondegenerate
He-core'' branch of evolution in Figure~\ref{figure:wd_flow}). One
cannot exclude that a Chandrasekhar mass may be accumulated by the WD in
this channel of evolution, but the probability of such a scenario seems to
be very low, $\sim 1\%$ of the inferred Galactic rate of
SNe~Ia~\cite{sol_yung_am05}\epubtkFootnote{The recent discovery of SN~Ia
  SN~2003fg with a mass estimate $\sim
  2\,M_\odot$~\cite{howell_2003fg_superch, jeffery_2003fg_superch} may
  support this scenario.}. If the He-star completes core He-burning
before RLOF, it becomes a CO-WD. In Figure~\ref{figure:wd_flow} it
``jumps'' into the ``double degenerate'' branch of evolution.

\begin{table}
  \renewcommand{\arraystretch}{1.2}
  \centering
  \begin{tabular}{l|c@{\quad}c@{\quad}c@{\quad}c@{\quad}c@{\quad}c}
    \hline \hline
    Donor & CO-WD & MS/SG & He-star & He-WD & RG \\
    \hline
    Counterpart & Close binary WD & Supersoft XRS & Blue sd & AM~CVn &
    Symbiotic star \\
    Mass transfer mode & Merger & RLOF & RLOF & RLOF & Wind \\
    Young population & $\bf 10^{-3}$ & $10^{-4}$ & $10^{-4}$ & $10^{-5}$ & $10^{-6}$ \\
    Old population & $\bf 10^{-3}$ & --- & --- & $10^{-5}$ & $10^{-6}$ \\
    \hline \hline
  \end{tabular}
  \caption{\it Occurrence rates of SNe~Ia in the candidate progenitor
    systems (in $ \mathrm{yr}^{-1} $), after~\cite{yungelson05a}. SG
    stands for sub-giant, RG for red giant, and XRS for X-ray source.}
  \label{table:snoccurr}
  \renewcommand{\arraystretch}{1.0}
\end{table}

\noindent
{\bf Type~Ia supernovae.~~}
Table~\ref{table:snoccurr} summarizes order
of magnitude model estimates of the occurrence rate of SNe~Ia produced
via different channels. For comparison, the rate of SNe~Ia from wide
binaries (symbiotic stars) is also given. The estimates are obtained
by a population synthesis code used before in, e.g., \cite{ty94, yl98,
  ty02} for the value of common envelope parameter
$\alpha_\mathrm{ce}=1$. The differences in the assumptions with other
population synthesis codes or in the assumed parameters of the models
result in numbers that vary by a factor of several; this is the reason
for giving only order of magnitude estimates. The estimates are
shown for $T=10 \mathrm{\ Gyr}$ after the beginning of star formation.

A ``young'' population had a constant star formation rate for $10
\mathrm{\ Gyr}$; in the ``old'' one the same amount of gas was
converted into stars in $1 \mathrm{\ Gyr}$. Both populations have a mass
comparable to the mass of the Galactic disk. We also list in the table
the types of observed systems associated with a certain channel and the
mode of mass transfer. These numbers have to be compared to the
inferred Galactic occurrence rate of SNe~Ia: $(4\pm2)\times 10^{-3}
\mathrm{\ yr}^{-1}$~\cite{captur01}. Table~\ref{table:snoccurr} shows
that, say, for elliptical galaxies where star formation occurred in a
burst, the DD scenario is the only one able to respond to the occurrence
of SNe~Ia, while in giant disk galaxies with continuing star formation
other scenarios may contribute as well.

For about two decades since the prediction of the possibility of the
merger of pairs of white dwarfs with total mass $\geq M_\mathrm{Ch}$, the
apparent absence of observed DDs with proper mass and merger times
shorter than Hubble time was considered as the major ``observational''
difficulty for the DD scenario. Theoretical models predicted that it may
be necessary to investigate for binarity up to 1,000 field WDs with $V
\lesssim 16\mbox{\,--\,}17$ for finding a proper
candidate~\cite{nyp_01}. Currently, it is likely that this problem is
resolved (see Section~\ref{section:observations}).

The merger of pairs of WDs occurs via an intermediate stage in which the
lighter of the two dwarfs transforms into a disc~\cite{ty79a, bbc_90,
  mochko90, loren_gwr05} from which the matter accretes onto the
central object. It was shown for one-dimensional non-rotating models
that the central C-ignition and SN~Ia explosion are possible only for
$\dot{M}_{\mathrm a}\lesssim(0.1\mbox{\,--\,}0.2)
\dot{M}_{{\mathrm{Edd}}}$~\cite{nomoto_iben85, timmes_sn94}. But it was
expected that in the merger products of binary dwarfs $\dot{M}_{\mathrm
  a}$ is close to $\dot{M}_{\mathrm{Edd}}\sim 10^{-5}\,M_\odot
\mathrm{\ yr}^{-1}$~\cite{mochko90} because of high viscosity in the
transition layer
between the core and the disk. For such an $\dot{M}_{\mathrm a}$, the nuclear
burning will start at the core edge, propagate inward and convert the
dwarf into an ONeMg one. The latter will collapse without a
SN~Ia~\cite{isern_83}. However, an analysis of the role of deposition
of angular momentum into a central object by Piersanti and
coauthors~\cite{piers_03b, piers_03a} led them to conclusion that, as
a result of the spin-up of rotation of the WD, instabilities associated with
rotation, deformation of the WD, and AML by a distorted
configuration via GWR, an $\dot{M}_{\mathrm a}$ that is initially $\sim
10^{-5}\,M_\odot \mathrm{\ yr}^{-1}$ decreases to $\simeq 4\times
10^{-7}\,M_\odot \mathrm{\ yr}^{-1}$. For this
$\dot{M}_{\mathrm a}$ close-to-center ignition of carbon becomes
possible. The efficiency of the mechanism suggested
in~\cite{piers_03b,piers_03a} is disputed, for instance, by Saio and
Nomoto~\cite{saio_rot_no_dd04} who found that an off-center carbon
ignition occurs even when the effect of stellar rotation is included,
if $\dot{M}_{\mathrm a} > 3 \times 10^{-6}\,M_\odot \mathrm{\
  yr}^{-1}$. The latter authors find
that the critical accretion rate for the off-center ignition is hardly
changed by the effect of rotation. The problem has to be considered as
unsettled until a better understanding of redistribution of angular
momentum during the merger process will become available.

Because of a long absence of apparent candidates for the DD scenario
and its theoretical problems, the SD scenario is often considered as the
most promising one. However, it also encounters severe problems. Even
stably burning white dwarfs must have radiatively driven winds. At
$\dot{M}_\mathrm{accr} \lesssim 10^{-8}\,M_\odot \mathrm{\ yr}^{-1}$
all accumulated mass is lost
in nova explosions~\cite{prkov95, yaron05}. Even if
$ \dot{M}_\mathrm{accr} $ allows
accumulation of a He-layer, most of the latter is lost after the
He-flash~\cite{it96symb, cassisi_etal98, piers_99}, dynamically or via
the frictional interaction of the binary components with the giant-size
common envelope. As a result, mass accumulation efficiency is always
$<1$ and may be even negative. On the other hand, it was noted that
the flashes become less violent and more effective accumulation of
matter may occur if mass is transferred on a rate close to the
thermal one or the dwarf is rapidly rotating~\cite{it84a, yl98,
  ivanova_taam03, hp04, fty04, yoon_stab04, yoon_lang_rotsn05}. Thus,
crucial for this SN~Ia scenario are the range of donor masses that may
support mass-loss rates ``efficient'' for the growth of WD, mechanisms
for stabilizing mass loss in the necessary range, convection and
angular momentum transfer in the accreted layer that define the amount
of mass loss in the outbursts, and the amount of matter that escape in
the wind\epubtkFootnote{All these caveats date back as far as to
  MacDonalds' 1984 paper~\cite{macdonald_sn84} but are still not
  resolved.}. If the diversity of SNe~Ia is associated with the spread
of mass of the exploding objects, it would be more easily explained in
a SD scenario, since the latter allows white dwarfs to grow
efficiently in mass by shell burning, which is stabilized by accretion-induced
spin-up. This inference may be supported by the discovery of the
``super-Chandrasekhar'' mass SN~Ia SN~2003fg (mass estimate $\sim
2\,M_\odot$ \cite{howell_2003fg_superch, jeffery_2003fg_superch}).
Even under assumption of the most favourable conditions for a SN~Ia in
the SD scenario, the estimates of the current Galactic
occurrence rate for this channel do not exceed
$1\times10^{-3} \mathrm{\ yr}^{-1} $~\cite{hp04}, i.e.\ they may contribute up to
50\% of the lowest estimate of the inferred Galactic SN~Ia occurrence
rate.

On the observational side, the major objection to the
SD scenario comes from the fact that no hydrogen is
observed in the spectra of SNe~Ia, while it is expected that $\sim
0.15\,M_\odot$ of H-rich matter may be stripped from the companion by the
SN shell~\cite{marietta00}\epubtkFootnote{The recently discovered
  hydrogen-rich SN~Ia 2001ic and similarly 1997cy~\cite{hamuy03} may
  belong to the so-called SN\,1.5 type or occur in a symbiotic
  system~\cite{cy04}.}. Hydrogen may be discovered both in very early
and late optical spectra of SNe and in radio- and X-ray
ranges~\cite{eck_95, marietta00, lentz_02}. Panagia et
al.~\cite{panagia_radio06} find a firm upper limit to a steady
mass-loss rate for individual SN systems of $\sim3 \times
10^{-8}\,M_\odot \mathrm{\ yr}^{-1}$. As well, no
expected~\cite{marietta00, canal_01} high
luminosity and/or high velocity former companions to exploding WD were
discovered as yet\epubtkFootnote{The reported discovery of Tycho Brahe's
  1572 SN~Ia companion~\cite{rlapuente04} is not confirmed as
  yet.}. The SD scenario also predicts the existence of many more supersoft
X-ray sources than are expected from observations, even considering
severe problems in estimating incompleteness of the samples of the
latter (see for instance~\cite{stefrap95}).

To summarize, the problem of progenitors of SNe~Ia is still unsettled. Large
uncertainties in the model parameters involved in the computation of
the evolution leading to a SN~Ia and in computations of the explosions
themselves, do not allow to
exclude any type of progenitors. The existence of at least two families of
progenitors is suggested by observations (see, e.g.,
\cite{mannucci_sn05}). A high
proportion of ``peculiar'' SN~Ia of $(36\pm9)\%$~\cite{li_01} suggests a
large spread in the ignition conditions in the exploding objects that also may
be attributed to the diversity of progenitors. Note that a high proportion of
``peculiar''  SNe~Ia casts a certain doubt to their use as standard candles
for cosmology. 

As shown in the flowchart in Figure~\ref{figure:wd_flow}, there are configurations for
which it is expected that stable accretion of He onto a CO-WD occurs: in
AM~CVn systems in the double-degenerate formation channel and in
precursors of AM~CVn systems in the helium-star channel. In the latter
systems the mass exchange rate is close to $(1\mbox{\,--\,}3) \times
10^{-8}\,M_\odot \mathrm{\ yr}^{-1}$, practically irrespective of
the combination of donor and
accretor mass. It was suggested that in such systems the accumulation of
a $\sim 0.1\,M_\odot$ degenerate He-layer onto a
$(0.6\mbox{\,--\,}0.9)\,M_\odot$ accretor is possible prior to
He-detonation and that the latter initiates a compressional wave that
results in the central detonation of carbon~\cite{livne90, lg91, ww94,
  la95}; even if central carbon ignition does not occur, the scale of
the event is comparable to weak SNe~\cite{lt91, it91,
  tk_eld92}. For a certain time these events involving
sub-Chandrasekhar mass accretors (nicknamed ``edge-lit detonations'',
ELD) that may occur at the rates of $\sim 10^{-3} \mathrm{\ yr}^{-1} $ were
considered as one of the alternative mechanisms for SNe~Ia, although it
was shown by H\"offlich and Khokhlov~\cite{hk96} that the behaviour of
light-curves produced by them does not resemble any of the known
SNe~Ia. Thus, until recently, the real identification of these events
remained a problem. However, it was shown recently by Yoon and
Langer~\cite{yoon_langer04}, who considered angular-momentum accretion
effects, that the helium envelope is heated efficiently by friction in the
differentially rotating spun-up layers. As a result, helium ignites
much earlier and under much less degenerate conditions compared to the
corresponding non-rotating case. If the efficiency of energy dissipation
is high enough, detonation may be avoided and, instead of a SN,
recurrent helium novae may occur. The outburst, typically, happens
after accumulation of $0.02\,M_\odot$. Currently, there is one object
known, identified as He-nova -- V445~Pup~\cite{wagner_hen01a,
  wagner_hen01b, ashok03a, ashok03b}. If He-novae are really
associated with mass-transfer from low-mass helium stars to CO white
dwarfs, then the estimates of the birth rate of the latter systems
($\approx 0.6 \times 10^{-3} \mathrm{\ yr}^{-1}$)~\cite{ty05} and of the
amount of matter available for transfer ($\approx 0.2\,M_\odot$) give
an occurrence rate of He-novae of $\sim0.1 \mathrm{\ yr}^{-1}$, i.e.\ one He-nova per
several 100 ``ordinary'' hydrogen-rich novae.

As we mentioned above, intermediate mass donors, before becoming white
dwarfs, pass through the stage of a helium star. If the mass of the
latter is above $\simeq 0.8\,M_\odot$, it expands to giant dimensions
after exhaustion of He in the core and may overflow the Roche lobe and,
under proper conditions, transfer mass stably~\cite{it85}. For the
range of mass-accretion rates expected for these stars, both the
conditions for stable and unstable helium burning may be fulfilled. In
the former case the accumulation of $M_\mathrm{Ch}$ and a SN~Ia become
possible, as it was shown explicitly by Yoon and
Langer~\cite{yoon_lang_sn03}. However, the probability of such a SN~Ia
is only $\sim 10^{-5} \mathrm{\ yr}^{-1}$.

\noindent
{\bf Ultra-compact X-ray binaries.~~}
The suggested channels for formation of UCXBs in the field are, in fact,
``hybrids'' of scenarios presented in
Figures~\ref{figure:massive_flow} and~\ref{figure:wd_flow}. In
progenitors of these systems, the primary becomes a neutron star,
while the secondary is not massive enough. Then, several scenarios
similar to the scenarios for the systems with the first-formed white
dwarf are open. A white dwarf may overflow the Roche lobe due to systemic
AML via GWR. A low-mass
companion to a neutron star may overflow the Roche lobe at the end of the
main sequence and become a low-mass He-rich donor. A core
helium-burning star may be brought in contact by AML due to GWR; mass
loss quenches nuclear
burning and the donor becomes a helium ``semidegenerate''
object. An additional scenario is provided by the formation of a neutron-star
component by AIC of an accreting white dwarf. We refer the reader to
the pioneering papers~\cite{tfey85, nelson86, th86, skh86, tfey87,
  fe89, vwb90} and to more recent studies~\cite{nyp01, ynh02,
  prp_low02, pfahl_low03, bel_taam_ucb04, zand05, sluys_ucb05a,
  sluys_ucb05b, kim_coalesc04, kalogera_nswd05}. An analysis of the chemical
composition of donors in these systems seems to be a promising way for
discrimination between systems of different origin~\cite{nelemans04b,
  Nelemans:2006gs, zand05, werner06}: Helium dwarf donors should
display products of H-burning, while He-star descendants should
display products of He-burning
products. Actually, both carbon/oxygen and helium/nitrogen discs in
UCXBs are discovered~\cite{nelemans_jonker_ucxb06}.

In globular clusters, UCXBs are formed most probably by dynamical
interactions, as first suggested by Fabian et
al.~\cite{fpr_xray_gc75} (see, e.g., \cite{rasio_cb_gc00,
  ivanova_ultra_gc05, verbunt_xray_gc05, lombardi_ucb_gc06,
  Benacquista_LRR02} and references therein for the latest studies on
the topic).

\newpage


\section{Observations}
\label{section:observations}

The state of interrelations between observations and theoretical
interpretations are different for different groups of compact
binaries. Such cataclysmic variables as novae stars have been observed for
centuries, their lower-amplitude cousins (including AM~CVn-type stars)
for decades, and their origin and evolution found their theoretical
explanation after the role of common envelopes and gravitational waves
radiation and magnetic braking were recognized~\cite{pac67a, pac76,
  ty79a, vz81}. At present, about 2,000 CVs are known; see the online
catalogue by Downes et al.~\cite{downes06} at~\cite{url03}.

More than 600 of them have measured orbital periods; see the online
catalogue by Kolb and Ritter~\cite{ritter-kolb-cat03} at~\cite{url04}.

In particular, there are at present 17 confirmed AM~CVn-stars with
measured or estimated periods and two more candidate systems; see the
lists and references in~\cite{nelemans_amcvn05, stroeer06} and~\cite{url05}.
  
Ultracompact X-ray binaries were discovered with the advent of the X-ray
astronomy era in the late 1960s and may be found in the first published
catalogues of X-ray sources (see for instance~\cite{giacconi74}). Their
detailed optical investigation became possible only with 8~m-class
telescopes. Currently, 12 UCXB systems with measured or suspected
periods are known plus six candidates; six of the known systems reside in
globular clusters (see lists and references
in~\cite{nelemans_jonker_ucxb06, bassa-ucb06}). Contrary to
cataclysmic binaries, the place of UCXBs in the scenarios of evolution
of close binaries, their origin and evolution were studied already
before optical identification~\cite{ty81, rj84, tfey85}.

Unlike CVs and UCXBs, the existence of close detached white dwarfs (DDs) was
first deduced from the analysis of scenarios for the evolution of close
binaries~\cite{webbink-79, ty79a, ty81, it84a, Webbink84}. Since it was
also inferred that DDs may be precursors of SNe~Ia, this theoretical
prediction stimulated optical surveys for DDs and the first of them was
detected in 1988 by Saffer, Liebert, and
Olszewski~\cite{slo88}. However, a series of surveys for
DDs performed over a decade~\cite{rob_shaft87, bgr_90,
  fwg91, mdd95, sly98} resulted in only about a dozen of definite
detections~\cite{marsh00}.


\subsection{SPY project}
\label{section:spy}

The major effort to discover DDs was undertaken by the ``ESO Supernovae Ia
Progenitors surveY (SPY)'' project: a systematic radial velocity
survey for DDs with the UVES spectrograph at the ESO VLT (with PI R.~Napiwotzky;
see~\cite{nap_01, nap_spy_03, spn_wd04, nap_dubr, napiwotzki-05} for
the details of the project and~\cite{napiwotzki-05} for the latest
published results). The project was aimed at DDs as potential
progenitors of type~Ia supernovae, but brought as a by-product an
immense wealth of data on white dwarfs. More than 1,000 white dwarfs
and pre-white dwarfs were observed (practically all white dwarfs
brighter than $V\approx16.5$\, available for observations from the ESO
site in Chile). SPY tremendously increased the number of detected DDs
to more than 150. Their system parameters are continuously determined from
follow-up observations. Figure~\ref{figure:pm_spy} shows the total masses
of the currently known close DDs vs.\ orbital periods and compares them with
the Chandrasekhar mass and the critical periods necessary for merger of
components in Hubble time for given $M_\mathrm{tot}$ (data available in
the fall of 2005; R.~Napiwotzky, private communication) Altogether,
$\sim5$
super-Chandrasekhar total mass DDs are expected to be found by SPY. At
the moment, several systems with masses close to the Chandrasekhar
limit and merger time shorter than Hubble time, including a probable
SN~Ia progenitor candidate are already detected. The second candidate
super-Chandrasekhar mass binary white dwarf was discovered by
Tovmassian et al.~\cite{tovmassian-04}.

\epubtkImage{figure07.png}{
  \begin{figure}[htbp]
    \centerline{\includegraphics[scale=0.5]{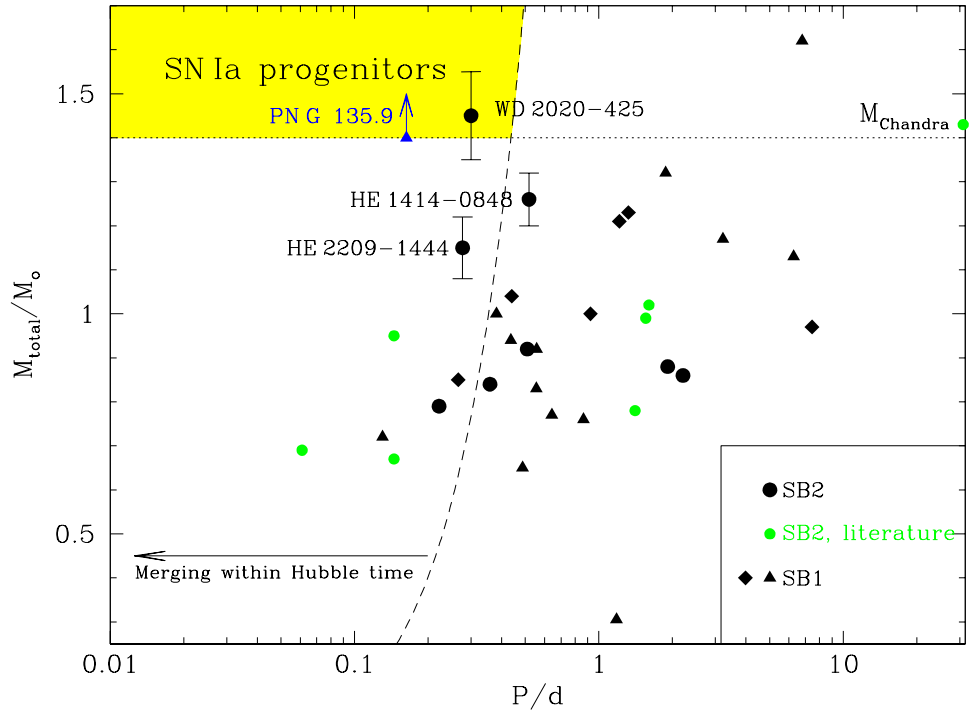}}
    \caption{\it Known close binaries with two WD components,
      or a WD and a sd component. Green circles mark
      systems known prior to the SPY project. Black filled symbols mark
     the positions of DDs and WD\,+\,sd systems detected in
      the SPY project. A blue triangle marks the positions of the WD
      component of the binary planetary nebula nucleus PN~G135.9+55.91
      detected by Tovmassian et al.~\cite{tovmassian-04}. (Courtesy
      R.~Napiwotzki.)}
    \label{figure:pm_spy}
  \end{figure}
}

\epubtkImage{figure08.png}{
  \begin{figure}[htbp]
    \centerline{\includegraphics[scale=0.4]{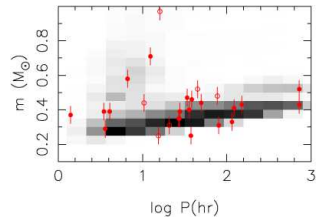}}
    \caption{\it The position of the primary components of known DDs in the
      ``orbital-period--mass'' diagram. The underlying gray scale plot
      is a model prediction from Nelemans et al.~\cite{nyp01}. (Figure
      from~\cite{url02}.)}
    \label{figure:pm_theory}
   \end{figure}
}

Figure~\ref{figure:pm_theory} shows the position of the observed
components of known DDs vs.\ the theoretical expectations with account for
observational selection effects in the $P_\mathrm{orb}$--$m$
diagram~\cite{nyp_01}. The agreement may be considered as quite
satisfactory.

\newpage


\section{Evolution of Interacting Double-Degenerate Systems}
\label{section:am-evol}

Angular momentum losses via GWR may bring detached double degenerates
into contact. The mass--radius relation for degenerate stars has a negative
power ($\simeq -1/3$ for WDs with a mass exceeding $0.1\,M_\odot$,
irrespective of their chemical composition and
temperature~\cite{del_bild_fin03}). Hence, the lower mass WD fills its
Roche lobe first.

In a binary with stable mass transfer the change of the radius of the
donor exactly matches the change of its Roche lobe. This condition
combined with an approximation to the size of the Roche lobe
valid for low $q$~\cite{kopal1959},
\begin{equation}
  R_{\mathrm{L}} \approx
  0.4622 \, a \left(\frac{m}{ M + m}\right)^{1/3}\!\!\!\!\!\!\!\!,
  \label{eq:r_L}
\end{equation}
where $m$ and $M$ are, respectively, the masses of the prospective donor and
accretor, renders two relations important for the study of compact
binaries.
\begin{enumerate}
\item It provides a relation between the orbital period of the
  binary $P_\mathrm{orb}$ and the mass $M_2$ and radius $R_{2}$ of the donor:
  \begin{equation}
    P_{\mathrm{orb}} \simeq 101 \mathrm{\ s}
    \left( \frac{R_2}{0.01\,R_\odot} \right)^{3/2}
    \left( \frac{0.1\,M_\odot}{M_2} \right)^{1/2}\!\!\!\!\!\!\!\!.
    \label{eq:pdrel}
  \end{equation}
\item It allows to derive the rate of mass transfer for
  a semidetached binary in which mass transfer is driven by angular
  momentum losses:
  \begin{equation}
    \frac{\dot{m}}{m} =
    \left ( \frac{\dot{J}}{J} \right )_\mathrm{GWR}
    \!\! \times \left( \frac{\zeta (m)}{2} + \frac{5}{6} -
    \frac{m}{M} \right)^{-1}\!\!\!\!\!\!\!,
    \label{eq:mdot}
  \end{equation}
  where $\zeta (m) = d \ln r/ d \ln m$.
\end{enumerate}
For the mass transfer to be stable, the term in the brackets must be
positive, i.e.\
\begin{equation}
 \label{eq:eqdmdt} 
 \frac{m}{M} < \frac{5}{6} + \frac{\zeta(m)}{2}. 
\end{equation}
Violation of this criterion results in mass loss by the donor on a
dynamical time scale and, most probably, merger of components.
Of course, Equation~(\ref{eq:eqdmdt}) clearly oversimplifies the
conditions for stable mass exchange. A rigorous treatment has to
include a consideration of tidal effects, angular momentum exchange,
and possible super-Eddington $\dot{M}$ immediately after RLOF by the donor and
associated common-envelope formation, possible ignition of accreted
helium~\cite{web_iben87a, hw99, marsh_nel_st04,
  gokhale_stab_am06}. However, the study of these effects is still in
the embryonic state and usually the evolution of IDDs or other binaries
with WD donors is calculated applying $M$--$R$ relations, without
considering detailed models of WDs (or low-mass helium stars in the
appropriate formation channel); see, e.g., \cite{pac67a, vil71,
  faulkner-71, ty79a, rj84, taam_wade_aml85, vr88, ty96, nyp01, ynh02,
  nyp04}. Figure~\ref{figure:dmdtam} shows examples of the evolution for
systems with a helium degenerate donor or a low-mass
``semidegenerate'' helium star donor and a carbon-oxygen accretor
with initial masses that are currently thought to be typical for
progenitors of AM~CVn systems -- $0.2$ and $0.6\,M_\odot$, respectively. For
Figure~\ref{figure:dmdtam} the mass--radius relation for zero-temperature
white dwarfs from the article by Verbunt and Rappaport~\cite{vr88} is
used; for the
low-mass He-star mass--radius relation approximating results of
evolutionary computations by Tutukov and Fedorova~\cite{tf89} were
used: $R \approx 0.043 \, m^{-0.062}$ (in solar units). The same
equations are applied for obtaining the model of the population of
AM~CVn-stars discussed in Sections~\ref{section:waves},
\ref{section:opt+x}, and~\ref{section:gwr+em}.
The mass of the donor in the system may be a discriminator between
the formation channels. For instance, a large mass of the donor $(0.18
\,M_\odot)$ found for the prototype of the class, AM~CVn
itself~\cite{roelofs_am06} favours the helium channel for this system.

\epubtkImage{figure09.png}{
  \begin{figure}[htbp]
    \centerline{\includegraphics[scale=0.5]{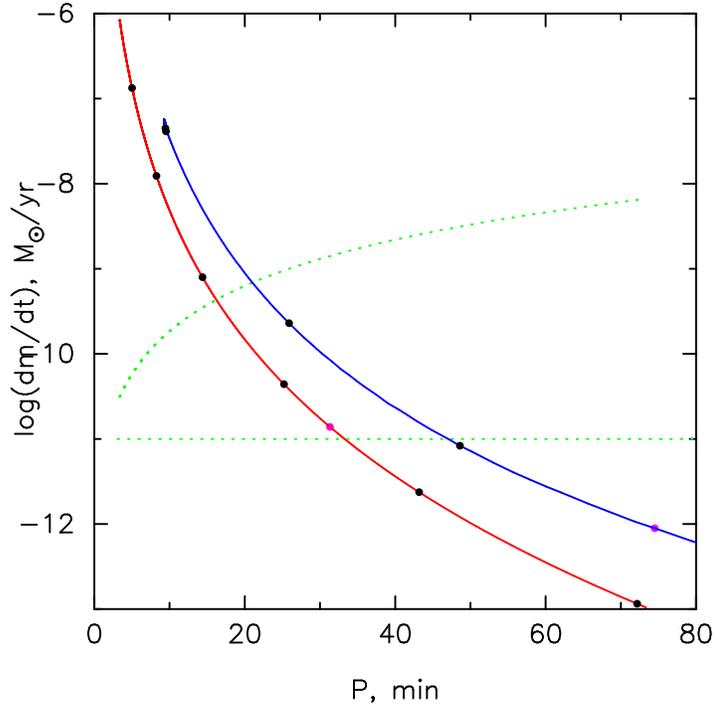}}
    \caption{\it Mass-loss rate vs.\ orbital period for ``typical''
      AM~CVn-stars: an interacting double degenerate system with initial
      masses of donor and accretor $0.2\,M_\odot$ and $0.6\,M_\odot$
      (red line) and a ``semidegenerate'' low-mass helium star donor
      plus white dwarf accretor of the same initial masses (blue
      line). Black dots on the red curve mark positions of the system
      at $\log T \mathrm{\ (yr)} = 5, 6, 7, 8, 9, 10$ from Roche-lobe
      overflow; on the blue curve they mark $\log T \mathrm{\ (yr)} =
      5, 6, 7, 8$. Green lines mark lower and upper limits of the disk
      instability region according to Tsugawa and
      Osaki~\cite{to97}. Below the magenta circles $q <0.02$ and
      conventional evolutionary computations may be not adequate for
      description of mass-transfer process (see the text).}
    \label{figure:dmdtam}
  \end{figure}
}

From the Equations~(\ref{A:GW:dEdt}, \ref{eq:r_L}, \ref{eq:mdot})
it follows that for $m \ll M$ the mass loss rate scales as $M^{1/3}$. As
a result, for all combinations of donor and accretor, the $P$--$\dot{m}$
lines form two rather narrow strips within which they converge with
decreasing $m$. We should note that the time span between formation
of a pair of WDs and contact may be from several Myr to several
Gyr~\cite{ty96}. This means that the approximation of zero-temperature
white dwarfs is not always valid.  Below we discuss the implications
of finite entropy of the donors for the population of AM~CVn-stars.

The ``theoretical'' model of evolution from shorter periods to
longer ones is supported by observations which found that the UV
luminosity of AM~CVn-stars is increasing as the orbital period gets
shorter, since shorter periods are associated with higher
$\dot{M}$~\cite{ramsay_am06}.
 
Note that there is a peculiar difference between white dwarf pairs that
merge and pairs that start stable mass exchange. The pairs that
coalesce stop emitting GWs in a relatively small time-scale (of the
order of the period of the last stable orbit, typically a few
minutes)~\cite{loren_gwr05}. Thus, if we would be lucky to
observe a chirping WD and a sudden disappearance of the signal, this
will manifest a merger. However, the chance of such event is small
since the Galactic occurrence rate of mergers of WDsis $\sim
10^{-2} \mathrm{\ yr}^{-1}$ only.

Apart from ``double-degenerate'' and ``helium-star'' channels for
the formation of AM~CVn-stars, there exists the third,
``CV''-channel~\cite{tfey85, tfey87, phr_am03}. In this channel,
the donor star fills its Roche lobe at the main-sequence stage or just
after its completion. For such donors the chemical inhomogeneity
inhibits complete mixing at $M\simeq0.3\,M_\odot$ typical for initially
non-evolved donors. The mixing is delayed to lower masses and as a
result the donors become helium dwarfs with some traces of
hydrogen. After reaching the minimum period they start to evolve to
the longer ones. The minimum of periods for these systems is $\simeq
5\mbox{\,--\,}7 \mathrm{\ min}$. However, the birth rate of systems
that can penetrate the region occupied by observed AM~CVn-stars is
much lower than the birth rate in ``double degenerate'' and
``helium-star'' channels and we do not take this channel into
account below.

\newpage


\section{Gravitational Waves from Compact Binaries with White-Dwarf Components}
\label{section:waves}

It was expected initially that contact W~UMa binaries will dominate
the gravitational wave spectrum at low
frequencies~\cite{mir65}. However, it was shown in~\cite{ty79a,
  mty_gwr81, it84a, eis87, lp87} 
that it is, most probably, totally dominated by detached and
semidetached double white dwarfs.

As soon as it was recognized that the birth rate of Galactic close
double white dwarfs may be rather high and even before the first close
DD was detected, Evans, Iben, and Smarr in 1987~\cite{eis87}
accomplished an analytical study of the detectability of the signal
from the Galactic ensemble of DDs, assuming certain average parameters
for DDs. Their main findings may be formulated as follows. Let us assume
that there exists a certain distribution of DDs over frequency of the
signal $f$ and strain amplitude $h$: $n(f, h)$. The weakest signal is
$h_\mathrm{w}$. For the time span of observations $\tau_\mathrm{int}$, the
elementary resolution bin of the detector is $\Delta f_\mathrm{int}
\approx 1/\tau_\mathrm{int}$. Then, integration of $n(f, h)$ over
amplitude down to a certain limiting $h$ and over $\Delta f$ gives the
mean number of sources per unit resolution bin for a volume defined by
$h$. If for a certain $h_n$
\begin{equation}
  \int^\infty_{h_n} \left(\frac{dn}{df\,dh}\right)\,\Delta
  \nu_\mathrm{int} = 1,
  \label{eq:confusion}
\end{equation}
then all sources with $h_n>h>h_\mathrm{w}$ overlap. If in a certain resolution
bin $h_n>h_\mathrm{w}$ does not exist, individual sources may be resolved in
this bin for a given integration time (if they are above the detector's
noise level). In the bins where binaries overlap they produce
so-called ``confusion noise'': an incoherent sum of signals; the
frequency, above which the resolution of individual sources becomes
possible got the name of ``confusion limit'', $f_\mathrm{c}$. Evans et~al.\
found $f_\mathrm{c} \approx 10 \mathrm{\ mHz}$ and $3 \mathrm{\ mHz}$ for
integration times $10^6 \mathrm{\ s}$ and $10^8 \mathrm{\ s}$,
respectively.

Independently, the effect of confusion of Galactic binaries was
demonstrated by Lipunov, Postnov, and Prokhorov~\cite{lpp87} who used
simple analytical estimates of the GW confusion background produced by
unresolved binaries whose evolution is driven by GWs only; in this
approximation, the expected level of the background depends solely on
the Galactic merger rate of binary WDs (see~\cite{Grishchuk_al01} for
more details). Later, more involved analytic studies of the
GW background produced by binary stars at low
frequencies were continued in~\cite{fbhhs89, hbw90, pp98, hb00}.

A more detailed approach to the estimate of the GW background is possible 
using population synthesis models~\cite{lp87, wh98, nyp01, nyp04}. 

Nelemans et~al.~\cite{nyp01} constructed a model of the gravitational
wave signal from the Galactic disk population of binaries containing
two compact objects. The model included detached DDs, semidetached DDs,
detached systems of NSs and BHs with WD companions,
binary NSs and BHs. For the details of the model we
refer the reader to the original paper and references
therein. Table~\ref{table:birthrates} shows the number of systems
with different combinations of components in the Nelemans
et al.\ model\epubtkFootnote{These numbers, like also for models
  discussed below, represent one random realization of the model and
  are subject to Poisson noise.}. Note that these numbers strongly depend
on the assumptions in the population synthesis code, especially on the
normalization of stellar birth rate, star formation history,
distributions of binaries over initial masses of components and their
orbital separations, treatment of stellar evolution, common envelope
formalism, etc. For binaries with relativistic components
(i.e.\ descending from massive stars) an additional uncertainty is
brought in by assumptions on stellar wind mass loss and natal
kicks. The factor of uncertainty in the estimated number of systems of
a specific type may be up to a factor $\sim 10$ (cf.~\cite{han98,
  nyp01, ty02, htp02}). Thus these numbers have to be taken with some
caution; we will show the effect of changing some of approximations
below. Table~\ref{table:birthrates} immediately shows that detached
DDs, as expected, dominate the population of compact binaries.

Population synthesis computations yield the ensemble of Galactic
binaries at a given epoch with their specific parameters $M_1$, $M_2$,
and $a$. Figure~\ref{figure:strain} shows examples of the relation
between frequency of emitted radiation and amplitude of the signals from a
``typical'' double degenerate system that evolves into contact and
merges, for an initially detached double degenerate system that stably
exchanges matter after contact, i.e.\ an AM~CVn-type star and its
progenitor, and for an UCXB and its progenitor. For the AM~CVn system
effective spin-orbital coupling is assumed~\cite{nyp01a, mns04}. For
the system with a NS, the mass exchange rate is limited by the
Eddington one and excess of the matter is ``re-ejected'' from the
system'' (see Section~\ref{sec:reemission} and~\cite{ynh02}). Note
that for an AM~CVn-type star it takes only $\sim 300 \mathrm{\ Myr}$
after contact to evolve
to $\log f =-3$ which explains their accumulation at lower $f$. For UCXBs
this time span is only $\sim 20 \mathrm{\ Myr}$. In the discussed model, the
systems are distributed randomly in the Galactic disk according to 
\begin{equation}
  \rho(R, z) = \rho_\mathrm{0} \, e^{-R/H} \,
  \mathrm{sech\,}(z/\beta)^2 \mathrm{\ pc}^{-3},
  \label{eq:rho_gal}
\end{equation}
where $H = 2.5 \mathrm{\ kpc}$~\cite{sac97} and $\beta = 200 \mathrm{\
  pc}$. The Sun is located at $R_\odot = 8.5 \mathrm{\ kpc}$ and
$z_\odot = -30 \mathrm{\ pc}$.

\begin{table}
  \renewcommand{\arraystretch}{1.2}
  \centering
  \begin{tabular}{l|ccc}
    \hline \hline
    Type & Birth rate & Merger rate & Number\\
    \hline
    Detached DD & $2.5 \times 10^{-2}$ & $1.1 \times 10^{-2}$ & $1.1 \times 10^{8}$ \\
    Semidetached DD & $3.3 \times 10^{-3}$ & --- & $4.2 \times 10^{7}$ \\
    NS\,+\,WD & $2.4 \times 10^{-4}$ & $1.4 \times 10^{-4}$ & $2.2 \times 10^{6}$ \\
    NS\,+\,NS & $5.7 \times 10^{-5}$ & $2.4 \times 10^{-5}$ & $7.5 \times 10^{5}$ \\
    BH\,+\,WD & $8.2 \times 10^{-5}$ & $1.9 \times 10^{-6}$ & $1.4 \times 10^{6}$ \\
    BH\,+\,NS & $2.6 \times 10^{-5}$ & $2.9 \times 10^{-6}$ & $4.7 \times 10^{5}$ \\
    BH\,+\,BH & $1.6 \times 10^{-4}$ & --- & $2.8 \times 10^{6}$ \\
    \hline \hline
  \end{tabular}
  \caption{\it Current birth rates and merger rates per year for
    Galactic disk binaries containing two compact objects and their
    total number in the Galactic disk~\cite{nyp01}.}
  \label{table:birthrates}
  \renewcommand{\arraystretch}{1.0}
\end{table}

\epubtkImage{figure10.png}{
  \begin{figure}[htbp]
    \centerline{\includegraphics[scale=0.5]{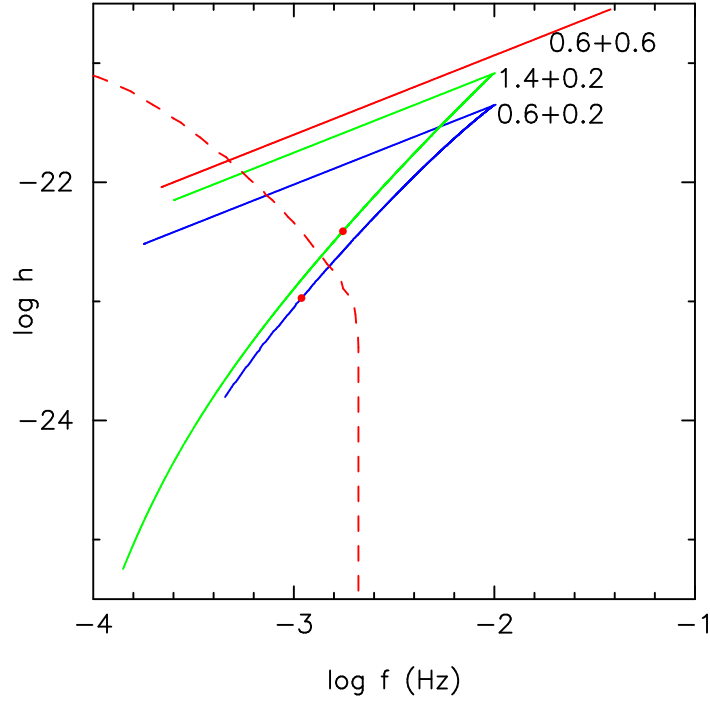}}
    \caption{\it Dependence of the dimensionless strain amplitude for
      a WD\,+\,WD detached system with initial masses of the
      components of $0.6\,M_\odot + 0.6\,M_\odot$ (red line), a
      WD\,+\,WD system with $0.6\,M_\odot + 0.2\,M_\odot$ (blue line)
      and a WD\,+\,NS system with $1.4\,M_\odot + 0.2\,M_\odot$ (green
      line). All systems have an initial separation of components
      $1\,R_\odot$ and are assumed to be at a distance of
      $1 \mathrm{\ kpc}$ (i.e.\ the actual strength of the signal has
      to be scaled with factor $1/d$, with $d$ in $\mathrm{kpc}$). For
      the DD system the line shows an evolution into contact, while
      for the two other systems the upper branches show pre-contact
      evolution and lower branches -- a post-contact evolution with
      mass exchange. The total time-span of evolution covered by the
      tracks is $13.5 \mathrm{\ Gyr}$. Red dots mark the positions of
      systems with mass-ratio of components $q=0.02$ below which the
      conventional picture of evolution with a mass exchange may be
      not valid. The red dashed line marks the position of the
      confusion limit as determined in~\cite{nyp04}.}
    \label{figure:strain}
  \end{figure}
}

\epubtkImage{figure11.png}{
  \begin{figure}[htbp]
    \centerline{\includegraphics[scale=0.6]{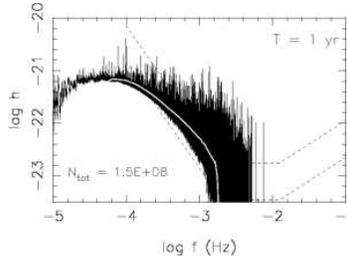}}
    \caption{\it GWR background produced by detached and semidetached
      double white dwarfs as it would be detected at the Earth. The
      assumed integration time is $1 \mathrm{\ yr}$. The `noisy' black line gives
      the total power spectrum, the white line the average. The dashed
      lines show the expected LISA sensitivity for a S/N of 1 and
      5~\cite{lhh00}. Semidetached  double white dwarfs contribute to
      the peak between $\log f \simeq -3.4$ and $-3.0$. (Figure
      from~\cite{nyp01}.)}
    \label{fig:GWR_bg}
  \end{figure}
}

\epubtkImage{figure12.png}{
  \begin{figure}[htbp]
    \centerline{\includegraphics[scale=0.55]{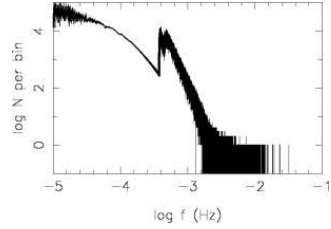}}
    \caption{\it The number of systems per bin on a logarithmic
      scale. Semidetached double white dwarfs contribute to the peak
      between $\log f \simeq -3.4$ and $-3.0$. (Figure
      from~\cite{nyp01}.)}
    \label{fig:GWR_bins}
  \end{figure}
}

\epubtkImage{figure13.png}{
  \begin{figure}[htbp]
    \centerline{\includegraphics[scale=0.45]{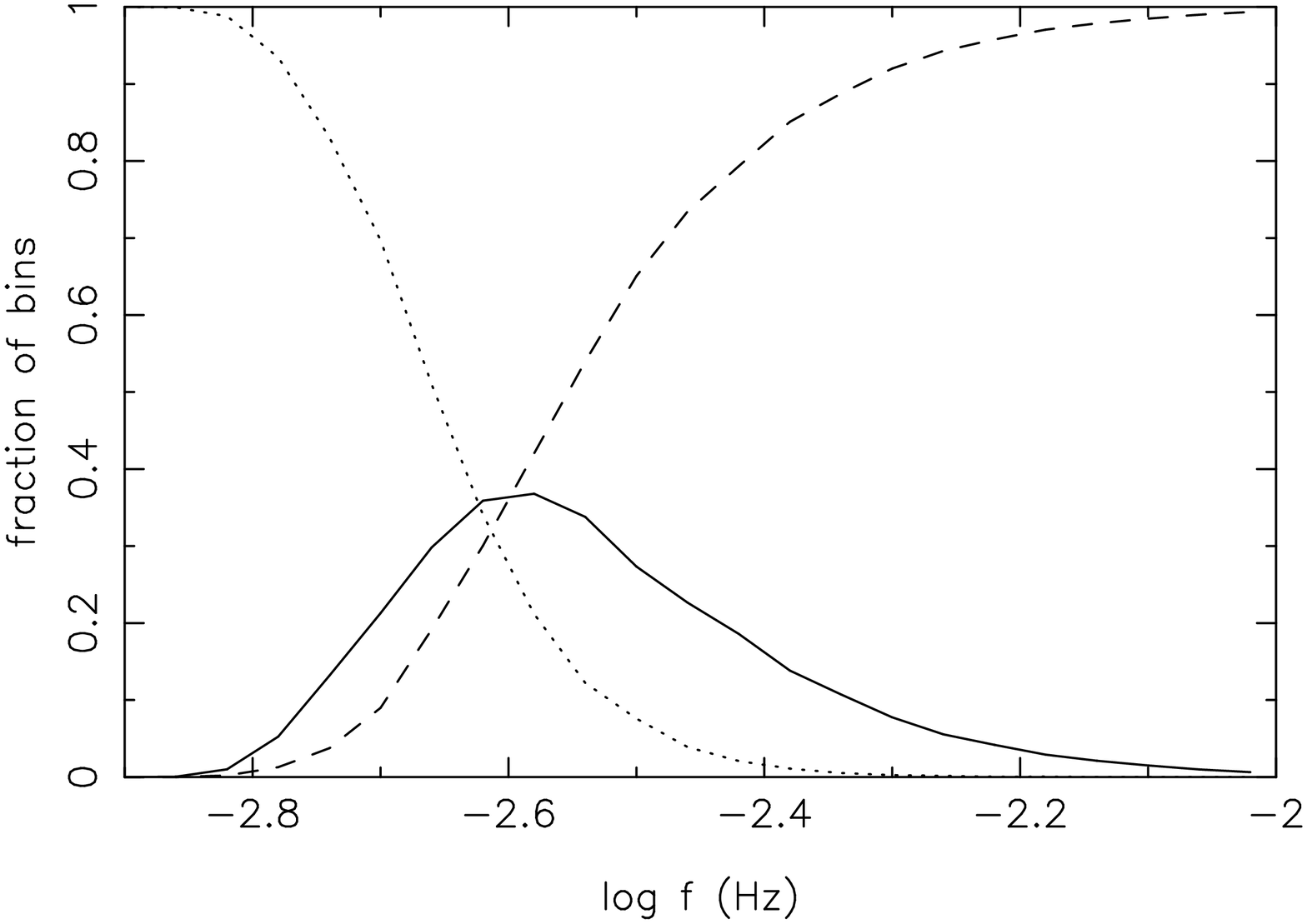}}
    \caption[]{\it Fraction of bins that contain exactly one system
      (solid line), empty bins (dashed line), and bins that contain
      more than one system (dotted line) as function of the frequency of
      the signals. (Figure from~\cite{nyp01}.)}
    \label{fig:GWR_fraction}
  \end{figure}
}

Then it is possible to compute strain amplitude for each system. The
power spectrum of the signal from the population of binaries as it
would be detected by a gravitational wave detector, may be simulated
by computation of the distribution of binaries over $\Delta f = 1/T$ wide
bins, with $T$ being the total integration time. Figure~\ref{fig:GWR_bg}
shows the resulting confusion limited background signal. In
Figure~\ref{fig:GWR_bins} the number of systems per bin is
plotted. The assumed integration time is $T=1 \mathrm{\ yr}$. Semidetached double
white dwarfs, which are less numerous than their detached cousins and
have lower strain amplitude dominate the number of systems per bin in
the frequency interval $-3.4 \lesssim \log f \mathrm{\ (Hz)} \lesssim
-3.0$ producing a peak there, as explained in the comment to
Figure~\ref{figure:strain}.

Figure~\ref{fig:GWR_bg} shows that there are many systems with a signal
amplitude much higher than the average in the bins with $f < f_\mathrm{c}$,
suggesting that even in the frequency range seized by confusion noise
some systems may be detectable above the noise.

Population synthesis also shows that the notion of a unique ``confusion
limit'' is an artifact of the assumption of a continuous distribution of
systems over their parameters. For a discrete population of sources it
appears that for a given integration time there is a range of
frequencies where there are both empty resolution bins and bins containing
more than one system (see Figure~\ref{fig:GWR_fraction}). For this
``statistical'' notion of $f_\mathrm{c}$, Nelemans et al.~\cite{nyp01} get the
first bin containing exactly one system at $\log f \mathrm{\ (Hz)} \approx
-2.84$, while up to $\log f \mathrm{\ (Hz)} \approx -2.3$ there are bins
containing more than one system.

\epubtkImage{figure14.png}{
  \begin{figure}[htbp]
    \centerline{\includegraphics[scale=0.55]{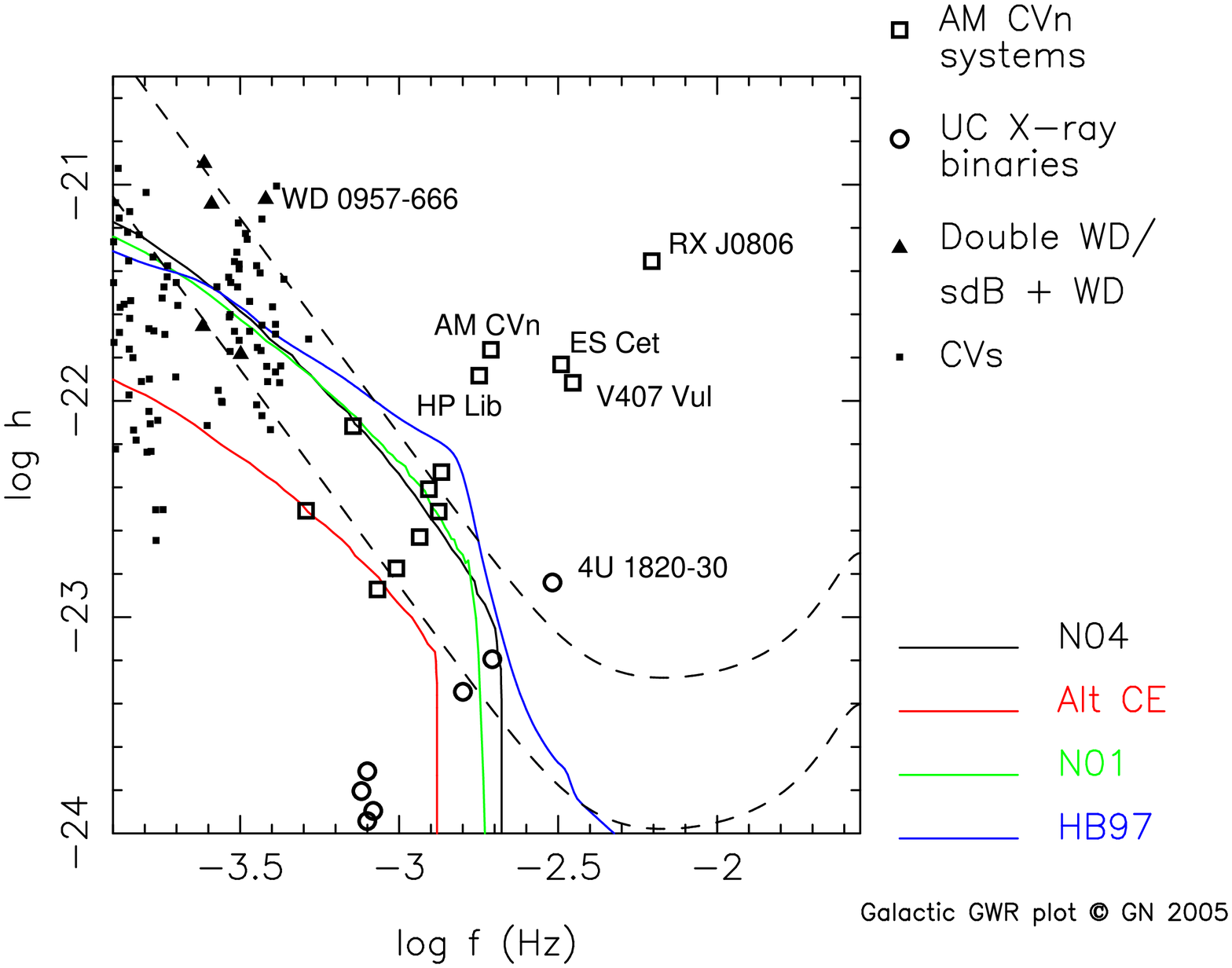}}
    \caption{\it Gravitational waves background formed by the Galactic
      population of white dwarfs and signal amplitudes produced by
      some of the most compact binaries with white dwarf
      components~(\cite{nelemans_amcvn05}). The green line presents
      results from~\cite{nyp01}, the black one from~\cite{nyp04},
      while the red
      one presents a model with all assumptions similar to~\cite{nyp04},
      but with the $\gamma$-formalism for the treatment of common
      envelopes~\cite{Nelemans_al00, nelemans_tout05} (see also
      Section~\ref{A:CE}). The blue line shows the background derived
      in~\cite{bh97}. (Figure from~\cite{url06}.)}
    \label{figure:gwr_back}
  \end{figure}
}

As we noted above, predictions of the population synthesis models are
sensitive to the assumptions of the model; one of the most important
is the treatment of common envelopes (see
Section~\ref{A:CE}). Figure~\ref{figure:gwr_back} compares the average
gravitational waves background formed by Galactic population of white
dwarfs under different assumptions on star formation history, IMF,
assumed Hubble time, and treatment of some details of stellar
evolution (cf.~\cite{nyp01, nyp04}). The comparison with the work of other
authors~\cite{hils98, wh98, seto_dwd02} shows that both the frequency
of the confusion limit and the level of confusion noise are uncertain
within a factor of $\sim 4$. This uncertainty is clearly high enough to
influence seriously the estimates of the possibilities for a detection of
compact binaries. Note, however, that there are systems expected to be
detected above the noise in all models (see below).

\epubtkImage{figure15.png}{
  \begin{figure}[htbp]
    \centerline{\includegraphics[scale=0.5]{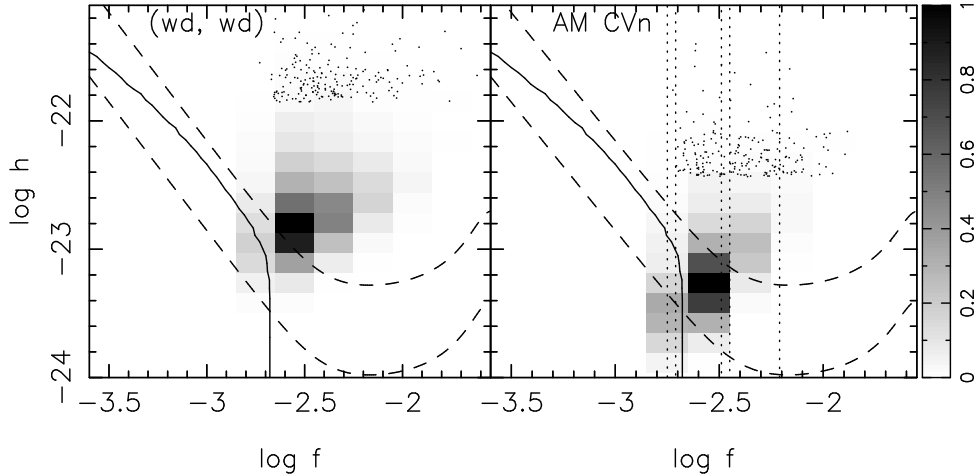}}
    \caption{\it Strain amplitude $h$ as a function of the frequency for
      the model populations of resolved DDs (left panel, $\simeq 10,\!700$
      objects) and AM~CVn systems (right panel, $\simeq 11,\!000$
      objects.). The gray shades give the density distribution of the
      resolved systems normalized to the maximum density in each panel
      ($1,\!548$ and $1,\!823$ per ``cell'' for the double white dwarfs and
      AM~CVn-stars panels, respectively). The 200 strongest sources in
      each sample are shown as dots to enhance their visibility. In
      the AM~CVn panel the periods of several observed short period
      systems are indicated by the vertical dotted lines. The solid line
      shows the average background noise due to detached white
      dwarfs. The LISA sensitivities for an integration time of
      one year and a signal-to-noise ratio of 1 and 5 are indicated by
      the dashed lines~\cite{lhh00}. (Figure from~\cite{nyp04}.)}
    \label{figure:fh}
  \end{figure}
}

Within the model of Nelemans et al.~\cite{nyp01} there are about
12,100 detached DD systems that can be resolved above $f_\mathrm{c}$ and
$\approx 6,\!100$ systems with $f < f_\mathrm{c}$ that are detectable above the
noise. This result was confirmed in a follow-up paper~\cite{nyp04}
which used a more up-to-date SFH and Galaxy model (this resulted in
a slight decrease of the number of ``detectable'' systems -- to
$\sim 11,\!000$). The frequency -- strain amplitude diagram for DD systems is
plotted in the left panel of Figure~\ref{figure:fh}. In the latter
study the following was noted. Previous studies of GW emission of the
AM~CVn systems~\cite{hb00, nyp01} have found that they hardly
contribute to the GW background noise, even despite at $f=(0.3 -
1.0) \mathrm{\ mHz}$ they outnumber the detached DDs. This happens because at
these $f$ their chirp mass $\mathcal{M}$ is well below that of a
typical detached system. But it was overlooked before that at higher
$f$, where the number of AM~CVn systems is much smaller, their
$\mathcal{M}$ is similar to that of the detached systems from which
they descend. It was shown that, out of the total population of
$\sim 140,\!000 $AM~CVn-stars with $P\leq1500 \mathrm{\ s}$, for
$T_{\mathrm{obs}}=1 \mathrm{\ yr}$, LISA may be expected to resolve
$\sim 11,\!000$ AM~CVn systems at $S/N \geq 1$ (or $\sim 3,\!000$ at $S/N
\geq 5$), i.e.\ the numbers of ``resolvable'' detached DDs and
interacting DDs are similar. Given all uncertainties in the input data,
these numbers are in reasonable agreement with the estimates of the
numbers of potentially resolved detached DDs obtained by other authors,
e.g., 3,000 to 6,000~\cite{seto_dwd02, cooray_ecl04}. These numbers
may be compared with expectation that $\sim 10$ Galactic NS\,+\,WD
binaries will be detected~\cite{cooray04}.

The population of potentially resolved AM~CVn-type stars is plotted in the
right panel of Figure~\ref{figure:fh}. Peculiarly enough, as a
comparison of Figures~\ref{figure:gwr_back} and~\ref{figure:fh} shows,
the AM~CVn-type systems appear, in fact, dominant among so-called
``verification binaries'' for LISA: binaries that are well known from
electromagnetic observations and whose radiation is estimated to be
sufficiently strong to be detected; see the list of 30 promising
candidates in~\cite{stroeer06} and references therein, and the
permanently updated (more rigorous) list of these binaries supported
by G.~Nelemans, G.~Ramsay, and T.~Marsh at~\cite{url05}.

Systems RXJ0806.3+1527, V407~Vul, ES~Cet, and AM~CVn are currently
considered as the best candidates. We must note that the most severe
``astronomical'' problems concerning ``verification binaries'' are
their distances, which for most systems are only estimates, and poorly
constrained component masses.

\newpage


\section{AM CVn-Type Stars as Sources of Optical and X-ray Emission}
\label{section:opt+x}

The circumstances mentioned in the previous paragraph stress the
importance of studying AM~CVn-stars in all possible
wavebands. LISA will measure a combination of all the
parameters that determine the GWR signal (frequency, chirp mass,
distance, position in the sky, and inclination angle; see,
e.g., \cite{hell03}), so if some of these parameters (period, position)
can be obtained from optical or X-ray observations, the other
parameters can be determined with higher accuracy. This is
particularly interesting for the distances, inclinations, and masses of
the systems, which are very difficult to measure with other methods.
 
In the optical, the total sample of AM~CVn-type stars is expected to
be dominated by long-period members of the class due to emission of
their disks. But the shortest periods AM~CVn-type stars that are
expected to be observed with LISA may be observed both in
optical and X-rays thanks to high mass-transfer rates (see
Figure~\ref{figure:dmdtam}). A model for electromagnetic-emission
properties of the ensemble of the shortest orbital period
$P\leq 1500 \mathrm{\ s}$ was constructed by Nelemans et al.~\cite{nyp04}.

In~\cite{nyp04}, only systems with He-WD or ``semidegenerate'' He-star
donors were considered (see Figure~\ref{figure:wd_flow}). Systems with
donors descending from strongly evolved MS-stars were excluded from
consideration, since their fraction in the orbital period range
interesting for LISA is negligible. The ``optimistic'' model
of~\cite{nyp01} was considered, which assumes efficient spin-orbital
coupling in the initial phase of mass-transfer and avoids edge-lit
detonations of helium accreted at low $\dot{M}$. Average temperature
and blackbody emission models in $V$-band and in the ROSAT
$0.1\mbox{\,--\,}2.4 \mathrm{\ keV}$ X-ray band were considered,
taking into account interstellar absorption. The ROSAT band
was chosen because of the discovery of AM~CVn itself~\cite{ull95a} and
two candidate AM~CVn systems as ROSAT sources
(RXJ0806.3+1527~\cite{ipc99} and V407~Vul~\cite{mhg_96}) and because
of the possibility for a comparison to the ROSAT all-sky survey.

One may identify four main emission sites: the accretion disc and
boundary layer between the disc and the accreting white dwarf, the
impact spot in the case of direct impact accretion, the accreting
star, and the donor star.

\noindent
{\bf Optical emission.~~}
The luminosity of the disk may be estimated as 
\begin{equation}
  L_\mathrm{disc} = \frac{1}{2} G M \dot{m}
  \left( \frac{1}{R} - \frac{1}{R_\mathrm{L1}} \right)
  \mathrm{\ erg\ s}^{-1},
  \label{eq:L_acc}
\end{equation}
with $M$ and $R$ being the mass and radius of the accretor, $R_\mathrm{L1}$
being the distance of the first Lagrangian point to the centre of the
accretor, and $\dot{m}$ being the mass transfer rate,
respectively. Optical emission of
the disk was modeled as that of a single temperature disc that extends
to 70\% of the Roche lobe of the accretor and radiates as a
blackbody~\cite{wad84}.

The emission from the donor was treated as the emission of a cooling white
dwarf, using approximations to the cooling models of
Hansen~\cite{hansen99}.

The emission from the accretor was treated  as the unperturbed cooling
luminosity of the white dwarf\epubtkFootnote{This is true for
  short-period systems, but becomes an oversimplification at longer
  periods and for $P_{\mathrm{orb}} \gtrsim 40 \mathrm{\ min}$ heating
  of the WD by
  accretion has to be taken into account~\cite{bild_accr06}.}.

A magnitude-limited sample was considered, with $V_\mathrm{lim}=20
\mathrm{\ mag}$,
typical for observed short-period AM~CVn-type stars. Interstellar
absorption was estimated using Sandage's model~\cite{san72} and
Equation~(\ref{eq:rho_gal}).

\noindent
{\bf X-ray emission.~~}
Most AM~CVn systems experience a short ($10^6\mbox{\,--\,}10^7
\mathrm{\ yr}$) ``direct impact'' stage in the beginning of
mass-transfer~\cite{hb00, nyp01, mns04}. Hence, in modeling the X-ray
emission of AM~CVn systems one has to distinguish two cases: direct
impact and disk accretion.

In the case of a direct impact a small area of the accretor's surface is
heated. One may assume that the total accretion luminosity is radiated
as a blackbody with a temperature given by
\begin{equation}
  \left( \frac{T_\mathrm{BB}}{T_{\odot}} \right)^4 =
  \frac{1}{s} R^{-2} L_\mathrm{acc},
\end{equation}
where $L_\mathrm{acc}$ and $R$ are in solar units and $L_\mathrm{acc}$ is
defined by Equation~(\ref{eq:L_acc}). The fraction $s$ of the surface that
is radiating depends on the details of the accretion. It was set to
0.001, consistent with expectations for a ballistic stream~\cite{ls75}
and the observed X-ray emission of V407~Vul, a known direct-impact
system~\cite{mar_ste02}.

In the presence of a disk, X-ray emission was assumed to be coming from
a boundary layer with temperature~\cite{pringle77}
\begin{equation}
  T_\mathrm{BL} = 5 \times 10^{5} 
  \left( \frac{\dot{m}}{10^{18} \mathrm{\ g\ s}^{-1}} \right)^\frac{2}{9} 
  \left( \frac{M}{M_\odot} \right)^\frac{1}{3}
  \left( \frac{R}{5 \times 10^8 \mathrm{\ cm}} \right)^{-\frac{7}{9}}
  \mathrm{\ K}.
\end{equation}
The systems with an X-ray flux in the ROSAT band higher than
$10^{-13} \mathrm{\ erg\ s}^{-1} \mathrm{\ cm}^{-2}$ were selected. Then, the intrinsic
flux in this band, the distance and an estimate of the Galactic
hydrogen absorption~\cite{morrison_abs83} provide an estimate of
detectable flux.

\epubtkImage{figure16.png}{
  \begin{figure}[htbp]
    \centerline{\includegraphics[scale=0.5]{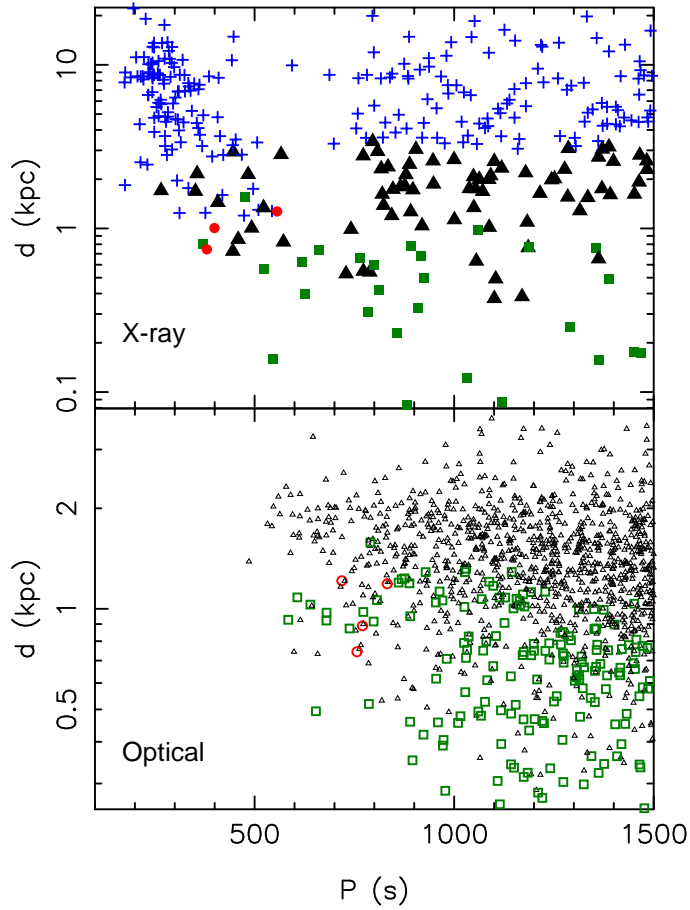}}
    \caption{\it Distribution of short period AM~CVn-type systems
      detectable in soft X-rays and as optical sources as a function of
      orbital period and distance. Top panel: systems detectable in
      X-rays only (blue pluses), direct impact systems observable in
      X-ray and $V$-band (red filled circles), systems detectable in
      X-ray with an optically visible donor (green squares), and systems
      detectable in X-rays and with an optically visible disc (large
      filled triangles). Bottom panel: direct impact systems (red open
      circles), systems with a visible donor (green squares), and systems
      with a visible accretion disc (small open triangles). The overlap of
      these systems with systems observable in gravitational waves is
      shown in Figure~\ref{figure:amcvn_hist}. (Updated figure
      from~\cite{nyp04}, see also~\cite{url07}.)}
    \label{figure:amcvn_obs}
  \end{figure}
}

Figure~\ref{figure:amcvn_obs} presents the resulting model. In the top panel there
are 220 systems only detectable in X-rays and 330 systems also detectable in the
$V$-band. One may distinguish two subpopulations in the top panel: In the
shortest period range there are systems with white-dwarf donors with such high
$\dot{M}$ that even sources close to the Galactic centre are detectable.
Spatially, these objects are concentrated in a small area on the sky. At longer
periods the X-rays get weaker (and softer) so only the systems close to the
Earth can be detected. They are more evenly distributed over the sky. Several of
these systems are also detectable in the optical (filled symbols). There are 30
systems that are close enough to the Earth that the donor stars can be seen as
well as the discs (filled squares). Above $P=600 \mathrm{\ s}$ the systems with
helium-star donors show up and have a high enough mass transfer rate to be X-ray
sources, the closer ones of which are also visible in the optical, as these systems
always have a disc. The bottom panel shows the 1,230 ``conventional'' AM~CVn
systems, detectable only by optical emission, which for most systems
emanates only from their accretion disc. Of this population 170 objects 
closest to the Earth also have a visible donor. The majority of the optically
detectable systems with orbital periods between $1,\!000$ and $1,\!500
\mathrm{\ s}$ are
expected to show outbursts due to the viscous-thermal disc
instability~\cite{to97} which could enhance the chance of their
discovery.

\newpage


\section{AM~CVn-Type Stars Detectable in GWR and Electromagnetic Spectrum}
\label{section:gwr+em} 

Figure~\ref{figure:amcvn_hist}~\cite{nyp04} shows the distributions
vs.\ orbital periods for the total number of AM~CVn systems with $P
\leq 1500 \mathrm{\ s}$ and for AM~CVn LISA sources that have optical
and/or X-ray counterparts. The interrelations between numbers of
sources emitting in different wavebands are shown in the legend to the
right of the figure. Out of 11,000 systems detectable in GWR, 2,060
are expected to be in the direct-impact (DI) stage and only 325 in the
mass-transfer via disk stage. Thus, the majority of the DI systems are
expected to be detectable in GWR; some 5\% of DI systems are expected
to emit X-rays. There are 1,336 systems detectable in the optical waveband
and 326 in X-rays; 106 members of the latter samples may be
detected in both spectral ranges.

\epubtkImage{figure17.png}{
  \begin{figure}[htbp]
    \centerline{\includegraphics[scale=0.47]{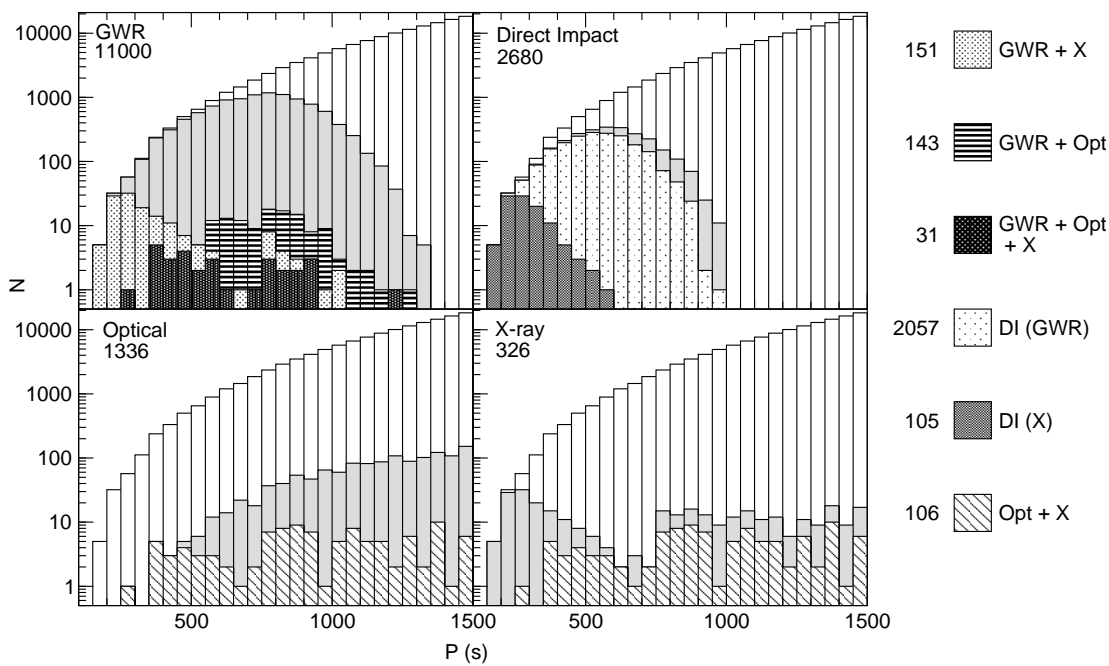}}
    \caption{\it Short-period AM~CVn systems, subdivided in different
      types. Each panel shows the total population as the white
      histogram. The top left panel shows 11,000 systems that can be
      resolved by LISA in gray, and they are subdivided into the
      ones that have optical counterparts (GWR\,+\,Opt), X-ray
      counterparts (GWR\,+\,X), and both (GWR\,+\,Opt\,+\,X). The top right
      panel shows the systems that are in the direct impact phase of
      accretion in gray, and they are subdivided in GWR and X-ray
      sources. The bottom two panels show (again in gray) the
      populations that are detectable in the optical band (left panel)
      and the X-ray band (right panel). The distribution of sources
      detectable both in optical and X-ray bands is shown as hatched
      bins in both lower panels (Opt\,+\,X). (Figure from~\cite{nyp04}.)}
    \label{figure:amcvn_hist}
  \end{figure}
}

An additional piece of information may be obtained from eclipsing
AM~CVn-stars: They would provide radii of the components and orbital
inclinations of the systems. A systematical study of the possibility
of eclipses was never carried out, but an estimate for a ``typical''
system with initial masses of components $(0.25 + 0.60)\,M_\odot$ shows
that the probability for eclipsing of the accretor is about 30\% at
$P=1000 \mathrm{\ s}$, and even higher for eclipsing (a part of) the accretion
disc. The first detection of an eclipsing AM~CVn-type star --
SDSS~J0926+3624 ($P_\mathrm{orb} = 28.3 \mathrm{\ min}$) -- was
recently reported by Anderson et~al.~\cite{anderson_sdss}.

For WD\,+\,WD pairs detectable by LISA the prospects of optical
identification are negligible, since for them cooling luminosity is
the only source of emission. Most of the potentially detectable dwarfs are
located close to the Galactic center and will be very faint. Estimates
based on the model~\cite{nyp04} predict for the bulk of them  $V\approx
35 \mathrm{\ mag}$, with only 75 objects detectable with $V <25 \mathrm{\ mag}$. Even
inclusion of brightening of the dwarfs close to contact under the
assumption of efficient tidal heating~\cite{itf98} increases this
number to $\approx 130$ only (G.~Nelemans, private communication).

In the discussion above, we considered the X-ray flux of AM~CVn-type systems
in the ROSAT waveband: $0.1\mbox{\,--\,}2.4 \mathrm{\
  keV}$. It may be compared with the expected flux in the Chandra
and XMM bands: $0.1\mbox{\,--\,}15 \mathrm{\ keV}$. Since
most of the  spectra of model AM~CVn-stars are rather soft, the flux
$\mathcal{F}$ in the latter band is generally not much larger than
in the ROSAT band: 80\% of systems have
$\mathcal{F}_\mathrm{XMM}/\mathcal{F}_\mathrm{ROSAT} < 1.5$; 96\%
have $\mathcal{F}_\mathrm{XMM}/\mathcal{F}_\mathrm{ROSAT} <
3$. However, Chandra and XMM have much higher
sensitivity. For instance, the Chandra observations of the
Galactic centre have a completeness limit of $3 \times 10^{-15}
\mathrm{\ erg\ cm}^{-2} \mathrm{\ s}^{-1}$, almost two orders of magnitude deeper than
our assumed ROSAT limit~\cite{muno_chandra_cat03}. The expected
number of X-ray sources in the $0.1\mbox{\,--\,}15 \mathrm{\ keV}$
band detectable down to $10^{-14} \mathrm{\ erg\ cm}^{-2} \mathrm{\ s}^{-1}$ is 644 and
it is 1085 down to $10^{-15} \mathrm{\ erg\ cm}^{-2} \mathrm{\ s}^{-1}$. In the
Chandra mosaic image of the Galactic
centre~\cite{wang_chandra02}, roughly down to $10^{-14} \mathrm{\ erg\
  cm}^{-2} \mathrm{\ s}^{-1}$ there are $\sim 1,\!000$ point sources, presumably
associated with accreting white dwarfs, neutron stars, and black
holes. Model~\cite{nyp04} predicts 16 X-ray systems in this region.


\subsection{Effects of finite entropy}
\label{section:fin_entr}

Results concerning the AM~CVn-stars population presented above were
obtained assuming a mass--radius relation for zero-temperature
WDs. Evidently, this is a quite crude approximation, having in mind that
in some cases the time span between the emergence of the second white dwarf
from the common envelope and contact may be as short as several
Myr~\cite{ty96}. As the first step to more realistic models, Deloye
and coauthors~\cite{del_arbitr05} considered the effects of finite
entropy of the donors by using their finite-entropy models for white
dwarfs~\cite{del_bild_fin03}. We illustrate some of these effects
following~\cite{del_arbitr05}.

\epubtkImage{figure18.png}{
  \begin{figure}[htbp]
    \centerline{\includegraphics[scale=1.1]{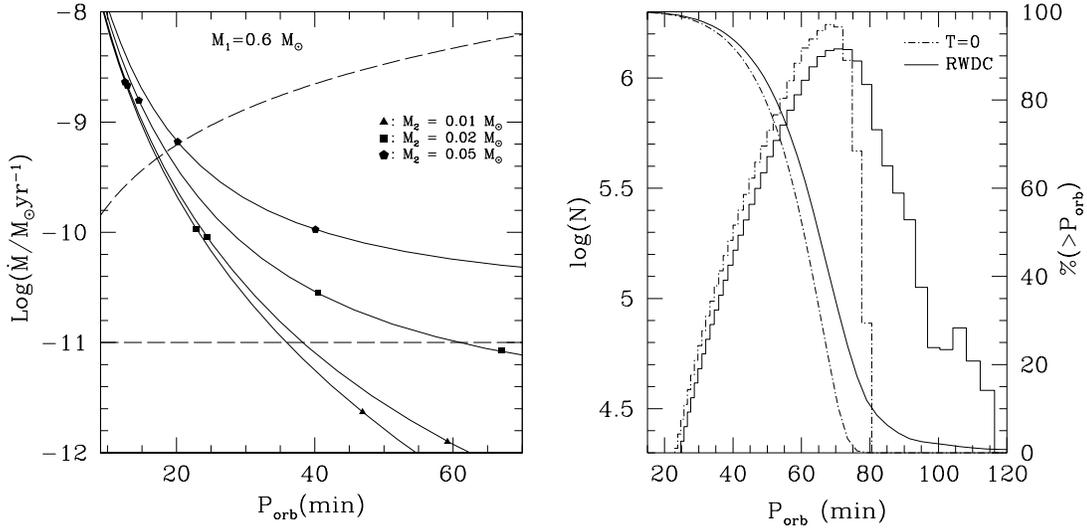}}
    \caption{\it Effects of a finite entropy of donors on the properties
      of AM~CVn-stars. The left panel shows the relation between $P_\mathrm{orb}$
      and $\dot{M}$ along tracks for a system with initial masses of
      components $0.2\,M_\odot$ and $0.6\,M_\odot$ (like in
      Figure~\ref{figure:dmdtam}). The solid lines show the evolution
      for donors with $T_\mathrm{c}=10^4 \mathrm{\ K}$, $10^6
      \mathrm{\ K}$, $5\times10^{6} \mathrm{\ K}$, and
      $10^{7} \mathrm{\ K}$ (left to right). The symbols show the positions of models
      with $M_2 = 0.01\,M_\odot$ (triangles), $0.02\,M_\odot$ (squares), and
      $0.05\,M_\odot$ (pentagons). The disk stability criteria (for $q=0.05$)
      are shown by the dashed lines (after~\cite{to97}). The right
      panel compares the numbers of systems as a function of
      $P_\mathrm{orb}$ for the model with a $T=0$ WD (dot-dashed line)
      and the model with ``realistic'' cooling (solid lines,
      RWDC). The smooth curves show the percentage of each population
      laying above a given $P_\mathrm{orb}$. (Figures
      from~\cite{del_arbitr05}.)}
    \label{figure:deloye1}
  \end{figure} 
}

The effects of finite entropy become noticeably important for
$M\lesssim 0.1\,M_\odot$. Isentropic WD with $T>0$ (i) have larger radii than
$T=0$ objects and (ii) the $M$--$R$ relations for them are steeper than
for $T=0$ (i.e.\ in the range of interest they are still negative but
have a lower absolute value). By virtue of Equations~(\ref{eq:pdrel},
\ref{eq:mdot}) this means that for a given orbital period they have
higher $\dot{M}$. This effect is illustrated in the left panel of
Figure~\ref{figure:deloye1}. (The period--mass relation is not
single-valued, since the $M$--$R$ relation has two branches: a branch where
the object is thermally supported and a branch where degenerate
electrons provide the dominant pressure support.) The right panel of
Figure~\ref{figure:deloye1} compares a model of the population of
AM~CVn-stars computed under assumptions that the donor white dwarfs have
$T=0$ and a model which takes into account cooling of the prospective
donors between formation and RLOF. The change in the rate of evolution
(shown in the left panel) shifts systems with ``realistic'' cooling to
longer orbital periods as compared to the $T=0$ population.

Finite entropy of the donors also influences the gravitational waves
signals from AM~CVn-stars. Again, by virtue of the requirement of
$R_\mathrm{donor}=R_\mathrm{L}$, the systems with $T=0$ donors and hot donors will
have a different $P_\mathrm{orb}$ for the same combination of component
masses, i.e.\ different radii at the contact and different relation
between chirp mass $\mathcal{M}$ and $P_\mathrm{orb}$. This alters the GW
amplitude, Equation~(\ref{A:meanh}.

For instance, if a $0.2\,M_\odot$ donor is fully degenerate, it
overflows its Roche lobe at $P_\mathrm{orb} \approx 3.5 \mathrm{\ min}$ and then
evolves to longer $P_\mathrm{orb}$. If ``realistic'' cooling is
considered, there are donors that make contact at $P_\mathrm{orb}$ up to
$\approx 25 \mathrm{\ min}$. Hotter donors at fixed $P_\mathrm{orb}$
are more massive, increasing $\mathcal{M}$ and increasing $h$. Thus, the
contribution of the individual systems to the integrated GW flux from the
total ensemble increases, but their higher rate of evolution decreases
the density of population of the sources detectable at low $f$, since they
are lost in the background confusion noise of Galactic WDs. But
altogether, the ensemble of sources detectable by LISA with $S/N>1$
diminishes by about 10\% only. Note that finite entropy of donors does not
significantly affect the properties of the $\sim 10,\!000$ systems that are
expected to be observed both in electromagnetic spectrum and
gravitational waves.

There are some more subtle effects related to finite entropy for which
we refer the reader to the original paper.


\subsection{Going further}
\label{section:further_am}

The major issue concerning compact binaries -- DDs, IDDs, UCXBs -- is
their number. Theoretical predictions strongly depend on the assumed
parameters and range within an order of magnitude (see references in
Section~\ref{section:waves}). The treatment of common envelopes and the
distribution of stars over $q$ are, perhaps, the crux. On the other hand,
observational estimates suffer from numerous selection effects. For
instance, the estimates of the local space density of white dwarfs may
be uncertain by a factor of $\sim5$:
cf.\ $4.2\times 10^{-3} \mathrm{\ pc}^{-3}$~\cite{khh99},
$(5\pm0.7)\times 10^{-3} \mathrm{\ pc}^{-3}$~\cite{holberg_wd02}, or
$(20\pm7)\times 10^{-3} \mathrm{\ pc}^{-3}$~\cite{fes98}. The problem with
AM~CVn-stars is their deficiency by a factor of several compared even to
the most pessimistic predictions, but in this case the situation improves:
about half of the known and candidate systems were found within the past
decade, while for another half it took about 30 years. In contrast to
DDs, systematic searches for AM~CVn-stars started only recently; in
fact, the majority of known objects were found serendipitously. Rather
successful was the search for AM~CVn-candidates in the SDSS catalogue and
follow-up observations which brought two confirmed and three candidate
stars~\cite{roelofs_am05, anderson_sdss}. Further progress may be
expected from dedicated surveys with wide-field cameras. ``RApid
Temporal Survey'' (RATS) by Ramsey and Hakala~\cite{rats05} is aimed
at discovering objects with periodicities down to $2 \mathrm{\ min}$ and is
sensitive to $V \sim 22.5$. Preliminary results~\cite{ramsay_rats2}
yielded no new AM~CVn-stars, though several objects were expected in
the 6 square degrees of the sky covered so far. This may be related to the fact
that initially the interstellar extinction for these objects, which have
to be located predominantly in the Galactic plane, was
underestimated. 

Another promising project is ``OmegaWhite'' (with PI P.~Groot,
University of Nijmegen in the Netherlands); see~\cite{url08}.

This project is also aimed at compact binaries, it will cover 400
square degrees of the sky, and will provide broad-band and narrow band
photometric information down to 21st magnitude or equivalent line
fluxes.

Another key issue in the studies of detached and interacting binary
white dwarfs is the determination of their distances. In this aspect,
the Gaia space probe which is aimed at the creation of a
three-dimensional chart of
our Galaxy by providing positional and radial velocity measurements
for about one billion stars (see~\cite{url09}) may appear the most
promising source of information. 

A failure to discover a significant number of detached and interacting
double degenerates or to confirm current ideas on their structure and
evolution will mean that serious drawbacks exist either in the
implementation of known stellar evolution physics and observational
statistical data in the population synthesis codes or in our
understanding of the processes occurring in compact binaries or in the
treatment of selection effects. Special attention in theoretical
  studies has to be paid to the onset of mass-transfer.

Above, we presented some of the current ideas on formation and compact
binaries that may be interesting for general relativity and cosmology
and on signals that may be expected from them in the LISA
waveband. There is another side of the problem -- the analysis of the
signal, would it be detected. This topic is out of the scope of this
brief review. We refer the reader only to several papers discussing
methods for detecting and subtracting individual binary signals from a
data stream with many overlapping signals~\cite{cornish_larson03},
inferring properties of the distribution of white-dwarf
binaries~\cite{edlund05}, the accuracy of parameter estimation of low-mass
binaries~\cite{takahashi_wd02, rogan_lisa06}, and the discussion of the
quality of information that may be provided by LISA~\cite{svn_lisa06}.

\newpage


\section{Conclusions}
\label{sec:concl}

The current understanding of the evolution of binary stars is
firmly based on observations of many types of binary systems, from
wide non-interacting pairs to very close compact binaries consisting
of remnants of stellar evolution. The largest uncertainties in the
specific parameters of the double compact binary formed at the end
of the evolution of a massive binary system are related to the physical
properties of the pre-supernovae: masses, magnetic fields, equation
of state (for NSs), spins, possible kick velocities, etc. This
situation is due to our limited understanding of both the late stages of
stellar evolution and especially of the supernovae
explosion mechanisms. The understanding of the origin and
evolution of compact white dwarf binaries also suffers from
incompleteness of our knowledge of white dwarf formation and, in
particular, on the common envelope treatment. The progress in these
fields, both observational and theoretical, will have a major effect
on the apprehension of the formation and evolution of compact binary
systems. On the other hand, the phenomenological approach used to
describe these uncertainties proves to be successful in explaining
many observed properties of various binary stars, so the
constraints derived from the studies of binary stars should be taken
into account in modeling stellar evolution and supernovae
explosions.

Of course, specifying and checking the initial distributions of
orbital parameters of binary stars and parameters of binary evolution
(such as evolution in the common envelopes) stay in the short-list of
the important actions to be done. Here an essential role belongs to
detailed numerical simulations.

\noindent
{\bf Further observations of compact binaries.~~}
Clearly, discoveries of new types of compact binary systems have provided the
largest impetus for studies of binary star evolution. The well
known examples include
the discovery of X-ray binaries, relativistic binary pulsars, and
millisecond recycled pulsars. In the nearest future we expect the
discovery of NS\,+\,BH binaries which are predicted by the massive binary
evolution scenario in the form of binary radio pulsars with
BH companions~\cite{Lipunov_al94, Lipunov_al05, Pfahl_al05}. It is
very likely that we already observe the coalescence of double NS/BH
systems as short gamma-ray bursts in other
galaxies~\cite{Gehrels_al05}. We also anticipate that coalescences of
NS\,+\,BH or BH\,+\,BH binaries can be found first in GW data
analysis~\cite{Grishchuk_al01}. The efforts of the LIGO collaboration
to put constraints on the compact binary coalescences from the
analysis of the existing GW observations are very
important~\cite{Abbott_al04, Abbott_al06a, Abbott_al06b}.

The formation and evolution of compact binaries is a very
interdisciplinary field of modern astrophysics, ranging from
studies of the equation of state for superdense matter inside neutron
stars and testing effects of strong gravity in relativistic compact
binaries to hydrodynamical effects in stellar winds and formation of
common envelopes. So, further progress in this field, which
will be made by means of traditional astronomical
observations and new tools, like gravitational wave
and neutrino detectors, will undoubtedly have an impact on
astronomy and astrophysics as a whole.

\newpage


\section{Acknowledgments}
\label{section:acknowledgements}

The authors acknowledge the referee for careful reading of the
manuscript and useful comments. The authors would like to thank
L.P.~Grishchuk, V.M.~Lipunov, M.E.~Prokhorov, A.G.~Kuranov,
A.~Tutukov, M.~Livio, G.~Dubus, J.-P.~Lasota, E.P.J.~van den Heuvel, and
R.~Napiwotzki for numerous useful discussions and joint research in
the evolution of binary stars. Writing of this review would have been
impossible without the long-term research cooperation with G.~Nelemans. We
acknowledge him also for useful discussions, help in collecting
information, and updating figures. Useful discussions with participants
of the First Nijmegen Workshop on AM~CVn-stars (July 2005) are
acknowledged. We thank P.~Groot for providing information on the
``OmegaWhite''-project. An intensive use of the Smithsonian/NASA ADS
Astronomy Abstract Service and arXiv is acknowledged. The work is
partially supported by the RFBR grants 06-02-16025, 04-02-16720 and by
Russian Academy of Sciences Basic Research Program ``Origin and
Evolution of Stars and Galaxies''.

\newpage


\bibliography{refs}

\begin{thebibliography}{100}

\bibitem{Abbott_al06a}
Abbott, B., Abbott, R., Adhikari, R., Ageev, A., Agresti, J., Ajith, P., Allen,
  B., Allen, J., Amin, R., Anderson, S.B., and coauthors, 422, ``Search for
  gravitational waves from binary black hole inspirals in LIGO data'', {\em
  Phys. Rev. D}, {\bf 73}, 1--17, 062001, (2006). Related online version (cited
  on 4 November 2006):
  \newline\url{http://adsabs.harvard.edu/abs/2006PhRvD..73f2001A}.
  \epubtkKeywords{Compact binaries, Gravitational wave sources, Gravitational
  wave data analysis}

\bibitem{Abbott_al06b}
Abbott, B., Abbott, R., Adhikari, R., Ageev, A., Agresti, J., Ajith, P., Allen,
  B., Allen, J., Amin, R., Anderson, S.B., and coauthors, 519, ``Joint LIGO and
  TAMA300 search for gravitational waves from inspiralling neutron star
  binaries'', {\em Phys. Rev. D}, {\bf 73}, 1--10, 102002, (2006). Related
  online version (cited on 4 November 2006):
  \newline\url{http://adsabs.harvard.edu/abs/2006PhRvD..73j2002A}.
  \epubtkKeywords{Compact binaries, Gravitational wave sources, Gravitational
  wave data analysis}

\bibitem{Abbott_al04}
Abbott, B. et al. (LIGO Scientific~Collaboration), ``Analysis of LIGO data for
  gravitational waves from binary neutron stars'', {\em Phys. Rev. D}, {\bf
  69}, 1--16, 122001, (2004). \epubtkKeywords{Gravitational wave observations,
  Relativistic binary systems}

\bibitem{Abt83}
Abt, H.A., ``Normal and abnormal binary frequencies'', {\em Annu. Rev. Astron.
  Astrophys.}, {\bf 21}, 343--372, (1983). \epubtkKeywords{Astronomical
  observations, Binary systems}

\bibitem{Acernese_al05}
Acernese, F. et al. (VIRGO~Collaboration), ``Status of Virgo'', {\em Class.
  Quantum Grav.}, {\bf 22}, S869--S880, (2005). Related online version (cited
  on 21 June 2006): \newline\url{http://arXiv.org/abs/gr-qc/0406123}.
  Proceedings of the 9th Gravitational Wave Data Analysis Workshop, Annecy,
  France, 15--18 December 2004. \epubtkKeywords{Gravitational waves,
  Gravitational wave detectors}

\bibitem{alexander-76}
Alexander, M.E., Chau, W.Y., and Henriksen, R.N., ``Orbital evolution of a
  singly condensed, close binary, by mass loss from the primary and by
  accretion drag on the condensed member'', {\em Astrophys. J.}, {\bf 204},
  879--888, (1976). Related online version (cited on 21 June 2006):
  \newline\url{http://adsabs.harvard.edu/abs/1976ApJ...204..879A}.
  \epubtkKeywords{Close binaries}

\bibitem{Anderson_al90}
Anderson, S.B., Gorham, P.W., Kulkarni, S.R., Prince, T.A., and Wolszczan, A.,
  ``Discovery of two radio pulsars in the globular cluster M15'', {\em Nature},
  {\bf 346}, 42--44, (1990). \epubtkKeywords{Pulsars, Relativistic binary
  systems}

\bibitem{anderson_sdss}
Anderson, S.F., Haggard, D., Homer, L., Joshi, N.R., Margon, B., Silvestri,
  N.M., Szkody, P., Wolfe, M.A., Agol, E., Becker, A.C., Henden, A., Hall,
  P.B., Knapp, G.R., Richmond, M.W., Schneider, D.P., Stinson, G., Barentine,
  J.C., Brewington, H.J., Brinkmann, J., Harvanek, M., Kleinman, S.J.,
  Krzesinski, J., Long, D., Neilsen, E.H., Nitta, A., and Snedden, S.A.,
  ``Ultracompact AM Canum Venaticorum Binaries from the Sloan Digital Sky
  Survey: Three Candidates Plus the First Confirmed Eclipsing System'', {\em
  Astron. J.}, {\bf 130}, 2230--2236, (2005). Related online version (cited on
  21 June 2006):
  \newline\url{http://adsabs.harvard.edu/abs/2005AJ....130.2230A}.
  \epubtkKeywords{Close binaries, White dwarfs}

\bibitem{ashok03a}
Ashok, N.M., ``Infrared study of the first identified helium nova V445
  Puppis'', {\em Bull. Astr. Soc. India}, {\bf 33}, 75, (2005). Related online
  version (cited on 21 June 2006):
  \newline\url{http://adsabs.harvard.edu/abs/2005BASI...33...75A}.
  \epubtkKeywords{Binary systems}

\bibitem{ashok03b}
Ashok, N.M., and Banerjee, D.P.K., ``The enigmatic outburst of V445 Puppis -- A
  possible helium nova?'', {\em Astron. Astrophys.}, {\bf 409}, 1007--1015,
  (2003). Related online version (cited on 21 June 2006):
  \newline\url{http://adsabs.harvard.edu/abs/2003A&A...409.1007A}.
  \epubtkKeywords{Close binaries, White dwarfs}

\bibitem{Babak_Grishchuk00}
Babak, S.V., and Grishchuk, L.P., ``Energy-momentum tensor for the
  gravitational field'', {\em Phys. Rev. D}, {\bf 61}, 024038, (2000).
  \epubtkKeywords{Energy and momentum}

\bibitem{Bagot96}
Bagot, P., ``Boost of the orbital motion in high mass X-ray binaries'', {\em
  Astron. Astrophys.}, {\bf 314}, 576--584, (1996). \epubtkKeywords{Binary
  stars}

\bibitem{Bagot97}
Bagot, P., ``On the progenitors of double neutron star systems'', {\em Astron.
  Astrophys.}, {\bf 322}, 533--544, (1997). \epubtkKeywords{Neutron stars,
  Relativistic binary systems}

\bibitem{bag_spz_lry_grb98}
Bagot, P., Portegies~Zwart, S.F., and Yungelson, L.R., ``Gamma-ray bursts and
  density evolution of neutron star binary mergers'', {\em Astron. Astrophys.},
  {\bf 332}, L57--L60, (1998). Related online version (cited on 4 November
  2006): \newline\url{http://adsabs.harvard.edu/abs/1998A&A...332L..57B}.
  \epubtkKeywords{Gamma-ray bursts, Compact binaries, Gravitational wave
  sources}

\bibitem{baraffe_upper01}
Baraffe, I., Heger, A., and Woosley, S.E., ``On the Stability of Very Massive
  Primordial Stars'', {\em Astrophys. J.}, {\bf 550}, 890--896, (2001). Related
  online version (cited on 21 June 2006):
  \newline\url{http://adsabs.harvard.edu/abs/2001ApJ...550..890B}.
  \epubtkKeywords{Stellar evolution}

\bibitem{Barish_Weiss99}
Barish, B.C., and Weiss, R., ``LIGO and the detection of gravitational waves'',
  {\em Phys. Today}, {\bf 52}, 44--50, (1999). Related online version (cited on
  21 June 2006):
  \newline\url{http://adsabs.harvard.edu/abs/1999PhT....52j..44B}.
  \epubtkKeywords{Gravitational waves, Gravitational wave detectors}

\bibitem{Barkov_al01}
Barkov, M.V., Bisnovatyi-Kogan, G.S., and Lamzin, S.A., ``The Thermal Evolution
  of Thorne--Zytkow Objects'', {\em Astron. Rep.}, {\bf 45}, 230--235, (2001).
  Related online version (cited on 21 June 2006):
  \newline\url{http://adsabs.harvard.edu/abs/2001ARep...45..230B}.
  \epubtkKeywords{Stellar evolution, Neutron stars}

\bibitem{Barthelmy_al05}
Barthelmy, S.D., Chincarini, G., Burrows, D.N., Gehrels, N., Covino, S.,
  Moretti, A., Romano, P., O'Brien, P.T., Sarazin, C.L., Kouveliotou, C., Goad,
  M., Vaughan, S., Tagliaferri, G., Zhang, B., Antonelli, L.A., Campana, S.,
  Cummings, J.R., D'Avanzo, P., Davies, M.B., Giommi, P., Grupe, D., Kaneko,
  Y., Kennea, J.A., King, A., Kobayashi, S., Melandri, A., M\'esz\'aros, P.,
  Nousek, J.A., Patel, S., Sakamoto, T., and Wijers, R.A.M.J., ``An origin for
  short {$\gamma$}-ray bursts unassociated with current star formation'', {\em
  Nature}, {\bf 438}, 994--996, (2005). Related online version (cited on 21
  June 2006): \newline\url{http://adsabs.harvard.edu/abs/2005Natur.438..994B}.
  \epubtkKeywords{Gamma-ray bursts, Relativistic binary stars}

\bibitem{Baryshev_Paturel01}
Baryshev, Y.V., and Paturel, G., ``Statistics of the detection rates for tensor
  and scalar gravitational waves from the Local Galaxy universe'', {\em Astron.
  Astrophys.}, {\bf 371}, 378--392, (2001). Related online version (cited on 4
  November 2006):
  \newline\url{http://adsabs.harvard.edu/abs/2001A&A...371..378B}.
  \epubtkKeywords{Gravitational wave sources, Extragalactic astronomy,
  Supernovae}

\bibitem{bassa-ucb06}
Bassa, C.G., Jonker, P.G., in't Zand, J.J.M., and Verbunt, F., ``Two new
  candidate ultra-compact X-ray binaries'', {\em Astron. Astrophys.}, {\bf
  446}, L17--L20, (2006). Related online version (cited on 21 June 2006):
  \newline\url{http://adsabs.harvard.edu/abs/2006A&A...446L..17B}.
  \epubtkKeywords{X-ray binaries, Neutron stars}

\bibitem{Belczynski_al02}
Belczynski, K., Bulik, T., and Kalogera, V., ``Merger Sites of Double Neutron
  Stars and Their Host Galaxies'', {\em Astrophys. J. Lett.}, {\bf 571},
  L147--L150, (2002). \epubtkKeywords{Relativistic binary systems}

\bibitem{bel_kal02}
Belczynski, K., Kalogera, V., and Bulik, T., ``A Comprehensive Study of Binary
  Compact Objects as Gravitational Wave Sources: Evolutionary Channels, Rates,
  and Physical Properties'', {\em Astrophys. J.}, {\bf 572}, 407--431, (2002).
  Related online version (cited on 21 June 2006):
  \newline\url{http://adsabs.harvard.edu/abs/2002ApJ...572..407B}.
  \epubtkKeywords{Compact binaries, Gravitational wave sources}

\bibitem{Belczhynski_al_startreck}
Belczynski, K., Kalogera, V., Rasio, F.A., Taam, R.E., Zezas, A., Bulik, T.,
  Maccarone, T.J., and Ivanova, N., ``Compact Object Modeling with the
  StarTrack Population Synthesis Code'', {\em Astrophys. J. Suppl. Ser.},
  submitted, (2005). Related online version (cited on 21 June 2006):
  \newline\url{http://arXiv.org/abs/astro-ph/0511811}. \epubtkKeywords{Binary
  stars, Neutron stars}

\bibitem{belcz_short_grb06}
Belczynski, K., Perna, R., Bulik, T., Kalogera, V., Ivanova, N., and Lamb,
  D.Q., ``A Study of Compact Object Mergers as Short Gamma-Ray Burst
  Progenitors'', {\em Astrophys. J.}, {\bf 648}, 1110--1116, (2006). Related
  online version (cited on 4 November 2006):
  \newline\url{http://adsabs.harvard.edu/abs/2006ApJ...648.1110B}.
  \epubtkKeywords{Gamma-ray bursts, Compact binaries}

\bibitem{bel_taam_ucb04}
Belczynski, K., and Taam, R.E., ``Galactic Populations of Ultracompact
  Binaries'', {\em Astrophys. J.}, {\bf 603}, 690--696, (2004). Related online
  version (cited on 21 June 2006):
  \newline\url{http://adsabs.harvard.edu/abs/2004ApJ...603..690B}.
  \epubtkKeywords{Close binaries, Compact binaries}

\bibitem{Benacquista_LRR02}
Benacquista, M., ``Relativistic Binaries in Globular Clusters'', {\em Living
  Rev. Relativity}, {\bf 9}, lrr-2006-2, (2006). URL (cited on 21 June 2006):
  \newline\url{http://www.livingreviews.org/lrr-2006-2}.
  \epubtkKeywords{Gravitational wave sources, Binary systems, Pulsars}

\bibitem{bh97}
Bender, P.L., and Hils, D., ``Confusion noise level due to galactic and
  extragalactic binaries'', {\em Class. Quantum Grav.}, {\bf 14}, 1439--1444,
  (1997). \epubtkKeywords{Compact binaries, Gravitational wave sources}

\bibitem{bbc_90}
Benz, W., Bowers, R.L., Cameron, A.G., and Press, W.H., ``Dynamic mass exchange
  in doubly degenerate binaries. I. 0.9 and 1.2 Mo stars'', {\em Astrophys.
  J.}, {\bf 348}, 647--667, (1990). \epubtkKeywords{Close binaries, White
  dwarfs}

\bibitem{Berger06}
Berger, E., ``The Afterglows and Host Galaxies of Short GRBs: An Overview'', in
  Holt, S.S., Gehrels, N., and Nousek, J.A., eds., {\em Gamma-Ray Bursts in the
  Swift Era}, Sixteenth Maryland Astrophysics Conference, Washington, DC, 29
  November -- 2 December 2005, vol. 836 of AIP Conference Proceedings,
  (American Institute of Physics, Melville, U.S.A., 2006). Related online
  version (cited on 21 June 2006):
  \newline\url{http://adsabs.harvard.edu/abs/2006astro.ph..2004B}.
  \epubtkKeywords{Gamma-ray bursts, Astronomical observations}

\bibitem{Berger_al06}
Berger, E., Fox, D.B., Price, P.A., Nakar, E., Gal-Yam, A., Holz, D.E.,
  Schmidt, B.P., Cucchiara, A., Cenko, S.B., Kulkarni, S.R., Soderberg, A.M.,
  Frail, D.A., Penprase, B.E., Rau, A., Ofek, E., Bell~Burnell, S.J., Cameron,
  P.B., Cowie, L.L., Dopita, M.A., Hook, I., Peterson, B.A., Podsiadlowski, P.,
  Roth, K.C., Rutledge, R.E., Sheppard, S.S., and Songaila, A., ``A New
  Population of High Redshift Short-Duration Gamma-Ray Bursts'', {\em
  Astrophys. J.}, submitted, (2006). Related online version (cited on 4
  November 2006): \newline\url{http://arXiv.org/abs/astro-ph/0611128}.
  \epubtkKeywords{Gamma-ray bursts, Gravitational wave sources}

\bibitem{Berger_al05}
Berger, E., Price, P.A., Cenko, S.B., Gal-Yam, A., Soderberg, A.M., Kasliwal,
  M., Leonard, D.C., Cameron, P.B., Frail, D.A., Kulkarni, S.R., Murphy, D.C.,
  Krzeminski, W., Piran, T., Lee, B.L., Roth, K.C., Moon, D.-S., Fox, D.B.,
  Harrison, F.A., Persson, S.E., Schmidt, B.P., Penprase, B.E., Rich, J.,
  Peterson, B.A., and Cowie, L.L., ``The afterglow and elliptical host galaxy
  of the short {$\gamma$}-ray burst GRB 050724'', {\em Nature}, {\bf 438},
  988--990, (2005). Related online version (cited on 21 June 2006):
  \newline\url{http://adsabs.harvard.edu/abs/2005Natur.438..988B}.
  \epubtkKeywords{Gamma-ray bursts, Astronomical observations}

\bibitem{Bethe_Brown98}
Bethe, H.A., and Brown, G.E., ``Evolution of Binary Compact Objects That
  Merge'', {\em Astrophys. J.}, {\bf 506}, 780--789, (1998).
  \epubtkKeywords{Relativistic binary systems}

\bibitem{Bhattacharya_vandenHeuvel91}
Bhattacharya, D., and van~den Heuvel, E.P.J., ``Formation and evolution of
  binary and millisecond radio pulsars'', {\em Phys. Rep.}, {\bf 203}, 1--124,
  (1991). \epubtkKeywords{Binary systems, Pulsars}

\bibitem{bild_accr06}
Bildsten, L., Townsley, D.M., Deloye, C.J., and Nelemans, G., ``The Thermal
  State of the Accreting White Dwarf in AM Canum Venaticorum Binaries'', {\em
  Astrophys. J.}, {\bf 640}, 466--473, (2006). Related online version (cited on
  21 June 2006):
  \newline\url{http://adsabs.harvard.edu/abs/2006ApJ...640..466B}.
  \epubtkKeywords{Close binaries, White dwarfs}

\bibitem{B-K02}
Bisnovatyi-Kogan, G.S., {\em Stellar Physics, Vol.~2: Stellar Evolution and
  Stability}, (Springer, Berlin, Germany; New York, U.S.A., 2002).
  \epubtkKeywords{White dwarfs, Neutron stars, Black holes, Supernovae}

\bibitem{B-K06}
Bisnovatyi-Kogan, G.S., ``Binary and recycled pulsars: 30 years after
  observational discovery'', {\em Phys. Usp.}, {\bf 176}, 53--68, (2006).
  Related online version (cited on 21 June 2006):
  \newline\url{http://adsabs.harvard.edu/abs/2005A&AT...24..151R}.
  \epubtkKeywords{Binary stars, Pulsars}

\bibitem{BK_Komberg74}
Bisnovatyi-Kogan, G.S., and Komberg, B.V., ``Pulsars and close binary
  systems'', {\em Sov. Astron.}, {\bf 18}, 217--221, (1974). Related online
  version (cited on 21 June 2006):
  \newline\url{http://adsabs.harvard.edu/abs/1974SvA....18..217B}.
  \epubtkKeywords{Neutron stars, Close binaries, Pulsars}

\bibitem{Blaauw61}
Blaauw, A., ``On the origin of the O- and B-type stars with high velocities
  (the ``run-away'' stars), and some related problems'', {\em Bull. Astron.
  Inst. Neth.}, {\bf 15}, 265--290, (1961). \epubtkKeywords{Binary stars,
  Supernovae}

\bibitem{Blinnikov_al84}
Blinnikov, S.I., Novikov, I.D., Perevodchikova, T.V., and Polnarev, A.G.,
  ``Exploding Neutron Stars in Close Binaries'', {\em Sov. Astron. Lett.}, {\bf
  10}, 177, (1984). \epubtkKeywords{Relativistic binary systems, Gamma-ray
  bursts}

\bibitem{Bloom_al05}
Bloom, J.S., Prochaska, J.X., Pooley, D., Blake, C.H., Foley, R.J., Jha, S.,
  Ramirez-Ruiz, E., Granot, J., Filippenko, A.V., Sigurdsson, S., Barth, A.J.,
  Chen, H.-W., Cooper, M.C., Falco, E.E., Gal, R.R., Gerke, B.F., Gladders,
  M.D., Greene, J.E., Hennanwi, J., Ho, L.C., Hurley, K., Koester, B.P., Li,
  W., Lubin, L., Newman, J., Perley, D.A., Squires, G.K., and Wood-Vasey, W.M.,
  ``Closing in on a Short-Hard Burst Progenitor: Constraints from Early-Time
  Optical Imaging and Spectroscopy of a Possible Host Galaxy of GRB 050509b'',
  {\em Astrophys. J.}, {\bf 638}, 354--368, (2006). Related online version
  (cited on 21 June 2006):
  \newline\url{http://adsabs.harvard.edu/abs/2006ApJ...638..354B}.
  \epubtkKeywords{Gamma-ray bursts, Astronomical observations}

\bibitem{bgr_90}
Bragaglia, A., Greggio, L., Renzini, A., and D'Odorico, S., ``Double
  Degenerates among DA white dwarfs'', {\em Astrophys. J.}, {\bf 365},
  L13--L17, (1990). \epubtkKeywords{Stars, White dwarfs}

\bibitem{bp95}
Brandt, N., and Podsiadlowski, P., ``The effects of high-velocity supernova
  kicks on the orbital properties and sky distribution of neutron'', {\em Mon.
  Not. R. Astron. Soc.}, {\bf 274}, 461--484, (1995). \epubtkKeywords{Close
  binaries, Supernovae}

\bibitem{Brandt_Podsiadlowski95}
Brandt, N., and Podsiadlowski, P., ``The effects of high-velocity supernova
  kicks on the orbital properties and sky distributions of neutron-star
  binaries'', {\em Mon. Not. R. Astron. Soc.}, {\bf 274}, 461--484, (1995).
  \epubtkKeywords{Supernovae, Binary systems}

\bibitem{brown_bh00}
Brown, G.E., Lee, C.-H., Wijers, R.A.M.J., and Bethe, H.A., ``Evolution of
  black holes in the Galaxy'', {\em Phys. Rep.}, {\bf 333}, 471--504, (2000).
  Related online version (cited on 4 November 2006):
  \newline\url{http://adsabs.harvard.edu/abs/2000PhR...333..471B}.
  \epubtkKeywords{Black hole formation, Compact binaries}

\bibitem{Brumberg_al75}
Brumberg, V.A., Zeldovich, I.B., Novikov, I.D., and Shakura, N.I., ``Component
  masses and inclination of binary systems containing a pulsar, determined from
  relativistic effects'', {\em Sov. Astron. Lett.}, {\bf 1}, 2--4, (1975).
  \epubtkKeywords{Neutron stars, Relativistic binary systems}

\bibitem{Bulik_Belczynski04}
Bulik, T., and Belczynski, K., ``Compact Object Mergers as Progenitors of Short
  Gamma-Ray Bursts'', {\em Baltic Astron.}, {\bf 13}, 280--283, (2004).
  \epubtkKeywords{Relativistic binary systems, Gamma-ray bursts}

\bibitem{bunning_ritter_me06}
B{\"u}ning, A., and Ritter, H., ``Numerical stability of mass transfer driven
  by Roche lobe overflow in close binaries'', {\em Astron. Astrophys.}, {\bf
  445}, 647--652, (2006). Related online version (cited on 21 June 2006):
  \newline\url{http://adsabs.harvard.edu/abs/2006A&A...445..647B}.
  \epubtkKeywords{Close binaries, Hydrodynamics}

\bibitem{Burgay_al03}
Burgay, M., D'Amico, N., Possenti, A., Manchester, R.N., Lyne, A.G., Joshi,
  B.C., McLaughlin, M.A., Kramer, M., Sarkissian, J.M., Camilo, F., Kalogera,
  V., Kim, C., and Lorimer, D.R., ``An increased estimate of the merger rate of
  double neutron stars from observations of a highly relativistic system'',
  {\em Nature}, {\bf 426}, 531--533, (2003). \epubtkKeywords{Double pulsars,
  Relativistic binary systems}

\bibitem{Burgay_al05}
Burgay, M., Possenti, A., Manchester, R.N., Kramer, M., McLaughlin, M.A.,
  Lorimer, D.R., Stairs, I.H., Joshi, B.C., Lyne, A.G., Camilo, F., D'Amico,
  N., Freire, P.C.C., Sarkissian, J.M., Hotan, A.W., and Hobbs, G.B.,
  ``Long-Term Variations in the Pulse Emission from PSR J0737--3039B'', {\em
  Astrophys. J. Lett.}, {\bf 624}, L113--L116, (2005). Related online version
  (cited on 21 June 2006):
  \newline\url{http://adsabs.harvard.edu/abs/2005ApJ...624L.113B}.
  \epubtkKeywords{Relativistic binary systems, Pulsars}

\bibitem{Burrows_al06}
Burrows, D.N., Grupe, D., Capalbi, M., Panaitescu, A., Patel, S.K.,
  Kouveliotou, C., Zhang, B., M\'esz\'aros, P., Chincarini, G., Gehrels, N.,
  and Wijers, R.A.M., ``Jet Breaks in Short Gamma-Ray Bursts. II: The
  Collimated Afterglow of GRB 051221A'', {\em Astrophys. J.}, submitted,
  (2006). Related online version (cited on 21 June 2006):
  \newline\url{http://arXiv.org/abs/astro-ph/0604320}.
  \epubtkKeywords{Gamma-ray bursts, Astronomical observations}

\bibitem{Camilo_Rasio05}
Camilo, F., and Rasio, F.A., ``Pulsars in Globular Clusters'', in Rasio, F.A.,
  and Stairs, I.H., eds., {\em Binary Radio Pulsars}, Meeting at the Aspen
  Center for Physics, Colorado, 12 -- 16 January 2004, vol. 328 of ASP
  Conference Series,  147--169, (Astronomical Society of the Pacific, San
  Francisco, U.S.A., 2005). Related online version (cited on 21 June 2006):
  \newline\url{http://arXiv.org/abs/astro-ph/0501226}. \epubtkKeywords{Double
  pulsars}

\bibitem{canal_01}
Canal, R., M{\'e}ndez, J., and Ruiz-Lapuente, P., ``Identification of the
  Companion Stars of Type Ia supernovae'', {\em Astrophys. J. Lett.}, {\bf
  550}, L53--L56, (2001). Related online version (cited on 21 June 2006):
  \newline\url{http://adsabs.harvard.edu/abs/2001ApJ...550L..53C}.
  \epubtkKeywords{Supernovae}

\bibitem{captur01}
Cappellaro, E., and Turatto, M., ``Supernova Types and Rates'', in Vanbeveren,
  D., ed., {\em The Influence of Binaries on Stellar Population Studies},
  Conference in Brussels, Belgium, 21 -- 25 August 2000, vol. 264 of
  Astrophysics and Space Science Library,  199, (Kluwer Academic Publishers,
  Dordrecht, Netherlands; Boston, U.S.A., 2001). Related online version (cited
  on 21 June 2006):
  \newline\url{http://adsabs.harvard.edu/abs/2001ibsp.conf..199C}.
  \epubtkKeywords{Supernovae, Astronomical observations}

\bibitem{cassisi_etal98}
Cassisi, S., Iben~Jr, I., and Tornamb\`e, A., ``Hydrogen-Accreting
  Carbon-Oxygen White Dwarfs'', {\em Astrophys. J.}, {\bf 496}, 376--385,
  (1998). \epubtkKeywords{White dwarfs, Accretion}

\bibitem{chevalier93}
Chevalier, R.A., ``Neutron star accretion in a stellar envelope'', {\em
  Astrophys. J. Lett.}, {\bf 411}, L33--L36, (1993). Related online version
  (cited on 4 November 2006):
  \newline\url{http://adsabs.harvard.edu/abs/1993ApJ...411L..33C}.
  \epubtkKeywords{Neutron stars, Accretion, Hydrodynamics}

\bibitem{Chugai84}
Chugai, N.N., ``Pulsar Space Velocities and Neutrino Chirality'', {\em Sov.
  Astron. Lett.}, {\bf 10}, 87, (1984). \epubtkKeywords{Supernovae, Neutron
  stars}

\bibitem{cy04}
Chugai, N.N., and Yungelson, L.R., ``Type Ia supernovae in dense circumstellar
  gas'', {\em Astron. Lett.}, {\bf 30}, 65--72, (2004). Related online version
  (cited on 10 November 2006):
  \newline\url{http://arXiv.org/abs/astro-ph/0308297}. \epubtkKeywords{Stars,
  Supernovae}

\bibitem{Clark_Eardley77}
Clark, J.P.A., and Eardley, D.M., ``Evolution of close neutron star binaries'',
  {\em Astrophys. J.}, {\bf 215}, 311--322, (1977).
  \epubtkKeywords{Relativistic binary systems, Astrophysics}

\bibitem{Clark_al79}
Clark, J.P.A., van~den Heuvel, E.P.J., and Sutantyo, W., ``Formation of neutron
  star binaries and their importance for gravitational radiation'', {\em
  Astron. Astrophys.}, {\bf 72}, 120--128, (1979).
  \epubtkKeywords{Gravitational wave sources, Relativistic binary systems}

\bibitem{colgate71}
Colgate, S.A., ``Neutron-Star Formation, Thermonuclear Supernovae, and
  Heavy-Element Reimplosion'', {\em Astrophys. J.}, {\bf 163}, 221, (1971).
  Related online version (cited on 4 November 2006):
  \newline\url{http://adsabs.harvard.edu/abs/1971ApJ...163..221C}.
  \epubtkKeywords{Supernovae, Neutron stars, White dwarfs}

\bibitem{cooray04}
Cooray, A., ``Gravitational-wave background of neutron star--white dwarf
  binaries'', {\em Mon. Not. R. Astron. Soc.}, {\bf 354}, 25--30, (2004).
  Related online version (cited on 21 June 2006):
  \newline\url{http://adsabs.harvard.edu/abs/2004MNRAS.354...25C}.
  \epubtkKeywords{Relativistic binary systems, Gravitational waves}

\bibitem{cooray_ecl04}
Cooray, A., Farmer, A.J., and Seto, N., ``The Optical Identification of Close
  White Dwarf Binaries in the Laser Interferometer Space Antenna Era'', {\em
  Astrophys. J. Lett.}, {\bf 601}, L47--L50, (2004). Related online version
  (cited on 21 June 2006):
  \newline\url{http://adsabs.harvard.edu/abs/2004ApJ...601L..47C}.
  \epubtkKeywords{White dwarfs, Binary systems, Gravitational wave detectors}

\bibitem{cornish_larson03}
Cornish, N.J., and Larson, S.L., ``LISA data analysis: Source identification
  and subtraction'', {\em Phys. Rev. D}, {\bf 67}, 103001, (2003). Related
  online version (cited on 21 June 2006):
  \newline\url{http://adsabs.harvard.edu/abs/2003PhRvD..67j3001C}.
  \epubtkKeywords{Gravitational wave data analysis}

\bibitem{Corongiu_al06}
Corongiu, A., Kramer, M., Stappers, B.W., Lyne, A.G., Jessner, A., Possenti,
  A., D'Amico, N., and L\"ohmer, O., ``The binary pulsar PSR J1811--1736:
  evidence of a low amplitude supernova kick'', {\em Astron. Astrophys.},
  accepted, (2006). Related online version (cited on 20 November 2006):
  \newline\url{http://arXiv.org/abs/astro-ph/0611436}. \epubtkKeywords{Double
  pulsars, Neutron stars}

\bibitem{Counselman73}
Counselman, C.C., ``Outcomes of Tidal Evolution'', {\em Astrophys. J.}, {\bf
  180}, 307--316, (1973). \epubtkKeywords{Binary stars}

\bibitem{Covino_al05}
Covino, S., Malesani, D., Israel, G.L., D'Avanzo, P., Antonelli, L.A.,
  Chincarini, G., Fugazza, D., Conciatore, M.L., Della~Valle, M., Fiore, F.,
  Guetta, D., Hurley, K., Lazzati, D., Stella, L., Tagliaferri, G., Vietri, M.,
  Campana, S., Burrows, D.N., D'Elia, V., Filliatre, P., Gehrels, N., Goldoni,
  P., Melandri, A., Mereghetti, S., Mirabel, I.F., Moretti, A., Nousek, J.,
  O'Brien, P.T., Pellizza, L.J., Perna, R., Piranomonte, S., Romano, P., and
  Zerbi, F.M., ``Optical emission from GRB 050709: A short/hard GRB in a
  star-forming galaxy'', {\em Astron. Astrophys.}, {\bf 447}, L5--L8, (2006).
  Related online version (cited on 21 June 2006):
  \newline\url{http://adsabs.harvard.edu/abs/2006A&A...447L...5C}.
  \epubtkKeywords{Gamma-ray bursts, Astronomical observations}

\bibitem{Cox_Giuli68}
Cox, J.P., and Giuli, R.T., {\em Principles of Stellar Structure}, (Gordon and
  Breach, New York, U.S.A., 1968). \epubtkKeywords{Astrophysics, Stellar
  evolution}

\bibitem{Crowder_Cornish05}
Crowder, J., and Cornish, N.J., ``Beyond LISA: Exploring future gravitational
  wave missions'', {\em Phys. Rev. D}, {\bf 72}, 083005, (2005). Related online
  version (cited on 21 June 2006):
  \newline\url{http://arXiv.org/abs/gr-qc/0506015}.
  \epubtkKeywords{Gravitational waves, Gravitational wave detectors}

\bibitem{crowther_araa06}
Crowther, P.A, ``Physical Properties of Wolf--Rayet Stars'', {\em Annu. Rev.
  Astron. Astrophys.}, submitted, (2006). Related online version (cited on 4
  November 2006): \newline\url{http://arXiv.org/abs/astro-ph/0610356}.
  \epubtkKeywords{Stars, Stellar winds, Astronomical observations}

\bibitem{Cutler_Thorne02}
Cutler, C., and Thorne, K.S., ``An Overview of Gravitational-Wave Sources'', in
  Bishop, N.T., and Maharaj, S.D., eds., {\em General Relativity and
  Gravitation}, Proceedings of the 16th International Conference on General
  Relativity and Gravitation, Durban, South Africa, 15--21 July 2001,  72--111,
  (World Scientific, Singapore; River Edge, U.S.A., 2002). Related online
  version (cited on 4 November 2006):
  \newline\url{http://arXiv.org/abs/gr-qc/0204090}.
  \epubtkKeywords{Gravitational wave sources}

\bibitem{freitas_ns06}
de~Freitas~Pacheco, J.A., Regimbau, T., Vincent, S., and Spallicci, A.,
  ``Expected Coalescence Rates of Ns--Ns Binaries for Laser Beam
  Interferometers'', {\em Int. J. Mod. Phys. D}, {\bf 15}, 235--249, (2006).
  Related online version (cited on 21 June 2006):
  \newline\url{http://adsabs.harvard.edu/abs/2006IJMPD..15..235D}.
  \epubtkKeywords{Compact binaries, Gravitational wave sources}

\bibitem{dek90}
de~Kool, M., ``Common envelope evolution and double cores of planetary
  nebulae'', {\em Astrophys. J.}, {\bf 358}, 189--195, (1990).
  \epubtkKeywords{Close binaries, White dwarfs, Hydrodynamics}

\bibitem{deMarco_Moe05}
de~Marco, O., and Moe, M., ``Common Envelope Evolution through Planetary Nebula
  Eyes'', in Szczerba, R., Stasi\'nska, G., and G\'orny, S.K., eds., {\em
  Planetary Nebulae as Astronomical Tools}, International Conference in
  Gda\'nsk, Poland, 28 June -- 2 July 2005, vol. 804 of AIP Conference
  Proceedings,  169--172, (American Institute of Physics, Melville, U.S.A.,
  2005). Related online version (cited on 21 June 2006):
  \newline\url{http://adsabs.harvard.edu/abs/2005AIPC..804..169D}.
  \epubtkKeywords{Binary stars}

\bibitem{Delgado-Donate_al04}
Delgado-Donate, E.J., Clarke, C.J., and Bate, M.R.and~Hodgkin, S.T., ``On the
  properties of young multiple stars'', {\em Mon. Not. R. Astron. Soc.}, {\bf
  351}, 617--629, (2004). \epubtkKeywords{Astronomical observations, Binary
  systems}

\bibitem{del_bild_fin03}
Deloye, C.J., and Bildsten, L., ``White Dwarf Donors in Ultracompact Binaries:
  The Stellar Structure of Finite-Entropy Objects'', {\em Astrophys. J.}, {\bf
  598}, 1217--1228, (2003). Related online version (cited on 21 June 2006):
  \newline\url{http://adsabs.harvard.edu/abs/2003ApJ...598.1217D}.
  \epubtkKeywords{Close binaries, White dwarfs}

\bibitem{del_arbitr05}
Deloye, C.J., Bildsten, L., and Nelemans, G., ``Arbitrarily Degenerate Helium
  White Dwarfs as Donors in AM Canum Venaticorum Binaries'', {\em Astrophys.
  J.}, {\bf 624}, 934--945, (2005). Related online version (cited on 21 June
  2006): \newline\url{http://adsabs.harvard.edu/abs/2005ApJ...624..934D}.
  \epubtkKeywords{Binary systems, White dwarfs}

\bibitem{dessart_onecoll06}
Dessart, L., Burrows, A., Ott, C., Livne, E., Yoon, S.-C., and Langer, N.,
  ``Multi-Dimensional Simulations of the Accretion-Induced Collapse of White
  Dwarfs to Neutron Stars'', {\em Astrophys. J.}, {\bf 644}, 1063--1084,
  (2006). Related online version (cited on 21 June 2006):
  \newline\url{http://adsabs.harvard.edu/abs/2006astro.ph..1603D}.
  \epubtkKeywords{White dwarfs, Accretion, Gravitational collapse}

\bibitem{Dewey_Cordes87}
Dewey, R.J., and Cordes, J.M., ``Monte Carlo simulations of radio pulsars and
  their progenitors'', {\em Astrophys. J.}, {\bf 321}, 780--798, (1987).
  \epubtkKeywords{Pulsars, Binary systems}

\bibitem{Dewi_Pols03}
Dewi, J.D.M., and Pols, O.R., ``The late stages of evolution of helium
  star-neutron star binaries and the formation of double neutron star
  systems'', {\em Mon. Not. R. Astron. Soc.}, {\bf 344}, 629--643, (2003).
  Related online version (cited on 20 November 2006):
  \newline\url{http://arXiv.org/abs/astro-ph/0306066}. \epubtkKeywords{Close
  binaries, Neutron stars}

\bibitem{Dewi_Tauris00}
Dewi, J.D.M., and Tauris, T.M., ``On the energy equation and efficiency
  parameter of the common envelope evolution'', {\em Astron. Astrophys.}, {\bf
  360}, 1043--1051, (2000). \epubtkKeywords{Binary systems}

\bibitem{Dewi_vdHeuvel04}
Dewi, J.D.M., and van~den Heuvel, E.P.J., ``The formation of the double neutron
  star pulsar J0737-3039'', {\em Mon. Not. R. Astron. Soc.}, {\bf 349},
  169--172, (2004). \epubtkKeywords{Pulsars, Relativistic binary systems}

\bibitem{stefrap95}
Di~Stefano, R., and Rappaport, S., ``How Large is the Population of Luminous
  Supersoft X-Ray Sources?'', in Fruchter, A.S., Tavani, M., and Backer, D.C.,
  eds., {\em Millisecond Pulsars: A Decade of Surprise}, Conference in Aspen,
  Colorado, 3 -- 7 January 1994, vol.~72 of ASP Conference Series,  155,
  (Astronomical Society of the Pacific, San Francisco, U.S.A., 1995).
  \epubtkKeywords{X-ray binaries, Astronomical observations}

\bibitem{Dorofeev_al85}
Dorofeev, O.F., Rodionov, V.N., and Ternov, I.M., ``Anisotropic Neutrino
  Emission from Beta-Decays in a Strong Magnetic Field'', {\em Sov. Astron.
  Lett.}, {\bf 11}, 123, (1985). \epubtkKeywords{Supernovae, Neutron stars}

\bibitem{downes06}
Downes, R.A., Webbink, R.F., Shara, M.M., Ritter, H., Kolb, U., and Duerbeck,
  H.W., ``Catalog of Cataclysmic Variables (Downes+ 2001--2006)'', web
  interface to database, Harvard-Smithsonian Center for Astrophysics. URL
  (cited on 21 June 2006):
  \newline\url{http://vizier.cfa.harvard.edu/viz-bin/VizieR?-source=V/123A}.
  \epubtkKeywords{Close binaries, White dwarfs}

\bibitem{Drake_Sarna03}
Drake, J.J., and Sarna, M.J., ``X-Ray Evidence of the Common Envelope Phase of
  V471 Tauri'', {\em Astrophys. J. Lett.}, {\bf 594}, L55--L58, (2003).
  \epubtkKeywords{Binary systems}

\bibitem{Duquennoy_Mayor91}
Duquennoy, A., and Mayor, M., ``Multiplicity among solar-type stars in the
  solar neighbourhood. II -- Distribution of the orbital elements in an
  unbiased sample'', {\em Astron. Astrophys.}, {\bf 248}, 485--524, (1991).
  \epubtkKeywords{Astronomical observations, Binary systems}

\bibitem{eck_95}
Eck, C.R., Cowan, J.J., Roberts, D.A., Boffi, F.R., and Branch, D., ``Radio
  Observations of the Type Ia Supernova 1986G as a Test of a Symbiotic-Star
  Progenitor'', {\em Astrophys. J. Lett.}, {\bf 451}, L53, (1995). Related
  online version (cited on 21 June 2006):
  \newline\url{http://adsabs.harvard.edu/abs/1995ApJ...451L..53E}.
  \epubtkKeywords{Supernovae, Astronomical observations}

\bibitem{edlund05}
Edlund, J.A., Tinto, M., Kr{\'o}lak, A., and Nelemans, G.,
  ``White-dwarf--white-dwarf galactic background in the LISA data'', {\em Phys.
  Rev. D}, {\bf 71}, 1--16, 122003, (2005). Related online version (cited on 21
  June 2006): \newline\url{http://adsabs.harvard.edu/abs/2005PhRvD..71l2003E}.
  \epubtkKeywords{Compact binaries, Gravitational wave sources, LISA}

\bibitem{Eggleton83}
Eggleton, P.P., ``Approximations to the radii of Roche lobes'', {\em Astrophys.
  J.}, {\bf 268}, 368--369, (1983). \epubtkKeywords{Binary stars}

\bibitem{Eichler_al89}
Eichler, D., Livio, M., Piran, T., and Schramm, D.N., ``Nucleosynthesis,
  Neutrino Bursts and $\gamma$-Rays from Coalescing Neutron Stars'', {\em
  Nature}, {\bf 340}, 126--128, (1989). \epubtkKeywords{Relativistic binary
  systems, Gamma-ray bursts}

\bibitem{url09}
European Space Agency, ``GAIA homepage'', project homepage, (2006). URL (cited
  on 21 June 2006):
  \newline\url{http://sci.esa.int/science-e/www/area/index.cfm?fareaid=26}.
  \epubtkKeywords{Astronomical observations}

\bibitem{eis87}
Evans, C.R., Iben~Jr, I., and Smarr, L.L., ``Degenerate dwarf binaries as
  promising, detectable sources of gravitational radiation'', {\em Astrophys.
  J.}, {\bf 323}, 129--139, (1987). Related online version (cited on 21 June
  2006): \newline\url{http://adsabs.harvard.edu/abs/1987ApJ...323..129E}.
  \epubtkKeywords{Binary systems, Gravitational wave detectors}

\bibitem{fpr_xray_gc75}
Fabian, A.C., Pringle, J.E., and Rees, M.J., ``Tidal capture formation of
  binary systems and X-ray sources in globular clusters'', {\em Mon. Not. R.
  Astron. Soc.}, {\bf 172}, 15--18, (1975). Related online version (cited on 21
  June 2006): \newline\url{http://adsabs.harvard.edu/abs/1975MNRAS.172P..15F}.
  \epubtkKeywords{Star clusters, X-ray binaries}

\bibitem{fadeyev03a}
Fadeyev, Y.A., and Novikova, M.F., ``Radial Pulsations of Helium Stars with
  Masses from 1 to $10 M_\odot$'', {\em Astron. Lett.}, {\bf 29}, 522--529,
  (2003). Related online version (cited on 21 June 2006):
  \newline\url{http://adsabs.harvard.edu/abs/2003AstL...29..522F}.
  \epubtkKeywords{Stellar evolution}

\bibitem{fadeyev04}
Fadeyev, Y.A., and Novikova, M.F., ``Radial Pulsations of Helium Stars with
  Masses from 10 to $50 M_\odot$'', {\em Astron. Lett.}, {\bf 30}, 707--714,
  (2004). Related online version (cited on 21 June 2006):
  \newline\url{http://adsabs.harvard.edu/abs/2004AstL...30..707F}.
  \epubtkKeywords{Stellar evolution}

\bibitem{fbhhs89}
Faller, J.E., Bender, P.L., Hall, J.L., Hils, D., Stebbins, R.T., and Vincent,
  M.A., ``An antenna for laser gravitational-wave observations in space'', {\em
  Adv. Space Res.}, {\bf 9}, 107--111, (1989). Related online version (cited on
  21 June 2006):
  \newline\url{http://adsabs.harvard.edu/abs/1989AdSpR...9..107F}.
  \epubtkKeywords{LISA}

\bibitem{Faulkner_al05}
Faulkner, A.J., Kramer, M., Lyne, A.G., Manchester, R.N., McLaughlin, M.A.,
  Stairs, I.H., Hobbs, G., Possenti, A., Lorimer, D.R., D'Amico, N., Camilo,
  F., and Burgay, M., ``PSR J1756--2251: A New Relativistic Double Neutron Star
  System'', {\em Astrophys. J. Lett.}, {\bf 618}, L119--L122, (2005).
  \epubtkKeywords{Pulsars, Relativistic binary systems}

\bibitem{Faulkner_al04}
Faulkner, A.J., Stairs, I.H., Kramer, M., Lyne, A.G., Hobbs, G., Possenti, A.,
  Lorimer, D.R., Manchester, R.N., McLaughlin, M.A., D'Amico, N., Camilo, F.,
  and Burgay, M., ``The Parkes Multibeam Pulsar Survey -- V. Finding binary and
  millisecond pulsars'', {\em Mon. Not. R. Astron. Soc.}, {\bf 355}, 147--158,
  (2004). \epubtkKeywords{Pulsars, Binary stars, Astronomical observations}

\bibitem{faulkner-71}
Faulkner, J., ``Ultrashort-Period Binaries, Gravitational Radiation, and Mass
  Transfer. I. The Standard Model, with Applications to WZ Sagittae and Z
  Camelopardalis'', {\em Astrophys. J.}, {\bf 170}, L99, (1971). Related online
  version (cited on 21 June 2006):
  \newline\url{http://adsabs.harvard.edu/abs/1971ApJ...170L..99F}.
  \epubtkKeywords{Close binaries, Gravitational waves}

\bibitem{ffw72}
Faulkner, J., Flannery, B.P., and Warner, B., ``Ultrashort-period binaries. II.
  HZ 29 (= AM CVn): A double-white-dwarf semidetached postcataclysmic nova?'',
  {\em Astrophys. J.}, {\bf 175}, L79, (1972). \epubtkKeywords{White dwarfs,
  Close binaries}

\bibitem{fe89}
Fedorova, A.V., and Ergma, E.V., ``Evolution of binaries with ultra-short
  periods -- Systematic study'', {\em Astrophys. J. Suppl. Ser.}, {\bf 151},
  125--134, (1989). Related online version (cited on 21 June 2006):
  \newline\url{http://adsabs.harvard.edu/abs/1989Ap&SS.151..125F}.
  \epubtkKeywords{Close binaries, Neutron stars}

\bibitem{fty04}
Fedorova, A.V., Tutukov, A.V., and Yungelson, L.R., ``Type Ia Supernovae in
  semi-detached binary systems'', {\em Astron. Lett.}, {\bf 30}, 73--85,
  (2004). Related online version (cited on 21 June 2006):
  \newline\url{http://arXiv.org/abs/astro-ph/0309052}.
  \epubtkKeywords{Supernovae, Binary stars, White dwarfs}

\bibitem{fes98}
Festin, L, ``The luminosity function of white dwarfs and M dwarfs using dark
  nebulae as opaque outer screens'', {\em Astron. Astrophys.}, {\bf 336},
  883--894, (1998). \epubtkKeywords{White dwarfs, Astronomical observations}

\bibitem{figer_upper05}
Figer, D.F., ``An upper limit to the masses of stars'', {\em Nature}, {\bf
  434}, 192--194, (2005). Related online version (cited on 21 June 2006):
  \newline\url{http://adsabs.harvard.edu/abs/2005Natur.434..192F}.
  \epubtkKeywords{Stars}

\bibitem{Flanagan_Hughes98}
Flanagan, {\'{E}}.{\'{E}}., and Hughes, S.A., ``Measuring gravitational waves
  from binary black hole coalescences. I. Signal to noise for inspiral, merger,
  and ringdown'', {\em Phys. Rev. D}, {\bf 57}, 4535--4565, (1998).
  \epubtkKeywords{Relativistic binary systems, Data analysis}

\bibitem{Flannery_vdHeuvel75}
Flannery, B.P., and van~den Heuvel, E.P.J., ``On the origin of the binary
  pulsar PSR 1913+16'', {\em Astron. Astrophys.}, {\bf 39}, 61--67, (1975).
  \epubtkKeywords{Relativistic binary systems, Pulsars}

\bibitem{forster_delay06}
F{\"o}rster, F., Wolf, C., Podsiadlowski, P., and Han, Z., ``Constraints on
  Type Ia supernova progenitor time delays from high-$z$ supernovae and the
  star formation history'', {\em Mon. Not. R. Astron. Soc.}, {\bf 368},
  1893--1904, (2006). Related online version (cited on 21 June 2006):
  \newline\url{http://adsabs.harvard.edu/abs/2006MNRAS.368.1893F}.
  \epubtkKeywords{Supernovae, Astronomical observations}

\bibitem{fwg91}
Foss, D., Wade, R.A., and Green, R.F., ``Limits on the space density of double
  degenerates as type Ia supernova progenitors'', {\em Astrophys. J.}, {\bf
  374}, 281--287, (1991). Related online version (cited on 21 June 2006):
  \newline\url{http://adsabs.harvard.edu/abs/1991ApJ...374..281F}.
  \epubtkKeywords{Compact binaries, Supernovae}

\bibitem{Fox_al05}
Fox, D.B., Frail, D.A., Price, P.A., Kulkarni, S.R., Berger, E., Piran, T.,
  Soderberg, A.M., Cenko, S.B., Cameron, P.B., Gal-Yam, A., Kasliwal, M.M.,
  Moon, D.-S., Harrison, F.A., Nakar, E., Schmidt, B.P., Penprase, B.,
  Chevalier, R.A., Kumar, P., Roth, K., Watson, D., Lee, B.L., Shectman, S.,
  Phillips, M.M., Roth, M., McCarthy, P.J., Rauch, M., Cowie, L., Peterson,
  B.A., Rich, J., Kawai, N., Aoki, K., Kosugi, G., Totani, T., Park, H.-S.,
  {MacFadyen}, A., and Hurley, K.C., ``The afterglow of GRB 050709 and the
  nature of the short-hard {$\gamma$}-ray bursts'', {\em Nature}, {\bf 437},
  845--850, (2005). Related online version (cited on 21 June 2006):
  \newline\url{http://adsabs.harvard.edu/abs/2005Natur.437..845F}.
  \epubtkKeywords{Gamma-ray bursts, Astronomical observations}

\bibitem{Fryer99}
Fryer, C.L., ``Mass Limits For Black Hole Formation'', {\em Astrophys. J.},
  {\bf 522}, 413--418, (1999). \epubtkKeywords{Supernovae, Black holes}

\bibitem{Fryer04}
Fryer, C.L., ed., {\em Stellar Collapse}, Proceedings of ``Core Collapse of
  Massive Stars'', 200th AAS meeting, Albuquerque, NM, June 2002, vol. 302 of
  Astrophysics and Space Science Library, (Kluwer Academic Publishers,
  Dordrecht, Netherlands; Boston, U.S.A., 2004). \epubtkKeywords{Neutron stars,
  Supernovae}

\bibitem{fbb98}
Fryer, C.L., Burrows, A., and Benz, W., ``Population Syntheses for Neutron Star
  Systems with Intrinsic Kicks'', {\em Astrophys. J.}, {\bf 496}, 333, (1998).
  URL (cited on 4 November 2006):
  \newline\url{http://adsabs.harvard.edu/abs/1998ApJ...496..333F}.
  \epubtkKeywords{Close binaries, Supernovae, Neutron stars}

\bibitem{Fryer_Kalogera01}
Fryer, C.L., and Kalogera, V., ``Theoretical Black Hole Mass Distributions'',
  {\em Astrophys. J.}, {\bf 554}, 548--560, (2001). \epubtkKeywords{Supernovae,
  Black holes}

\bibitem{Fukugita_Peebles04}
Fukugita, M., and Peebles, P.J.E., ``The Cosmic Energy Inventory'', {\em
  Astrophys. J.}, {\bf 616}, 643--668, (2004). Related online version (cited on
  21 June 2006): \newline\url{http://arXiv.org/abs/gr-qc/0406095}.
  \epubtkKeywords{Cosmology}

\bibitem{berro_ritossa_one97}
Garc\'ia-Berro, E., Ritossa, C., and Iben~Jr, I., ``On the Evolution of Stars
  That Form Electron-Degenerate Cores Processed by Carbon Burning. III. The
  Inward Propagation of a Carbon-burning Flame and Other Properties of a $9
  M_\odot$ Model Star'', {\em Astrophys. J.}, {\bf 485}, 765, (1997). Related
  online version (cited on 21 June 2006):
  \newline\url{http://adsabs.harvard.edu/abs/1997ApJ...485..765G}.
  \epubtkKeywords{Stellar evolution}

\bibitem{Gehrels_al05}
Gehrels, N., Sarazin, C.L., O'Brien, P.T., Zhang, B., Barbier, L., Barthelmy,
  S.D., Blustin, A., Burrows, D.N., Cannizzo, J., Cummings, J.R., Goad, M.,
  Holland, S.T., Hurkett, C.P., Kennea, J.A., Levan, A., Markwardt, C.B.,
  Mason, K.O., M\'esz\'aros, P., Page, M., Palmer, D.M., Rol, E., Sakamoto, T.,
  Willingale, R., Angelini, L., Beardmore, A., Boyd, P.T., Breeveld, A.,
  Campana, S., Chester, M.M., Chincarini, G., Cominsky, L.R., Cusumano, G.,
  de~Pasquale, M., Fenimore, E.E., Giommi, P., Gronwall, C., Grupe, D., Hill,
  J.E., Hinshaw, D., Hjorth, J., Hullinger, D., Hurley, K.C., Klose, S.,
  Kobayashi, S., Kouveliotou, C., Krimm, H.A., Mangano, V., Marshall, F.E.,
  McGowan, K., Moretti, A., Mushotzky, R.F., Nakazawa, K., Norris, J.P.,
  Nousek, J.A., Osborne, J.P., Page, K., Parsons, A.M., Patel, S., Perri, M.,
  Poole, T., Romano, P., Roming, P.W.A., Rosen, S., Sato, G., Schady, P.,
  Smale, A.P., Sollerman, J., Starling, R., Still, M., Suzuki, M., Tagliaferri,
  G., Takahashi, T., Tashiro, M., Tueller, J., Wells, A.A., White, N.E., and
  Wijers, R.A.M.J., ``A short $\gamma$-ray burst apparently associated with an
  elliptical galaxy at redshift $z = 0.225$'', {\em Nature}, {\bf 437},
  851--854, (2005). Related online version (cited on 21 June 2006):
  \newline\url{http://adsabs.harvard.edu/abs/2005Natur.437..851G}.
  \epubtkKeywords{Astronomical observations, Relativistic binary systems}

\bibitem{ghosh_cyg3analog06}
Ghosh, K.K., Rappaport, S., Tennant, A.F., Swartz, D.A., Pooley, D., and
  Madhusudhan, N., ``Discovery of a 3.6 hr Eclipsing Luminous X-Ray Binary in
  the Galaxy NGC 4214'', {\em Astrophys. J.}, {\bf 650}, 872--878, (2006).
  Related online version (cited on 21 June 2006):
  \newline\url{http://arXiv.org/abs/astro-ph/0604466}. \epubtkKeywords{X-ray
  binaries}

\bibitem{giacconi74}
Giacconi, R., Murray, S., Gursky, H., Kellogg, E., Schreier, E., Matilsky, T.,
  Koch, D., and Tananbaum, H., ``The Third UHURU Catalog of X-Ray Sources'',
  {\em Astrophys. J. Suppl. Ser.}, {\bf 27}, 37, (1974). Related online version
  (cited on 21 June 2006):
  \newline\url{http://adsabs.harvard.edu/abs/1974ApJS...27...37G}.
  \epubtkKeywords{X-ray binaries, Astronomical observations}

\bibitem{gilponsetal03}
Gil-Pons, P., Garc{\'{i}}a-Berro, E., Jos{\' e}, J., Hernanz, M., and Truran,
  J.W., ``The frequency of occurrence of novae hosting an ONe white dwarf'',
  {\em Astron. Astrophys.}, {\bf 407}, 1021--1028, (2003). Related online
  version (cited on 21 June 2006):
  \newline\url{http://adsabs.harvard.edu/abs/2003A&A...407.1021G}.
  \epubtkKeywords{Binary systems, White dwarfs}

\bibitem{gokhale_stab_am06}
Gokhale, V., Meng~Peng, X., and Frank, J., ``Evolution of Close White Dwarf
  Binaries'', (October, 2006). URL (cited on 4 November 2006):
  \newline\url{http://arXiv.org/abs/astro-ph/0610919}. \epubtkKeywords{Compact
  binaries, White dwarfs}

\bibitem{Grindlay_al06}
Grindlay, J., Portegies~Zwart, S.F., and McMillan, S.L.W., ``Short gamma-ray
  bursts from binary neutron star mergers in globular clusters'', {\em Nature
  Phys.}, {\bf 2}, 116--119, (2006). Related online version (cited on 21 June
  2006): \newline\url{http://adsabs.harvard.edu/abs/2006NatPh...2..116G}.
  \epubtkKeywords{Relativistic binary systems, Gamma-ray bursts}

\bibitem{Grishchuk_al01}
Grishchuk, L.P., Lipunov, V.M., Postnov, K.A., Prokhorov, M.E., and
  Sathyaprakash, B.S., ``Gravitational Wave Astronomy: In Anticipation of First
  Sources to be Detected'', {\em Phys. Usp.}, {\bf 44}, 1--51, (2001). Related
  online version (cited on 21 June 2006):
  \newline\url{http://arXiv.org/abs/astro-ph/0008481}.
  \epubtkKeywords{Cosmology, Gravitational waves, Gravitational wave sources,
  Relativistic binary systems}

\bibitem{url08}
Groot, P.J., ``OmegaWhite and VPHAS+'', other, Astro-Wise, (2006). URL (cited
  on 21 June 2006):
  \newline\url{http://www.astro-wise.org/Presentations/LC2005course/astrowisel%
eiden_groot.pdf}. \epubtkKeywords{Close binaries, Astronomical observations,
  Gravitational waves}

\bibitem{guerrero04}
Guerrero, J., Garc{\'{i}}a-Berro, E., and Isern, J., ``Smoothed Particle
  Hydrodynamics simulations of merging white dwarfs'', {\em Astron.
  Astrophys.}, {\bf 413}, 257--272, (2004). \epubtkKeywords{Binary systems,
  White dwarfs, Supernovae}

\bibitem{Gunn_Ostriker70}
Gunn, J.E., and Ostriker, J.P., ``On the Nature of Pulsars. III. Analysis of
  Observations'', {\em Astrophys. J.}, {\bf 160}, 979--1002, (1970).
  \epubtkKeywords{Pulsars, Neutron stars}

\bibitem{hachisu_rec01}
Hachisu, I., and Kato, M., ``Recurrent Novae as a Progenitor System of Type Ia
  Supernovae. I. RS Ophiuchi Subclass: Systems with a Red Giant Companion'',
  {\em Astrophys. J.}, {\bf 558}, 323--350, (2001). Related online version
  (cited on 21 June 2006):
  \newline\url{http://adsabs.harvard.edu/abs/2001ApJ...558..323H}.
  \epubtkKeywords{Supernovae}

\bibitem{hknu99}
Hachisu, I., Kato, M., Nomoto, K., and Umeda, H., ``A New Evolutionary Path to
  Type Ia Supernovae: A Helium-rich Supersoft X-Ray Source Channel'', {\em
  Astrophys. J.}, {\bf 519}, 314--323, (1999). Related online version (cited on
  21 June 2006):
  \newline\url{http://adsabs.harvard.edu/abs/1999ApJ...519..314H}.
  \epubtkKeywords{Supernovae}

\bibitem{Halbwachs_al03}
Halbwachs, J.L., Mayor, M., Udry, S., and Arenou, F., ``Multiplicity among
  solar-type stars. III. Statistical properties of the F7-K binaries with
  periods up to 10 years'', {\em Astron. Astrophys.}, {\bf 397}, 159--175,
  (2003). \epubtkKeywords{Astronomical observations, Binary systems}

\bibitem{hamann_wn06}
Hamann, W.-R., Gr{\"a}fener, G., and Liermann, A., ``The Galactic WN stars.
  Spectral analyses with line-blanketed model atmospheres versus stellar
  evolution models with and without rotation'', {\em Astron. Astrophys.}, {\bf
  457}, 1015--1031, (2006). Related online version (cited on 4 November 2006):
  \newline\url{http://adsabs.harvard.edu/abs/2006A&A...457.1015H}.
  \epubtkKeywords{Stars, Stellar winds}

\bibitem{hamuy03}
Hamuy, M., Phillips, M.M., Suntzeff, N.B., Maza, J., Gonzalez, L.E., Roth, M.,
  Krisciunas, K., Morrell, N., Green, E.M., Persson, S.E., and McCarthy, P.E.,
  ``An asymptotic-giant-branch star in the progenitor system of a type Ia
  supernova'', {\em Nature}, {\bf 424}, 651--654, (2003). Related online
  version (cited on 21 June 2006):
  \newline\url{http://adsabs.harvard.edu/abs/2003Natur.424..651H}.
  \epubtkKeywords{Supernovae}

\bibitem{han98}
Han, Z., ``The formation of double degenerates and related objects'', {\em Mon.
  Not. R. Astron. Soc.}, {\bf 296}, 1019--1040, (1998). Related online version
  (cited on 21 June 2006):
  \newline\url{http://adsabs.harvard.edu/abs/1998MNRAS.296.1019H}.
  \epubtkKeywords{Close binaries, White dwarfs}

\bibitem{hp04}
Han, Z., and Podsiadlowski, P., ``The single-degenerate channel for the
  progenitors of Type Ia supernovae'', {\em Mon. Not. R. Astron. Soc.}, {\bf
  350}, 1301--1309, (2004). Related online version (cited on 21 June 2006):
  \newline\url{http://adsabs.harvard.edu/abs/2004MNRAS.350.1301H}.
  \epubtkKeywords{Supernovae, Stellar evolution}

\bibitem{hp06}
Han, Z., and Podsiadlowski, P., ``A single-degenerate model for the progenitor
  of the Type Ia supernova 2002ic'', {\em Mon. Not. R. Astron. Soc.}, {\bf
  368}, 1095--1100, (2006). Related online version (cited on 21 June 2006):
  \newline\url{http://adsabs.harvard.edu/abs/2006MNRAS.368.1095H}.
  \epubtkKeywords{Supernovae}

\bibitem{Han_al02}
Han, Z., Podsiadlowski, P., Maxted, P.F.L., Marsh, T.R., and Ivanova, N., ``The
  origin of subdwarf B stars -- I. The formation channels'', {\em Mon. Not. R.
  Astron. Soc.}, {\bf 336}, 449--466, (2002). \epubtkKeywords{Binary systems}

\bibitem{hpmmi02}
Han, Z., Podsiadlowski, P., Maxted, P.F.L., Marsh, T.R., and Ivanova, N., ``The
  origin of subdwarf B stars -- I. The formation channels'', {\em Mon. Not. R.
  Astron. Soc.}, {\bf 336}, 449--466, (2002). Related online version (cited on
  21 June 2006):
  \newline\url{http://adsabs.harvard.edu/abs/2002MNRAS.336..449H}.
  \epubtkKeywords{Binary systems, Stellar evolution}

\bibitem{hw99}
Han, Z., and Webbink, R.F., ``Stability and energetics of mass transfer in
  double white dwarfs'', {\em Astron. Astrophys.}, {\bf 349}, L17--L20, (1999).
  \epubtkKeywords{Compact binaries, White dwarfs, Hydrodynamics}

\bibitem{hansen99}
Hansen, B.M.S., ``Cooling models for old white dwarfs'', {\em Astrophys. J.},
  {\bf 520}, 680--695, (1999). \epubtkKeywords{White dwarfs}

\bibitem{heger_death03}
Heger, A., Fryer, C.L., Woosley, S.E., Langer, N., and Hartmann, D.H., ``How
  Massive Single Stars End Their Life'', {\em Astrophys. J.}, {\bf 591},
  288--300, (2003). Related online version (cited on 4 November 2006):
  \newline\url{http://adsabs.harvard.edu/abs/2003ApJ...591..288H}.
  \epubtkKeywords{Stellar evolution, Supernovae, Gravitational collapse}

\bibitem{hell03}
Hellings, R.W., ``LISA data analysis: The detection and initial guess problems
  for monochromatic binaries'', {\em Class. Quantum Grav.}, {\bf 20},
  1019--1029, (2003). Related online version (cited on 21 June 2006):
  \newline\url{http://adsabs.harvard.edu/abs/2003CQGra..20.1019H}.
  \epubtkKeywords{Gravitational wave data analysis, Close binaries}

\bibitem{hills83}
Hills, J.G., ``The effects of sudden mass loss and a random kick velocity
  produced in a supernova explosion on the dynamics of a binary star of
  arbitrary orbital eccentricity -- Applications to X-ray binaries and to the
  binary pulsars'', {\em Astrophys. J.}, {\bf 267}, 322--333, (1983). URL
  (cited on 4 November 2006):
  \newline\url{http://adsabs.harvard.edu/abs/1983ApJ...267..322H}.
  \epubtkKeywords{Close binaries, Supernovae}

\bibitem{hils98}
Hils, D., ``Confusion Noise Estimate for Gravitational Wave Measurements in
  Space'', in Folkner, W.M., ed., {\em Laser Interferometer Space Antenna},
  Second International LISA Symposium on the Detection and Observation of
  Gravitational Waves in Space, Pasadena, California, July 1998, vol. 456 of
  AIP Conference Proceedings,  68--78, (American Institure of Physics,
  Woodbury, U.S.A., 1998). Related online version (cited on 21 June 2006):
  \newline\url{http://adsabs.harvard.edu/abs/1998AIPC..456...68H}.
  \epubtkKeywords{Compact binaries, Gravitational wave sources}

\bibitem{hb00}
Hils, D., and Bender, P.L., ``Gravitational Radiation from Helium
  Cataclysmics'', {\em Astrophys. J.}, {\bf 537}, 334--341, (2000). Related
  online version (cited on 21 June 2006):
  \newline\url{http://adsabs.harvard.edu/abs/2000ApJ...537..334H}.
  \epubtkKeywords{Close binaries, Gravitational wave sources}

\bibitem{hbw90}
Hils, D, Bender, P.L., and Webbink, R.F., ``Gravitational radiation from the
  Galaxy'', {\em Astrophys. J.}, {\bf 360}, 75--94, (1990).
  \epubtkKeywords{Binary systems, Gravitational radiation}

\bibitem{hw87}
Hjellming, M.S., and Webbink, R.F., ``Thresholds for rapid mass transfer in
  binary systems. I. Polytropic models'', {\em Astrophys. J.}, {\bf 318},
  794--808, (1987). \epubtkKeywords{Close binaries}

\bibitem{Hjorth_al03}
Hjorth, J., Sollerman, J., {M{\o}ller}, P., Fynbo, J.P.U., Woosley, S.E.,
  Kouveliotou, C., Tanvir, N.R., Greiner, J., Andersen, M.I., Castro-Tirado,
  A.J., Castro~Cer{\'o}n, J.M., Fruchter, A.S., Gorosabel, J.and~Jakobsson, P.,
  Kaper, L., Klose, S., Masetti, N.and~Pedersen, H., Pedersen, K., Pian, E.,
  Palazzi, E., Rhoads, J.E., Rol, E., van~den Heuvel, E.P.J., Vreeswijk, P.M.,
  Watson, D., and Wijers, R.A.M.J., ``A very energetic supernova associated
  with the {$\gamma$}-ray burst of 29 March 2003'', {\em Nature}, {\bf 423},
  847--850, (2003). \epubtkKeywords{Astronomical observations, Supernovae}

\bibitem{Hjorth_al05}
Hjorth, J., Watson, D., Fynbo, J.P.U., Price, P.A., Jensen, B.L., J{\o}rgensen,
  U.G., Kubas, D., Gorosabel, J., Jakobsson, P., Sollerman, J., Pedersen, K.,
  and Kouveliotou, C., ``The optical afterglow of the short {$\gamma$}-ray
  burst GRB 050709'', {\em Nature}, {\bf 437}, 859--861, (2005). Related online
  version (cited on 21 June 2006):
  \newline\url{http://adsabs.harvard.edu/abs/2005Natur.437..859H}.
  \epubtkKeywords{Gamma-ray bursts, Astronomical observations}

\bibitem{Hobbs_al05}
Hobbs, G., Lorimer, D.R., Lyne, A.G., and Kramer, M., ``A statistical study of
  233 pulsar proper motions'', {\em Mon. Not. R. Astron. Soc.}, {\bf 360},
  974--992, (2005). Related online version (cited on 21 June 2006):
  \newline\url{http://arXiv.org/abs/astro-ph/0504584}.
  \epubtkKeywords{Astronomical observations, Pulsars}

\bibitem{hk96}
H\"oflich, P., Khokhlov, A., Wheeler, J.C., Phillips, M.M., Suntzeff, N.B., and
  Hamuy, M., ``Maximum Brightness and Postmaximum Decline of Light Curves of
  Type Ia Supernovae: A Comparison of Theory and Observations'', {\em
  Astrophys. J. Lett.}, {\bf 472}, L81--L84, (1996). Related online version
  (cited on 21 June 2006):
  \newline\url{http://adsabs.harvard.edu/abs/1996ApJ...472L..81H}.
  \epubtkKeywords{Supernovae}

\bibitem{holberg_wd02}
Holberg, J.B., Oswalt, T.D., and Sion, E.M., ``A Determination of the Local
  Density of White Dwarf Stars'', {\em Astrophys. J.}, {\bf 571}, 512--518,
  (2002). Related online version (cited on 21 June 2006):
  \newline\url{http://adsabs.harvard.edu/abs/2002ApJ...571..512H}.
  \epubtkKeywords{White dwarfs}

\bibitem{Hopman_al06}
Hopman, C., Guetta, D., Waxman, E., and Portegies~Zwart, S., ``The Redshift
  Distribution of Short Gamma-Ray Bursts from Dynamically Formed Neutron Star
  Binaries'', {\em Astrophys. J. Lett.}, {\bf 643}, L91--L94, (2006). Related
  online version (cited on 21 June 2006):
  \newline\url{http://adsabs.harvard.edu/abs/2006ApJ...643L..91H}.
  \epubtkKeywords{Gamma-ray bursts, Relativistic binary stars}

\bibitem{Hotan_al05}
Hotan, A.W., Bailes, M., and Ord, S.M., ``Geodetic Precession in PSR
  J1141--6545'', {\em Astrophys. J.}, {\bf 624}, 906--913, (2005). Related
  online version (cited on 21 June 2006):
  \newline\url{http://adsabs.harvard.edu/abs/2005ApJ...624..906H}.
  \epubtkKeywords{Relativistic binary systems, Pulsars}

\bibitem{howell_2003fg_superch}
Howell, D.A., Sullivan, M., Nugent, P.E., Ellis, R.S., Conley, A.J., Le~Borgne,
  D., Carlberg, R.G., Guy, J., Balam, D., Basa, S., Fouchez, D., Hook, I.M.,
  Hsiao, E.Y., Neill, J.D., Pain, R., Perrett, K.M., and Pritchet, C.J., ``The
  type Ia supernova SNLS--03D3bb from a super-Chandrasekhar-mass white dwarf
  star'', {\em Nature}, {\bf 443}, 308--311, (2006). Related online version
  (cited on 4 November 2006):
  \newline\url{http://adsabs.harvard.edu/abs/2006Natur.443..308H}.
  \epubtkKeywords{Close binaries, White dwarfs, Supernovae}

\bibitem{Hulse_Taylor75}
Hulse, R.A., and Taylor, J.H., ``Discovery of a pulsar in a binary system'',
  {\em Astrophys. J. Lett.}, {\bf 195}, L51--L53, (1975).
  \epubtkKeywords{Double pulsars, Relativistic binary systems}

\bibitem{Hurley_al02}
Hurley, J.R., Tout, C.A., and Pols, O.R., ``Evolution of binary stars and the
  effect of tides on binary populations'', {\em Mon. Not. R. Astron. Soc.},
  {\bf 329}, 897--928, (2002). Related online version (cited on 21 June 2006):
  \newline\url{http://adsabs.harvard.edu/abs/2002MNRAS.329..897H}.
  \epubtkKeywords{Binary stars}

\bibitem{htp02}
Hurley, J.R., Tout, C.A., and Pols, O.R., ``Evolution of binary stars and the
  effect of tides on binary populations'', {\em Mon. Not. R. Astron. Soc.},
  {\bf 329}, 897--928, (2002). Related online version (cited on 21 June 2006):
  \newline\url{http://adsabs.harvard.edu/abs/2002MNRAS.329..897H}.
  \epubtkKeywords{Close binaries}

\bibitem{Hyman_al05}
Hyman, S.D., Lazio, T.J.W., Kassim, N.E., Ray, P.S., Markwardt, C.B., and
  Yusef-Zadeh, F., ``A powerful bursting radio source towards the Galactic
  Centre'', {\em Nature}, {\bf 434}, 50--52, (2005). \epubtkKeywords{Pulsars}

\bibitem{iben86}
Iben~Jr, I., ``On the evolution of binary components which first fill their
  Roche lobes after the exhaustion of central helium'', {\em Astrophys. J.},
  {\bf 304}, 201--216, (1986). Related online version (cited on 21 June 2006):
  \newline\url{http://adsabs.harvard.edu/abs/1986ApJ...304..201I}.
  \epubtkKeywords{Close binaries, Stellar evolution}

\bibitem{il93}
Iben~Jr, I., and Livio, M., ``Common envelopes in binary star evolution'', {\em
  Publ. Astron. Soc. Pac.}, {\bf 105}, 1373--1406, (1993). Related online
  version (cited on 08 December 2006):
  \newline\url{http://adsabs.harvard.edu/abs/1993PASP..105.1373I}.
  \epubtkKeywords{Close binaries}

\bibitem{iben_ritossa_one97}
Iben~Jr, I., Ritossa, C., and Garc\'ia-Berro, E., ``On the Evolution of Stars
  That Form Electron-Degenerate Cores Processed by Carbon Burning. IV. Outward
  Mixing during the Second Dredge-Up Phase and Other Properties of a $10.5
  M_\odot$ Model Star'', {\em Astrophys. J.}, {\bf 489}, 772, (1997). Related
  online version (cited on 21 June 2006):
  \newline\url{http://adsabs.harvard.edu/abs/1997ApJ...489..772I}.
  \epubtkKeywords{Stellar evolution}

\bibitem{it84a}
Iben~Jr, I., and Tutukov, A.V., ``Supernovae of type I as end products of the
  evolution of binaries with components of moderate initial mass ($M \lesssim 9
  M_\odot$)'', {\em Astrophys. J. Suppl. Ser.}, {\bf 54}, 335--372, (1984).
  Related online version (cited on 21 June 2006):
  \newline\url{http://adsabs.harvard.edu/abs/1984ApJS...54..335I}.
  \epubtkKeywords{Supernovae, Stellar evolution, Close binaries}

\bibitem{it85}
Iben~Jr, I., and Tutukov, A.V., ``On the evolution of close binaries with
  components of initial mass between 3 solar masses and 12 solar masses'', {\em
  Astrophys. J. Suppl. Ser.}, {\bf 58}, 661--710, (1985). \epubtkKeywords{Close
  binaries, Stellar evolution}

\bibitem{it91}
Iben~Jr, I., and Tutukov, A.V., ``Helium star cataclysmics'', {\em Astrophys.
  J.}, {\bf 370}, 615--629, (1991). Related online version (cited on 21 June
  2006): \newline\url{http://adsabs.harvard.edu/abs/1991ApJ...370..615I}.
  \epubtkKeywords{Close binaries, White dwarfs}

\bibitem{it96symb}
Iben~Jr, I., and Tutukov, A.V., ``On the Evolution of Symbiotic Stars and Other
  Binaries with Accreting Degenerate Dwarfs'', {\em Astrophys. J. Suppl. Ser.},
  {\bf 105}, 145, (1996). Related online version (cited on 21 June 2006):
  \newline\url{http://adsabs.harvard.edu/abs/1996ApJS..105..145I}.
  \epubtkKeywords{Close binaries, White dwarfs}

\bibitem{itf98}
Iben~Jr, I., Tutukov, A.V., and Fedorova, A.V., ``On the Luminosity of White
  Dwarfs in Close Binaries Merging under the Influence of Gravitational Wave
  Radiation'', {\em Astrophys. J.}, {\bf 503}, 344, (1998). Related online
  version (cited on 21 June 2006):
  \newline\url{http://adsabs.harvard.edu/abs/1998ApJ...503..344I}.
  \epubtkKeywords{Close binaries, White dwarfs}

\bibitem{ity95b}
Iben~Jr, I., Tutukov, A.V., and Yungelson, L.R., ``A Model of the Galactic
  X-Ray Binary Population. II. Low-Mass X-Ray Binaries in the Galactic Disk'',
  {\em Astrophys. J. Suppl. Ser.}, {\bf 100}, 233, (1995).
  \epubtkKeywords{X-ray binaries}

\bibitem{zand05}
in't Zand, J.J.M., Cumming, A., van~der Sluys, M.V., Verbunt, F., and Pols,
  O.R., ``On the possibility of a helium white dwarf donor in the presumed
  ultracompact binary 2S 0918--549'', {\em Astron. Astrophys.}, {\bf 441},
  675--684, (2005). Related online version (cited on 4 November 2006):
  \newline\url{http://adsabs.harvard.edu/abs/2005A&A...441..675I}.
  \epubtkKeywords{X-ray binaries, White dwarfs}

\bibitem{isern_83}
Isern, J., Labay, J., Hernanz, M., and Canal, R., ``Collapse and explosion of
  white dwarfs. I. Precollapse evolution'', {\em Astrophys. J.}, {\bf 273},
  320--329, (1983). Related online version (cited on 21 June 2006):
  \newline\url{http://adsabs.harvard.edu/abs/1983ApJ...273..320I}.
  \epubtkKeywords{White dwarfs, Gravitational collapse}

\bibitem{ipc99}
Israel, G.L., Panzera, M.R., Campana, S., Lazzati, D., Covino, S., Tagliaferri,
  G., and Stella, L., ``The discovery of 321 S pulsations in the ROSAT HRI
  light curves of 1BMW J080622.8+152732 = RX J0806.3+1527'', {\em Astron.
  Astrophys.}, {\bf 349}, L1--L4, (1999). Related online version (cited on 21
  June 2006): \newline\url{http://adsabs.harvard.edu/abs/1999A&A...349L...1I}.
  \epubtkKeywords{X-ray binaries, Astronomical observations}

\bibitem{Ivanova_al03}
Ivanova, N., Belczynski, K., Kalogera, V., Rasio, F.A., and Taam, R.E., ``The
  Role of Helium Stars in the Formation of Double Neutron Stars'', {\em
  Astrophys. J.}, {\bf 592}, 475--485, (2003). \epubtkKeywords{Neutron stars,
  Relativistic binary systems}

\bibitem{ivanova_ultra_gc05}
Ivanova, N., Rasio, F.A., Lombardi, J.C., Dooley, K.L., and Proulx, Z.F.,
  ``Formation of Ultracompact X-Ray Binaries in Dense Star Clusters'', {\em
  Astrophys. J. Lett.}, {\bf 621}, L109--L112, (2005). Related online version
  (cited on 21 June 2006):
  \newline\url{http://adsabs.harvard.edu/abs/2005ApJ...621L.109I}.
  \epubtkKeywords{X-ray binaries, Star clusters}

\bibitem{ivanova_taam03}
Ivanova, N., and Taam, R.E., ``Thermal Timescale Mass Transfer and the
  Evolution of White Dwarf Binaries'', {\em Astrophys. J.}, {\bf 601},
  1058--1066, (2004). Related online version (cited on 21 June 2006):
  \newline\url{http://adsabs.harvard.edu/abs/2004ApJ...601.1058I}.
  \epubtkKeywords{Compact binaries}

\bibitem{jeffery_2003fg_superch}
Jeffery, D.J., Branch, D., and Baron, E., ``On SN 2003fg: The Probable
  Super-Chandrasekhar-Mass SN Ia'', (September, 2006). URL (cited on 4 November
  2006): \newline\url{http://arXiv.org/abs/astro-ph/0609804}.
  \epubtkKeywords{White dwarfs, Supernovae}

\bibitem{Johnston_al05}
Johnston, S., Hobbs, G., Vigeland, S., Kramer, M., Weisberg, J.M., and Lyne,
  A.G., ``Evidence for alignment of the rotation and velocity vectors in
  pulsars'', {\em Mon. Not. R. Astron. Soc.}, {\bf 364}, 1397--1412, (2005).
  Related online version (cited on 21 June 2006):
  \newline\url{http://adsabs.harvard.edu/abs/2005MNRAS.364.1397J}.
  \epubtkKeywords{Pulsars}

\bibitem{jonker_nelemans04}
Jonker, P.G., and Nelemans, G., ``The distances to Galactic low-mass X-ray
  binaries: Consequences for black hole luminosities and kicks'', {\em Mon.
  Not. R. Astron. Soc.}, {\bf 354}, 355--366, (2004). Related online version
  (cited on 21 June 2006):
  \newline\url{http://adsabs.harvard.edu/abs/2004MNRAS.354..355J}.
  \epubtkKeywords{X-ray binaries, Black hole formation}

\bibitem{kahabka_sss95}
Kahabka, P., ``Recurrent supersoft X-ray sources'', {\em Astron. Astrophys.},
  {\bf 304}, 227, (1995). Related online version (cited on 21 June 2006):
  \newline\url{http://adsabs.harvard.edu/abs/1995A&A...304..227K}.
  \epubtkKeywords{X-ray binaries, Astronomical observations}

\bibitem{kahabka02}
Kahabka, P., ``Super Soft Sources'', in Lewin, W.H.G., and van~der Klis, M.,
  eds., {\em Compact Stellar X-Ray Sources}, Cambridge Astrophysics Series,
  (Cambridge University Press, Cambridge, U.K., 2006). Related online version
  (cited on 21 June 2006): \newline\url{http://arXiv.org/abs/astro-ph/0212037}.
  \epubtkKeywords{X-ray binaries, Astronomical observations}

\bibitem{kh97}
Kahabka, P., and van~den Heuvel, P.J., ``Luminous supersoft X-ray sources'',
  {\em Annu. Rev. Astron. Astrophys.}, {\bf 35}, 69--100, (1997).
  \epubtkKeywords{Binary systems, X-ray binaries}

\bibitem{kalogera96}
Kalogera, V., ``Orbital Characteristics of Binary Systems after Asymmetric
  Supernova Explosions'', {\em Astrophys. J.}, {\bf 471}, 352, (1996). Related
  online version (cited on 4 November 2006):
  \newline\url{http://adsabs.harvard.edu/abs/1996ApJ...471..352K}.
  \epubtkKeywords{Close binaries, Supernovae}

\bibitem{Kalogera00}
Kalogera, V., ``Spin-Orbit Misalignment in Close Binaries with Two Compact
  Objects'', {\em Astrophys. J.}, {\bf 541}, 319--328, (2000).
  \epubtkKeywords{Relativistic binary systems}

\bibitem{Kalogera_al04}
Kalogera, V., Kim, C., Lorimer, D.R., Burgay, M., D'Amico, N., Possenti, A.,
  Manchester, R.N., Lyne, A.G., Joshi, B.C., McLaughlin, M.A., Kramer, M.,
  Sarkissian, J.M., and Camilo, F., ``The Cosmic Coalescence Rates for Double
  Neutron Star Binaries'', {\em Astrophys. J. Lett.}, {\bf 601}, L179--L182,
  (2004). \epubtkKeywords{Relativistic binary systems}

\bibitem{kalogera_nswd05}
Kalogera, V., Kim, C., Lorimer, D.R., Ihm, M., and Belczynski, K., ``The
  Galactic Formation Rate of Eccentric Neutron Star -- White Dwarf Binaries'',
  in Rasio, F.A., and Stairs, I.H., eds., {\em Binary Radio Pulsars}, Meeting
  at the Aspen Center for Physics, Colorado, 12 -- 16 January 2004, vol. 328 of
  ASP Conference Series,  261--267, (Astronomical Society of the Pacific, San
  Francisco, U.S.A., 2005). Related online version (cited on 21 June 2006):
  \newline\url{http://adsabs.harvard.edu/abs/2005ASPC..328..261K}.
  \epubtkKeywords{Pulsars, Binary systems}

\bibitem{Kalogera_al01}
Kalogera, V., Narayan, R., Spergel, D.N., and Taylor, J.H., ``The Coalescence
  Rate of Double Neutron Star Systems'', {\em Astrophys. J.}, {\bf 556},
  340--356, (2001). \epubtkKeywords{Relativistic binary systems}

\bibitem{kw96}
Kalogera, V., and Webbink, R.F., ``Formation of Low-Mass X-Ray Binaries. I.
  Constraints on Hydrogen-rich Donors at the Onset of the X-Ray Phase'', {\em
  Astrophys. J.}, {\bf 458}, 301--311, (1996). \epubtkKeywords{X-ray binaries,
  Close binaries}

\bibitem{kw98}
Kalogera, V., and Webbink, R.F., ``Formation of Low-Mass X-Ray Binaries. II.
  Common Envelope Evolution of Primordial Binaries with Extreme Mass Ratios'',
  {\em Astrophys. J.}, {\bf 493}, 351--368, (1998). \epubtkKeywords{X-ray
  binaries, Close binaries, Hydrodynamics}

\bibitem{Karachentsev_al04}
Karachentsev, I.D., Karachentseva, V.E., Huchtmeier, W.K., and Makarov, D.I.,
  ``A Catalog of Neighboring Galaxies'', {\em Astron. J.}, {\bf 127},
  2031--2068, (2004). Related online version (cited on 4 November 2006):
  \newline\url{http://adsabs.harvard.edu/abs/2004AJ....127.2031K}.
  \epubtkKeywords{Astronomical observations, Extragalactic astronomy}

\bibitem{Kaspi_al96}
Kaspi, V.M., Bailes, M., Manchester, R.N., Stappers, B.W., and Bell, J.F.,
  ``Evidence from a precessing pulsar orbit for a neutron-star birth kick'',
  {\em Nature}, {\bf 381}, 584--586, (1996). \epubtkKeywords{Supernovae,
  Neutron stars}

\bibitem{kathac94}
Kato, M., and Hachisu, I., ``Optically thick winds in nova outbursts'', {\em
  Astrophys. J.}, {\bf 437}, 802--826, (1994). Related online version (cited on
  21 June 2006):
  \newline\url{http://adsabs.harvard.edu/abs/1994ApJ...437..802K}.
  \epubtkKeywords{Close binaries, White dwarfs}

\bibitem{Katz_Canel96}
Katz, J.I., and Canel, L.M., ``The Long and the Short of Gamma-Ray Bursts'',
  {\em Astrophys. J.}, {\bf 471}, 915, (1996). \epubtkKeywords{Gamma-ray
  bursts, Relativistic binary systems}

\bibitem{Kim_al03}
Kim, C., Kalogera, V., and Lorimer, D.R., ``The Probability Distribution of
  Binary Pulsar Coalescence Rates. I. Double Neutron Star Systems in the
  Galactic Field'', {\em Astrophys. J.}, {\bf 584}, 985--995, (2003).
  \epubtkKeywords{Pulsars, Relativistic binary systems}

\bibitem{kim_etal06}
Kim, C., Kalogera, V., and Lorimer, D.R., ``Effect of PSR J0737--3039 on the
  DNS Merger Rate and Implications for GW Detection'', (August, 2006). URL
  (cited on 4 November 2006):
  \newline\url{http://arXiv.org/abs/astro-ph/0608280}. \epubtkKeywords{Double
  pulsars, Gravitational wave sources}

\bibitem{kim_coalesc04}
Kim, C., Kalogera, V., Lorimer, D.R., and White, T., ``The Probability
  Distribution Of Binary Pulsar Coalescence Rates. II. Neutron Star--White
  Dwarf Binaries'', {\em Astrophys. J.}, {\bf 616}, 1109--1117, (2004). Related
  online version (cited on 21 June 2006):
  \newline\url{http://adsabs.harvard.edu/abs/2004ApJ...616.1109K}.
  \epubtkKeywords{Compact binaries, Binary pulsars}

\bibitem{kitaura_one06}
Kitaura, F.S., Janka, H.-T., and Hillebrandt, W., ``Explosions of O-Ne-Mg
  cores, the Crab supernova, and subluminous type II-P supernovae'', {\em
  Astron. Astrophys.}, {\bf 450}, 345--350, (2006). Related online version
  (cited on 21 June 2006):
  \newline\url{http://adsabs.harvard.edu/abs/2006A&A...450..345K}.
  \epubtkKeywords{Supernovae}

\bibitem{khh99}
Knox, R.A., Hawkins, M.R.S., and Hambly, N.C., ``A survey for cool white dwarfs
  and the age of the Galactic disc'', {\em Mon. Not. R. Astron. Soc.}, {\bf
  306}, 736--752, (1999). \epubtkKeywords{White dwarfs, Astronomical
  observations}

\bibitem{kopal1959}
Kopal, Z., {\em Close Binary Systems}, vol.~5 of The International Astrophysics
  Series, (Chapman \& Hall; Wiley, London, U.K.; New York, U.S.A., 1959).
  \epubtkKeywords{Close binaries}

\bibitem{Kornilov_Lipunov83a}
Kornilov, V.G., and Lipunov, V.M., ``Neutron Stars in Massive Binary Systems --
  Part One -- Classification and Evolution'', {\em Sov. Astron.}, {\bf 27},
  163, (1983). \epubtkKeywords{Binary systems, Neutron stars}

\bibitem{Kornilov_Lipunov83b}
Kornilov, V.G., and Lipunov, V.M., ``Neutron Stars in Massive Binary Systems --
  Part Two -- Numerical Modeling'', {\em Sov. Astron.}, {\bf 27}, 334, (1983).
  \epubtkKeywords{Binary systems, Neutron stars}

\bibitem{Kotake_al06}
Kotake, K., Sato, K., and Takahashi, K., ``Explosion mechanism, neutrino burst
  and gravitational wave in core-collapse supernovae'', {\em Rep. Prog. Phys.},
  {\bf 69}, 971--1143, (2006). Related online version (cited on 21 June 2006):
  \newline\url{http://adsabs.harvard.edu/abs/2006RPPh...69..971K}.
  \epubtkKeywords{Supernovae, Gravitational collapse}

\bibitem{kmg62}
Kraft, R.P., Mathews, J., and Greenstein, J.L., ``Binary Stars among
  Cataclysmic Variables. II. Nova WZ Sagittae: A Possible Radiator of
  Gravitational Waves'', {\em Astrophys. J.}, {\bf 136}, 312, (1962). Related
  online version (cited on 21 June 2006):
  \newline\url{http://adsabs.harvard.edu/abs/1962ApJ...136..312K}.
  \epubtkKeywords{Close binaries, White dwarfs}

\bibitem{kroupa_etal93}
Kroupa, P., Tout, C.A., and Gilmore, G., ``The distribution of low-mass stars
  in the Galactic disc'', {\em Mon. Not. R. Astron. Soc.}, {\bf 262}, 545--587,
  (1993). Related online version (cited on 4 November 2006):
  \newline\url{http://adsabs.harvard.edu/abs/1993MNRAS.262..545K}.
  \epubtkKeywords{Stars, Astronomical observations}

\bibitem{kroupa_weidner_03}
Kroupa, P., and Weidner, C., ``Galactic-Field Initial Mass Functions of Massive
  Stars'', {\em Astrophys. J.}, {\bf 598}, 1076--1078, (2003). Related online
  version (cited on 4 November 2006):
  \newline\url{http://adsabs.harvard.edu/abs/2003ApJ...598.1076K}.
  \epubtkKeywords{Stars, Astronomical observations}

\bibitem{kudr_urban_winds06}
Kudritzki, R.-P., and Urbaneja, M.A., ``Parameters and Winds of Hot Massive
  Stars'', (July, 2006). URL (cited on 4 November 2006):
  \newline\url{http://arXiv.org/abs/astro-ph/0607460}. \epubtkKeywords{Stellar
  winds}

\bibitem{Kuiper35}
Kuiper, G.P., ``Problems of Double-Star Astronomy. I'', {\em Publ. Astron. Soc.
  Pac.}, {\bf 47}, 15--42, (1935). Related online version (cited on 21 June
  2006): \newline\url{http://adsabs.harvard.edu/abs/1935PASP...47...15K}.
  \epubtkKeywords{Astronomical observations, Binary systems}

\bibitem{Kumar63}
Kumar, S.S., ``The Structure of Stars of Very Low Mass'', {\em Astrophys. J.},
  {\bf 137}, 1121, (1963). \epubtkKeywords{Astrophysics, Stars}

\bibitem{Kuranov_Postnov06}
Kuranov, A.G., and Postnov, K.A., ``Neutron stars in globular clusters:
  Formation and observational manifestations'', {\em Astron. Lett.}, {\bf 32},
  393--405, (2006). Related online version (cited on 21 June 2006):
  \newline\url{http://arXiv.org/abs/astro-ph/0605115}. \epubtkKeywords{Neutron
  stars, Binary stars}

\bibitem{Kusenko04}
Kusenko, A., ``Pulsar Kicks from Neutrino Oscillations'', {\em Int. J. Mod.
  Phys. D}, {\bf 13}, 2065--2084, (2004). \epubtkKeywords{Neutron stars}

\bibitem{Lai01}
Lai, D., ``Neutron Star Kicks and Asymmetric Supernovae'', in Blaschke, D.,
  Glendenning, N.K., and Sedrakian, A., eds., {\em Physics of Neutron Star
  Interiors}, vol. 578 of Lecture Notes in Physics,  424, (Springer, Berlin,
  Germany; New York, U.S.A., 2001). Related online version (cited on 21 June
  2006): \newline\url{http://adsabs.harvard.edu/abs/2001LNP...578..424L}.
  \epubtkKeywords{Pulsars, Supernovae}

\bibitem{Lai_al01}
Lai, D., Chernoff, D.F., and Cordes, J.M., ``Pulsar Jets: Implications for
  Neutron Star Kicks and Initial Spins'', {\em Astrophys. J.}, {\bf 549},
  1111--1118, (2001). \epubtkKeywords{Pulsars, Supernovae}

\bibitem{L_L_v1}
Landau, L.D., and Lifshitz, E.M., {\em Mechanics}, vol.~1 of Course of
  Theoretical Physics, (Pergamon Press, Oxford, U.K.; New York, U.S.A., 1969),
  2nd edition. \epubtkKeywords{Dynamics}

\bibitem{L_L_v2}
Landau, L.D., and Lifshitz, E.M., {\em The Classical Theory of Fields}, vol.~2
  of Course of Theoretical Physics, (Pergamon Press, Oxford, U.K.; New York,
  U.S.A., 1975), 4th edition. \epubtkKeywords{Theories of gravity}

\bibitem{lhh00}
Larson, S.L., Hiscock, W.A., and Hellings, R.W., ``Sensitivity curves for
  spaceborne gravitational wave interferometers'', {\em Phys. Rev. D}, {\bf
  62}, 062001, (2000). \epubtkKeywords{Gravitational wave detectors}

\bibitem{Lattimer_Prakash04}
Lattimer, J.M., and Prakash, M., ``The Physics of Neutron Stars'', {\em
  Science}, {\bf 304}, 536--542, (2004). \epubtkKeywords{Neutron stars}

\bibitem{Lee_al05a}
Lee, W.H., Ramirez-Ruiz, E., and Granot, J., ``A Compact Binary Merger Model
  for the Short, Hard GRB 050509b'', {\em Astrophys. J. Lett.}, {\bf 630},
  L165--L168, (2005). Related online version (cited on 21 June 2006):
  \newline\url{http://adsabs.harvard.edu/abs/2005ApJ...630L.165L}.
  \epubtkKeywords{Relativistic binary stars, Gamma-ray bursts}

\bibitem{Lee_al05b}
Lee, W.H., Ramirez-Ruiz, E., and Page, D., ``Dynamical Evolution of
  Neutrino-Cooled Accretion Disks: Detailed Microphysics, Lepton-Driven
  Convection, and Global Energetics'', {\em Astrophys. J.}, {\bf 632},
  421--437, (2005). Related online version (cited on 21 June 2006):
  \newline\url{http://adsabs.harvard.edu/abs/2005ApJ...632..421L}.
  \epubtkKeywords{Relativistic binary stars, Accretion disks}

\bibitem{lentz_02}
Lentz, E.J., Baron, E., Hauschildt, P.H., and Branch, D., ``Detectability of
  Hydrogen Mixing in Type Ia Supernova Premaximum Spectra'', {\em Astrophys.
  J.}, {\bf 580}, 374--379, (2002). \epubtkKeywords{Supernovae}

\bibitem{lesaffre_sn06}
Lesaffre, P., Han, Z., Tout, C.A., Podsiadlowski, P., and Martin, R.G., ``The C
  flash and the ignition conditions of Type Ia supernovae'', {\em Mon. Not. R.
  Astron. Soc.}, {\bf 368}, 187--195, (2006). Related online version (cited on
  21 June 2006):
  \newline\url{http://adsabs.harvard.edu/abs/2006MNRAS.368..187L}.
  \epubtkKeywords{Supernovae}

\bibitem{li_01}
Li, W., Filippenko, A.V., Treffers, R.R., Riess, A.G., Hu, J., and Qiu, Y., ``A
  High Intrinsic Peculiarity Rate among Type Ia Supernovae'', {\em Astrophys.
  J.}, {\bf 546}, 734--743, (2001). Related online version (cited on 21 June
  2006): \newline\url{http://adsabs.harvard.edu/abs/2001ApJ...546..734L}.
  \epubtkKeywords{Supernovae}

\bibitem{lh97}
Li, X.-D., and van~den Heuvel, P.J., ``Evolution of white dwarf binaries:
  Supersoft X-ray sources and progenitors of type Ia upernovae'', {\em Astron.
  Astrophys.}, {\bf 322}, L9--L12, (1997). \epubtkKeywords{Compact binaries,
  X-ray binaries, Supernovae}

\bibitem{lt91}
Limongi, M., and Tornamb\`{e}, A., ``He stars and He-accreting CO white
  dwarfs'', {\em Astrophys. J.}, {\bf 371}, 317--331, (1991).
  \epubtkKeywords{Close binaries, Accretion, White dwarfs}

\bibitem{Lipunov92}
Lipunov, V.M., {\em Astrophysics of Neutron Stars}, (Springer, Berlin, Germany;
  New York, U.S.A., 1992). \epubtkKeywords{Neutron stars}

\bibitem{Lipunov_al05}
Lipunov, V.M., Bogomazov, A.I., and Abubekerov, M.K., ``How abundant is the
  population of binary radio pulsars with black holes?'', {\em Mon. Not. R.
  Astron. Soc.}, {\bf 359}, 1517--1523, (2005). \epubtkKeywords{Pulsars, Black
  holes, Binary systems}

\bibitem{Lipunov_al95_gwsky}
Lipunov, V.M., Nazin, S.N., Panchenko, I.E., Postnov, K.A., and Prokhorov,
  M.E., ``The gravitational wave sky'', {\em Astron. Astrophys.}, {\bf 298},
  677, (1995). Related online version (cited on 4 November 2006):
  \newline\url{http://adsabs.harvard.edu/abs/1995A&A...298..677L}.
  \epubtkKeywords{Gravitational wave sources, Extragalactic astronomy}

\bibitem{lp87}
Lipunov, V.M., and Postnov, K.A., ``Spectrum of gravitational radiation of
  binary systems'', {\em Sov. Astron.}, {\bf 31}, 228--235, (1988).
  \epubtkKeywords{Binary stars, Gravitational wave sources}

\bibitem{lpp87}
Lipunov, V.M., Postnov, K.A., and Prokhorov, M.E., ``The sources of
  gravitaional waves with continuous and discrete spectra'', {\em Astron.
  Astrophys.}, {\bf 176}, L1--L4, (1987). \epubtkKeywords{Binary stars,
  Gravitational wave sources}

\bibitem{Lipunov_al96}
Lipunov, V.M., Postnov, K.A., and Prokhorov, M.E., ``The Scenario Machine:
  Binary Star Population Synthesis'', {\em Astrophys. Space Phys. Rev.}, {\bf
  9}, 1--178, (1996). Related online version (cited on 13 November 2006):
  \newline\url{http://xray.sai.msu.ru/~mystery/articles/review/}.
  \epubtkKeywords{Astrophysics, Binary systems}

\bibitem{LPP96}
Lipunov, V.M., Postnov, K.A., and Prokhorov, M.E., ``The Scenario Machine:
  Restrictions on key parameters of binary evolution'', {\em Astron.
  Astrophys.}, {\bf 310}, 489--507, (1996). \epubtkKeywords{Astrophysics,
  Binary systems}

\bibitem{LPP97}
Lipunov, V.M., Postnov, K.A., and Prokhorov, M.E., ``Black holes and
  gravitational waves: Possibilities for simultaneous detection using
  first-generation laser interferometers'', {\em Astron. Lett.}, {\bf 23},
  492--497, (1997). \epubtkKeywords{Black holes, Relativistic binary systems}

\bibitem{Lipunov_al97}
Lipunov, V.M., Postnov, K.A., and Prokhorov, M.E., ``Formation and coalescence
  of relativistic binary stars: The effect of kick velocity'', {\em Mon. Not.
  R. Astron. Soc.}, {\bf 288}, 245--259, (1997). \epubtkKeywords{Relativistic
  binary stars, Neutron stars, Black holes}

\bibitem{Lipunov_al94}
Lipunov, V.M., Postnov, K.A., Prokhorov, M.E., and Osminkin, E.Y., ``Binary
  Radiopulsars with Black Holes'', {\em Astrophys. J. Lett.}, {\bf 423},
  L121--L124, (1994). Related online version (cited on 08 December 2006):
  \newline\url{http://adsabs.harvard.edu/abs/1994ApJ...423L.121L}.
  \epubtkKeywords{Pulsars, Black holes, Binary systems}

\bibitem{Lipunov_al95_evol}
Lipunov, V.M., Postnov, K.A., Prokhorov, M.E., Panchenko, I.E., and
  J{\o}rgensen, H.E., ``Evolution of the Double Neutron Star Merging Rate and
  the Cosmological Origin of Gamma-Ray Burst Sources'', {\em Astrophys. J.},
  {\bf 454}, 593, (1995). Related online version (cited on 4 November 2006):
  \newline\url{http://adsabs.harvard.edu/abs/1995ApJ...454..593L}.
  \epubtkKeywords{Gamma-ray bursts, Compact binaries}

\bibitem{livne90}
Livne, E., ``Successive detonations in accreting white dwarfs as an alternative
  mechanism for type I supernovae'', {\em Astrophys. J.}, {\bf 354}, L53--L55,
  (1990). Related online version (cited on 21 June 2006):
  \newline\url{http://adsabs.harvard.edu/abs/1990ApJ...354L..53L}.
  \epubtkKeywords{White dwarfs, Supernovae}

\bibitem{la95}
Livne, E., and Arnett, D., ``Explosions of Sub-Chandrasekhar Mass White Dwarfs
  in Two Dimensions'', {\em Astrophys. J.}, {\bf 452}, 62, (1995). Related
  online version (cited on 21 June 2006):
  \newline\url{http://adsabs.harvard.edu/abs/1995ApJ...452...62L}.
  \epubtkKeywords{White dwarfs, Supernovae}

\bibitem{lg91}
Livne, E., and Glasner, A., ``Numerical simulations of off-center detonations
  in helium shells'', {\em Astrophys. J.}, {\bf 370}, 272--281, (1991).
  \epubtkKeywords{Supernovae}

\bibitem{lombardi_ucb_gc06}
Lombardi, J.C., Proulx, Z.F., Dooley, K.L., Theriault, E.M., Ivanova, N., and
  Rasio, F.A., ``Stellar Collisions and Ultracompact X-Ray Binary Formation'',
  {\em Astrophys. J.}, {\bf 640}, 441--458, (2006). Related online version
  (cited on 21 June 2006):
  \newline\url{http://adsabs.harvard.edu/abs/2006ApJ...640..441L}.
  \epubtkKeywords{X-ray binaries}

\bibitem{lyhnp05}
Lommen, D., Yungelson, L.R., van~den Heuvel, E., Nelemans, G., and
  Portegies~Zwart, S., ``Cygnus X-3 and the problem of the missing Wolf--Rayet
  X-ray binaries'', {\em Astron. Astrophys.}, {\bf 443}, 231--241, (2005).
  Related online version (cited on 21 June 2006):
  \newline\url{http://adsabs.harvard.edu/abs/2005A&A...443..231L}.
  \epubtkKeywords{X-ray binaries}

\bibitem{loren_gwr05}
Lor{\'e}n-Aguilar, P., Guerrero, J., Isern, J., Lobo, J.A., and
  Garc{\'{i}}a-Berro, E., ``Gravitational wave radiation from the coalescence
  of white dwarfs'', {\em Mon. Not. R. Astron. Soc.}, {\bf 356}, 627--636,
  (2005). Related online version (cited on 21 June 2006):
  \newline\url{http://adsabs.harvard.edu/abs/2005MNRAS.356..627L}.
  \epubtkKeywords{Close binaries, White dwarfs, Gravitational wave sources}

\bibitem{Lorimer_LRR01}
Lorimer, D.R., ``Binary and Millisecond Pulsars at the New Millennium'', {\em
  Living Rev. Relativity}, {\bf 8}, lrr-2005-7, (2005). URL (cited on 21 June
  2006): \newline\url{http://www.livingreviews.org/lrr-2005-7}.
  \epubtkKeywords{Pulsars, Binary systems, Relativistic binary systems}

\bibitem{Lorimer_al06}
Lorimer, D.R., Stairs, I.H., Freire, P.C., Cordes, J.M., Camilo, F., Faulkner,
  A.J., Lyne, A.G., Nice, D.J., Ransom, S.M., Arzoumanian, Z., Manchester,
  R.N., Champion, D.J., van Leeuwen, J., Mclaughlin, M.A., Ramachandran, R.,
  Hessels, J.W., Vlemmings, W., Deshpande, A.A., Bhat, N.D., Chatterjee, S.,
  Han, J.L., Gaensler, B.M., Kasian, L., Deneva, J.S., Reid, B., Lazio, T.J.,
  Kaspi, V.M., Crawford, F., Lommen, A.N., Backer, D.C., Kramer, M., Stappers,
  B.W., Hobbs, G.B., Possenti, A., D'Amico, N., and Burgay, M., ``Arecibo
  Pulsar Survey Using ALFA. II. The Young, Highly Relativistic Binary Pulsar
  J1906+0746'', {\em Astrophys. J.}, {\bf 640}, 428--434, (2006). Related
  online version (cited on 21 June 2006):
  \newline\url{http://adsabs.harvard.edu/abs/2006ApJ...640..428L}.
  \epubtkKeywords{Relativistic binary systems, Pulsars}

\bibitem{ls75}
Lubow, S.H., and Shu, F.H., ``Gas dynamics of semidetached binaries'', {\em
  Astrophys. J.}, {\bf 198}, 383, (1975). \epubtkKeywords{Close binaries,
  Hydrodynamics}

\bibitem{Lyne_al04}
Lyne, A.G., Burgay, M., Kramer, M., Possenti, A., Manchester, R.N., Camilo, F.,
  McLaughlin, M.A., Lorimer, D.R., D'Amico, N., Joshi, B.C., Reynolds, J., and
  Freire, P.C.C., ``A Double-Pulsar System: A Rare Laboratory for Relativistic
  Gravity and Plasma Physics'', {\em Science}, {\bf 303}, 1153--1157, (2004).
  \epubtkKeywords{Double pulsars, Relativistic binary systems}

\bibitem{Lyne_al00}
Lyne, A.G., Camilo, F., Manchester, R.N., Bell, J.F., Kaspi, V.M., D'Amico, N.,
  McKay, N.P.F., Crawford, F., Morris, D.J., Sheppard, D.C., and Stairs, I.H.,
  ``The Parkes Multibeam Pulsar Survey: PSR J1811--1736, a pulsar in a highly
  eccentric binary system'', {\em Mon. Not. R. Astron. Soc.}, {\bf 312},
  698--702, (2000). \epubtkKeywords{Pulsars, Astronomical observations}

\bibitem{Lyne_Lorimer94}
Lyne, A.G., and Lorimer, D.R., ``High Birth Velocities of Radio Pulsars'', {\em
  Nature}, {\bf 369}, 127--129, (1994). \epubtkKeywords{Pulsars}

\bibitem{macdonald_sn84}
MacDonald, J., ``Are cataclysmic variables the progenitors of Type I
  supernovae?'', {\em Astrophys. J.}, {\bf 283}, 241--248, (1984). Related
  online version (cited on 21 June 2006):
  \newline\url{http://adsabs.harvard.edu/abs/1984ApJ...283..241M}.
  \epubtkKeywords{Close binaries, Supernovae}

\bibitem{madau_prog98}
Madau, P., ``Galaxy Evolution and the Cosmic Rate of Supernovae'', in
  D'Odorico, S., Fontana, A., and Giallongo, E., eds., {\em The Young Universe:
  Galaxy Formation and Evolution at Intermediate and High Redshift}, Meeting at
  the Rome Astronomical Observatory, Monteporzio, Italy, 29 September -- 3
  October 1997, vol. 146 of ASP Conference Series,  289, (Astronomical Society
  of the Pacific, San Francisco, U.S.A., 1998). Related online version (cited
  on 21 June 2006):
  \newline\url{http://adsabs.harvard.edu/abs/1998ASPC..146..289M}.
  \epubtkKeywords{Extragalactic astronomy, Galaxies}

\bibitem{Manchester_al01}
Manchester, R.N., Lyne, A.G., Camilo, F., Bell, J.F., Kaspi, V.M., D'Amico, N.,
  McKay, N.P.F., Crawford, F., Stairs, I.H., Possenti, A., Kramer, M., and
  Sheppard, D.C., ``The Parkes multi-beam pulsar survey -- I. Observing and
  data analysis systems, discovery and timing of 100 pulsars'', {\em Mon. Not.
  R. Astron. Soc.}, {\bf 328}, 17--35, (2001). \epubtkKeywords{Neutron stars,
  Pulsars, Astronomical observations}

\bibitem{mannucci_sn05}
Mannucci, F., Della~Valle, M., and Panagia, N., ``Two populations of
  progenitors for Type Ia supernovae?'', {\em Mon. Not. R. Astron. Soc.}, {\bf
  370}, 773--783, (2006). Related online version (cited on 21 June 2006):
  \newline\url{http://adsabs.harvard.edu/abs/2005astro.ph.10315M}.
  \epubtkKeywords{Supernovae}

\bibitem{marietta00}
Marietta, E., Burrows, A., and Fryxell, B., ``Type Ia Supernova Explosions in
  Binary Systems: The Impact on the Secondary Star and its Consequences'', {\em
  Astrophys. J. Suppl. Ser.}, {\bf 128}, 615--650, (2000). Related online
  version (cited on 21 June 2006):
  \newline\url{http://adsabs.harvard.edu/abs/2000ApJS..128..615M}.
  \epubtkKeywords{Binary systems, Supernovae}

\bibitem{marsh00}
Marsh, T.R., ``Detached white-dwarf close-binary stars -- CV's extended
  family'', {\em New Astron. Rev.}, {\bf 44}, 119--124, (2000). Related online
  version (cited on 21 June 2006):
  \newline\url{http://adsabs.harvard.edu/abs/2000NewAR..44..119M}.
  \epubtkKeywords{Compact binaries, White dwarfs, Astronomical observations}

\bibitem{mdd95}
Marsh, T.R., Dhillon, V.S., and Duck, S.R., ``Low-mass white dwarfs need
  friends: Five new double-degenerate close binary stars'', {\em Mon. Not. R.
  Astron. Soc.}, {\bf 275}, 828, (1995). \epubtkKeywords{Compact binaries,
  White dwarfs, Astronomical observations}

\bibitem{mns04}
Marsh, T.R., Nelemans, G., and Steeghs, D., ``Mass transfer between double
  white dwarfs'', {\em Mon. Not. R. Astron. Soc.}, {\bf 350}, 113--128, (2004).
  Related online version (cited on 21 June 2006):
  \newline\url{http://adsabs.harvard.edu/abs/2004MNRAS.350..113M}.
  \epubtkKeywords{Close binaries, White dwarfs, Hydrodynamics}

\bibitem{marsh_nel_st04}
Marsh, T.R., Nelemans, G., and Steeghs, D., ``Mass transfer between double
  white dwarfs'', {\em Mon. Not. R. Astron. Soc.}, {\bf 350}, 113--128, (2004).
  Related online version (cited on 21 June 2006):
  \newline\url{http://adsabs.harvard.edu/abs/2004MNRAS.350..113M}.
  \epubtkKeywords{Compact binaries, White dwarfs, Hydrodynamics}

\bibitem{mar_ste02}
Marsh, T.R., and Steeghs, D., ``V407 Vul: A direct impact accretor'', {\em Mon.
  Not. R. Astron. Soc.}, {\bf 331}, L7--L11, (2002). Related online version
  (cited on 21 June 2006):
  \newline\url{http://adsabs.harvard.edu/abs/2002MNRAS.331L...7M}.
  \epubtkKeywords{Close binaries, White dwarfs}

\bibitem{mty_gwr81}
Masevich, A.G., Tutukov, A.V., and Yungelson, L.R., ``Gravitational radiation
  and the evolution of dwarf binaries'', {\em Priroda}, {\bf -}, 68--76,
  (1981). Related online version (cited on 21 June 2006):
  \newline\url{http://adsabs.harvard.edu/abs/1981Prir........68M}. In Russian.
  \epubtkKeywords{Compact binaries, Gravitational radiation}

\bibitem{mty76}
Massevitch, A.G., Tutukov, A.V., and Yungelson, L.R., ``Evolution of massive
  close binaries and formation of neutron stars and black holes'', {\em
  Astrophys. Space Sci.}, {\bf 40}, 115--133, (1976).
  \epubtkKeywords{Astrophysics, Close binaries, Neutron stars, Black holes}

\bibitem{McClintock_Remillard03}
McClintock, J.E., and Remillard, R.A., ``Black Hole Binaries'', in Lewin,
  W.H.G., and van~der Klis, M., eds., {\em Compact Stellar X-Ray Sources},
  vol.~39 of Cambridge Astrophysics Series, chapter~4, (Cambridge University
  Press, Cambridge, U.K., 2006). \epubtkKeywords{Black holes, Binary systems}

\bibitem{Miller_Scalo79}
Miller, G.E., and Scalo, J.M., ``The initial mass function and stellar
  birthrate in the solar neighborhood'', {\em Astrophys. J. Suppl. Ser.}, {\bf
  41}, 513--547, (1979). \epubtkKeywords{Astronomical observations}

\bibitem{mir65}
Mironovskii, V.N., ``Gravitational Radiation of Double Stars'', {\em Sov.
  Astron.}, {\bf 9}, 752, (1965). Related online version (cited on 21 June
  2006): \newline\url{http://adsabs.harvard.edu/abs/1965SvA.....9..752M}.
  \epubtkKeywords{Binary systems, Gravitational wave sources}

\bibitem{mitalas76}
Mitalas, R., ``Effect of Asymmetric Explosion on Orbital Elements of Circular
  Binaries'', {\em Astron. Astrophys.}, {\bf 46}, 323, (1976). Related online
  version (cited on 4 November 2006):
  \newline\url{http://adsabs.harvard.edu/abs/1976A&A....46..323M}.
  \epubtkKeywords{Close binaries, Supernovae}

\bibitem{Mochkovitch_al93}
Mochkovitch, R., Hernanz, M., Isern, J., and Martin, X., ``Gamma-ray bursts as
  collimated jets from neutron star/black hole mergers'', {\em Nature}, {\bf
  361}, 236--238, (1993). \epubtkKeywords{Relativistic binary systems,
  Gamma-ray bursts}

\bibitem{mochko90}
Mochkovitch, R., and Livio, M., ``The coalescence of white dwarfs and type I
  supernovae. The merged configuration'', {\em Astron. Astrophys.}, {\bf 236},
  378--384, (1990). Related online version (cited on 21 June 2006):
  \newline\url{http://adsabs.harvard.edu/abs/1990A&A...236..378M}.
  \epubtkKeywords{Close binaries, White dwarfs, Supernovae}

\bibitem{morrison_abs83}
Morrison, R., and McCammon, D., ``Interstellar photoelectric absorption cross
  sections, 0.03--10 keV'', {\em Astrophys. J.}, {\bf 270}, 119--122, (1983).
  Related online version (cited on 21 June 2006):
  \newline\url{http://adsabs.harvard.edu/abs/1983ApJ...270..119M}.
  \epubtkKeywords{Astrophysics}

\bibitem{mhg_96}
Motch, C., Haberl, F., Guillout, P., Pakull, M., Reinsch, K., and Krautter, J.,
  ``New cataclysmic variables from the ROSAT All-Sky Survey'', {\em Astron.
  Astrophys.}, {\bf 307}, 459--469, (1996). Related online version (cited on 21
  June 2006): \newline\url{http://adsabs.harvard.edu/abs/1996A&A...307..459M}.
  \epubtkKeywords{X-ray binaries, White dwarfs}

\bibitem{muno_chandra_cat03}
Muno, M.P., Baganoff, F.K., Bautz, M.W., Brandt, W.N., Broos, P.S., Feigelson,
  E.D., Garmire, G.P., Morris, M.R., Ricker, G.R., and Townsley, L.K., ``A Deep
  Chandra Catalog of X-Ray Point Sources toward the Galactic Center'', {\em
  Astrophys. J.}, {\bf 589}, 225--241, (2003). Related online version (cited on
  21 June 2006):
  \newline\url{http://adsabs.harvard.edu/abs/2003ApJ...589..225M}.
  \epubtkKeywords{X-ray astronomy}

\bibitem{muno_wester_magn06}
Muno, M.P., Clark, J.S., Crowther, P.A., Dougherty, S.M., de~Grijs, R., Law,
  C., McMillan, S.L.W., Morris, M.R., Negueruela, I., Pooley, D.,
  Portegies~Zwart, S.F., and Yusef-Zadeh, F., ``A Neutron Star with a Massive
  Progenitor in Westerlund 1'', {\em Astrophys. J. Lett.}, {\bf 636}, L41--L44,
  (2006). Related online version (cited on 4 November 2006):
  \newline\url{http://adsabs.harvard.edu/abs/2006ApJ...636L..41M}.
  \epubtkKeywords{Neutron stars, Supernovae}

\bibitem{Nakar_al06}
Nakar, E., Gal-Yam, A., and Fox, D.B., ``The Local Rate and the Progenitor
  Lifetimes of Short-Hard Gamma-Ray Bursts: Synthesis and Predictions for the
  Laser Interferometer Gravitational-Wave Observatory'', {\em Astrophys. J.},
  {\bf 650}, 281--290, (2006). Related online version (cited on 4 November
  2006): \newline\url{http://adsabs.harvard.edu/abs/2006ApJ...650..281N}.
  \epubtkKeywords{Gamma-ray bursts, Gravitational wave sources}

\bibitem{nap_01}
Napiwotzki, R., Christlieb, N., Drechsel, H., Hagen, H.-J., Heber, U., Homeier,
  D., Karl, C., Koester, D., Leibundgut, B., Marsh, T.R., Moehler, S.,
  Nelemans, G., Pauli, E.-M., Reimers, D., Renzini, A., and Yungelson, L.R.,
  ``Search for progenitors of supernovae type Ia with SPY'', {\em Astron.
  Nachr.}, {\bf 322}, 411--418, (2001). \epubtkKeywords{Supernovae,
  Astronomical observations}

\bibitem{nap_spy_03}
Napiwotzki, R., Christlieb, N., Drechsel, H., Hagen, H.-J., Heber, U., Homeier,
  D., Karl, C., Koester, D., Leibundgut, B., Marsh, T.R., Moehler, S.,
  Nelemans, G., Pauli, E.-M., Reimers, D., Renzini, A., and Yungelson, L.R.,
  ``SPY -- The Eso Supernovae Type Ia Progenitor Survey'', {\em Messenger},
  {\bf 2003}(112), 25--30, (2003). Related online version (cited on 21 June
  2006): \newline\url{http://adsabs.harvard.edu/abs/2003Msngr.112...25N}.
  \epubtkKeywords{Supernovae, Astronomical observations}

\bibitem{napiwotzki-05}
Napiwotzki, R., Karl, C.A., Nelemans, G., Yungelson, L.R., Christlieb, N.,
  Drechsel, H., Heber, U., Homeier, D., Koester, D., Kruk, J., Leibundgut, B.,
  Marsh, T.R., Moehler, S., Renzini, A., and Reimers, D., ``New Results from
  the Supernova Ia Progenitor Survey'', in Koester, D., and Moehler, S., eds.,
  {\em 14th European Workshop on White Dwarfs}, Workshop in Kiel, Germany, 19
  -- 23 July 2004, vol. 334 of ASP Conference Series,  375, (Astronomical
  Society of the Pacific, San Francisco, U.S.A., 2005). Related online version
  (cited on 21 June 2006):
  \newline\url{http://adsabs.harvard.edu/abs/2005ASPC..334..375N}.
  \epubtkKeywords{Supernovae, Astronomical observations}

\bibitem{nap_dubr}
Napiwotzki, R., Yungelson, L.R., Nelemans, G., Marsh, T.R., Leibundgut, B.,
  Renzini, R., Homeier, D., Koester, D., Moehler, S., Christlieb, N., Reimers,
  D., Drechsel, H., Heber, U., Karl, C., and Pauli, E.-M., ``Double Degenerates
  and Progenitors of Supernovae Type Ia'', in Hilditch, R.W., Hensberge, H.,
  and Pavlovski, K., eds., {\em Spectroscopically and Spatially Resolving the
  Components of Close Binary Stars}, Meeting in Dubrovnik, Croatia, 20 -- 24
  October 2003, vol. 318 of ASP Conference Series,  402--410, (Astronomical
  Society of the Pacific, San Francisco, U.S.A., 2004). Related online version
  (cited on 21 June 2006):
  \newline\url{http://adsabs.harvard.edu/abs/2004ASPC..318..402N}.
  \epubtkKeywords{Supernovae, Binary systems, White dwarfs}

\bibitem{Narayan87}
Narayan, R., ``The birthrate and initial spin period of single radio pulsars'',
  {\em Astrophys. J.}, {\bf 319}, 162--179, (1987). \epubtkKeywords{Pulsars}

\bibitem{nps91}
Narayan, R., Piran, T., and Shemi, A., ``Neutron star and black hole binaries
  in the Galaxy'', {\em Astrophys. J. Lett.}, {\bf 379}, L17--L20, (1991).
  Related online version (cited on 4 November 2006):
  \newline\url{http://adsabs.harvard.edu/abs/1991ApJ...379L..17N}.
  \epubtkKeywords{Compact binaries, Neutron stars, Black holes, Relativistic
  binary systems, Astrophysics}

\bibitem{url01}
NASA Marshall Space Flight Center, ``BATSE Web'', project homepage, (2006). URL
  (cited on 21 June 2006): \newline\url{http://www.batse.msfc.nasa.gov/batse/}.
  \epubtkKeywords{Astronomical observations, Gamma-ray bursts, Pulsars}

\bibitem{nelemans_amcvn05}
Nelemans, G., ``AM CVn stars'', in Hameury, J.-M., and Lasota, J.-P., eds.,
  {\em The Astrophysics of Cataclysmic Variables and Related Objects}, Meeting
  in Strasbourg, France, 11 -- 16 July 2004, vol. 330 of ASP Conference Series,
  ~27, (Astronomical Society of the Pacific, San Francisco, U.S.A., 2005).
  Related online version (cited on 21 June 2006):
  \newline\url{http://adsabs.harvard.edu/abs/2005ASPC..330...27N}.
  \epubtkKeywords{Close binaries, White dwarfs}

\bibitem{url07}
Nelemans, G., ``AM CVn stars'', personal homepage, Radboud Universiteit
  Nijmegen, (2006). URL (cited on 21 June 2006):
  \newline\url{http://www.astro.ru.nl/~nelemans/Research/AMCVn.html}.
  \epubtkKeywords{Close binaries, White dwarfs}

\bibitem{url06}
Nelemans, G., ``Gravitational waves'', personal homepage, Radboud Universiteit
  Nijmegen, (2006). URL (cited on 21 June 2006):
  \newline\url{http://www.astro.ru.nl/~nelemans/Research/GWR.html}.
  \epubtkKeywords{LISA, Gravitational waves, Gravitational wave detectors}

\bibitem{url05}
Nelemans, G., ``LISA Wiki Page'', personal homepage, Radboud Universiteit
  Nijmegen, (2006). URL (cited on 21 June 2006):
  \newline\url{http://www.astro.kun.nl/~nelemans/dokuwiki/doku.php}.
  \epubtkKeywords{LISA, Gravitational waves, Gravitational wave detectors}

\bibitem{url02}
Nelemans, G., ``Preparing for the start of gravitational wave astrophysics'',
  other, Kernfysisch Versneller Instituut, (2006). URL (cited on 21 June 2006):
  \newline\url{http://www.kvi.nl/~berg/appsym4/Nelemans.pdf}.
  \epubtkKeywords{Gravitational waves, Gravitational wave detectors}

\bibitem{nelemans_jonker_ucxb06}
Nelemans, G., and Jonker, P.G., ``Ultra-compact (X-ray) binaries'', {\em New
  Astron. Rev.}, submitted, (2006). Related online version (cited on 21 June
  2006): \newline\url{http://adsabs.harvard.edu/abs/2006astro.ph..5722N}.
  \epubtkKeywords{X-ray binaries}

\bibitem{nelemans04b}
Nelemans, G., Jonker, P.G., Marsh, T.R., and van~der Klis, M., ``Optical
  spectra of the carbon-oxygen accretion discs in the ultra-compact X-ray
  binaries 4U 0614+09, 4U 1543--624 and 2S 0918--549'', {\em Mon. Not. R.
  Astron. Soc.}, {\bf 348}, L7--L11, (2004). Related online version (cited on
  21 June 2006):
  \newline\url{http://adsabs.harvard.edu/abs/2004MNRAS.348L...7N}.
  \epubtkKeywords{X-ray binaries, White dwarfs, Accretion disks}

\bibitem{Nelemans:2006gs}
Nelemans, G., Jonker, P.G., and Steeghs, D., ``Optical spectroscopy of
  (candidate) ultra-compact X-ray binaries: Constraints on the composition of
  the donor stars'', {\em Mon. Not. R. Astron. Soc.}, {\bf 370}, 255--262,
  (2006). Related online version (cited on 21 June 2006):
  \newline\url{http://arXiv.org/abs/astro-ph/0604597}. \epubtkKeywords{X-ray
  binaries, White dwarfs}

\bibitem{nyp01a}
Nelemans, G., Portegies~Zwart, S.F., Verbunt, F., and Yungelson, L.R.,
  ``Population synthesis for double white dwarfs. II. Semi-detached systems: AM
  CVn stars'', {\em Astron. Astrophys.}, {\bf 368}, 939--949, (2001). Related
  online version (cited on 21 June 2006):
  \newline\url{http://adsabs.harvard.edu/abs/2001A&A...368..939N}.
  \epubtkKeywords{Binary systems, White dwarfs, Gravitational wave sources}

\bibitem{nsg_am00}
Nelemans, G., Steeghs, D., and Groot, P.J., ``Spectroscopic evidence for the
  binary nature of AM CVn'', {\em Mon. Not. R. Astron. Soc.}, {\bf 326},
  621--627, (2001). Related online version (cited on 21 June 2006):
  \newline\url{http://adsabs.harvard.edu/abs/2001MNRAS.326..621N}.
  \epubtkKeywords{Close binaries, White dwarfs}

\bibitem{Nelemans_al99}
Nelemans, G., Tauris, T.M., and van~den Heuvel, E.P.J., ``Constraints on mass
  ejection in black hole formation derived from black hole X-ray binaries'',
  {\em Astron. Astrophys.}, {\bf 352}, L87--L90, (1999).
  \epubtkKeywords{Astrophysics, Black holes}

\bibitem{nelemans_tout05}
Nelemans, G., and Tout, C.A., ``Reconstructing the evolution of white dwarf
  binaries: Further evidence for an alternative algorithm for the outcome of
  the common-envelope phase in close binaries'', {\em Mon. Not. R. Astron.
  Soc.}, {\bf 356}, 753--764, (2005). Related online version (cited on 21 June
  2006): \newline\url{http://adsabs.harvard.edu/abs/2005MNRAS.356..753N}.
  \epubtkKeywords{Close binaries, White dwarfs}

\bibitem{Nelemans_al00}
Nelemans, G., Verbunt, F., Yungelson, L.R., and Portegies~Zwart, S.F.,
  ``Reconstructing the evolution of double helium white dwarfs: Envelope loss
  without spiral-in'', {\em Astron. Astrophys.}, {\bf 360}, 1011--1018, (2000).
  Related online version (cited on 13 November 2006):
  \newline\url{http://adsabs.harvard.edu/abs/2000A&A...360.1011N}.
  \epubtkKeywords{Close binaries, Hydrodynamics, Binary systems, White dwarfs}

\bibitem{nyp01}
Nelemans, G., Yungelson, L.R., and Portegies~Zwart, S.F., ``The gravitational
  wave signal from the Galactic disk population of binaries containing two
  compact objects'', {\em Astron. Astrophys.}, {\bf 375}, 890--898, (2001).
  Related online version (cited on 21 June 2006):
  \newline\url{http://adsabs.harvard.edu/abs/2001A&A...375..890N}.
  \epubtkKeywords{Binary systems, White dwarfs, Gravitational wave sources}

\bibitem{nyp04}
Nelemans, G., Yungelson, L.R., and Portegies~Zwart, S.F., ``Short-Period AM CVn
  systems as optical, X-ray and gravitational-wave sources'', {\em Mon. Not. R.
  Astron. Soc.}, {\bf 349}, 181--192, (2004). Related online version (cited on
  21 June 2006):
  \newline\url{http://adsabs.harvard.edu/abs/2004MNRAS.349..181N}.
  \epubtkKeywords{Close binaries, White dwarfs, Gravitational wave sources}

\bibitem{nyp_01}
Nelemans, G., Yungelson, L.R., Portegies~Zwart, S.F., and Verbunt, F.,
  ``Population synthesis for double white dwarfs. I. Detached systems'', {\em
  Astron. Astrophys.}, {\bf 365}, 491--507, (2001). \epubtkKeywords{Close
  binaries, White dwarfs, Gravitational wave sources}

\bibitem{nelson86}
Nelson, L.A., Rappaport, S.A., and Joss, P.C., ``The evolution of ultrashort
  period binary systems'', {\em Astrophys. J.}, {\bf 304}, 231--240, (1986).
  Related online version (cited on 21 June 2006):
  \newline\url{http://adsabs.harvard.edu/abs/1986ApJ...304..231N}.
  \epubtkKeywords{Close binaries}

\bibitem{Nice_al96}
Nice, D.J., Sayer, R.W., and Taylor, J.H., ``PSR J1518+4904: A Mildly
  Relativistic Binary Pulsar System'', {\em Astrophys. J. Lett.}, {\bf 466},
  L87--L90, (1996). \epubtkKeywords{Pulsars, Relativistic binary systems}

\bibitem{Nice_al04}
Nice, D.J., Splaver, E.M., and Stairs, I.H., ``Arecibo Measurements of
  Pulsar--White Dwarf Binaries: Evidence for Heavy Neutron Stars'', in Rasio,
  F.A., and Stairs, I.H., eds., {\em Binary Radio Pulsars}, Meeting at the
  Aspen Center for Physics, Colorado, 12 -- 16 January 2004, vol. 328 of ASP
  Conference Series,  371, (Astronomical Society of the Pacific, San Francisco,
  U.S.A., 2004). Related online version (cited on 21 June 2006):
  \newline\url{http://adsabs.harvard.edu/abs/2005ASPC..328..371N}.
  \epubtkKeywords{Double pulsars}

\bibitem{nomoto_iben85}
Nomoto, K., and Iben~Jr, I., ``Carbon ignition in a rapidly accreting
  degenerate dwarf: A clue to the nature of the merging process in close
  binaries'', {\em Astrophys. J.}, {\bf 297}, 531--537, (1985). Related online
  version (cited on 21 June 2006):
  \newline\url{http://adsabs.harvard.edu/abs/1985ApJ...297..531N}.
  \epubtkKeywords{Close binaries, Accretion, White dwarfs}

\bibitem{nomkondo91}
Nomoto, K., and Kondo, Y., ``Conditions for accretion-induced collapse of white
  dwarfs'', {\em Astrophys. J. Lett.}, {\bf 367}, L19--L22, (1991). Related
  online version (cited on 21 June 2006):
  \newline\url{http://adsabs.harvard.edu/abs/1991ApJ...367L..19N}.
  \epubtkKeywords{Accretion, White dwarfs, Gravitational collapse}

\bibitem{nl00}
Nugis, T., and Lamers, H.J.G.L.M., ``Mass-loss rates of Wolf--Rayet stars as a
  function of stellar parameters'', {\em Astron. Astrophys.}, {\bf 360},
  227--244, (2000). \epubtkKeywords{Stellar evolution}

\bibitem{Shore_al94}
Nussbaumer, H., and Orr, A., eds., {\em Star Clusters: Saas-Fee Advanced Course
  22}, Lecture Notes of the Saas-Fee Advanced Course 22, Les Diablerets,
  Switzerland, April 6--11, 1992, Saas-Fee Advanced Courses, (Springer, Berlin,
  Germany; New York, U.S.A., 1994). \epubtkKeywords{Binary systems}

\bibitem{Oechslin_Janka06}
Oechslin, R., and Janka, T., ``Short Gamma-Ray Bursts from Binary Neutron Star
  Mergers'', (April, 2006). URL (cited on 21 June 2006):
  \newline\url{http://arXiv.org/abs/astro-ph/0604562}.
  \epubtkKeywords{Gamma-ray bursts, Relativistic binary stars}

\bibitem{orosz_bh03}
Orosz, J.A, ``Inventory of black hole binaries'', in van~der Hucht, K.A.,
  Herrero, A., and Esteban, C., eds., {\em A Massive Star Odyssey: From Main
  Sequence to Supernova}, Conference in Lanzarote, Spain, 24--28 June 2002,
  vol. 212 of IAU Symposia,  365, (Astronomical Society of the Pacific, San
  Francisco, U.S.A., 2003). Related online version (cited on 4 November 2006):
  \newline\url{http://arXiv.org/abs/astro-ph/0209041}. \epubtkKeywords{Stellar
  evolution, Black holes}

\bibitem{Oshaughnessy_al05}
O'Shaughnessy, R., Kim, C., Fragos, T., Kalogera, V., and Belczynski, K.,
  ``Constraining Population Synthesis Models via the Binary Neutron Star
  Population'', {\em Astrophys. J.}, {\bf 633}, 1076--1084, (2005). Related
  online version (cited on 21 June 2006):
  \newline\url{http://adsabs.harvard.edu/abs/2005ApJ...633.1076O}.
  \epubtkKeywords{Relativistic binary systems}

\bibitem{Ostriker75}
Ostriker, J.P., ``Common Envelope Binaries'', Structure and Evolution of Close
  Binary Systems (IAU Symposium 73), Cambridge, England, 28 July -- 1 August,
  1975, conference paper, (1976). \epubtkKeywords{Binary systems}

\bibitem{Paciesas_al99}
Paciesas, W.S., Meegan, C.A., Pendleton, G.N.and~Briggs, M.S., Kouveliotou, C.,
  Koshut, T.M., Lestrade, J.P., McCollough, M.L., Brainerd, J.J., Hakkila, J.,
  Henze, W., Preece, R.D., Connaughton, V., Kippen, R.M., Mallozzi, R.S.,
  Fishman, G.J., Richardson, G.A., and Sahi, M., ``The Fourth BATSE Gamma-Ray
  Burst Catalog (Revised)'', {\em Astrophys. J. Suppl. Ser.}, {\bf 122},
  465--495, (1999). \epubtkKeywords{Gamma-ray bursts}

\bibitem{pac67a}
Paczy\'nski, B., ``Gravitational Waves and the Evolution of Close Binaries'',
  {\em Acta Astron.}, {\bf 17}, 287, (1967). Related online version (cited on
  21 June 2006):
  \newline\url{http://adsabs.harvard.edu/abs/1967AcA....17..287P}.
  \epubtkKeywords{Close binaries, Gravitational wave sources}

\bibitem{pac76}
Paczy\'nski, B., ``Common envelope binaries'', in Eggleton, P.P., Mitton, S.,
  and Whelan, J., eds., {\em Structure and Evolution of Close Binary Systems
  (IAU Symposium 73)}, Conference in Cambridge, U.K., 28 July -- 1 August 1975,
   75--80, (D. Reidel, Dordrecht, Netherlands; Boston, U.S.A., 1976).
  \epubtkKeywords{Close binaries, Hydrodynamics}

\bibitem{Paczynski86}
Paczy\'nski, B., ``Gamma-ray bursters at cosmological distances'', {\em
  Astrophys. J. Lett.}, {\bf 308}, L43--L46, (1986).
  \epubtkKeywords{Relativistic binary systems, Gamma-ray bursts}

\bibitem{Paczynski91}
Paczy\'nski, B., ``Cosmological gamma-ray bursts'', {\em Acta Astron.}, {\bf
  41}, 257--267, (1991). \epubtkKeywords{Relativistic binary systems, Gamma-ray
  bursts}

\bibitem{panagia_radio06}
Panagia, N., Van~Dyk, S.D., Weiler, K.W., Sramek, R.A., Stockdale, C.J., and
  Murata, K.P., ``A Search for Radio Emission from Type Ia Supernovae'', {\em
  Astrophys. J.}, submitted, (2006). Related online version (cited on 21 June
  2006): \newline\url{http://arXiv.org/abs/astro-ph/0603808}.
  \epubtkKeywords{Supernovae, Radio emission}

\bibitem{Panchenko_al99}
Panchenko, I.E., Lipunov, V.M., Postnov, K.A., and Prokhorov, M.E., ``Stellar
  evolution, GRB and their hosts'', {\em Astron. Astrophys. Suppl.}, {\bf 138},
  517--518, (1999). \epubtkKeywords{Relativistic binary systems, Gamma-ray
  bursts}

\bibitem{Perez-Gonzalez_al03}
P\'erez-Gonz\'alez, P.G., Zamorano, J., Gallego, J., Arag\'on-Salamanca, A.,
  and Gil~de Paz, A., ``Spatial Analysis of the H{$\alpha$} Emission in the
  Local Star-Forming UCM Galaxies'', {\em Astrophys. J.}, {\bf 591}, 827--842,
  (2003). \epubtkKeywords{Astronomical observations}

\bibitem{perlmutter_cosm99}
Perlmutter, S., Aldering, G., Goldhaber, G., Knop, R.A., Nugent, P.E., Castro,
  P.G., Deustua, S., Fabbro, S., Goobar, A., Groom, D.E., Hook, I.M., Kim,
  A.G., Kim, M.Y., Lee, J.C., Nunes, N.J., Pain, R., Pennypacker, C.R., Quimby,
  R., Lidman, C., Ellis, R.S., Irwin, M., McMahon, R.G., Ruiz-Lapuente, P.,
  Walton, N., Schaefer, B., Boyle, B.J., Filippenko, A.V., Matheson, T.,
  Fruchter, A.S., Panagia, N., Newberg, H.J.M., Couch, W.J., and Project, The
  Supernova~Cosmology, ``Measurements of Omega and Lambda from 42 High-Redshift
  Supernovae'', {\em Astrophys. J.}, {\bf 517}, 565--586, (1999). Related
  online version (cited on 21 June 2006):
  \newline\url{http://adsabs.harvard.edu/abs/1999ApJ...517..565P}.
  \epubtkKeywords{Supernovae, Cosmology}

\bibitem{Peters64}
Peters, P.C., ``Gravitational Radiation and the Motion of Two Point Masses'',
  {\em Phys. Rev.}, {\bf 136}, B1224--B1232, (1964). \epubtkKeywords{Binary
  systems, Gravitational radiation}

\bibitem{Pfahl_al05}
Pfahl, E., Podsiadlowski, P., and Rappaport, S., ``Relativistic Binary Pulsars
  with Black Hole Companions'', {\em Astrophys. J.}, {\bf 628}, 343--352,
  (2005). Related online version (cited on 21 June 2006):
  \newline\url{http://adsabs.harvard.edu/abs/2005ApJ...628..343P}.
  \epubtkKeywords{Pulsars, Black holes, Binary systems}

\bibitem{Pfahl_al02a}
Pfahl, E., Rappaport, S., and Podsiadlowski, P., ``A Comprehensive Study of
  Neutron Star Retention in Globular Clusters'', {\em Astrophys. J.}, {\bf
  573}, 283--305, (2002). Related online version (cited on 4 November 2006):
  \newline\url{http://adsabs.harvard.edu/abs/2002ApJ...573..283P}.
  \epubtkKeywords{Neutron stars, Star clusters, Supernovae}

\bibitem{pfahl_low03}
Pfahl, E., Rappaport, S., and Podsiadlowski, P., ``The Galactic Population of
  Low- and Intermediate-Mass X-Ray Binaries'', {\em Astrophys. J.}, {\bf 597},
  1036--1048, (2003). Related online version (cited on 21 June 2006):
  \newline\url{http://adsabs.harvard.edu/abs/2003ApJ...597.1036P}.
  \epubtkKeywords{X-ray binaries}

\bibitem{prps02}
Pfahl, E., Rappaport, S., Podsiadlowski, P., and Spruit, H., ``A New Class of
  High-Mass X-Ray Binaries: Implications for Core Collapse and Neutron Star
  Recoil'', {\em Astrophys. J.}, {\bf 574}, 364--376, (2002). Related online
  version (cited on 4 November 2006):
  \newline\url{http://adsabs.harvard.edu/abs/2002ApJ...574..364P}.
  \epubtkKeywords{X-ray binaries, Neutron stars, Supernovae}

\bibitem{Phinney91}
Phinney, E.S., ``The rate of neutron star binary mergers in the universe:
  Minimal predictions for gravity wave detectors'', {\em Astrophys. J. Lett.},
  {\bf 380}, L17--L21, (1991). \epubtkKeywords{Relativistic binary systems}

\bibitem{piers_99}
Piersanti, L., Cassisi, S., Iben~Jr, I., and Tornamb\'e, A., ``On the Very Long
  Term Evolutionary Behavior of Hydrogen-accreting Low-Mass CO White Dwarfs'',
  {\em Astrophys. J. Lett.}, {\bf 521}, L59--L62, (1999). \epubtkKeywords{Close
  binaries, Accretion, White dwarfs}

\bibitem{piers_03b}
Piersanti, L., Gagliardi, S., Iben~Jr, I., and Tornamb{\' e}, A.,
  ``Carbon-Oxygen White Dwarf Accreting CO-Rich Matter. II. Self-Regulating
  Accretion Process up to the Explosive Stage'', {\em Astrophys. J.}, {\bf
  598}, 1229--1238, (2003). Related online version (cited on 21 June 2006):
  \newline\url{http://adsabs.harvard.edu/abs/2003ApJ...598.1229P}.
  \epubtkKeywords{Close binaries, Accretion, White dwarfs}

\bibitem{piers_03a}
Piersanti, L., Gagliardi, S., Iben~Jr, I., and Tornamb{\' e}, A.,
  ``Carbon-Oxygen White Dwarfs Accreting CO-Rich Matter. I. A Comparison
  between Rotating and Nonrotating Models'', {\em Astrophys. J.}, {\bf 583},
  885--901, (2003). Related online version (cited on 21 June 2006):
  \newline\url{http://adsabs.harvard.edu/abs/2003ApJ...583..885P}.
  \epubtkKeywords{Close binaries, Accretion, White dwarfs}

\bibitem{Piran_Guetta06}
Piran, T., and Guetta, D., ``The rate and luminosity function of Short GRBs'',
  in Holt, S.S., Gehrels, N., and Nousek, J.A., eds., {\em Gamma-Ray Bursts in
  the Swift Era}, Sixteenth Maryland Astrophysics Conference, Washington, DC,
  29 November -- 2 December 2005, vol. 836 of AIP Conference Proceedings,
  (American Institute of Physics, Melville, U.S.A., 2006). Related online
  version (cited on 21 June 2006):
  \newline\url{http://adsabs.harvard.edu/abs/2006astro.ph..2208P}.
  \epubtkKeywords{Gamma-ray bursts, Relativistic binary stars}

\bibitem{phr_am03}
Podsiadlowski, P., Han, Z., and Rappaport, S.A., ``Cataclysmic variables with
  evolved secondaries and the progenitors of AM CVn stars'', {\em Mon. Not. R.
  Astron. Soc.}, {\bf 340}, 1214--1228, (2003). Related online version (cited
  on 4 November 2006):
  \newline\url{http://adsabs.harvard.edu/abs/2003MNRAS.340.1214P}.
  \epubtkKeywords{Close binaries, White dwarfs}

\bibitem{Podsiadlowski_al04}
Podsiadlowski, P., Langer, N., Poelarends, A.J.T., Rappaport, S., Heger, A.,
  and Pfahl, E., ``The Effects of Binary Evolution on the Dynamics of Core
  Collapse and Neutron Star Kicks'', {\em Astrophys. J.}, {\bf 612},
  1044--1051, (2004). \epubtkKeywords{Binary systems, Supernovae}

\bibitem{Podsiadlowski_al03}
Podsiadlowski, P., Rappaport, S., and Han, Z., ``On the formation and evolution
  of black hole binaries'', {\em Mon. Not. R. Astron. Soc.}, {\bf 341},
  385--404, (2003). \epubtkKeywords{Binary systems, Black holes}

\bibitem{prp_low02}
Podsiadlowski, P., Rappaport, S., and Pfahl, E.D., ``Evolutionary Sequences for
  Low- and Intermediate-Mass X-Ray Binaries'', {\em Astrophys. J.}, {\bf 565},
  1107--1133, (2002). Related online version (cited on 21 June 2006):
  \newline\url{http://adsabs.harvard.edu/abs/2002ApJ...565.1107P}.
  \epubtkKeywords{Stellar evolution, Binary systems}

\bibitem{Podsiadlowski_al05}
Podsiadlowski, Ph., Dewi, J.D.M., Lesaffre, P., Miller, J.C., Newton, W.G., and
  Stone, J.R., ``The Double Pulsar J0737--3039: Testing the Neutron Star
  Equation of State'', {\em Mon. Not. R. Astron. Soc.}, {\bf 361}, 1243--1249,
  (2005). Related online version (cited on 21 June 2006):
  \newline\url{http://arXiv.org/abs/astro-ph/0506566}. \epubtkKeywords{Pulsars,
  Supernovae}

\bibitem{Politano_Weiler06}
Politano, M., and Weiler, K.P., ``The Distribution of Secondary Masses in
  Post-Common-Envelope Binaries: A Potential Test of Disrupted Magnetic
  Braking'', {\em Astrophys. J. Lett.}, {\bf 641}, L137--L140, (2006). Related
  online version (cited on 21 June 2006):
  \newline\url{http://adsabs.harvard.edu/abs/2006ApJ...641L.137P}.
  \epubtkKeywords{Binary stars}

\bibitem{pol_98}
Pols, O.R., Schr\"oder, K.-P., Hurley, J.R., Tout, C.A., and Eggleton, P.P.,
  ``Stellar evolution models for $Z = 0.0001$ to 0.03'', {\em Mon. Not. R.
  Astron. Soc.}, {\bf 298}, 525--536, (1998). Related online version (cited on
  21 June 2006):
  \newline\url{http://adsabs.harvard.edu/abs/1998MNRAS.298..525P}.
  \epubtkKeywords{Stellar evolution}

\bibitem{Popov_Prokhorov04}
Popov, S.B., and Prokhorov, M.E., ``Population synthesis in astrophysics'',
  (November, 2004). URL (cited on 21 June 2006):
  \newline\url{http://arXiv.org/abs/astro-ph/0411792}. \epubtkKeywords{Binary
  stars, Neutron stars}

\bibitem{Popova_al82}
Popova, E.I., Tutukov, A.V., and Yungelson, L.R., ``Study of physical
  properties of spectroscopic binary stars'', {\em Astrophys. J. Suppl. Ser.},
  {\bf 88}, 55--80, (1982). \epubtkKeywords{Astronomical observations, Binary
  systems}

\bibitem{PortegiesZwart_al97}
Portegies~Zwart, S.F., Verbunt, F., and Ergma, E., ``The formation of
  black-holes in low-mass X-ray binaries'', {\em Astron. Astrophys.}, {\bf
  321}, 207--212, (1997). \epubtkKeywords{Astrophysics, Black holes}

\bibitem{py98}
Portegies~Zwart, S.F., and Yungelson, L.R., ``Formation and Evolution of Binary
  Neutron Stars'', {\em Astron. Astrophys.}, {\bf 332}, 173--188, (1998).
  \epubtkKeywords{Close binaries, Double pulsars, Neutron stars, Relativistic
  binary systems}

\bibitem{pp98}
Postnov, K.A., and Prokhorov, M.E., ``Galactic binary gravitational wave noise
  within LISA frequency band'', {\em Astrophys. J.}, {\bf 494}, 674--679,
  (1998). Related online version (cited on 21 June 2006):
  \newline\url{http://arXiv.org/abs/astro-ph/9801034}. \epubtkKeywords{Compact
  binaries, Gravitational radiation, LISA}

\bibitem{Postnov_Prokhorov00}
Postnov, K.A., and Prokhorov, M.E., ``Binary black hole formaiton and
  merging'', in Tr{\^{a}}n Than~V{\^{a}}n, J., Dumarchez, J., Raynoud, S.,
  Salomon, C., Thorsett, S., and Vinet, J.Y., eds., {\em Gravitational Waves
  and Experimental Gravity}, XXXIVth Rencontres de Moriond, Les Arcs, France,
  23 -- 30 January 1999,  113--118, (World Publishers, Hanoi, Vietnam, 2000).
  \epubtkKeywords{Relativistic binary stars, Black holes}

\bibitem{prkov95}
Prialnik, D., and Kovetz, A., ``An extended grid of multicycle nova evolution
  models'', {\em Astrophys. J.}, {\bf 445}, 789--810, (1995). Related online
  version (cited on 21 June 2006):
  \newline\url{http://adsabs.harvard.edu/abs/1995ApJ...445..789P}.
  \epubtkKeywords{Close binaries, White dwarfs}

\bibitem{Prince_al91}
Prince, T.A., Anderson, S.B., Kulkarni, S.R., and Wolszczan, A., ``Timing
  observations of the 8 hour binary pulsar 2127+11C in the globular cluster
  M15'', {\em Astrophys. J. Lett.}, {\bf 374}, L41--L44, (1991).
  \epubtkKeywords{Double pulsars}

\bibitem{pringle77}
Pringle, J.E., ``Soft X-ray emission from dwarf novae'', {\em Mon. Not. R.
  Astron. Soc.}, {\bf 178}, 195--202, (1977). Related online version (cited on
  21 June 2006):
  \newline\url{http://adsabs.harvard.edu/abs/1977MNRAS.178..195P}.
  \epubtkKeywords{White dwarfs, X-ray binaries}

\bibitem{Prokhorov_Postnov97}
Prokhorov, M.E., and Postnov, K.A., ``Direct observation of the kick during the
  birth of a neutron star in the binary pulsar PSR B1259--63'', {\em Astron.
  Lett.}, {\bf 23}, 439--444, (1997). \epubtkKeywords{Supernovae, Neutron
  stars}

\bibitem{Raguzova_Popov05}
Raguzova, N.V., and Popov, S.B., ``Be X-ray binaries and candidates'', {\em
  Astron. Astrophys. Trans.}, {\bf 24}, 151--185, (2005). Related online
  version (cited on 21 June 2006):
  \newline\url{http://adsabs.harvard.edu/abs/2005A&AT...24..151R}.
  \epubtkKeywords{Binary stars}

\bibitem{ramsay_rats2}
Ramsay, G., Brocksopp, C., Groot, P.J., Hakala, P., Lehto, H.J., Marsh, T.R.,
  Napiwotzki, R., Nelemans, G., Potter, S., Slee, B., Steeghs, D., and Wu, K.,
  ``Recent observational progress in AM CVn binaries'', (October, 2006). URL
  (cited on 4 November 2006):
  \newline\url{http://arXiv.org/abs/astro-ph/0610357}. \epubtkKeywords{Close
  binaries, White dwarfs, Astronomical observations}

\bibitem{ramsay_am06}
Ramsay, G., Groot, P.J., Marsh, T.R., Nelemans, G., Steeghs, D., and Hakala,
  P., ``XMM-Newton observations of AM CVn binaries: V396 Hya and SDSS
  J1240-01'', {\em Astron. Astrophys.}, {\bf 457}, 623--627, (2006). Related
  online version (cited on 4 November 2006):
  \newline\url{http://adsabs.harvard.edu/abs/2006A&A...457..623R}.
  \epubtkKeywords{X-ray astronomy, Close binaries, White dwarfs}

\bibitem{rats05}
Ramsay, G., and Hakala, P., ``RApid Temporal Survey (RATS) -- I. Overview and
  first results'', {\em Mon. Not. R. Astron. Soc.}, {\bf 360}, 314--321,
  (2005). Related online version (cited on 21 June 2006):
  \newline\url{http://adsabs.harvard.edu/abs/2005MNRAS.360..314R}.
  \epubtkKeywords{Astronomical observations}

\bibitem{rdss94}
Rappaport, S., {Di Stefano}, R., and Smith, J.D., ``Formation and evolution of
  luminous supersoft X-ray sources'', {\em Astrophys. J.}, {\bf 426}, 692--703,
  (1994). Related online version (cited on 21 June 2006):
  \newline\url{http://adsabs.harvard.edu/abs/1994ApJ...426..692R}.
  \epubtkKeywords{X-ray astronomy}

\bibitem{rj84}
Rappaport, S., and Joss, P.C., ``The lower main sequence and the nature of
  secondary stars in ultracompact binaries'', {\em Astrophys. J.}, {\bf 283},
  232--240, (1984). Related online version (cited on 21 June 2006):
  \newline\url{http://adsabs.harvard.edu/abs/1984ApJ...283..232R}.
  \epubtkKeywords{Close binaries, White dwarfs}

\bibitem{Rasio_Livio96}
Rasio, F.A., and Livio, M., ``On the Formation and Evolution of Common Envelope
  Systems'', {\em Astrophys. J.}, {\bf 471}, 366, (1996).
  \epubtkKeywords{Binary systems}

\bibitem{rasio_cb_gc00}
Rasio, F.A., Pfahl, E.D., and Rappaport, S., ``Formation of Short-Period Binary
  Pulsars in Globular Clusters'', {\em Astrophys. J. Lett.}, {\bf 532},
  L47--L50, (2000). Related online version (cited on 21 June 2006):
  \newline\url{http://adsabs.harvard.edu/abs/2000ApJ...532L..47R}.
  \epubtkKeywords{Binary pulsars, Star clusters}

\bibitem{Rees76}
Rees, M.J., ``Opacity-limited hierarchical fragmentation and the masses of
  protostars'', {\em Mon. Not. R. Astron. Soc.}, {\bf 176}, 483--486, (1976).
  \epubtkKeywords{Astrophysics, Stars}

\bibitem{rrw74}
Refsdal, S., Roth, M.L., and Weigert, A., ``On the binary system AS Eri'', {\em
  Astron. Astrophys.}, {\bf 36}, 113--122, (1974). Related online version
  (cited on 21 June 2006):
  \newline\url{http://adsabs.harvard.edu/abs/1974A&A....36..113R}.
  \epubtkKeywords{Close binaries, White dwarfs}

\bibitem{Remillard_McClintock06}
Remillard, R.A., and McClintock, J.E., ``X-Ray Properties of Black-Hole
  Binaries'', {\em Annu. Rev. Astron. Astrophys.}, {\bf 44}, (2006). Related
  online version (cited on 21 June 2006):
  \newline\url{http://adsabs.harvard.edu/abs/2006astro.ph..6352R}.
  \epubtkKeywords{Black holes, Astronomical observations}

\bibitem{Ricci_Brillet97}
Ricci, F., and Brillet, A., ``A Review of Gravitational Wave Detectors'', {\em
  Annu. Rev. Nucl. Part. Sci.}, {\bf 47}, 111--156, (1997). Related online
  version (cited on 21 June 2006):
  \newline\url{http://adsabs.harvard.edu/abs/1997ARNPS..47..111R}.
  \epubtkKeywords{Gravitational waves, Gravitational wave detectors}

\bibitem{riess_cosm98}
Riess, A.G., Filippenko, A.V., Challis, P., Clocchiatti, A., Diercks, A.,
  Garnavich, P.M., Gilliland, R.L., Hogan, C.J., Jha, S., Kirshner, R.P.,
  Leibundgut, B., Phillips, M.M., Reiss, D., Schmidt, B.P., Schommer, R.A.,
  Smith, R.C., Spyromilio, J., Stubbs, C., Suntzeff, N.B., and Tonry, J.,
  ``Observational Evidence from Supernovae for an Accelerating Universe and a
  Cosmological Constant'', {\em Astron. J.}, {\bf 116}, 1009--1038, (1998).
  Related online version (cited on 21 June 2006):
  \newline\url{http://adsabs.harvard.edu/abs/1998AJ....116.1009R}.
  \epubtkKeywords{Supernovae, Cosmology}

\bibitem{ritossa_berro_one96}
Ritossa, C., Garc{\'{i}}a-Berro, E., and Iben~Jr, I., ``On the Evolution of
  Stars That Form Electron-Degenerate Cores Processed by Carbon Burning. II.
  Isotope Abundances and Thermal Pulses in a $10 M_\odot$ Model with an ONe
  Core and Applications to Long-Period Variables, Classical Novae, and
  Accretion-Induced Collapse'', {\em Astrophys. J.}, {\bf 460}, 489, (1996).
  Related online version (cited on 21 June 2006):
  \newline\url{http://adsabs.harvard.edu/abs/1996ApJ...460..489R}.
  \epubtkKeywords{Stellar evolution, White dwarfs}

\bibitem{ritossa_berro_one99}
Ritossa, C., Garc\'ia-Berro, E., and Iben~Jr, I., ``On the Evolution of Stars
  that Form Electron-degenerate Cores Processed by Carbon Burning. V. Shell
  Convection Sustained by Helium Burning, Transient Neon Burning, Dredge-Out,
  URCA Cooling, and Other Properties of an $11 M_\odot$ Population I Model
  Star'', {\em Astrophys. J.}, {\bf 515}, 381--397, (1999). Related online
  version (cited on 21 June 2006):
  \newline\url{http://adsabs.harvard.edu/abs/1999ApJ...515..381R}.
  \epubtkKeywords{Stellar evolution, White dwarfs}

\bibitem{ritter-kolb-cat03}
Ritter, H., and Kolb, U., ``Catalogue of cataclysmic binaries, low-mass X-ray
  binaries and related objects (Seventh edition)'', {\em Astron. Astrophys.},
  {\bf 404}, 301--303, (2003). Related online version (cited on 21 June 2006):
  \newline\url{http://adsabs.harvard.edu/abs/2003A&A...404..301R}.
  \epubtkKeywords{Close binaries, White dwarfs, X-ray binaries}

\bibitem{url04}
Ritter, H., and Kolb, U., ``Catalogue of Cataclysmic Binaries, Low-Mass X-Ray
  Binaries and Related Objects'', project homepage, Open University, (2006).
  URL (cited on 21 June 2006): \newline\url{http://physics.open.ac.uk/RKcat/}.
  \epubtkKeywords{Close binaries, White dwarfs, X-ray binaries}

\bibitem{rob_shaft87}
Robinson, E.L., and Shafter, A.W., ``An Upper Limit to the Space Density of
  Short-period Noninteracting Binary White Dwarfs'', {\em Astrophys. J.}, {\bf
  322}, 296, (1987). Related online version (cited on 21 June 2006):
  \newline\url{http://adsabs.harvard.edu/abs/1987ApJ...322..296R}.
  \epubtkKeywords{Compact binaries, White dwarfs}

\bibitem{roelofs_am05}
Roelofs, G.H.A., Groot, P.J., Marsh, T.R., Steeghs, D., Barros, S.C.C., and
  Nelemans, G., ``SDSS J124058.03-015919.2: A new AM CVn star with a 37-min
  orbital period'', {\em Mon. Not. R. Astron. Soc.}, {\bf 361}, 487--494,
  (2005). Related online version (cited on 21 June 2006):
  \newline\url{http://adsabs.harvard.edu/abs/2005MNRAS.361..487R}.
  \epubtkKeywords{Close binaries, White dwarfs}

\bibitem{roelofs_am06}
Roelofs, G.H.A., Groot, P.J., Nelemans, G., Marsh, T.R., and Steeghs, D.,
  ``Kinematics of the ultra-compact helium accretor AM Canum Venaticorum'',
  (June, 2006). URL (cited on 4 November 2006):
  \newline\url{http://arXiv.org/abs/astro-ph/0606327}. \epubtkKeywords{Close
  binaries, White dwarfs, Accretion}

\bibitem{rogan_lisa06}
Rogan, A., and Bose, S., ``Parameter estimation of binary compact objects with
  LISA: Effects of time-delay interferometry, Doppler modulation, and frequency
  evolution'', (May, 2006). URL (cited on 21 June 2006):
  \newline\url{http://arXiv.org/abs/astro-ph/0605034}. \epubtkKeywords{Compact
  binaries, LISA}

\bibitem{Romani90}
Romani, R.W., ``A unified model of neutron-star magnetic fields'', {\em
  Nature}, {\bf 347}, 741--743, (1990). \epubtkKeywords{Neutron stars}

\bibitem{Romani95}
Romani, R.W., ``Neutron Star Magnetic Fields'', in Fruchter, A.S., Tavani, M.,
  and Backer, D.C., eds., {\em Millisecond Pulsars: A Decade of Surprise},
  Conference in Aspen, Colorado, 3 -- 7 January 1994, vol.~72 of ASP Conference
  Series,  288, (Astronomical Society of the Pacific, San Francisco, U.S.A.,
  1995). \epubtkKeywords{Neutron stars}

\bibitem{rs83}
Ruderman, M.A., and Shaham, J., ``Fate of very low-mass secondaries in
  accreting binaries and the 1.5-ms pulsar'', {\em Nature}, {\bf 304},
  425--427, (1983). \epubtkKeywords{Close binaries, Accretion, Pulsars}

\bibitem{ruiz_canal_cosm98}
Ruiz-Lapuente, P., and Canal, R., ``Type Ia Supernova Counts at High $z$:
  Signatures of Cosmological Models and Progenitors'', {\em Astrophys. J.
  Lett.}, {\bf 497}, L57--L60, (1998). Related online version (cited on 21 June
  2006): \newline\url{http://adsabs.harvard.edu/abs/1998ApJ...497L..57R}.
  \epubtkKeywords{Supernovae, Cosmology}

\bibitem{rlapuente04}
Ruiz-Lapuente, P., Comeron, F., M\'endez, J., Canal, R., Smartt, S.J.,
  Filippenko, A.V., Kurucz, R.L., Chornock, R., Foley, R.J., Stanishev, V., and
  Ibata, R., ``The binary progenitor of Tycho Brahe's 1572 supernova'', {\em
  Nature}, {\bf 431}, 1069--1072, (2004). \epubtkKeywords{Binary systems,
  Supernovae}

\bibitem{sac97}
Sackett, P.D., ``Does the Milky Way Have a Maximal Disk?'', {\em Astrophys.
  J.}, {\bf 483}, 103, (1997). Related online version (cited on 21 June 2006):
  \newline\url{http://adsabs.harvard.edu/abs/1997ApJ...483..103S}.
  \epubtkKeywords{Galactic structure}

\bibitem{slo88}
Saffer, R.A., Liebert, J., and Olszewski, E.W., ``Discovery of a close detached
  binary DA white dwarf system'', {\em Astrophys. J.}, {\bf 334}, 947--957,
  (1988). \epubtkKeywords{Close binaries, White dwarfs, Astronomical
  observations}

\bibitem{sly98}
Saffer, R.A., Livio, M., and Yungelson, L.R., ``Close Binary White Dwarf
  Systems: Numerous New Detections and Their Interpretation'', {\em Astrophys.
  J.}, {\bf 502}, 394, (1998). \epubtkKeywords{Close binaries, White dwarfs,
  Astronomical observations}

\bibitem{saio_rot_no_dd04}
Saio, H., and Nomoto, K., ``Off-Center Carbon Ignition in Rapidly Rotating,
  Accreting Carbon-Oxygen White Dwarfs'', {\em Astrophys. J.}, {\bf 615},
  444--449, (2004). Related online version (cited on 21 June 2006):
  \newline\url{http://adsabs.harvard.edu/abs/2004ApJ...615..444S}.
  \epubtkKeywords{Close binaries, Accretion, White dwarfs}

\bibitem{salpeter55}
Salpeter, E.E., ``The Luminosity Function and Stellar Evolution'', {\em
  Astrophys. J.}, {\bf 121}, 161--167, (1955). Related online version (cited on
  21 June 2006):
  \newline\url{http://adsabs.harvard.edu/abs/1955ApJ...121..161S}.
  \epubtkKeywords{Stars, Astronomical observations}

\bibitem{san72}
Sandage, A., ``The redshift-distance relation. II. The Hubble diagram and its
  scatter for first-ranked cluster galaxies: A formal value for $q_0$'', {\em
  Astrophys. J.}, {\bf 178}, 1--24, (1972). Related online version (cited on 21
  June 2006): \newline\url{http://adsabs.harvard.edu/abs/1972ApJ...178....1S}.
  \epubtkKeywords{Galaxy clustering, Cosmology}

\bibitem{sandq_binhedw00}
Sandquist, E.L., Taam, R.E., and Burkert, A., ``On the Formation of Helium
  Double Degenerate Stars and Pre-Cataclysmic Variables'', {\em Astrophys. J.},
  {\bf 533}, 984--997, (2000). Related online version (cited on 21 June 2006):
  \newline\url{http://adsabs.harvard.edu/abs/2000ApJ...533..984S}.
  \epubtkKeywords{Close binaries, White dwarfs}

\bibitem{sandq_ce98}
Sandquist, E.L., Taam, R.E., Chen, X., Bodenheimer, P., and Burkert, A.,
  ``Double Core Evolution. X. Through the Envelope Ejection Phase'', {\em
  Astrophys. J.}, {\bf 500}, 909, (1998). Related online version (cited on 21
  June 2006): \newline\url{http://adsabs.harvard.edu/abs/1998ApJ...500..909S}.
  \epubtkKeywords{Close binaries, Hydrodynamics}

\bibitem{skh86}
Savonije, G.J., de~Kool, M., and van~den Heuvel, E.P.J., ``The minimum orbital
  period for ultra-compact binaries with helium burning secondaries'', {\em
  Astron. Astrophys.}, {\bf 155}, 51--57, (1986). Related online version (cited
  on 21 June 2006):
  \newline\url{http://adsabs.harvard.edu/abs/1986A&A...155...51S}.
  \epubtkKeywords{Close binaries, Stellar evolution}

\bibitem{schatz_msw62}
Schatzman, E., ``A theory of the role of magnetic activity during star
  formation'', {\em Ann. Astrophys.}, {\bf 25}, 18, (1962). Related online
  version (cited on 21 June 2006):
  \newline\url{http://adsabs.harvard.edu/abs/1962AnAp...25...18S}.
  \epubtkKeywords{Stars}

\bibitem{Schiminovich_al05}
Schiminovich, D., Ilbert, O., Arnouts, S., Milliard, B., Tresse, L.,
  Le~F\`evre, O., Treyer, M., Wyder, T.K., Budav\'ari, T., Zucca, E., Zamorani,
  G., Martin, D.C., Adami, C., Arnaboldi, M., Bardelli, S., Barlow, T.,
  Bianchi, L., Bolzonella, M., Bottini, D., Byun, Y.-I., Cappi, A., Contini,
  T., Charlot, S., Donas, J., Forster, K., Foucaud, S., Franzetti, P.,
  Friedman, P.G., Garilli, B., Gavignaud, I., Guzzo, L., Heckman, T.M., Hoopes,
  C., Iovino, A., Jelinsky, P., Le~Brun, V.and~Lee, Y.-W., Maccagni, D.,
  Madore, B.F., Malina, R.and~Marano, B., Marinoni, C., McCracken, H.J.,
  Mazure, A., Meneux, B., Morrissey, P., Neff, S., Paltani, S.and~Pell{\`o},
  R., Picat, J.P., Pollo, A., Pozzetti, L., Radovich, M., Rich, R.M.,
  Scaramella, R., Scodeggio, M., Seibert, M., Siegmund, O., Small, T., Szalay,
  A.S., Vettolani, G., Welsh, B., Xu, C.K., and Zanichelli, A., ``The
  GALEX-VVDS Measurement of the Evolution of the Far-Ultraviolet Luminosity
  Density and the Cosmic Star Formation Rate'', {\em Astrophys. J. Lett.}, {\bf
  619}, L47--L50, (2005). \epubtkKeywords{Astronomical observations}

\bibitem{spn_wd04}
Schr{\"o}der, K.-P., Pauli, E.-M., and Napiwotzki, R., ``A population model of
  the solar neighbourhood thin disc white dwarfs'', {\em Mon. Not. R. Astron.
  Soc.}, {\bf 354}, 727--736, (2004). Related online version (cited on 21 June
  2006): \newline\url{http://adsabs.harvard.edu/abs/2004MNRAS.354..727S}.
  \epubtkKeywords{White dwarfs}

\bibitem{Schutz03}
Schutz, B.F., ``Observational evidence for supermassive black hole binaries'',
  in Centrella, J.M., ed., {\em The Astrophysics of Gravitational Wave
  Sources}, College Park, Maryland, 24 -- 26 April 2003, vol. 686 of AIP
  Conference Proceeedings,  3--29, (American Institute of Physics, Melville,
  U.S.A., 2003). \epubtkKeywords{Gravitational wave sources}

\bibitem{seto_dwd02}
Seto, N., ``Long-term operation of the Laser Interferometer Space Antenna and
  Galactic close white dwarf binaries'', {\em Mon. Not. R. Astron. Soc.}, {\bf
  333}, 469--474, (2002). Related online version (cited on 21 June 2006):
  \newline\url{http://adsabs.harvard.edu/abs/2002MNRAS.333..469S}.
  \epubtkKeywords{Close binaries, Gravitational wave sources, LISA}

\bibitem{Shapiro_Teukolsky83}
Shapiro, S.L., and Teukolsky, S.A., {\em Black Holes, White Dwarfs, and Neutron
  Stars}, (Wiley, New York, U.S.A., 1983). \epubtkKeywords{White dwarfs,
  Neutron stars, Black holes}

\bibitem{Shklovskii70}
Shklovskii, I.S., ``Possible causes of the secular increase in pulsar
  periods'', {\em Sov. Astron.}, {\bf 13}, 562--565, (1970).
  \epubtkKeywords{Pulsars}

\bibitem{siess_agb06}
Siess, L., ``Evolution of massive AGB stars. I. Carbon burning phase'', {\em
  Astron. Astrophys.}, {\bf 448}, 717--729, (2006). Related online version
  (cited on 21 June 2006):
  \newline\url{http://adsabs.harvard.edu/abs/2006A&A...448..717S}.
  \epubtkKeywords{Stellar evolution}

\bibitem{Skumanich72}
Skumanich, A., ``Time Scales for Ca II Emission Decay, Rotational Braking, and
  Lithium Depletion'', {\em Astrophys. J.}, {\bf 171}, 565--567, (1972).
  Related online version (cited on 21 June 2006):
  \newline\url{http://adsabs.harvard.edu/abs/1972ApJ...171..565S}.
  \epubtkKeywords{Stellar rotation}

\bibitem{sma67}
Smak, J., ``Light variability of HZ 29'', {\em Acta Astron.}, {\bf 17},
  255--270, (1967). Related online version (cited on 21 June 2006):
  \newline\url{http://adsabs.harvard.edu/abs/1967AcA....17..255S}.
  \epubtkKeywords{Close binaries, White dwarfs}

\bibitem{Soberman_al97}
Soberman, G.E., Phinney, E.S., and van~den Heuvel, E.P.J., ``Stability criteria
  for mass transfer in binary stellar evolution'', {\em Astron. Astrophys.},
  {\bf 327}, 620--635, (1997). \epubtkKeywords{Binary stars}

\bibitem{Soker_Harpaz03}
Soker, N., and Harpaz, A., ``Criticism of recent calculations of common
  envelope ejection'', {\em Mon. Not. R. Astron. Soc.}, {\bf 343}, 456--458,
  (2003). \epubtkKeywords{Binary systems}

\bibitem{sol_yung_am05}
Solheim, J.-E., and Yungelson, L.R., ``The White Dwarfs in AM CVn Systems --
  Candidates for SN Ia?'', in Koester, D., and Moehler, S., eds., {\em 14th
  European Workshop on White Dwarfs}, Workshop in Kiel, Germany, 19 -- 23 July
  2004, vol. 334 of ASP Conference Series,  387, (Astronomical Society of the
  Pacific, San Francisco, U.S.A., 2005). Related online version (cited on 21
  June 2006): \newline\url{http://adsabs.harvard.edu/abs/2005ASPC..334..387S}.
  \epubtkKeywords{Compact binaries, White dwarfs, Supernovae}

\bibitem{url03}
Space Telescope Science Institute, ``A Catalog and Atlas of Cataclysmic
  Variables'', project homepage, (2006). URL (cited on 21 June 2006):
  \newline\url{http://archive.stsci.edu/prepds/cvcat/index.html}.
  \epubtkKeywords{Close binaries, White dwarfs}

\bibitem{sparks-74}
Sparks, W.M., and Stecher, T.P., ``Supernova: The Result of the Death Spiral of
  a White Dwarf into a Red Giant'', {\em Astrophys. J.}, {\bf 188}, 149,
  (1974). Related online version (cited on 21 June 2006):
  \newline\url{http://adsabs.harvard.edu/abs/1974ApJ...188..149S}.
  \epubtkKeywords{Supernovae, White dwarfs}

\bibitem{Spruit_Phinney98}
Spruit, H.C., and Phinney, E.S., ``Birth kicks as the origin of pulsar
  rotation'', {\em Nature}, {\bf 393}, 139--141, (1998). Related online version
  (cited on 21 June 2006):
  \newline\url{http://adsabs.harvard.edu/abs/1998Natur.393..139S}.
  \epubtkKeywords{Neutron stars, Pulsars}

\bibitem{Stairs_LRR03}
Stairs, I.H., ``Testing General Relativity with Pulsar Timing'', {\em Living
  Rev. Relativity}, {\bf 6}, lrr-2003-5, (2003). URL (cited on 15 Jan 2005):
  \newline\url{http://www.livingreviews.org/lrr-2003-5}.
  \epubtkKeywords{Pulsars, Tests of relativistic gravity}

\bibitem{Stairs_al04}
Stairs, I.H., Thorsett, S.E., and Arzoumanian, Z., ``Measurement of
  Gravitational Spin-Orbit Coupling in a Binary-Pulsar System'', {\em Phys.
  Rev. Lett.}, {\bf 93}, 141101, (2004). Related online version (cited on 21
  June 2006): \newline\url{http://adsabs.harvard.edu/abs/2004PhRvL..93n1101S}.
  \epubtkKeywords{Relativistic binary systems, Pulsars}

\bibitem{Stairs_al02}
Stairs, I.H., Thorsett, S.E., Taylor, J.H., and Wolszczan, A., ``Studies of the
  Relativistic Binary Pulsar PSR B1534+12. I. Timing Analysis'', {\em
  Astrophys. J.}, {\bf 581}, 501--508, (2002). \epubtkKeywords{Pulsars,
  Relativistic binary systems}

\bibitem{Stanek_al03}
Stanek, K.Z., Matheson, T., Garnavich, P.M., Martini, P., Berlind, P.,
  Caldwell, N., Challis, P.and~Brown, W.R., Schild, R., Krisciunas, K.,
  Calkins, M.L., Lee, J.C., Hathi, N., Jansen, R.A., Windhorst,
  R.and~Echevarria, L., Eisenstein, D.J., Pindor, B., Olszewski, E.W., Harding,
  P., Holland, S.T., and Bersier, D., ``Spectroscopic Discovery of the
  Supernova 2003dh Associated with GRB 030329'', {\em Astrophys. J. Lett.},
  {\bf 591}, L17--L20, (2003). \epubtkKeywords{Astronomical observations,
  Supernovae}

\bibitem{stroeer06}
Stroeer, A., and Vecchio, A., ``The LISA verification binaries'', {\em Class.
  Quantum Grav.}, {\bf 23}, S809--S818, (2006). Related online version (cited
  on 21 June 2006): \newline\url{http://arXiv.org/abs/astro-ph/0605227}.
  \epubtkKeywords{Close binaries, Gravitational wave sources, LISA}

\bibitem{svn_lisa06}
Stroeer, A., Vecchio, A., and Nelemans, G., ``LISA Astronomy of Double White
  Dwarf Binary Systems'', {\em Astrophys. J. Lett.}, {\bf 633}, L33--L36,
  (2005). Related online version (cited on 21 June 2006):
  \newline\url{http://adsabs.harvard.edu/abs/2005ApJ...633L..33S}.
  \epubtkKeywords{Compact binaries, White dwarfs, LISA}

\bibitem{Sutantyo78}
Sutantyo, W., ``Asymmetric supernova explosions and the origin of binary
  pulsars'', {\em Astrophys. Space Sci.}, {\bf 54}, 479--488, (1978). Related
  online version (cited on 4 November 2006):
  \newline\url{http://adsabs.harvard.edu/abs/1978Ap&SS..54..479S}.
  \epubtkKeywords{Close binaries, Supernovae}

\bibitem{taam_ce00}
Taam, R.E., and Sandquist, E.L., ``Common Envelope Evolution of Massive Binary
  Stars'', {\em Annu. Rev. Astron. Astrophys.}, {\bf 38}, 113--141, (2000).
  Related online version (cited on 4 November 2006):
  \newline\url{http://adsabs.harvard.edu/abs/2000ARA&A..38..113T}.
  \epubtkKeywords{Stellar evolution, Close binaries, Hydrodynamics}

\bibitem{th86}
Taam, R.E., and van~den Heuvel, E.P.J., ``Magnetic field decay and the origin
  of neutron star binaries'', {\em Astrophys. J.}, {\bf 305}, 235--245, (1986).
  Related online version (cited on 21 June 2006):
  \newline\url{http://adsabs.harvard.edu/abs/1986ApJ...305..235T}.
  \epubtkKeywords{Close binaries, Neutron stars}

\bibitem{taam_wade_aml85}
Taam, R.E., and Wade, R.A., ``Angular momentum loss and the evolution of
  binaries of extreme mass ratio'', {\em Astrophys. J.}, {\bf 293}, 504--507,
  (1985). Related online version (cited on 21 June 2006):
  \newline\url{http://adsabs.harvard.edu/abs/1985ApJ...293..504T}.
  \epubtkKeywords{Close binaries}

\bibitem{takahashi_wd02}
Takahashi, R., and Seto, N., ``Parameter Estimation for Galactic Binaries by
  the Laser Interferometer Space Antenna'', {\em Astrophys. J.}, {\bf 575},
  1030--1036, (2002). Related online version (cited on 21 June 2006):
  \newline\url{http://adsabs.harvard.edu/abs/2002ApJ...575.1030T}.
  \epubtkKeywords{Binary systems, LISA}

\bibitem{tautak98}
Tauris, T.M., and Takens, R.J., ``Runaway velocities of stellar components
  originating from disrupted binaries via asymmetric supernova explosions'',
  {\em Astron. Astrophys.}, {\bf 330}, 1047--1059, (1998). URL (cited on 4
  November 2006):
  \newline\url{http://ads.inasan.rssi.ru/abs/1998A&A...330.1047T}.
  \epubtkKeywords{Close binaries, Supernovae}

\bibitem{Tauris_vdHeuvel00}
Tauris, T.M., and van~den Heuvel, E.P.J., ``New Direct Observational Evidence
  for Kicks in SNe'', in Kramer, M., Wex, N., and Wielebinski, R., eds., {\em
  Pulsar Astronomy: 2000 and Beyond (IAU Colloquium 177)}, 177th Colloquium of
  the IAU at the Max-Planck-Institut f\"ur Radioastronomie, Bonn, Germany, 30
  August -- 3 September 1999, ASP Conference Series,  595, (Astronomical
  Society of the Pacific, San Francisco, U.S.A., 2000).
  \epubtkKeywords{Supernovae, Neutron stars, Astrophysics}

\bibitem{Thorne87}
Thorne, K.S., ``Gravitational radiation'', in Hawking, S.W., and Israel, W.,
  eds., {\em Three Hundred Years of Gravitation},  330--458, (Cambridge
  University Press, Cambridge, U.K.; New York, U.S.A., 1987).
  \epubtkKeywords{Gravitational radiation}

\bibitem{Thorsett_Chakrabarty99}
Thorsett, S.E., and Chakrabarty, D., ``Neutron Star Mass Measurements. I. Radio
  Pulsars'', {\em Astrophys. J.}, {\bf 512}, 288--299, (1999).
  \epubtkKeywords{Neutron stars, Pulsars}

\bibitem{Timmes_al97}
Timmes, F.X., Diehl, R., and Hartmann, D.H., ``Constraints from 26Al
  Measurements on the Galaxy's Recent Global Star Formation Rate and
  Core-Collapse Supernovae Rate'', {\em Astrophys. J.}, {\bf 479}, 760, (1997).
  \epubtkKeywords{Astronomical observations, Supernovae}

\bibitem{timmes_sn94}
Timmes, F.X., Woosley, S.E., and Taam, R.E., ``The conductive propagation of
  nuclear flames. II. Convectively bounded flames in C+O and O+NE+MG cores'',
  {\em Astrophys. J.}, {\bf 420}, 348--363, (1994). Related online version
  (cited on 21 June 2006):
  \newline\url{http://adsabs.harvard.edu/abs/1994ApJ...420..348T}.
  \epubtkKeywords{White dwarfs, Supernovae}

\bibitem{Timmes_al96}
Timmes, F.X., Woosley, S.E., and Weaver, T.A., ``The Neutron Star and Black
  Hole Initial Mass Function'', {\em Astrophys. J.}, {\bf 457}, 834--843,
  (1996). \epubtkKeywords{Supernovae, Neutron stars, Black holes}

\bibitem{tovmassian-04}
Tovmassian, G.H., Napiwotzki, R., Richer, M.G., Stasi{\'n}ska, G., Fullerton,
  A.W., and Rauch, T., ``A Close Binary Nucleus in the MOST Oxygen-Poor
  Planetary Nebula PN G135.9+55.9'', {\em Astrophys. J.}, {\bf 616}, 485--497,
  (2004). Related online version (cited on 21 June 2006):
  \newline\url{http://adsabs.harvard.edu/abs/2004ApJ...616..485T}.
  \epubtkKeywords{Close binaries, White dwarfs}

\bibitem{truran_livio_one86}
Truran, J.W., and Livio, M., ``On the frequency of occurrence of
  oxygen-neon-magnesium white dwarfs in classical nova systems'', {\em
  Astrophys. J.}, {\bf 308}, 721--727, (1986). Related online version (cited on
  21 June 2006):
  \newline\url{http://adsabs.harvard.edu/abs/1986ApJ...308..721T}.
  \epubtkKeywords{Close binaries, White dwarfs}

\bibitem{to97}
Tsugawa, M., and Osaki, Y., ``Disk instability model for the AM Canum
  Venaticorum stars'', {\em Publ. Astron. Soc. Japan}, {\bf 49}, 75--84,
  (1997). Related online version (cited on 21 June 2006):
  \newline\url{http://adsabs.harvard.edu/abs/1997PASJ...49...75T}.
  \epubtkKeywords{Compact binaries, White dwarfs, Accretion}

\bibitem{Turolla_al05}
Turolla, R., Possenti, A., and Treves, A., ``Is the bursting radio-source
  J1745-3009 a double neutron star binary?'', {\em Astrophys. J. Lett.}, {\bf
  628}, L49--L52, (2005). Related online version (cited on 21 June 2006):
  \newline\url{http://adsabs.harvard.edu/abs/2005ApJ...628L..49T}.
  \epubtkKeywords{Pulsars, Relativistic binary systems}

\bibitem{Tutukov_Yungelson73b}
Tutukov, A., and Yungelson, L.R., ``Evolution of close binary with relativistic
  component'', {\em Nauchn. Inform.}, {\bf 27}, 86, (1973). Related online
  version (cited on 21 June 2006):
  \newline\url{http://adsabs.harvard.edu/abs/1973NInfo..27...86T}. In Russian.
  \epubtkKeywords{Astrophysics, Binary systems}

\bibitem{tf89}
Tutukov, A.V., and Fedorova, A.V., ``Formation and evolution of close binaries
  containing helium donors'', {\em Sov. Astron.}, {\bf 33}, 606--613, (1989).
  \epubtkKeywords{Close binaries, White dwarfs}

\bibitem{tf_bh02}
Tutukov, A.V., and Fedorova, A.V., ``Evolution of Close Stellar Binaries with
  Black Holes under the Action of Gravitational Radiation and Magnetic and
  Induced Stellar Winds of the Donor'', {\em Astron. Rep.}, {\bf 46}, 765--778,
  (2002). Related online version (cited on 21 June 2006):
  \newline\url{http://adsabs.harvard.edu/abs/2002ARep...46..765T}.
  \epubtkKeywords{Close binaries, Black holes}

\bibitem{tfey85}
Tutukov, A.V., Fedorova, A.V., Ergma, E.V., and Yungelson, L.R., ``Evolution of
  Low-Mass Close Binaries - The Minimum Orbital Period'', {\em Sov. Astron.
  Lett.}, {\bf 11}, 52, (1985). Related online version (cited on 21 June 2006):
  \newline\url{http://adsabs.harvard.edu/abs/1985SvAL...11...52T}.
  \epubtkKeywords{Close binaries, Stellar evolution}

\bibitem{tfey87}
Tutukov, A.V., Fedorova, A.V., Ergma, E.V., and Yungelson, L.R., ``The
  evolutionary status of MXB 1820-30 and other short-period low-mass X-ray
  sources'', {\em Sov. Astron. Lett.}, {\bf 13}, 328, (1987). Related online
  version (cited on 21 June 2006):
  \newline\url{http://adsabs.harvard.edu/abs/1987SvAL...13..328T}.
  \epubtkKeywords{X-ray binaries}

\bibitem{tfy82}
Tutukov, A.V., Fedorova, A.V., and Yungelson, L.R., ``The evolution of dwarf
  binaries'', {\em Sov. Astron. Lett.}, {\bf 8}, 365--370, (1982).
  \epubtkKeywords{Close binaries, White dwarfs}

\bibitem{tk_eld92}
Tutukov, A.V., and Khokhlov, A.M., ``Helium Shell Explosions in Accreting
  Carbon-Oxygen Dwarfs'', {\em Sov. Astron.}, {\bf 36}, 401, (1992). Related
  online version (cited on 21 June 2006):
  \newline\url{http://adsabs.harvard.edu/abs/1992SvA....36..401T}.
  \epubtkKeywords{Compact binaries, White dwarfs, Accretion}

\bibitem{ty73}
Tutukov, A.V., and Yungelson, L.R., ``Evolution of massive close binaries'',
  {\em Nauchn. Inform.}, {\bf 27}, 70, (1973). Related online version (cited on
  21 June 2006):
  \newline\url{http://adsabs.harvard.edu/abs/1973NInfo..27...70T}.
  \epubtkKeywords{Compact binaries, Neutron stars}

\bibitem{ty79}
Tutukov, A.V., and Yungelson, L.R., ``Evolution of massive common envelope
  binaries and mass loss'', in de~Loore, C., and Conti, P.S., eds., {\em Mass
  loss and evolution of O-type stars}, IAU Symposium 83 in Vancouver Island,
  Canada, 5 -- 9 June, 1978,  401, (D. Reidel, Dordrecht, Netherlands; Boston,
  U.S.A., 1979). \epubtkKeywords{Close binaries, Hydrodynamics}

\bibitem{ty79a}
Tutukov, A.V., and Yungelson, L.R., ``On the influence of emission of
  gravitational waves on the evolution of low-mass close binary stars'', {\em
  Acta Astron.}, {\bf 29}, 665--680, (1979). Related online version (cited on
  21 June 2006):
  \newline\url{http://adsabs.harvard.edu/abs/1979AcA....29..665T}.
  \epubtkKeywords{Close binaries, Gravitational radiation}

\bibitem{ty81}
Tutukov, A.V., and Yungelson, L.R., ``Evolutionary scenario for close binary
  systems of low and moderate masses'', {\em Nauchn. Inform.}, {\bf 49}, 3,
  (1981). \epubtkKeywords{Close binaries, White dwarfs}

\bibitem{Tutukov_Yungelson93a}
Tutukov, A.V., and Yungelson, L.R., ``Formation of neutron stars in binary
  systems'', {\em Astron. Rep.}, {\bf 37}, 411--431, (1993).
  \epubtkKeywords{Astrophysics, Neutron stars, Binary systems}

\bibitem{ty93b}
Tutukov, A.V., and Yungelson, L.R., ``The merger rate of neutron star and black
  hole binaries'', {\em Mon. Not. R. Astron. Soc.}, {\bf 260}, 675--678,
  (1993). \epubtkKeywords{Compact binaries, Gravitational wave sources, Neutron
  stars, Black holes, Relativistic binary systems}

\bibitem{ty94}
Tutukov, A.V., and Yungelson, L.R., ``Merging of Binary White Dwarfs, Neutron
  Stars and Black-Holes Under the Influence of Gravitational Wave Radiation'',
  {\em Mon. Not. R. Astron. Soc.}, {\bf 268}, 871, (1994).
  \epubtkKeywords{Compact binaries, Gravitational radiation}

\bibitem{ty96}
Tutukov, A.V., and Yungelson, L.R., ``Double-degenerate semidetached binaries
  with helium secondaries: cataclismic variables, supersoft X-ray sources,
  supernovae and accretion-induced collapses'', {\em Mon. Not. R. Astron.
  Soc.}, {\bf 280}, 1035--1045, (1996). \epubtkKeywords{Close binaries,
  Accretion, Supernovae}

\bibitem{ty02}
Tutukov, A.V., and Yungelson, L.R., ``A Model for the Population of Binary
  Stars in the Galaxy'', {\em Astron. Rep.}, {\bf 46}, 667--683, (2002).
  Related online version (cited on 21 June 2006):
  \newline\url{http://adsabs.harvard.edu/abs/2002ARep...46..667T}.
  \epubtkKeywords{Binary systems}

\bibitem{ull95a}
Ulla, A., ``The X-ray properties of AM Canum Venaticorum'', {\em Astron.
  Astrophys.}, {\bf 301}, 469, (1995). Related online version (cited on 21 June
  2006): \newline\url{http://adsabs.harvard.edu/abs/1995A&A...301..469U}.
  \epubtkKeywords{X-ray binaries}

\bibitem{vandenHeuvel83}
van~den Heuvel, E.P.J., ``Formation and evolution of X-ray binaries'', in
  Lewin, W.H.G., and van~den Heuvel, E.P.J., eds., {\em Accretion-Driven
  Stellar X-ray Sources}, vol.~4 of Cambridge Astrophysics Series,  303--341,
  (Cambridge University Press, Cambridge, U.K.; New York, U.S.A., 1983).
  \epubtkKeywords{Astrophysics, Binary stars}

\bibitem{hbnr92}
van~den Heuvel, E.P.J., Bhattacharya, D., Nomoto, K., and Rappaport, S.A.,
  ``Accreting white dwarf models for CAL 83, CAL 87 and other ultrasoft X-ray
  sources in the LMC'', {\em Astron. Astrophys.}, {\bf 262}, 97--105, (1992).
  Related online version (cited on 21 June 2006):
  \newline\url{http://adsabs.harvard.edu/abs/1992A&A...262...97V}.
  \epubtkKeywords{Close binaries, White dwarfs, Accretion, X-ray binaries}

\bibitem{vdHeuvel_deLoore73}
van~den Heuvel, E.P.J., and de~Loore, C., ``The nature of X-ray binaries. III.
  Evolution of massive close binarieswith one collapsed component, with a
  possible application to Cygnus X3'', {\em Astron. Astrophys.}, {\bf 25}, 387,
  (1973). \epubtkKeywords{Astrophysics, Binary systems}

\bibitem{Heuvel_Habets84}
van~den Heuvel, E.P.J., and Habets, G.M.H.J., ``Observational lower mass limit
  for black hole formation derived from massive X-ray binaries'', {\em Nature},
  {\bf 309}, 598--600, (1984). \epubtkKeywords{Astrophysics, Black holes}

\bibitem{sluys_ucb05a}
van~der Sluys, M.V., Verbunt, F., and Pols, O.R., ``Creating ultra-compact
  binaries through stable mass transfer'', in Burderi, L., Antonelli, L.A.,
  D'Antona, F., di~Salvo, T., Israel, G.L., Piersanti, L., Tornamb{\`e}, A.,
  and Straniero, O., eds., {\em Interacting Binaries: Accretion, Evolution, and
  Outcomes}, Conference in Cefal\`u, Italy, 4 -- 10 July 2004, vol. 797 of AIP
  Conference Proceedings,  627--630, (American Institute of Physics, Melville,
  U.S.A., 2005). Related online version (cited on 21 June 2006):
  \newline\url{http://adsabs.harvard.edu/abs/2005AIPC..797..627V}.
  \epubtkKeywords{X-ray binaries, Compact binaries}

\bibitem{sluys_ucb05b}
van~der Sluys, M.V., Verbunt, F., and Pols, O.R., ``Reduced magnetic braking
  and the magnetic capture model for the formation of ultra-compact binaries'',
  {\em Astron. Astrophys.}, {\bf 440}, 973--979, (2005). Related online version
  (cited on 4 November 2006):
  \newline\url{http://adsabs.harvard.edu/abs/2005A&A...440..973V}.
  \epubtkKeywords{X-ray binaries}

\bibitem{sluys_wd06}
van~der Sluys, M.V., Verbunt, F., and Pols, O.R., ``Modelling the formation of
  double white dwarfs'', {\em Astron. Astrophys.}, {\bf 460}, 209--228, (2006).
  Related online version (cited on 4 November 2006):
  \newline\url{http://adsabs.harvard.edu/abs/2006A&A...460..209V}.
  \epubtkKeywords{Compact binaries, White dwarfs}

\bibitem{verbunt_xray_gc05}
Verbunt, F., ``X-ray sources in globular clusters'', in Burderi, L., Antonelli,
  L.A., D'Antona, F., di~Salvo, T., Israel, G.L., Piersanti, L., Tornamb{\`e},
  A., and Straniero, O., eds., {\em Interacting Binaries: Accretion, Evolution,
  and Outcomes}, Conference in Cefal\`u, Italy, 4 -- 10 July 2004, vol. 797 of
  AIP Conference Proceedings,  30--39, (American Institute of Physics,
  Melville, U.S.A., 2005). Related online version (cited on 21 June 2006):
  \newline\url{http://adsabs.harvard.edu/abs/2005AIPC..797...30V}.
  \epubtkKeywords{X-ray binaries, Star clusters}

\bibitem{vr88}
Verbunt, F., and Rappaport, S., ``Mass transfer instabilities due to angular
  momentum flows in close binaries'', {\em Astrophys. J.}, {\bf 332}, 193--198,
  (1988). \epubtkKeywords{Close binaries}

\bibitem{vwb90}
Verbunt, F., Wijers, R.A.M.J., and Burm, H.M.G., ``Evolutionary scenarios for
  the X-ray binary pulsars 4U 1626-67 and Hercules X-1, and their implications
  for the decay of neutron star magnetic fields'', {\em Astron. Astrophys.},
  {\bf 234}, 195--202, (1990). Related online version (cited on 21 June 2006):
  \newline\url{http://adsabs.harvard.edu/abs/1990A&A...234..195V}.
  \epubtkKeywords{X-ray binaries, Neutron stars}

\bibitem{vz81}
Verbunt, F., and Zwaan, C., ``Magnetic braking in low-mass X-ray binaries'',
  {\em Astron. Astrophys.}, {\bf 100}, L7--L9, (1981). Related online version
  (cited on 21 June 2006):
  \newline\url{http://adsabs.harvard.edu/abs/1981A&A...100L...7V}.
  \epubtkKeywords{X-ray binaries}

\bibitem{vil71}
Vila, S.C., ``Late Evolution of Close Binaries'', {\em Astrophys. J.}, {\bf
  168}, 217--223, (1971). Related online version (cited on 21 June 2006):
  \newline\url{http://adsabs.harvard.edu/abs/1971ApJ...168..217V}.
  \epubtkKeywords{Close binaries}

\bibitem{voss_taurisns03}
Voss, R., and Tauris, T.M., ``Galactic distribution of merging neutron stars
  and black holes -- Prospects for short gamma-ray burst progenitors and
  LIGO/VIRGO'', {\em Mon. Not. R. Astron. Soc.}, {\bf 342}, 1169--1184, (2003).
  Related online version (cited on 21 June 2006):
  \newline\url{http://adsabs.harvard.edu/abs/2003MNRAS.342.1169V}.
  \epubtkKeywords{Compact binaries, Neutron stars, Black holes, Gravitational
  wave sources}

\bibitem{wad84}
Wade, R.A., ``A double grid of accretion disc model spectra for cataclysmic
  variable stars'', {\em Mon. Not. R. Astron. Soc.}, {\bf 208}, 381--398,
  (1984). Related online version (cited on 21 June 2006):
  \newline\url{http://adsabs.harvard.edu/abs/1984MNRAS.208..381W}.
  \epubtkKeywords{White dwarfs, Accretion disks}

\bibitem{wagner_hen01a}
Wagner, R.M., Foltz, C.B., and Starrfield, S.G., ``Possible Nova in Puppis'',
  {\em IAU Circ.}, {\bf 2001}(7556), (January, 2001). Related online version
  (cited on 21 June 2006):
  \newline\url{http://cfa-www.harvard.edu/iauc/07500/07556.html}.
  \epubtkKeywords{Close binaries, White dwarfs}

\bibitem{wagner_hen01b}
Wagner, R.M., Schwarz, G., Starrfield, S.G., Foltz, C.B., Howell, S., and
  Szkody, P., ``V445 Puppis'', {\em IAU Circ.}, {\bf 2001}(7717), (September,
  2001). Related online version (cited on 21 June 2006):
  \newline\url{http://cfa-www.harvard.edu/iauc/07700/07717.html}.
  \epubtkKeywords{Close binaries, White dwarfs}

\bibitem{wang_chandra02}
Wang, Q.D., Gotthelf, E.V., and Lang, C.C., ``A faint discrete source origin
  for the highly ionized iron emission from the Galactic Centre region'', {\em
  Nature}, {\bf 415}, 148--150, (2002). Related online version (cited on 21
  June 2006): \newline\url{http://adsabs.harvard.edu/abs/2002Natur.415..148W}.
  \epubtkKeywords{X-ray astronomy}

\bibitem{wr72}
Warner, B., and Robinson, E.L., ``Observations of rapid blue variables -- IX.
  AM CVn (HZ 29)'', {\em Mon. Not. R. Astron. Soc.}, {\bf 159}, 101--111,
  (1972). Related online version (cited on 21 June 2006):
  \newline\url{http://adsabs.harvard.edu/abs/1972MNRAS.159..101W}.
  \epubtkKeywords{Close binaries, White dwarfs, Astronomical observations}

\bibitem{webbink-79}
Webbink, R.F., ``The formation of the white dwarfs in close binary systems'',
  in van Horn, H.M., and Weidemann, V., eds., {\em White Dwarfs and Variable
  Degenerate Stars (IAU Colloquium 53)}, Colloquium, July 30 -- August 2 1979
  and Fourth Annual Workshop on Novae, Dwarf Novae and Other Cataclysmic
  Variables at the University of Rochester in Rochester, N.Y., 3 August 1979,
  426--447, (University of Rochester, Rochester, U.S.A., 1979).
  \epubtkKeywords{Close binaries, White dwarfs}

\bibitem{Webbink84}
Webbink, R.F., ``Double white dwarfs as progenitors of R Coronae Borealis stars
  and Type I supernovae'', {\em Astrophys. J.}, {\bf 277}, 355--360, (1984).
  \epubtkKeywords{Compact binaries, Binary stars, White dwarfs, Supernovae}

\bibitem{wh98}
Webbink, R.F., and Han, Z., ``Gravitational Radiation from Close Double White
  Dwarfs'', in Folkner, W.M., ed., {\em Laser Interferometer Space Antenna},
  Second International LISA Symposium on the Detection and Observation of
  Gravitational Waves in Space, Pasadena, California, July 1998, vol. 456 of
  AIP Conference Proceedings, ~61, (American Institure of Physics, Woodbury,
  U.S.A., 1998). Related online version (cited on 21 June 2006):
  \newline\url{http://adsabs.harvard.edu/abs/1998lain.conf...61W}.
  \epubtkKeywords{Close binaries, White dwarfs, Gravitational wave sources}

\bibitem{web_iben87a}
Webbink, R.F., and Iben~Jr, I., ``Tidal interaction and coalescence of close
  binary white dwarfs'', in Philip, A.G.D., Hayes, D.S., and Liebert, J.W.,
  eds., {\em The Second Conference on Faint Blue Stars (IAU Colloquium 95)},
  Conference in Tucson, Arizona, 1 -- 5 June 1987,  445--456, (L. Davis Press,
  Schenectady, U.S.A., 1987). Related online version (cited on 4 November
  2006): \newline\url{http://adsabs.harvard.edu/abs/1987fbs..conf..445W}.
  \epubtkKeywords{Close binaries, White dwarfs}

\bibitem{Weisberg_Taylor02}
Weisberg, J.M., and Taylor, J.H., ``General Relativistic Geodetic Spin
  Precession in Binary Pulsar B1913+16: Mapping the Emission Beam in Two
  Dimensions'', {\em Astrophys. J.}, {\bf 576}, 942--949, (2002).
  \epubtkKeywords{Pulsars, Binary systems}

\bibitem{werner06}
Werner, K., Nagel, T., Rauch, T., Hammer, N.J., and Dreizler, S., ``VLT
  spectroscopy and non-LTE modeling of the C/O-dominated accretion disks in two
  ultracompact X-ray binaries'', {\em Astron. Astrophys.}, {\bf 450}, 725--733,
  (2006). Related online version (cited on 21 June 2006):
  \newline\url{http://adsabs.harvard.edu/abs/2006A&A...450..725W}.
  \epubtkKeywords{X-ray binaries, Accretion disks}

\bibitem{Wettig_Brown96}
Wettig, T., and Brown, G.E., ``The evolution of relativistic binary pulsars'',
  {\em New Astronomy}, {\bf 1}, 17--34, (1996). \epubtkKeywords{Neutron stars,
  Pulsars, Relativistic binary systems}

\bibitem{Wex_al00}
Wex, N., Kalogera, V., and Kramer, M., ``Constraints on Supernova Kicks from
  the Double Neutron Star System PSR B1913+16'', {\em Astrophys. J.}, {\bf
  528}, 401--409, (2000). \epubtkKeywords{Supernovae, Neutron stars,
  Relativistic binary systems}

\bibitem{wi73}
Whelan, J., and Iben~Jr, I., ``Binaries and Supernovae of Type I'', {\em
  Astrophys. J.}, {\bf 186}, 1007--1014, (1973). Related online version (cited
  on 21 June 2006):
  \newline\url{http://adsabs.harvard.edu/abs/1973ApJ...186.1007W}.
  \epubtkKeywords{Close binaries, Supernovae}

\bibitem{Willems_al05}
Willems, B., Henninger, M., Levin, T., Ivanova, N., Kalogera, V., McGhee, K.,
  Timmes, F.X., and Fryer, C.L., ``Understanding Compact Object Formation and
  Natal Kicks. I. Calculation Methods and the Case of GRO J1655-40'', {\em
  Astrophys. J.}, {\bf 625}, 324--346, (2005). \epubtkKeywords{Black holes,
  Binary systems}

\bibitem{Willems_Kalogera04}
Willems, B., and Kalogera, V., ``Constraints on the Formation of PSR
  J0737--3039: The Most Probable Isotropic Kick Magnitude'', {\em Astrophys. J.
  Lett.}, {\bf 603}, L101--L104, (2004). \epubtkKeywords{Pulsars, Relativistic
  binary systems}

\bibitem{Willems_al04}
Willems, B., Kalogera, V., and Henninger, M., ``Pulsar Kicks and Spin Tilts in
  the Close Double Neutron Stars PSR J0737--3039, PSR B1534+12, and PSR
  B1913+16'', {\em Astrophys. J.}, {\bf 616}, 414--438, (2004).
  \epubtkKeywords{Pulsars, Relativistic binary systems}

\bibitem{Wolszczan90}
Wolszczan, A., ``PSR 1257+12 and PSR 1534+12'', {\em IAU Circ.}, {\bf
  1990}(5073), (August, 1990). Related online version (cited on 13 November
  2006): \newline\url{http://cfa-www.harvard.edu/iauc/05000/05073.html}.
  \epubtkKeywords{Double pulsars, Relativistic binary systems}

\bibitem{Woosley_Heger06}
Woosley, S.E., and Heger, A., ``The Progenitor Stars of Gamma-Ray Bursts'',
  {\em Astrophys. J.}, {\bf 637}, 914--921, (2006). Related online version
  (cited on 21 June 2006):
  \newline\url{http://adsabs.harvard.edu/abs/2006ApJ...637..914W}.
  \epubtkKeywords{Stellar evolution, Gamma-ray bursts}

\bibitem{Woosley_al02}
Woosley, S.E., Heger, A., and Weaver, T.A., ``The evolution and explosion of
  massive stars'', {\em Rev. Mod. Phys.}, {\bf 74}, 1015--1071, (2002).
  \epubtkKeywords{Stellar evolution, Supernovae}

\bibitem{Woosley_al95}
Woosley, S.E., Langer, N., and Weaver, T.A., ``The Presupernova Evolution and
  Explosion of Helium Stars That Experience Mass Loss'', {\em Astrophys. J.},
  {\bf 448}, 315, (1995). \epubtkKeywords{Neutron stars, Black holes,
  Supernovae}

\bibitem{ww94}
Woosley, S.E., and Weaver, T.A., ``Sub-Chandrasekhar mass models for Type Ia
  supernovae'', {\em Astrophys. J.}, {\bf 423}, 371--379, (1994). Related
  online version (cited on 21 June 2006):
  \newline\url{http://adsabs.harvard.edu/abs/1994ApJ...423..371W}.
  \epubtkKeywords{Supernovae, White dwarfs}

\bibitem{Yamaoka_al93}
Yamaoka, H., Shigeyama, T., and Nomoto, K., ``Formation of double neutron star
  systems and asymmetric supernova explosions'', {\em Astron. Astrophys.}, {\bf
  267}, 433--438, (1993). \epubtkKeywords{Binary stars, Supernovae,
  Relativistic binary systems}

\bibitem{yaron05}
Yaron, O., Prialnik, D., Shara, M.M., and Kovetz, A., ``An Extended Grid of
  Nova Models. II. The Parameter Space of Nova Outbursts'', {\em Astrophys.
  J.}, {\bf 623}, 398--410, (2005). Related online version (cited on 21 June
  2006): \newline\url{http://adsabs.harvard.edu/abs/2005ApJ...623..398Y}.
  \epubtkKeywords{Close binaries, White dwarfs, Accretion}

\bibitem{yoon_lang_sn03}
Yoon, S.-C., and Langer, N., ``The first binary star evolution model producing
  a Chandrasekhar mass white dwarf'', {\em Astron. Astrophys.}, {\bf 412},
  L53--L56, (2003). Related online version (cited on 21 June 2006):
  \newline\url{http://adsabs.harvard.edu/abs/2003A&A...412L..53Y}.
  \epubtkKeywords{White dwarfs, Binary systems}

\bibitem{yoon_langer04}
Yoon, S.-C., and Langer, N., ``Helium accreting CO white dwarfs with rotation:
  Helium novae instead of double detonation'', {\em Astron. Astrophys.}, {\bf
  419}, 645--652, (2004). \epubtkKeywords{White dwarfs, Accretion}

\bibitem{yoon_lang_rotsn05}
Yoon, S.-C., and Langer, N., ``On the evolution of rapidly rotating massive
  white dwarfs towards supernovae or collapses'', {\em Astron. Astrophys.},
  {\bf 435}, 967--985, (2005). Related online version (cited on 4 November
  2006): \newline\url{http://adsabs.harvard.edu/abs/2005A&A...435..967Y}.
  \epubtkKeywords{White dwarfs, Supernovae}

\bibitem{yoon_stab04}
Yoon, S.-C., Langer, N., and van~der Sluys, M.V., ``On the stability of
  thermonuclear shell sources in stars'', {\em Astron. Astrophys.}, {\bf 425},
  207--216, (2004). Related online version (cited on 4 November 2006):
  \newline\url{http://adsabs.harvard.edu/abs/2004A&A...425..207Y}.
  \epubtkKeywords{Stellar evolution, Stellar structure}

\bibitem{y73a}
Yungelson, L.R., ``Evolution of close binaries with mass loss from the system.
  III. Systems containing white dwarfs'', {\em Nauchn. Inform.}, {\bf 26},
  71--82, (1973). Related online version (cited on 4 November 2006):
  \newline\url{http://adsabs.harvard.edu/abs/1973NInfo..26...71Y}. In Russian.
  \epubtkKeywords{Close binaries, White dwarfs}

\bibitem{yungelson05b}
Yungelson, L.R., ``Population synthesis for low and intermediate mass
  binaries'', in Burderi, L., Antonelli, L.A., D'Antona, F., di~Salvo, T.,
  Israel, G.L., Piersanti, L., Tornamb{\`e}, A., and Straniero, O., eds., {\em
  Interacting Binaries: Accretion, Evolution, and Outcomes}, Conference in
  Cefal\`u, Italy, 4 -- 10 July 2004, vol. 797 of AIP Conference Proceedings,
  1--10, (American Institute of Physics, Melville, U.S.A., 2005). Related
  online version (cited on 21 June 2006):
  \newline\url{http://adsabs.harvard.edu/abs/2005AIPC..797....1Y}.
  \epubtkKeywords{Close binaries}

\bibitem{yungelson05a}
Yungelson, L.R., ``Population Synthesis for Progenitors of Type Ia
  Supernovae'', in Sion, E.M., Vennes, S., and H.L., Shipman, eds., {\em White
  Dwarfs: Galactic and Cosmological Probes}, Astrophysics and Space Science
  Library, (Springer, Dordrecht, Netherlands, 2005). Related online version
  (cited on 4 November 2006):
  \newline\url{http://arXiv.org/abs/astro-ph/0409677}. \epubtkKeywords{Binary
  systems, Supernovae}

\bibitem{yungelson_bh06}
Yungelson, L.R., Lasota, J.-P., Nelemans, G., Dubus, G., van~den Heuvel,
  E.P.J., Dewi, J., and Portegies~Zwart, S., ``The origin and fate of
  short-period low-mass black-hole binaries'', {\em Astron. Astrophys.}, {\bf
  454}, 559--569, (2006). Related online version (cited on 21 June 2006):
  \newline\url{http://adsabs.harvard.edu/abs/2006A&A...454..559Y}.
  \epubtkKeywords{Close binaries, Black holes}

\bibitem{yl98}
Yungelson, L.R., and Livio, M., ``Type Ia Supernovae: An Examination of
  Potential Progenitors and the Redshift Distribution'', {\em Astrophys. J.},
  {\bf 497}, 168, (1998). \epubtkKeywords{Supernovae, White dwarfs,
  Extragalactic astronomy}

\bibitem{ylttf96}
Yungelson, L.R., Livio, M., Truran, J.W., Tutukov, A., and Fedorova, A., ``A
  Model for the Galactic Population of Binary Supersoft X-Ray Sources'', {\em
  Astrophys. J.}, {\bf 466}, 890, (1996). \epubtkKeywords{X-ray binaries}

\bibitem{ynh02}
Yungelson, L.R., Nelemans, G., and van~den Heuvel, E.P.J., ``On the formation
  of neon-enriched donor stars in ultracompact X-ray binaries'', {\em Astron.
  Astrophys.}, {\bf 388}, 546--551, (2002). \epubtkKeywords{X-ray binaries}

\bibitem{ty05}
Yungelson, L.R., and Tutukov, A.V., ``A Model for the Population of Helium
  Stars in the Galaxy: Low-Mass Stars'', {\em Astron. Rep.}, {\bf 49},
  871--883, (2005). Related online version (cited on 21 June 2006):
  \newline\url{http://adsabs.harvard.edu/abs/2005ARep...49..871Y}.
  \epubtkKeywords{Stellar evolution}

\bibitem{ytl93}
Yungelson, L.R., Tutukov, A.V., and Livio, M., ``The formation of binary and
  single nuclei of Planetary Nebulae'', {\em Astrophys. J.}, {\bf 418},
  794--803, (1993). \epubtkKeywords{Close binaries, White dwarfs}

\bibitem{zeld72}
Zel'dovich, Y.B., Ivanova, L.N., and Nadezhin, D.K., ``Nonstationary
  Hydrodynamical Accretion onto a Neutron Star'', {\em Sov. Astron.}, {\bf 16},
  209, (1972). Related online version (cited on 4 November 2006):
  \newline\url{http://adsabs.harvard.edu/abs/1972SvA....16..209Z}.
  \epubtkKeywords{Neutron stars, Accretion, Hydrodynamics}

\end{thebibliography}

\end{document}